# Planetary Exploration Horizon 2061 – Report

# Chapter 3: From science questions to Solar System exploration

Véronique Dehant[1,2], Michel Blanc[3], Steve Mackwell[4], Krista M. Soderlund[5], Pierre Beck[6], Emma Bunce[7], Sébastien Charnoz[8], Bernard Foing[9], Valerio Filice[1], Leigh N. Fletcher[7], François Forget[10], Léa Griton[3], Heidi Hammel[11], Dennis Höning[12,13], Takeshi Imamura[14], Caitriona Jackman[15], Yohai Kaspi[16], Oleg Korablev[17], Jérémy Leconte[18], Emmanuel Lellouch[19], Bernard Marty[20], Nicolas Mangold[21], Patrick Michel[3], Alessandro Morbidelli[22], Olivier Mousis[23], Olga Prieto-Ballesteros[24], Tilman Spohn[13], Jürgen Schmidt[25], Veerle J. Sterken[26], Nicola Tosi[13], Ann C. Vandaele[27], Pierre Vernazza[23], Allona Vazan[28], Frances Westall[29]

[1] Royal observatory of Belgium, 3 avenue Circulaire, B1180 Brussels, Belgium
[2] Université catholique de Louvain, Belgium
[3] Institut de Recherche en Astrophysique et Planétologie, Observatoire Midi-Pyrénées Toulouse, France
[4] American Institute of Physics, Melville, USA
[5] Institute for Geophysics, Jackson School of Geosciences, University of Texas at Austin, Austin, TX, USA
[6] Institut d'astrophysique et de planétologie de Grenoble/ISTerre, Université Grenoble Alpes, France
[7] School of Physics and Astronomy, University of Leicester, UK
[8] Institut de Physique du Globe de Paris, France
[9] Leiden University, The Netherlands
[10] Laboratoire de Météorologie Dynamique, Institut Pierre Simon Laplace, Paris, France
[11] Association of Universities for Research in Astronomy, Washington, D.C., USA
[12] Vrije Universiteit Amsterdam, The Netherlands
[13] German Aerospace Centre (DLR), Berlin, Germany
[14] The University of Tokyo, Bunkyo-ku, Japan
[15] Dublin Institute for Advanced Studies, Dublin, Ireland
[16] Weizmann Institute of Science, Rehovot, Israel
[17] IKI, Russian Academy of Sciences, Moscow, Russia
[18] Laboratoire d'astrophysique de Bordeaux, Univ. Bordeaux, CNRS, B18N, Pessac, France
[19] LESIA, Observatoire de Paris, France
[20] Université de Lorraine, Vandoeuvre lès Nancy, France
[21] Laboratoire de Planétologie et Géodynamique, UMR 6112, CNRS, Université de Nantes, Nantes, France
[22] Université Côte d'Azur, CNRS, Observatoire de la Côte d'Azur, Nice, France
[23] Laboratoire d'Astrophysique de Marseille, Aix-Marseille Université, France
[24] Centro de Astrobiología-CSIC-INTA, Madrid, Spain
[25] University of Oulu, Finland
[26] ETH Zürich, Switzerland
[27] Royal Belgian Institute for Space Aeronomy, Brussels, Belgium
[28] The Open University of Israël, Israël
[29] CNRS-Centre de Biophysique Moléculaire, Orléans, France

# Contents

# 1. Introduction

The analysis of planetary systems and their different classes presented in Chapter 1, (Blanc et al., 2022), based on community contributions to the Horizon 2061 foresight exercise, led us to identify six key science questions, Q1 to Q6, concerning the diversity, origins, workings and habitability of planetary systems:

Q1- How well do we understand the diversity of planetary systems objects?

Q2- How well do we understand the diversity of planetary system architectures?

Q3- What are the origins and formation scenarios for planetary systems?

Q4- How do planetary systems work?

Q5- Do planetary systems host potential habitats?

Q6- Where and how to search for life?

Chapter 2 (Rauer et al., 2022) described how the broad variety of astronomical techniques for the study of extrasolar planetary systems can help partly to address these science questions. Nevertheless, future studies of the Solar System will keep a unique role in addressing these six key questions, for mainly three main reasons:

- It will remain by far the system that can be observed with the highest spatial resolution: in situ, close-up investigations of its objects by space probes will combine with the enhanced resolving power of Earth-based telescopes to scrutinize even its most distant objects (giant planet systems, Trans-Neptunian objects, etc.); it is also the only planetary system we can ever hope to sample in situ via descent probes and landers.
- Because the Solar System can be observed with this unique combination of orbiting spacecraft, in situ landers and probes, and giant ground- and space-based telescopes, it enables scrutiny of the broadest diversity of objects that a planetary system may be expected to host.
- The Solar System is, for the time being, the only planetary system where we can observe secondary systems within it, formed of an equally diverse collection of objects: numerous systems of satellites that vary in complexity, rings and plasma tori formed around its giant planets, and our Earth-Moon system where life is known to have evolved. Thus, studied together with the Solar System as a whole, these secondary systems provide us with unique insight into the diversity of planetary system architectures.
- Finally, the combination of space missions and telescope observations makes it possible to study in detail its interface with the local interstellar medium surrounding it, the heliopause, and offers the perspective that, very soon, we will take the first steps outside of our own home planetary system to explore its surroundings.

Fig. 3.1 illustrates three important characteristics of the Solar System: (1) the distribution of its objects over nearly three orders of magnitude of distance from the Sun (from a fraction of an Astronomical Unit (AU) to more than 100 AU), with the small terrestrial planets grouped near the Sun, the giant planets grouped outside, and several populations of small bodies distributed throughout this range of distances; (2) the diversity of its secondary systems, with their host of regular satellites mimicking terrestrial and dwarf planets, and their irregular satellites connected to different families of small bodies; and (3) the location of its interfaces with the interstellar medium at a distance short enough for in situ exploration spacecraft.

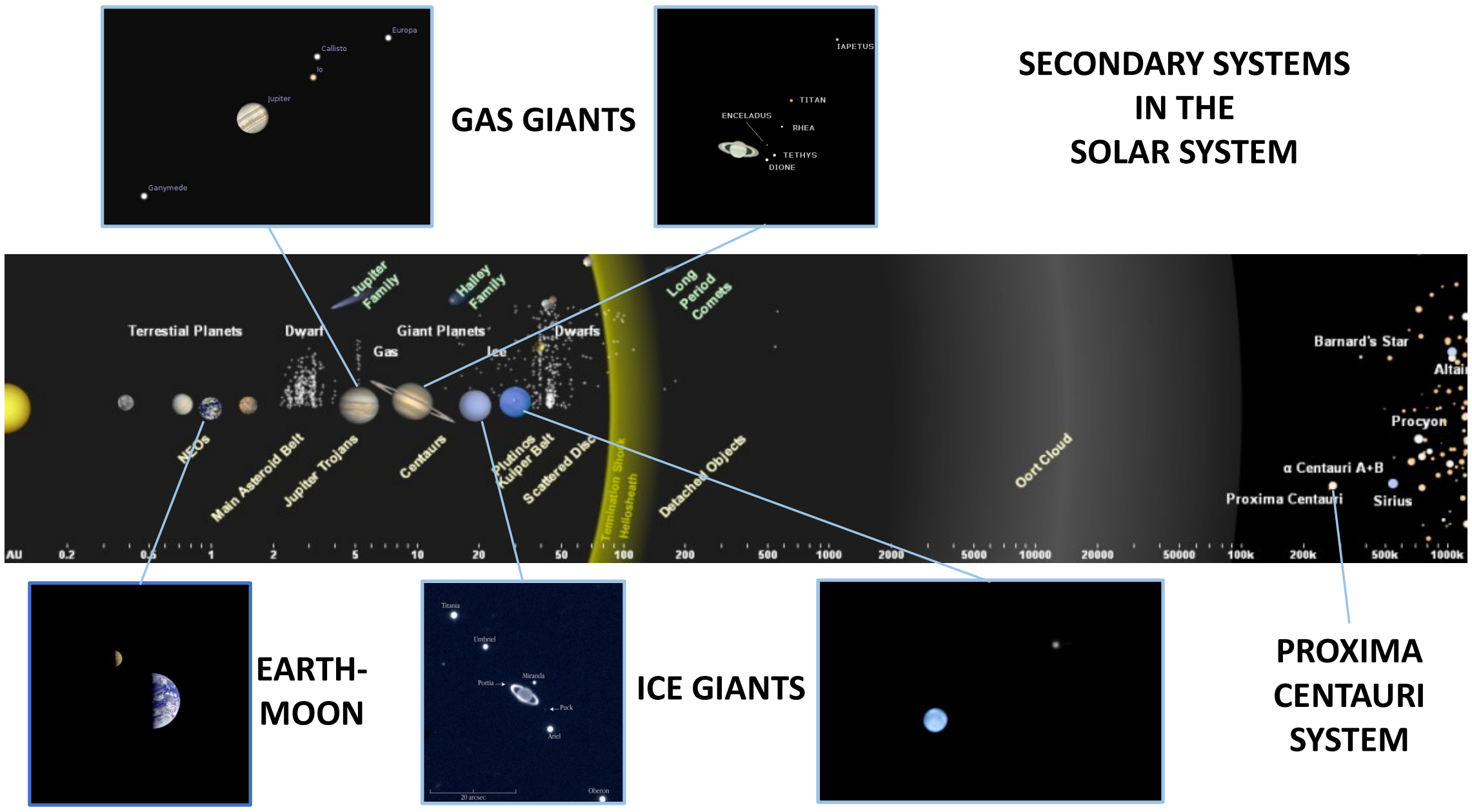

Figure 3.1. (Centre) Radial distribution of Solar System objects (Logarithmic scale in AU) showing their diversity (planets, small bodies, dwarf planets and moons) and their distribution over 5 orders of magnitude in distance from the Sun, from slightly more than 0.1 AU to over $10^4$ AU – still only a few percent of our distance to the closest stars. (Upper and lower panels) The main secondary systems of the Solar System display an equally broad diversity of architectures: gas giant systems (top), ice giant systems (bottom) contrast with the very different configuration of the Earth-Moon system. © NASA and ESO.

In this chapter, we take advantage of these unique properties of the Solar System to ask how its future exploration will provide partial answers to our six key questions, reformulated in terms that will address more specifically the planets and their secondary systems. For each of these questions, we will identify a set of key measurements needed to address them, together with the different places where they must be performed, and the types of scientific instruments needed. We will also provide preliminary ideas on the types of space missions required.

This scientific exploration of the Solar System, guided by our six questions, will be organized as follows. In Section 2, we will review what we know of the diversity of Solar System objects and identify the most important “knowledge gaps” to be filled in their inventory (question Q1). Then in Section 3, we will explore the diversity of architectures of primary and secondary planetary systems contained in the Solar System (question Q2).

Analysis of these two questions will set the stage for the next two questions: we will study in Section 4 the origin and formation scenarios of the Solar System (question Q3) and, in Section 5, some of its most important working mechanisms (question Q4).

Starting from the initial conditions that shaped its emergence from the Solar Nebula and driven by its working mechanisms operating at multiple and interconnected scales, the Solar System evolved towards the configuration that we can observe today. Exploration of the diverse environments in our system makes it possible to address perhaps the two most challenging of our six questions: does the Solar System host potential habitats where the conditions for the emergence of life, as we understand them, are fulfilled, or have been fulfilled in the past? This question (Q5) will be addressed in Section 6. Finally, Section 7 will address question Q6: where and how to search for life in the Solar System?

# 2. Diversity of Solar System Objects (Q1)

Thanks to planetary exploration and telescopic observations, we know that the Solar System offers a very broad diversity of objects to our inquiry:

- three categories of planets (gas giants, ice giants and terrestrial planets),
- dwarf planets and moons by the hundreds,
- a host of families of small bodies (asteroids, comets, centaurs and Trojans, Trans-Neptunian Objects (TNOs), Oort cloud objects,
- an extended population of small size interplanetary and interstellar dust particles,
- ring particles in the four giant planet systems.

The sizes of these objects range from about $10^8$ m for the largest one, Jupiter, to about $10^{-7}$m for the smallest of dust particles. Some of these objects, like the Moon and Mars, have been extensively studied and deserve to be studied more. However, for the specific purpose of exploring the diversity of Solar System objects, we will focus more attention in this section on the classes of objects that have been poorly covered by space exploration to date and are difficult to characterize from telescopes (for example, the most distant populations of small bodies and the ice giants).

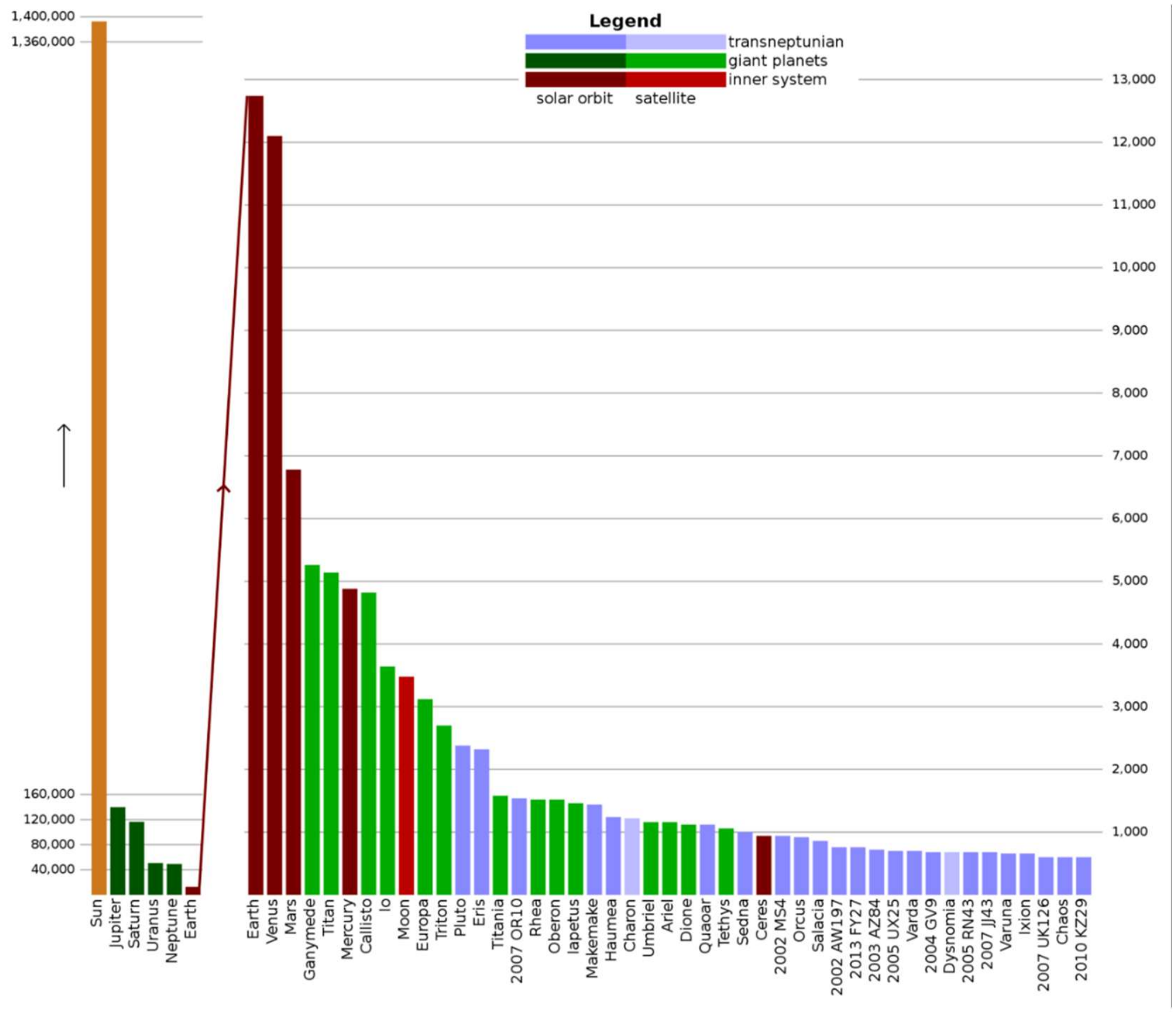

Figure 3.2. Relative sizes of the fifty largest bodies in the Solar System, coloured by orbital region. Values are diameters in kilometres. Scale is linear. © Tbayboy – Own work, CC BY-SA 4.0, https://commons.wikimedia.org/w/index.php?curid=40797976.

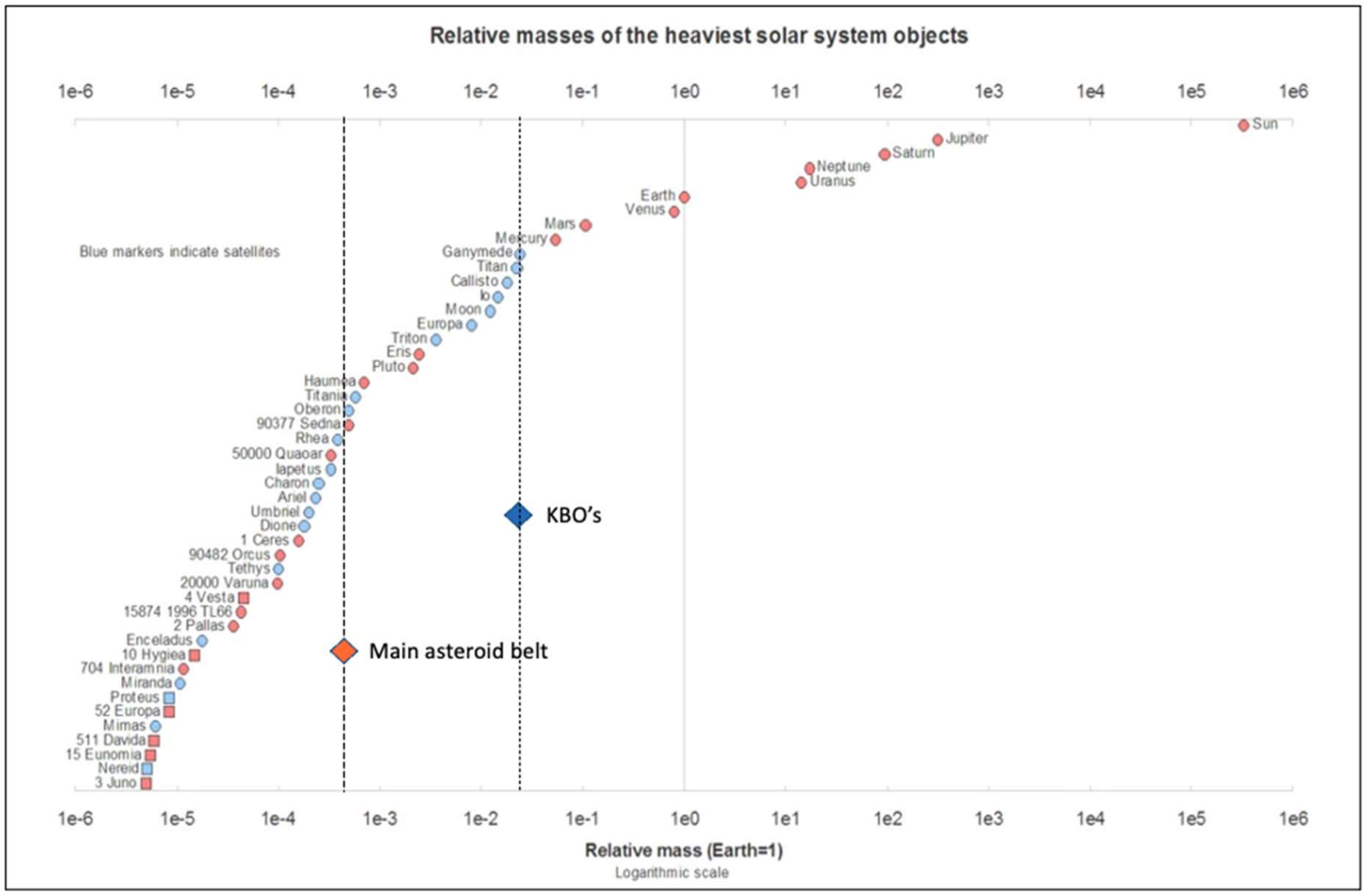

Figure 3.3. Relative masses of the heaviest Solar System objects. The total masses of KBO's deduced by Pitjeva & Pitjev (2018), blue diamond, and of the main asteroid belt from Pitjeva & Pitjev (2016), red diamond, are overlaid.
© https://space.fandom.com/wiki/List_of_solar_system_objects_by_mass.

Fig. 3.1 illustrates the range of distances we need to cover, combining remote observations from ground- and space-based telescopes, remote sensing from orbiters and flyby spacecraft, and in situ exploration by landers and descent probes, to establish a comprehensive inventory of these objects. As an introduction to this inventory, Fig. 3.2 and Fig. 3.3 show the size and mass spectra of Solar System objects.

Fig. 3.3 illustrates how the total mass of Solar System objects (excluding the Sun) is distributed between their different populations. Compared to the total mass of the Solar System (approximately 446.6 Earth masses, $M_E$), Jupiter alone represents 71% of this mass, the gas giants Jupiter and Saturn 92%. Adding the ice giants, giant planets gather 99.5% of the mass of the Solar System, whereas all smaller objects, terrestrial planets, small bodies, and moons amount to 0.5%.

Considering the two main populations of small bodies, Pitjeva & Pitjev (2018) estimated from astrometry, telescopic and flyby data (when available) that the total mass of TNOs amounts to $2.10^{-2}$ $M_E$, i.e. approximately the size of a Galilean moon of Jupiter, to be compared to approximately $4.10^{-4}$ $M_E$ for the main asteroid belt (Pitjeva & Pitjev, 2016), which is on the order of the mass of one of the small regular satellites of Uranus (Titania or Oberon). One can see that outer Solar System objects, those that likely formed beyond the ice line of the Solar Nebula, represent the vast majority of the mass of Solar System objects, including planets as well as small bodies.

Finally, it is interesting to note that, if we sum up from Fig. 3.3 the total mass of bodies that are suspected to be "habitable" (see Section 6), i.e. Earth, Mars and the main icy satellites of giant planets, they represent about 1.17 $M_E$ (including the Earth), or a fraction on the order of 0.26% of the total mass of Solar System objects.

Let us now explore the different categories of objects.

## 2.1. Planets

The eight planets of the Solar System offer a broad diversity of characteristics, whatever property we consider: internal structure, magnetic field, chemical composition, surface morphology and dynamics, atmospheric composition and dynamics.

### 2.1.1. Diversity in internal structure, chemical composition, magnetic fields

Our limited current knowledge of the internal structure of the eight Solar System planets is illustrated in Fig. 3.4. Further information can be found in Bennett et al. (2019).

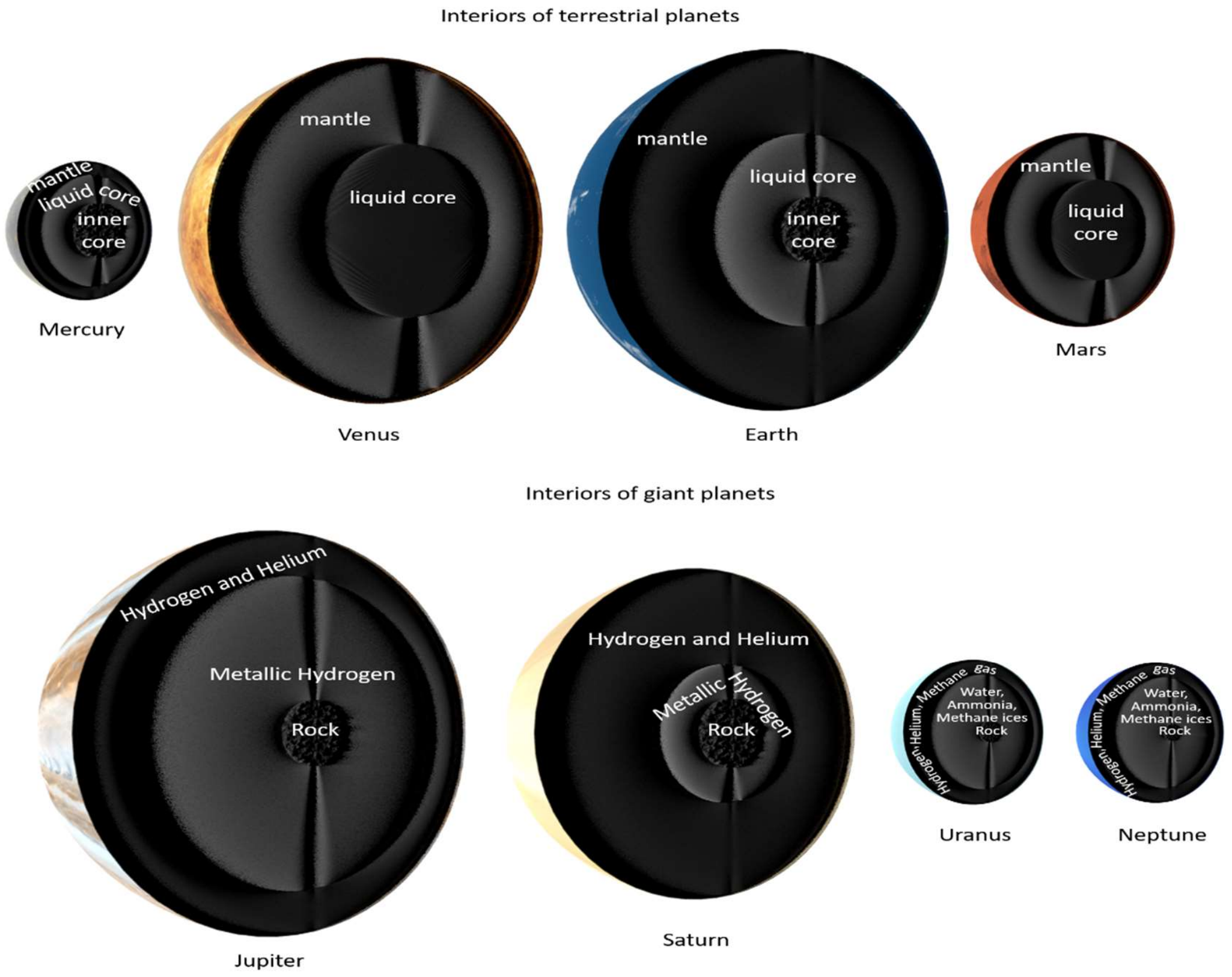

Figure 3.4. The standard classification of Solar System planets into terrestrial planets, gas giants and ice giants reflects distinctly different bulk compositions, densities and likely internal layering: while terrestrial planet compositions are dominated by silicates and metals with only a few percent of volatiles, gas giants are dominated by hydrogen and helium captured during their formation out of the Solar Nebula and ice giants are likely dominated by water and other ices surrounded by a shallower $H_2$/He outer mantle and atmosphere. Our knowledge of their internal structure, degree of differentiation and of the coupling mechanisms between their layers decreases drastically with increasing distance from the Sun.

All planets' radial structures are described to first order by a succession of layers from the centre to the exterior: core, mantle and crust for terrestrial planets, possibly surrounded by an atmosphere, whose density decreases outwards. At giant planets, however, recent gravity measurements (Wahl et al., 2017) and ring seismology observations (Mankovich and Fuller, 2021) showed that transitions between a rock/ice core, mantle and atmosphere are much smoother and more gradual than described in this simple picture (see Section 5.3). What we know of this characteristic chemical differentiation of planets comes from a very limited number of sources:

- The initial composition of the Solar Nebula, from which all planetary materials come;
- Gravimetry information, which indicates the bulk density, constrains the average composition and provides partial (though ambiguous) information on the internal structure. In the case of the most distant planets like the Ice Giants, direct gravimetry data is limited to the radio science experiment performed by Voyager 2 in 1986 and 1989, respectively;
- Ring seismology, which can identify regions that are stably stratified through pulsations within the planet;
- The thermodynamic and rheological properties of the different types of materials that contribute to the different layers, sometimes informed by high-pressure experiments performed in Earth laboratories;
- Finally, internal structure models fed by this diversity of data and constraints, applied to each planet.

It is from this limited and incomplete information that the current consensus on the internal diverse composition of the planets is based:

- For gas giants, a rock-and-ice core, surrounded by a hydrogen mantle with a small fraction of helium and metallic hydrogen conditions at high pressures;
- For ice giants, a rock-and-ice core, surrounded by a lower mantle composed of a mixture of heavy molecular species (possibly water-rich, depending on the poorly-known ice-to-rock ratio) and an upper mantle and atmosphere of hydrogen and helium;
- For terrestrial planets, a metallic core surrounded by a silicate mantle and crust, with an atmosphere of varying composition.

We also know that the radial structure of planets is controlled by several key mechanisms that operated through time:

- Their accretion history and the materials that contributed to this accretion;
- Their thermal evolution history, which saw chemical differentiation driven by partial or total melting, convection and migration of materials under the effect of buoyancy forces, progressive cooling and sometimes solidification of certain layers;
- The build-up and maintenance of internal phase transition boundaries between the solid (amorphous or crystalline), liquid and gas phases, which creates a particularly sharp boundary between solid and gas layers at terrestrial planets: the crust and its interface with the atmosphere in the form of a relatively narrow surface boundary;
- Finally, the contemporary working of each planet's "thermal engine" under the effect of its specific internal heat sources: decay of radionuclides for terrestrial planets, continuous contraction of the planetary body and possible phase separation for giant planets.

Beyond this generic description, many elements remain poorly known:

- our limited knowledge of the basic composition (the rock-to-ice ratio) of the least-explored of Solar System planets (ice giants and to a lesser extent gas giants) and with it, our poor understanding of radial transport mechanisms of mass and heat in their interiors;
- the internal structure prevailing right after formation and its early evolution, which for giant planets (unlike terrestrial planets) is crucial for the long-term evolution of their internal structure;
- the actual degree of differentiation and therefore also the chemical composition contrasts and degrees of mixing between layers, promising many surprises when more accurate measurements become available.

The recent results of Jovian gravity field measurements by Juno's radio science experiment are a perfect illustration of the major leap forward that an extensive radio science experiment onboard a low periapsis orbiter can produce: they revealed that chemical transitions in Jupiter's interior are much smoother than predicted by pre-Juno models and transformed our view of Jupiter's interior, leading to the revised internal structure model shown in Fig. 3.5 (from Wahl et al., 2017).

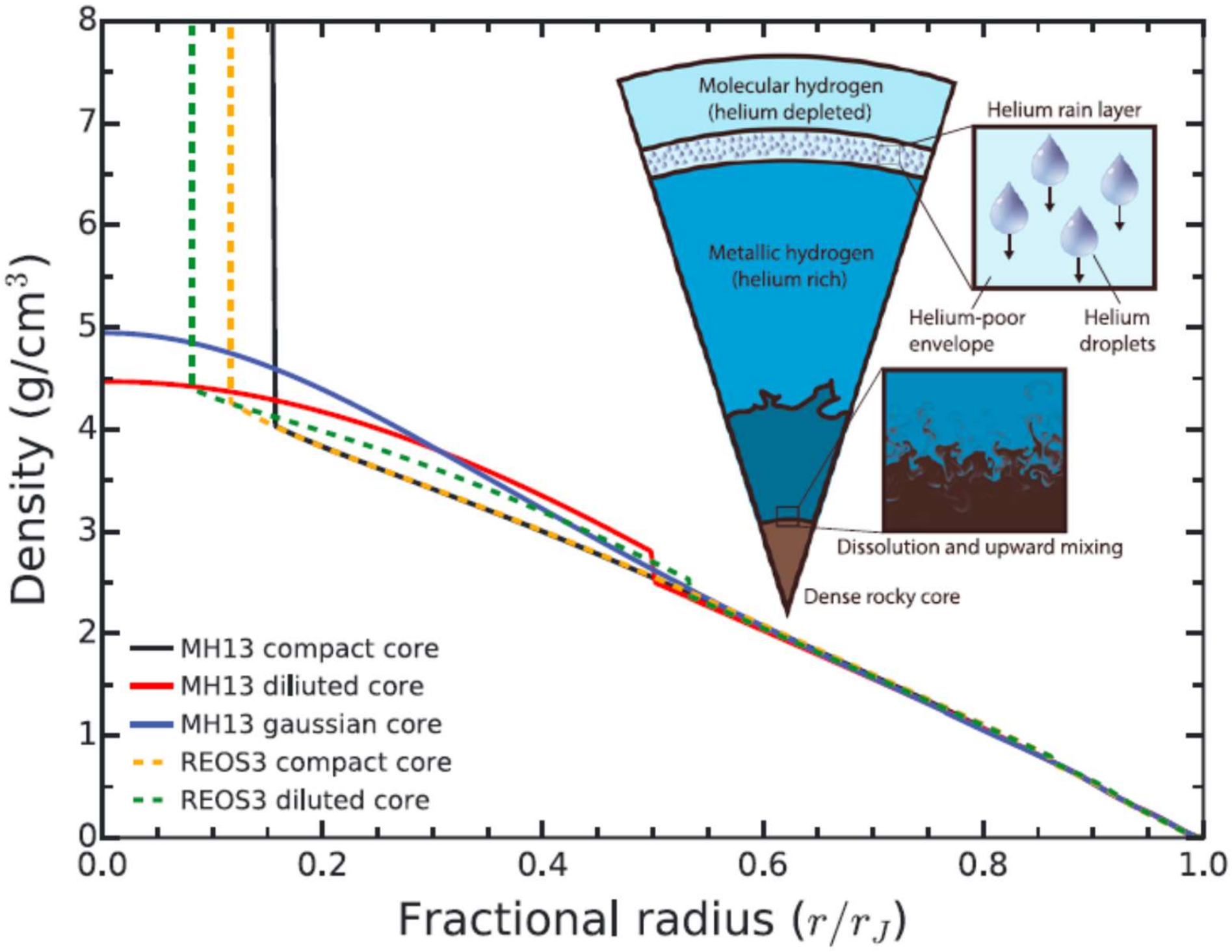

Figure 3.5. Radial density profiles from a selection of interior models based on ab initio computer simulations of hydrogen-helium mixtures are used to reproduce Juno's unprecedentedly accurate models of the even moments of Juno's gravity field up to order 8 in the spherical harmonic expansion. Adjustment of models to observations suggests that a dilute core, extending to a significant fraction of the planet's radius, represented by the continuous curves in blue and red, is helpful in reconciling the calculated gravity coefficients $J_n$ with Juno's observations. The inset illustrates what could be the corresponding internal structure of Jupiter, with a dense rocky core at the centre and part of its heavy element components partly dissolved and mixed inside the surrounding lower mantle. Helium "rains down" at the transition between the shallow molecular hydrogen envelope and the deeper metallic hydrogen mantle. From Wahl et al. (2017).

Complementing the gravity fields, planetary magnetic fields are another important product of the differentiation and internal dynamics of each planet. They are also very diverse and the mechanisms producing them are often considered to be a type of $\alpha$–$\Omega$ dynamo (the $\alpha$ effect transforms toroidal fields into poloidal ones through helical fluid motions; the $\Omega$ effect transforms poloidal fields into toroidal ones through differential rotation) operating in an electrically conducting convective layer at each magnetised planet: a liquid iron alloy core for terrestrial planets, the metallic hydrogen layer in gas giants and perhaps an ionic/superionic water ocean inside Uranus and Neptune. Magnetic fields provide constraints on the radial distribution and properties of these layers and on their ability to maintain an internal dynamo. Fig. 3.6, adapted from Soderlund & Stanley (2020), illustrates our current knowledge of the surface distribution of the radial components of the magnetic fields for planets with active dynamos (excluding the Jovian satellite Ganymede, not shown), a criterion that excludes Venus and to a lesser extent Mars despite its weak fossil magnetic field. It shows a broad diversity of magnetic field surface distributions, with variable relative intensities of the dipole component compared to higher-order components and variable orientations in space of their dipole axis. Secular variations, or changes in the field over time, have been modelled for Earth and Jupiter and can further provide constraints on interior flows. Over long timescales, evidence of remnant crustal magnetisation on Mercury suggests its dynamo is long-lived with a lifetime of billions of years. Our best understanding of magnetic field evolution over time beyond Earth is derived from paleomagnetic studies of lunar samples returned by the Apollo missions, which demonstrates an ancient high-field intensity epoch and a subsequent weak-field epoch that persisted until relatively recently.

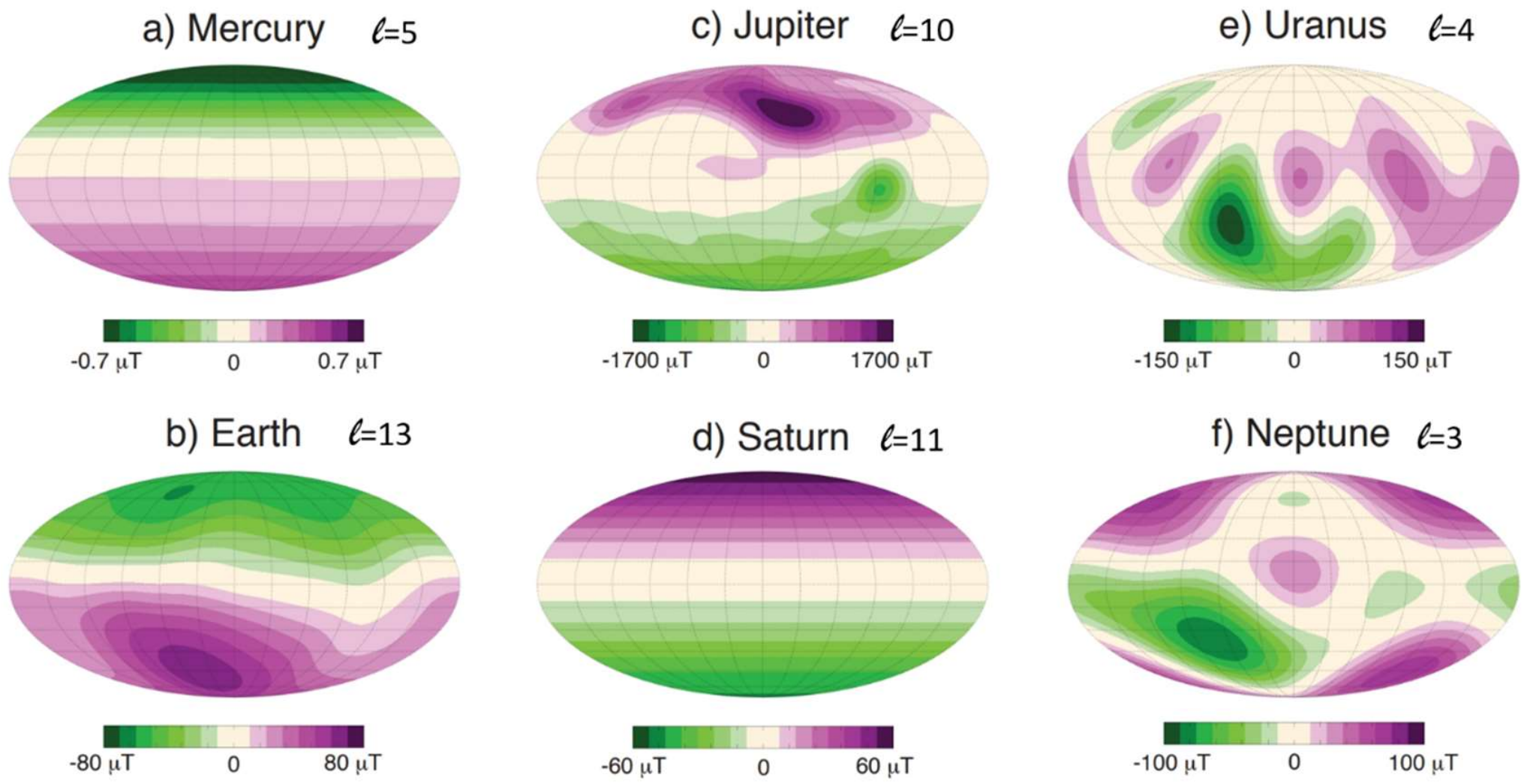

Figure 3.6. Radial magnetic field at the surfaces of a) Mercury, b) Earth, c) Jupiter, d) Saturn, e) Uranus and f) Neptune. The colours represent field intensity, where purple (green) indicates outward (inward) directed field. Spectral resolution of the observations in terms of spherical harmonic degree l for each planet is denoted above each representation. This comparison illustrates the ice giants' unique magnetic field morphologies. Adapted from Soderlund & Stanley (2020).

Despite the observational and experimental constraints described above, there are still many fundamental questions regarding the diversity of planetary interiors that are only partly answered, or sometimes just unsolved:

- What are the bulk compositions of Solar System planets?
- Are their interiors separated into well-distinguished chemically differentiated layers, or only partially differentiated with smooth composition and dynamical transitions?
- What are the dominant heat transport mechanisms within each planet?
- What are the characteristics of their intrinsic magnetic fields, how and where are they generated and what are the relationships of the magnetic field generation regions to the different layers of magnetised planets?

Our degree of information on these questions, mainly driven by our incomplete knowledge of planetary gravity and magnetic fields and of the bulk, interior, surface and atmospheric composition of planets, is extremely diverse and uneven. This is the direct result of the huge differences of their distances from our Earth-based telescopes and of the low number and observation capabilities of space missions that have visited them. The two ice giants, Uranus and Neptune, are by far the most poorly known. Characterising them with an accuracy comparable to the other two classes of planets, via orbital exploration and in situ descent probes, is an urgent task to reach a significantly better and more uniform knowledge of planetary interiors, dynamics, and compositions across the Solar System.

## 2.1.2. Diversity in surface morphology and geology of terrestrial planets and the Moon

Planetary surfaces represent the archives that enable deciphering the evolution of planets. The processes shaping the surface of terrestrial planets are diverse: internal processes build the topography and modify the surface; surface processes include a large number of processes, such as aeolian, fluvial or glacial erosion, provided the presence of an atmosphere; cratering and volcanism are ubiquitous processes that deeply modify the topography. The diversity of these processes explains why each planet has its own geological signature. By being so different, terrestrial planets and the Moon are complementary and provide fundamental data for understanding the evolution of the inner Solar System and thus Earth's evolution. The divergence of evolution between Venus, Earth and Mars remains a key question in relation to the combined roles of geodynamical and atmospheric evolution that left Earth as the only currently habitable body. Although terrestrial planets have triggered a huge number of space missions, especially to the Moon and Mars, key observations are still missing to understand fully their differences in evolution in the context of the unique terrestrial planet on which life evolved, Earth.

### The Moon

Earth-based telescopes, lunar orbiters, or landers on the Moon have allowed studying the geology and geomorphology of the Moon, especially the side facing the Earth. The lunar surface has been formed by a combination of processes, especially impact cratering and volcanism, and has also undergone space weathering due to high-energy particles and micrometeorites. The lunar landscape is characterized by impact craters, their ejecta, a few volcanoes, hills, lava flows and depressions filled by magma (see Fig. 3.7). There are bright and dark zones typical of the lunar surface. Light surfaces are lunar highlands (of anorthositic composition) and the dark plains are called maria (of basaltic composition). Since they are more heavily cratered, the highlands are older than the maria. On the far side of the Moon, there are only a few maria. These maria have lava flow patterns and collapse structures attributed to lava tubes. There are also grabens and tectonic features within the lunar maria near the edges of large impact basins. Impact cratering is the most evident process on the Moon as it appears everywhere on the lunar surface. A timescale of the surface processes on the lunar surface has been established from analysing the samples returned from the Moon by the Apollo missions. A summary of the lander sites is presented in panel (a) of Fig. 3.7.

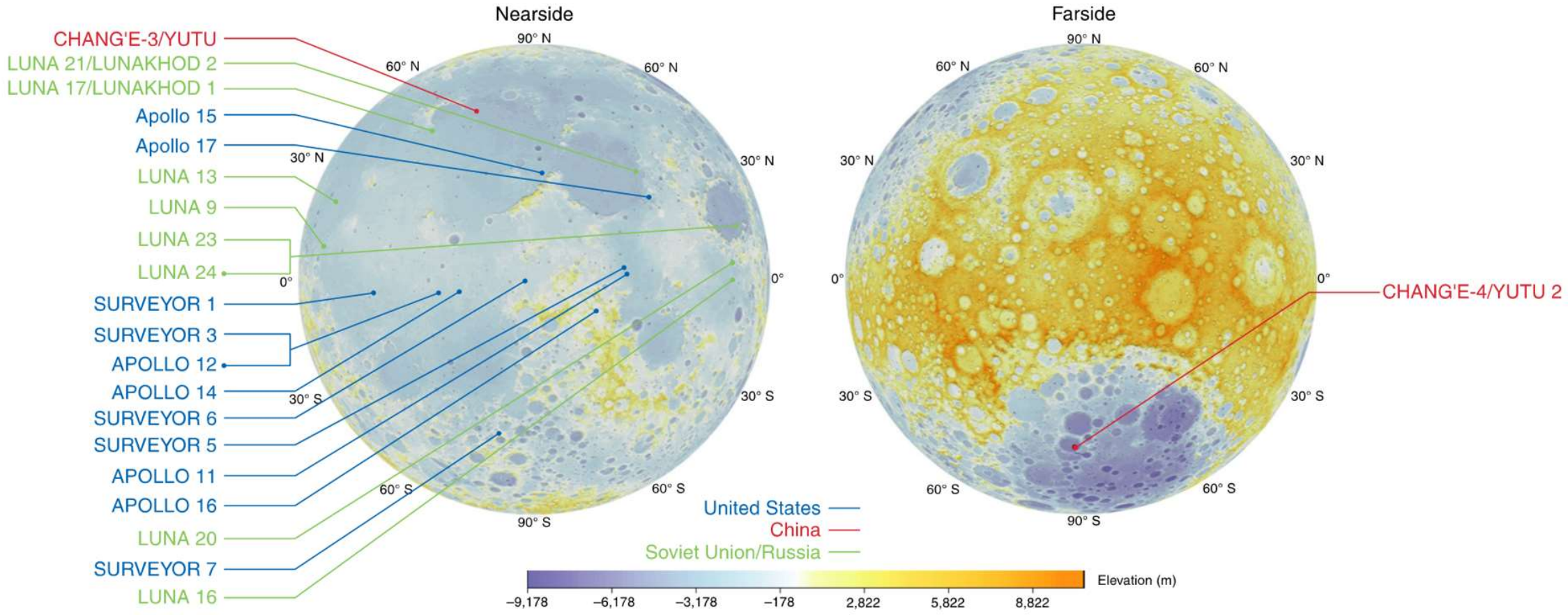

Figure 3.7 Map of all sites visited by robotic or human-tended missions on the Moon (note: Chang'e 5 should be added). From Wu et al. (2019).

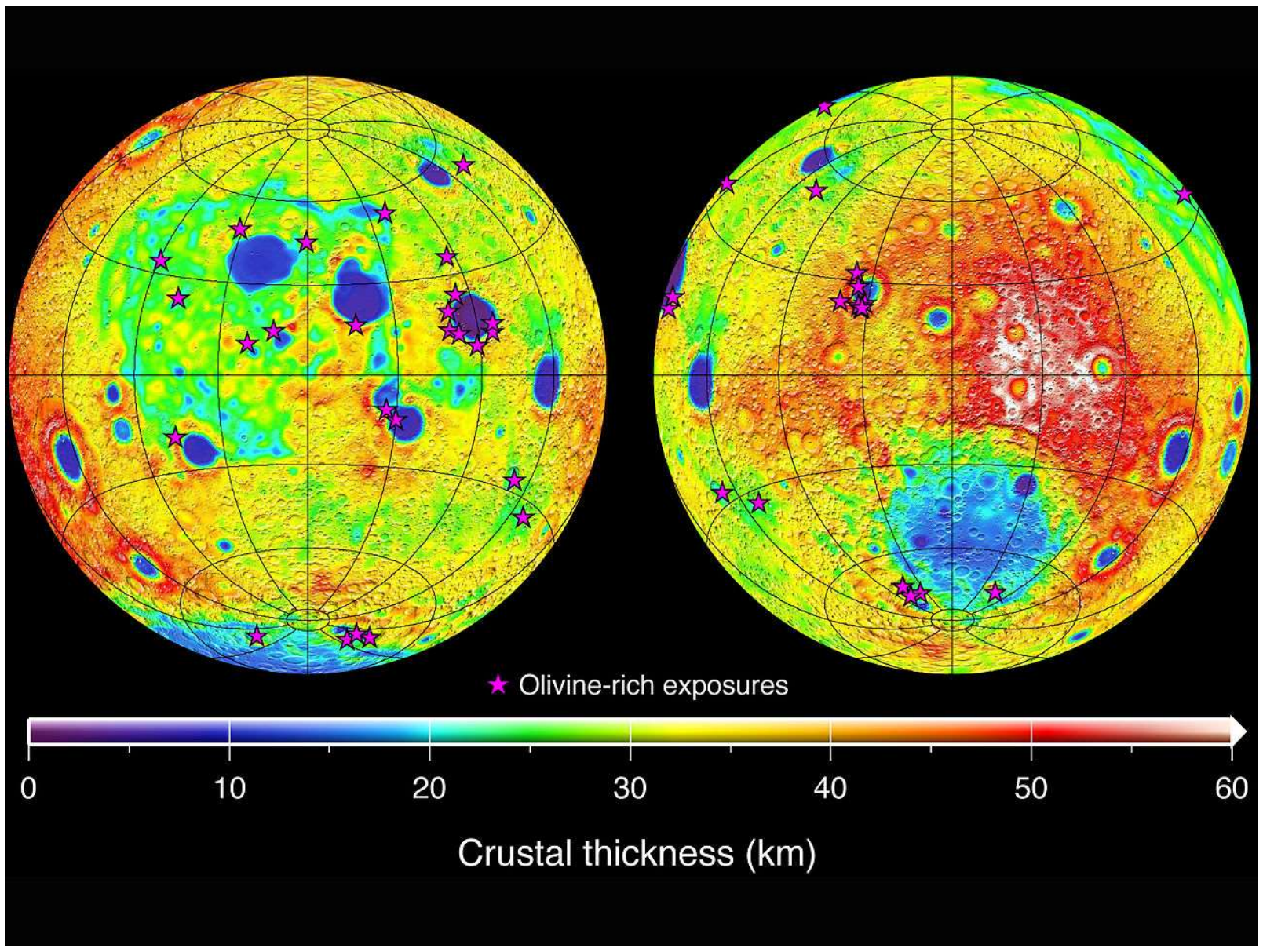

Figure 3.8. Map of geological terrains and crustal thickness of the Moon established by GRAIL and LOLA/LRO. One can recognize through the different topographical levels the division of lunar terrains between highlands (predominantly red), maria (predominantly green), and the major impact basins (blue to deep blue). The south-pole Aitkin impact basin, the largest of lunar impact basins, stands out on the far side. It bears a particular interest as one possible place where the crust might have been excavated to the point of uncovering upper mantle material. © NASA/JPL-Caltech/IPGP https://www.nasa.gov/mission_pages/grail/multimedia/pia16589.html#.YiNt1ejMKPo or https://www.planetary.org/space-images/moon-crust-thickness-grail; original from Wieczorek et al. (2013).

Using spacecraft-to-spacecraft tracking observations from the Gravity Recovery and Interior Laboratory (GRAIL), Zuber et al. (2013) have constructed the lunar gravitational field with the help of a spherical harmonic development up to degree and order 420 and have revealed features not previously resolved, including tectonic structures, volcanic landforms, basin rings, crater central peaks and numerous simple craters (see Fig. 3.8). This high-resolution gravity data combined with remote sensing and sample data, have allowed Wieczorek et al. (2013) to show the existence of a low crustal density with a porosity of ~12% to depths of at least a few kilometres (average crustal thickness between 34 and 43 kilometres). Matching the density, moment and Love number, Williams et al. (2014) have further calculated models including a fluid outer core with radius of 200–380 km, a solid inner core with radius of 0–280 km and mass fraction of 0–1% and a deep mantle zone of low seismic shear velocity.

The installation of seismometers on the Moon's surface during the Apollo era provided information on the lunar structure, formation and evolution. Seismic events were detected and used to constrain the structure of the Moon's crust and mantle down to a depth of about 1000 km. These data have been revisited recently (Weber et al., 2011) and have characterized the lunar core. The core size has also been resolved from Lunar Laser Ranging Data (LLR) (Viswanathan et al., 2019). The so-determined structure has resulted from the fractional crystallisation of a magma ocean shortly after its formation about 4.5 billion years ago, which has further been used to infer the Moon's history of differentiation. However, the uncertainties are still quite large. See Fig. 3.9.

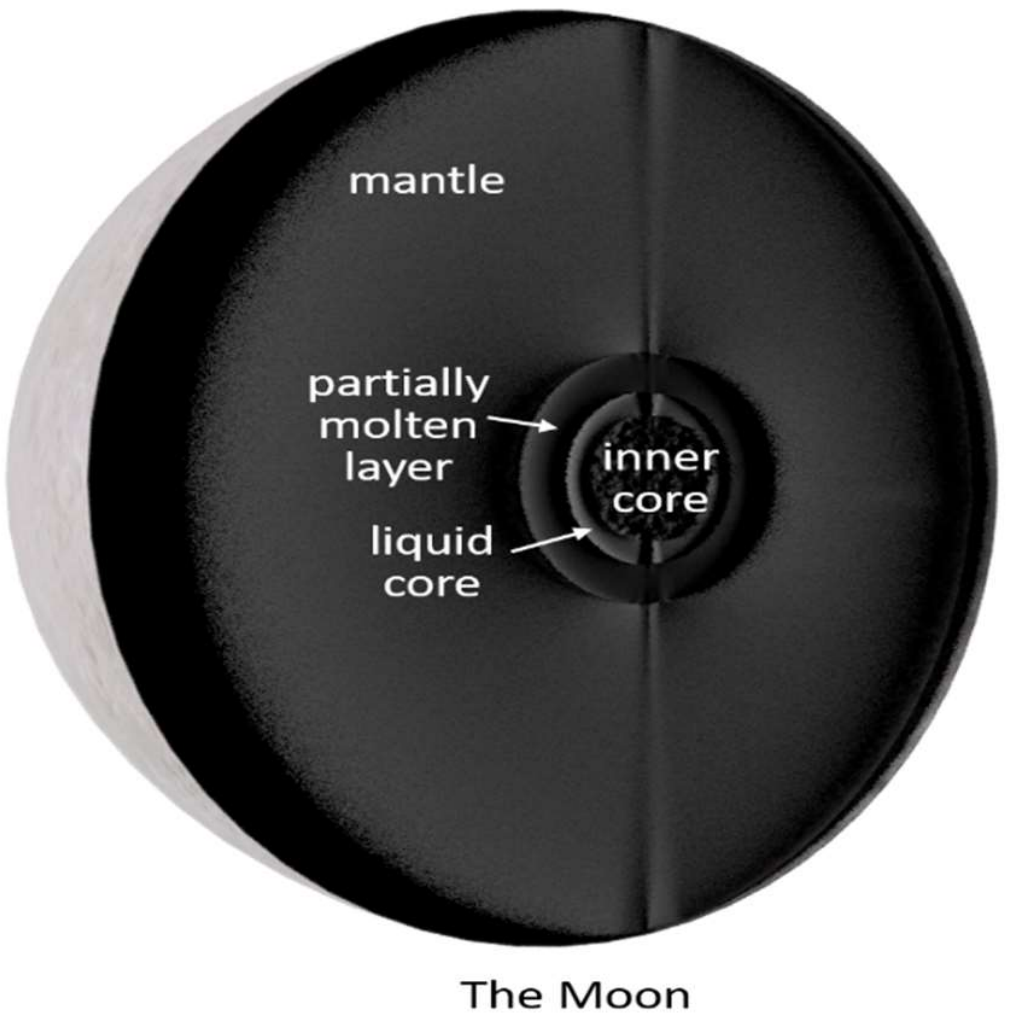

Figure 3.9. Internal structure model for the Moon from seismology. Adapted from Weber et al. (2011).

To solve these questions, a second wave of exploration of the Moon has started after the historical achievements made in the 1960s and 1970s by the Apollo and Luna missions. The Chinese lander Chang'e 4 has successfully landed on the far side for the first time offering new chemical data for the Moon. Regions where the mantle is suspected to outcrop are currently under assessment for future landed missions such as for the European mission Heracles, superseded by the European Large Logistics Lander (E3L). NASA is currently developing the Commercial Lunar Payload Services and the ambitious Artemis program that combines human missions with technological and scientific objectives, including the preparation of future manned missions to Mars (see Chapter 4 (Section 3.1) for a complete discussion). In this context, the next decade might lead to many important discoveries. One challenge will be to find and utilize in-situ resources for astronauts, including water and energy. This challenge will include technology-driven missions, but will also result in new scientific findings. For instance, a lander in the regions where water ice has been detected would be valuable to estimate the proportion of water and evaluate its use as a resource for astronauts. In parallel, the study of its isotopic composition, such as D/H ratio or of associated volatiles, could give fundamental inputs to understand better the origin of this water (cometary?). Future landed robotic and human missions will address a broad range of science questions.

## Mercury

Only reached by two space missions as of today, Mercury is often forgotten in the comparative planetology of the interior system. One reason is that Mercury's surface has been viewed as a Moon-like surface for a long time because of its cratered surface observed by Mariner 10. In the years 2011-2015, the NASA MESSENGER (Mercury Surface, Space Environment, Geochemistry and Ranging) mission confirmed the old age of most of the surface, but this characteristic should not make Mercury a planet of little interest. On one hand, Mercury's surface displays specific volcanic and tectonic landforms, such as dense networks of wrinkle ridges and vents formed by explosive volcanism. On the other hand, Mercury's surface can be used as a point of comparison to the Moon for cratering processes and related chronology, as well as for all processes linked to space weathering because of its proximity to the Sun. Space weathering is likely responsible for the fact that the orbital spectrometer onboard MESSENGER has not been able to characterise the mineralogy of the surface with the same success as spectrometers in lunar orbit. Some of the dark regions with low reflectance have been explained tentatively by the presence of graphite (Peplowski et al., 2016). In contrast, X-Ray and Gamma Ray spectrometry enabled global mapping of the chemistry of Mercury's surface, finding low FeO (<2 wt%) but high Sulphur (up to 4 wt% of elementary S) (Nittler et al., 2011). These results contrast with the usual view of Mercury as a metal-rich and volatile-poor body. It is also unclear whether fragments of a crystallized magma ocean are preserved at the surface, as is the case on the Moon.

The lack of better surface mineralogy remains a gap in knowledge to understand Mercury's evolution. The European mission Bepi-Colombo arriving at Mercury in 2025 may help to improve this situation, but the surface analyses will remain challenging for orbital spectrometry.

Ground truth from in-situ analyses would be welcome to enable comparisons with both spectral and gamma-ray data. A lander would thus enable a huge improvement in the knowledge of Mercury's surface, although the hot, high-radiation environment would certainly require several technological challenges to be solved.

A further intriguing observation of Mercury corresponds to measurements of high amounts of hydrogen made by MESSENGER neutron spectrometer in the Polar Regions, likely indicating water ice in permanently shadowed craters (Lawrence et al., 2013). Indeed, crater floors in regions near the poles are expected to have very cold temperatures (<100 K), especially when compared to the hot regions (>600 K) exposed to the Sun during the long Mercurian days. The origin and age of ice deposition are unknown.

Further studies of these deposits would be of major interest. Nevertheless, reaching these regions may be a huge technological challenge, especially regards to the huge temperature differences and space environment in the potential landing locations where ice has been detected.

## Mars

Mars is the terrestrial planet with the largest variety of surface processes after Earth, in combination with the presence of an atmosphere and volatiles at the surface, as well prolonged volcanic activity. While Mars' climate is currently cold and dry, the period during which Mars was warmer and wetter dates back to 3 Gy and before, with still unclear knowledge of the amplitude and duration of that past climate.

A major unknown of Mars' history is its early evolution, usually referred to as pre-Noachian (>4 Ga), roughly before the giant impacts that formed Hellas and Argyre. The surface of Mars does not preserve the morphology of that primordial period, limiting its analysis from orbital images. However, old crustal material may still be present locally in the cratered highlands. The finding of an ancient meteorite dating back to 4.4 Ga containing alkali feldspar (Hewins et al., 2017) and of felsic rocks at Gale crater by the Curiosity rover (Sautter et al., 2015) opened several questions regarding the composition of the primitive crust. Are they preserved pieces of primary crust formed by the crystallisation of a magma ocean as on the Moon? Is the ancient crust more felsic than the basaltic volcanism observed ubiquitously at the surface? Are there volatiles trapped in ancient magmatic rocks? The rover Perseverance from the Mars 2020 mission may enable analysis of the ancient crust on the rim of Jezero impact crater and may even enable sample return from such rock.

However, there are many exhumed outcrops of ancient crust at the surface of Mars, which would deserve a specific devoted mission, with in situ analysis of rocks with the help of landers or rovers and rock analysis techniques on these spacecraft, such as by camera, close up imager, various spectrometers (IR, Raman, Mossbauer), and/or molecular analysers, providing an accurate visual and spectral characterisation of the surface.

The primary objectives of rovers on Mars are the study of ancient sedimentary bodies that enable geologists to reconstitute the climate under which they formed, and that have potentially preserved biosignatures, if life ever appeared on the planet. To date, rovers have only been sent to sedimentary rocks of Late Noachian to Hesperian age (3.7 to 3 Ga), including the new rover Perseverance, which has landed in a Late Noachian to Hesperian paleolake (Mangold et al., 2020), although the ExoMars rover will sample older sediments, albeit of as yet uncertain origin (Quantin-Nataf et al., 2021). Spectral data collected from orbiters of Noachian regions show a widespread presence of phyllosilicates linked to aqueous alteration.

The analysis of residual sedimentary rocks in the ancient crust of the pre-Noachian period as observed from spectral data collected from orbiters could pave the way to a new view of Mars early history that cannot be traced from the current dataset. This would require landing in the Southern Highlands and therefore for instance use RHUs and precision landing.

In contrast to its ancient climate, the current climate of Mars is much better understood, but the occurrence of specific processes related to $CO_2$ frost, buried ice, or dust storms requires detailed studies as these processes will be of great importance for human exploration. In this regard, the presence of water ice in the near surface (few cm) is demonstrated by the presence of typical periglacial landforms (polygonal cracks, etc.), the abundance of hydrogen detected by neutron spectrometer and the local ground truth provided by the Phoenix lander at 69° latitude North. However, these high latitudes are not considered for early human exploration because of the seasonal occurrence of $CO_2$ frost, at temperatures below -120°C. For a long time it has been known that many potential periglacial landforms also exist in the mid-latitudes (25°-60°) (lobate aprons, sublimation lags, etc.) (e.g., Squyres, 1989), demonstrating that water ice is locally preserved as buried glaciers or relics of permafrost of potential interest as an in-situ resource.

Because water ice is thermodynamically unstable at the surface at these latitudes, and because sublimation landforms indicate that ice partly vanished, it would be worthwhile to explore these regions with subsurface sounding methods and drills (10 m scale) to provide local evidence for the presence of water ice before the start of human exploration of Mars.

### Venus

With its thick atmosphere and enigmatic surface, Venus remains a mysterious planet. Radar data from the Magellan spacecraft demonstrated that the whole surface of Venus was resurfaced ca. 500-700 Ma (e.g., Saunders & Pettengill, 1991), including many volcanic edifices but also structures rarely seen on other planets or moons, such as the circular coronae. Unlike the other terrestrial planets, Venus does not appear to preserve any terrains from its early history, raising questions concerning its evolution during its first 4 billion years. In theory, early Venus could have been more suitable for life than Mars, or even Earth, due to its closer distance to the presumably faint young Sun.

While this early history is difficult to reconstitute, the current composition of its crust, including the fate of volatiles, may give some hints of this past evolution, or, at least, elements to understand its geodynamics.

The surface of Venus has been reached by soviet landers in the 70s and 80s. During their short duration of activity (1 or 2 hours), X-fluorescence instruments onboard Venera 13, 14 and Vega 2, analysed the chemistry of rocks, suggesting that the Venusian rocks are dominated by basaltic compositions ranging from gabbro-norite to tholeiitic basalt (Surkov et al., 1986). However, those landers only sampled the surface at equatorial lowlands, far away from the highlands. The high concentration of Th, U and K (up to 5 wt% $K_2O$) measured by the Gamma-Ray Spectrometer at the Venera 8 landing site suggests that more differentiated rocks are also present (e.g., Basilevsky et al., 1992), thus questioning their role as internal heating for geodynamic processes. In addition, the identification of current thermal anomalies suggests that there is ongoing geodynamic activity, at least regionally (Smrekar et al., 2010). The composition of the mantle is only known from theoretical studies, although Venus' evolution requires a comprehensive understanding from the origin of volatiles from mantle to atmosphere (e.g., Gillmann & Tackley, 2014). Volatiles from the atmosphere play a strong role in the surface alteration.

The role of highly acidic gases such as HCl, HF, $H_2SO_4$, on basaltic rocks at >400°C, as is the case at the Venus' surface, needs to be understood better, especially because these processes can lead to the production of Sulphur-rich compounds in the subsurface, thus potentially trapping atmospheric Sulphur in the crust. The feedback of surface, crustal and mantle processes to the atmospheric evolution is an entire field requiring exploration.

ESA's Venus Express and JAXA's Akatsuki missions were the last to explore Venus from orbit, both in the last decade, while the soviet probe Vega 2 was the last to land on Venus in 1986, more than three decades ago. Several new missions to Venus are in development, such as a new European orbital mission (EnVision, scheduled for launch in 2032) with instruments such as interferometric radar, sounding radar, and thermal mapper. The NASA projects VERITAS (Venus Emissivity, Radio Science, InSAR, Topography and Spectroscopy) has been selected in Discovery and DAVINCI+ (Deep Atmosphere Venus Investigation of Noble gases, Chemistry, and Imaging, Plus) has been proposed but not selected in New Frontiers program. A new collaborative Russian-European mission (IKI Venera-D) is also under development for potential launch in 2026 or 2031, with an orbiter and a lander. However, these missions still face challenges. In particular, the use of orbital spectrometry for analysing surface composition is limited due to the thick cloud cover.

Spectroscopic analyses from balloons or during descent of landers are possible alternatives to enable, at least, a regional coverage. The in-situ exploration of the Venusian solid surface remains technologically challenging, especially the development of relatively long-lived electronic systems onboard landers, due to the high surface pressure, high temperature, and corrosive atmosphere. However, a better understanding of Venus' interior and surface evolution will not be possible without more in-situ data, if not returned samples from key areas. A mission coupling descent imagery, in-situ analyses, and sample return would certainly be the most ambitious type of mission for future Venus exploration. Sample return, however, requires new concepts for rockets because the gravity of Venus is almost the same as on Earth, thus requiring a huge energy for launch. Some of Venus' mysteries will likely take the whole century to be solved.

### 2.1.3 Diversity in Solar System atmospheres

Our exploration of the Solar System has revealed that there are two major classes of planet and moon atmospheres in the Solar System:

- **Primary atmospheres** formed at the time of planetary accretion before dissipation of the gas component of the Solar Nebula (up to 10 Myr maximum) that are mainly composed of the lightest and most dominant volatiles of the Solar Nebula: hydrogen, helium and trace volatiles in their reduced, hydrogenated forms. These have been preserved from such early times only on planets with a strong enough gravity to prevent escape of these light gases and avoid their loss to interplanetary space. Primary atmospheres prevail at giant planets.
- **Secondary atmospheres** formed on planets whose gravity is not large enough, and their temperatures too warm, to keep their primary atmospheres. This was the case for the terrestrial planets and Titan. Primordial light atmospheric gases then escape to space early in the planet's history and are replaced by volatiles degassing from the solid envelopes of the planet or moon and imported by comets or volatile-rich meteorites and micrometeorites, especially during the early phases of planetary evolution. This degassing was likely provided by two different processes over very different timescales through the planet's history (Forget & Leconte, 2014, and references therein; Grenfell et al., 2020; Forget et al., 2021): catastrophic outgassing, mainly of $H_2O$ and $CO_2$, was produced during the initial period of solidification of the magma ocean (500k to 3 million years), as these two constituents could not remain trapped in the solid phase of planetary silicate mantles; later on, over geologic timescales (billions of years) volcanic outgassing progressively released other volatiles trapped in the solid mantle, such as $H_2S$, $SO_2$, $CH_4$, $NH_3$, noble gases, in addition to $H_2O$ and $CO_2$ again, producing the diverse chemical composition observed in terrestrial planets.

Observation of the Solar System reveals a broad diversity among these two classes of atmospheres, to be likely surpassed only by the diversity of exoplanet atmospheres that will be characterised in the near future (see Grenfell et al., 2020).

**Terrestrial planets**

All processes governing the formation and evolution of planetary atmospheres (accretion, escape, outgassing, condensation, and weathering of atmospheric gases) are mainly controlled by the chemical composition of these gases and the environmental conditions reigning in each planet's atmosphere. The latter can, for convenience, be reduced to the planet's equilibrium temperature and mass, as shown in Fig. 3.10, which separates different domains of occurrence of planetary atmospheres in a notional temperature vs. mass diagram.

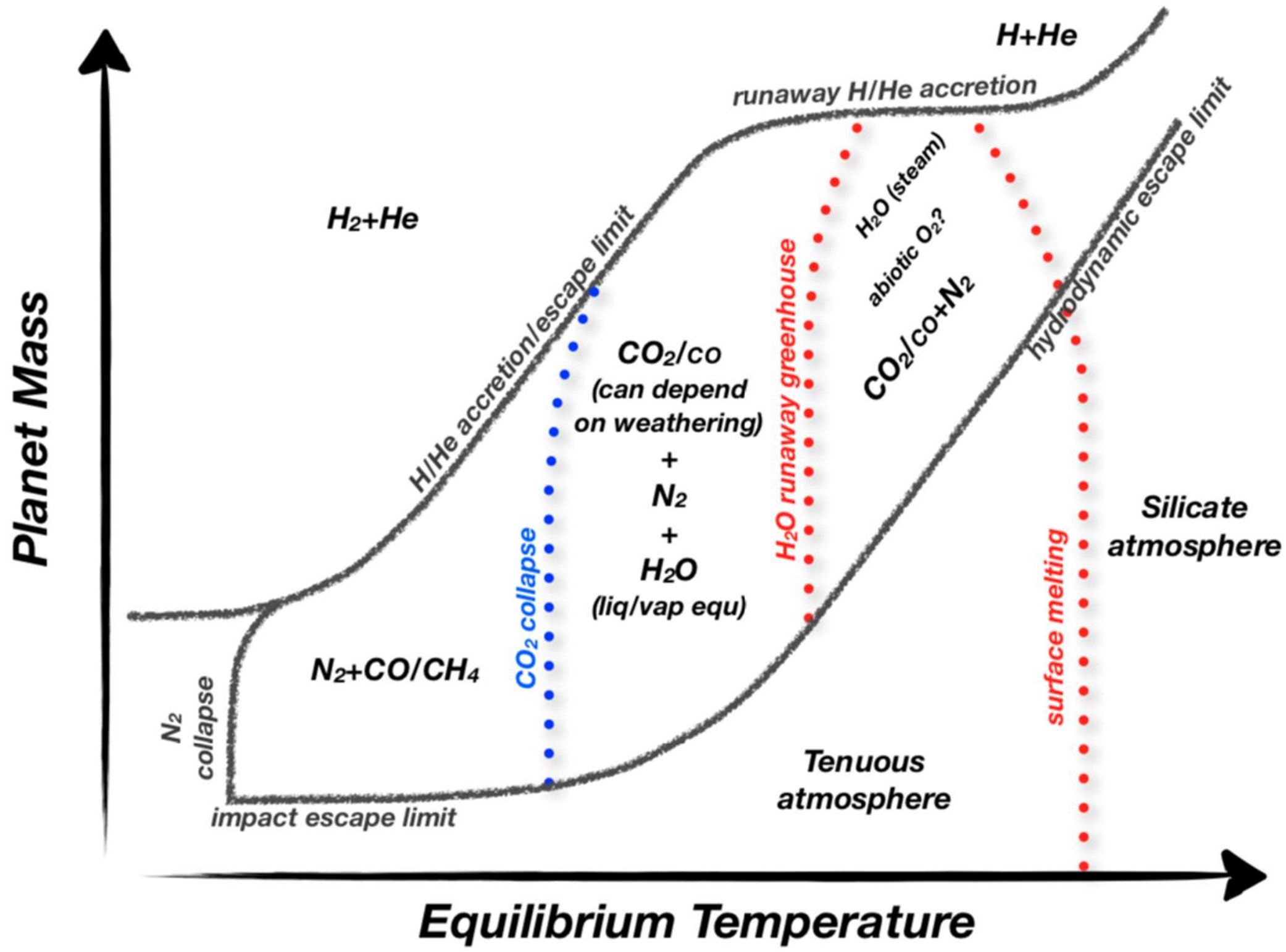

Figure 3.10. Schematic summary of the various classes of atmospheres. Each line represents a transition from one regime to another, but note that these "transitions" are in no way hard limits. Only the expected dominant species are indicated, but other trace gases would also be present. From Forget & Leconte (2014).

The two black curves separate the temperature versus mass parameter space in three domains. Above and to the left of the upper black curve, is the domain of primary H/He atmospheres, populated by giant planets that are massive enough to capture hydrogen during its accretion phase and to keep it from escaping to space over the age of the Solar System. Below and to the right of the lower black curve, the planet is not massive enough to retain even heavy gases and loses its dense atmosphere via hydrodynamic escape. At moderate to large temperatures, the mass limit for this to happen increases with increasing temperature. At low temperatures, the atmosphere of the lightest planets is lost just by impacts with other planetary or small bodies. In this domain, only tenuous atmospheres can be maintained, fed by the sublimation of ices on the low temperature side, as is the case for the tenuous atmospheres of most giant planet moons, or by the vaporisation of the silicate surface on the high temperature side (a case not encountered in the Solar System, but may be met among low-mass exoplanets orbiting close to their stars).

Thus, the domain between the upper and lower black curves is where terrestrial planets can preserve a stable dense atmosphere. This domain itself can be divided into several subdomains corresponding to very different compositions and potential evolution histories, bounded by the coloured near-vertical dotted curves:

- Left of the blue curve, temperatures are cold enough that $CO_2$ condenses (together with water). Nitrogen may then be left as the dominant constituent in the atmosphere (e.g., at Titan); at even lower temperatures to the left of the left-most vertical black curve, nitrogen condenses as well (e.g., at Triton).
- Between the blue and red curves, water is in equilibrium between the liquid and vapour phases, a water cycle may exist and possibly a carbonate-silicate cycle regulating the amount of $CO_2$ and the greenhouse effect in the atmosphere; this is the domain within which Earth has remained during most of its evolution.
- Right of the dotted red curve and up to the next dotted red curve corresponding to surface melting of silicates, water is vaporised into the atmosphere, no weathering limits the accumulation of $CO_2$ in the atmosphere. Here, the planet may experience a runaway greenhouse regime, as Venus likely does.
- Right of the second red curve, the temperature is high enough to melt the surface silicates and possibly generate a silicate secondary atmosphere.

It is interesting to note that many of the subdomains delimited in Fig. 3.10 are populated by one or several Solar System atmospheres: giant planet moons (except Titan) and Mercury on the low-mass side, giant planets on the high-mass side, and Venus, Earth, Mars and Titan in the intermediate domain. To conclude this section, let us focus on these four particularly important objects. Some of their typical parameters are displayed in Table 3.1 below.

| | Venus | Earth | Mars | Titan |
|---|---|---|---|---|
| Distance to the Sun, AU | 0.72 | 1 | 1.52 | 9.53 |
| Equatorial radius, km | 6052 | 6376 | 3380 | 2574 |
| Solar flux, W/m$^2$ | 2613 | 1364 | 589 | 15 |
| Surface pressure, bar | 92 | 1 | 0.006 | 1.45 |
| Main atmospheric gases | $CO_2$ 97%<br>$N_2$ 3%<br>$SO_2$ 0.015% | $N_2$ 79%<br>$0_2$ 18%<br>Ar 1%<br>$H_2O$ 2%<br>$CO_2$ 0.04% | $CO_2$ 96%<br>Ar 2%<br>$N_2$ 1.8% | $N_2$ 95.0%<br>$CH_4$ 4.9%<br>$H_2$0.2% |
| Bond albedo | 0.9 | 0.306 | 0.25 | 0.22 |
| Surface temperature, °C | 462° | 14°<br>(-90…+57) | -63°<br>(-40…130) | -179.5° |
| Greenhouse effect, K | +230 | +33 | +3 | +20 / -9 (net +12) |

Table 3.1. Main parameters characterising the atmospheres of the four terrestrial planetary objects with a dense atmosphere.

The remarkable number of similarities and differences offered by the atmospheres of these four objects located at very different distances from the Sun may guide our search for the causes of their origins, their diverging evolution paths, and the way they work, as well as teach us lessons about Earth's weather and climate:

- Venus and Mars possess $CO_2$-dominated atmospheres and experience very different greenhouse regimes: nearly non-existent greenhouse effect at Mars and runaway greenhouse at Venus, offering two opposite extreme cases to compare to Earth, which resides in the "comfortable zone" where liquid water is stable. A runaway greenhouse regime like that of Venus is the likely future of Earth.
- Venus and Titan both offer cases of global superrotation of their atmosphere (i.e., an additional differential rotation with respect to the rotation of the body itself), despite very different solar insolation and rotation rates, and this can be used to better understand the drivers of this phenomenon.
- Two important characteristics of Mars' atmosphere are the seasonal condensation of $CO_2$ on the winter polar cap and the importance of dust storms. Do they inform us about the "snowball-Earth" episodes of our climate and the severe atmospheric cooling episodes that past giant impacts induced on Earth's climate?
- The water condensation/evaporation cycle at Earth, coupled to a $CO_2$-driven greenhouse effect, offers some analogies with the condensation/cycle of methane coupled to molecular nitrogen greenhouse at Titan, producing similarities between the weather systems of these two planets despite very different temperatures and compositions.

Documenting further these similarities and differences by gathering the key data needed is essential for an in-depth understanding of the diversity of planetary atmospheres at large.

Key measurements towards this end are local measurements at the surfaces of the bodies and in their atmospheres, as well as remote sensing (from the UV to IR and even radio spectrum) from orbiters to determine atmosphere composition and properties and their changes over time.

**Giant planets**

Giant planets have preserved their primary atmospheres, dominated by hydrogen and helium and enriched by heavier volatiles (cosmogenically common elements C, O, N, S and others in their reduced and hydrogenated forms). In their case, there is no abrupt discontinuity, i.e. no "surface", between their interiors and their atmospheres. Indeed, the connection between the ever-changing meteorology and circulation of the 'visible' upper atmospheric layers to the processes at work in the deep hidden interiors remains a key driver for giant planet exploration today. The four giant planets serve as planetary-scale laboratories for atmospheric and oceanic processes at work on worlds spanning a broad range of parameter space – from rapidly-rotating, large-radii and hydrogen-dominated Jupiter and Saturn, to slow-rotating, intermediate-radii and ice-dominated Uranus and Neptune. Of these two classes of objects, Uranus and Neptune remain the least explored, to the extent that even their basic rock-to-ice ratio (i.e., their definition as "ice" giants) remains in question (e.g., Teanby et al., 2020).

What we know today of giant planet atmospheres mainly comes from their remote observation by ground- and space-based telescopes, alongside visiting planetary spacecraft (both flybys and orbiters). Remote sensing utilizes a variety of wavelengths: each particular spectral window provides a piece of the "puzzle" of these atmospheres, but even so, the puzzle remains to be assembled. Each spectral range, sensing either reflected sunlight (from the UV to the near IR) or thermal emission of the atmosphere (from mid-IR to submillimetre and a fraction of the radio spectrum), probes different depths and different constituents of the atmosphere. At infrared and radio wavelengths, the depth that we can probe is limited by the opacity of hydrogen and helium, with the exception of 'windows' into the deeper atmosphere near 5 microns (sensing the 2-6 bar cloud-forming region on Jupiter) and in the microwave range (1-100 cm, sensing 1-100 bars on Jupiter, approximately).

Fig. 3.11 illustrates the different and complementary multispectral information that can be gathered in this way on the atmospheres of the four giants. Visible and near-IR images use the strength of methane absorption bands to sound the clouds and aerosols throughout the upper troposphere. Infrared images use tropospheric absorption features and stratospheric emission features to determine atmospheric temperatures and gaseous composition in three dimensions. Aerosols are transparent at microwave (centimetre and millimetre) wavelengths, permitting sensing of the deeper tropospheres. Tracking of clouds as they move east and west with the prevailing zonal winds enables the study of giant planet meteorology. Furthermore, emissions from ions in the ionosphere and thermosphere reveal circulations in the upper atmosphere, including the redistribution of auroral energy.

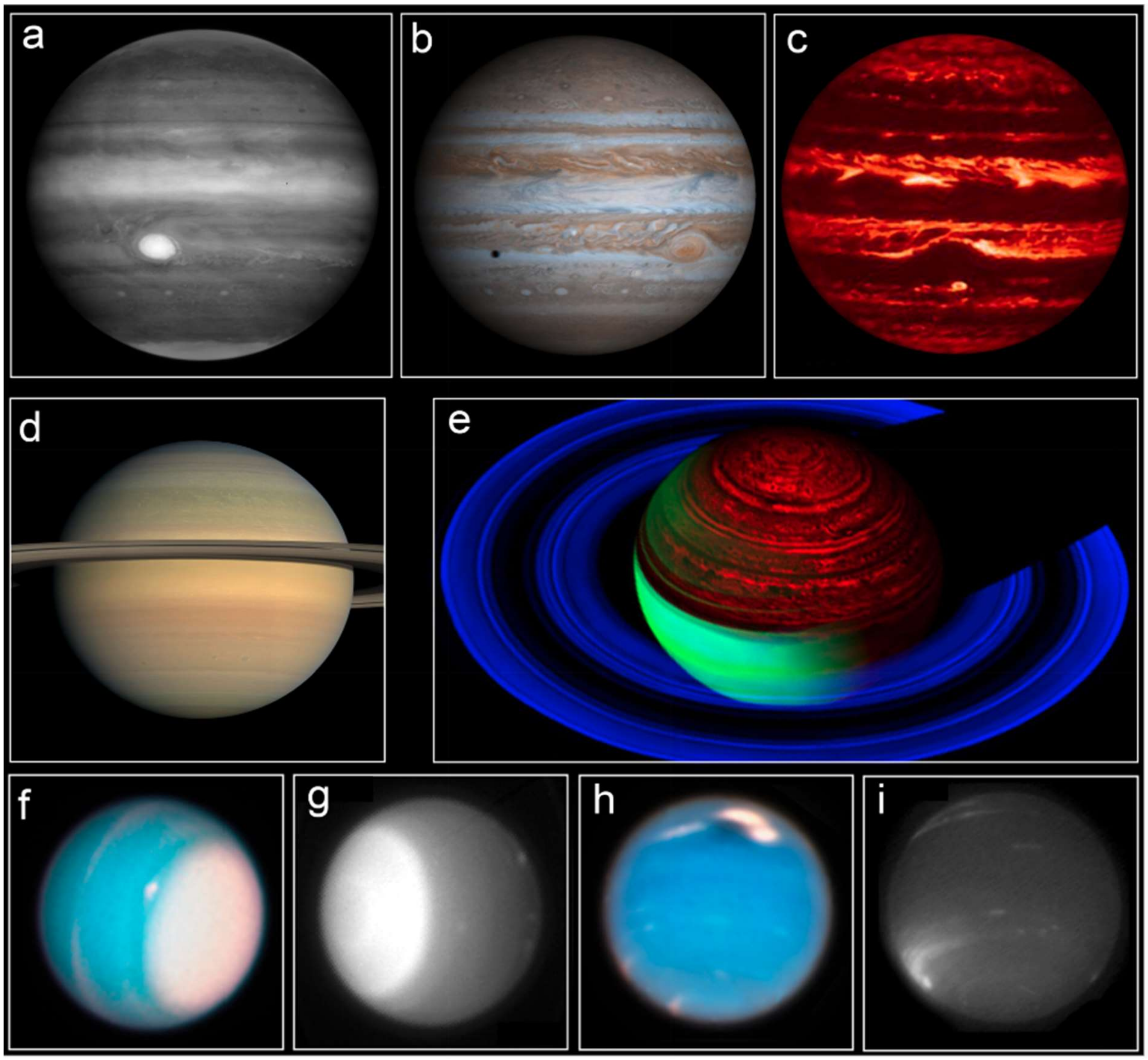

Figure 3.11. Multiwavelength images of Jupiter (upper row), Saturn (middle row) and Uranus and Neptune (bottom row). Images in the near-infrared in methane absorption bands (a, g, i) sample complex layers of hazes. Visible images (b, d, f, h) correspond to the top of the main upper cloud ($NH_3$ in Jupiter and Saturn and $CH_4$ in Uranus and Neptune). Infrared images at 4-5 µm (c, e) sample the opacity of a secondary and deeper cloud layer, most probably $NH_4SH$ in Jupiter and Saturn. From Mousis et al. (2021).

Remote sensing across these multiple wavelengths reveals two features common to giant planets atmospheres:

- Horizontally, these atmospheres are mainly organised in latitudinal bands interrupted by large-scale anticyclones, cyclones, convective storms and wave phenomena. These bands are associated with east-west jet streams and manifest as latitudinal contrasts in temperatures, aerosol albedo/colours and chemical compositions in the troposphere. As explicitly mentioned in Fletcher et al. (2020a), “on Jupiter, the reflective white bands of low temperatures, elevated aerosol opacities and enhancements of quasi-conserved chemical tracers are referred to as ‘zones.’ Conversely, the darker bands of warmer temperatures, depleted aerosols and reductions of chemical tracers are known as ‘belts’.” This banded structure has been extensively studied on all four giant planets at altitudes above the main clouds, but recent microwave observations from the ground (de Pater et al., 2021; Molter et al., 2021), Cassini (Janssen et al., 2013; 2017) and Juno (Bolton et al., 2017; Janssen et al., 2017) reveal that the banded structure extends much deeper into the tropospheres of the four giants. The bands are variable with time, changing over quasi-predictable nonseasonal cycles (multiple years) that remain poorly understood. Finally, the banded structure at low- and mid-latitudes gives way to fascinating dynamical regimes at high latitudes, including turbulence and organised circumpolar cyclones on Jupiter, polar vortices on Saturn and Neptune, seasonal clouds on Uranus and a hexagonal wave on Saturn.

- Vertically, giant planet atmospheres are dominated by a succession of clouds and hazes produced by the condensation of volatiles at different atmospheric levels. At each of the altitudes where they prevail, there exist clearings in the cloud layers opening latitude gaps, thus creating a band structure seen from above. Fig. 3.12 summarises current estimates of the vertical structure of these different layers of condensable species at the four giant planets based on thermochemical equilibrium. The top-most clouds are further contaminated by the products of methane photochemistry, which sediment downwards from stratospheric altitudes and form discrete haze layers. The relative concentrations of these condensable species increase from left to right, from Jupiter to Saturn and then to Uranus and Neptune, reflecting the increase with decreasing planet mass of the bulk relative concentration of ices with respect to hydrogen. Reality is likely to be significantly more complex, as cloud microphysics, storms and precipitation processes modify the cloud decks away from these equilibrium expectations.

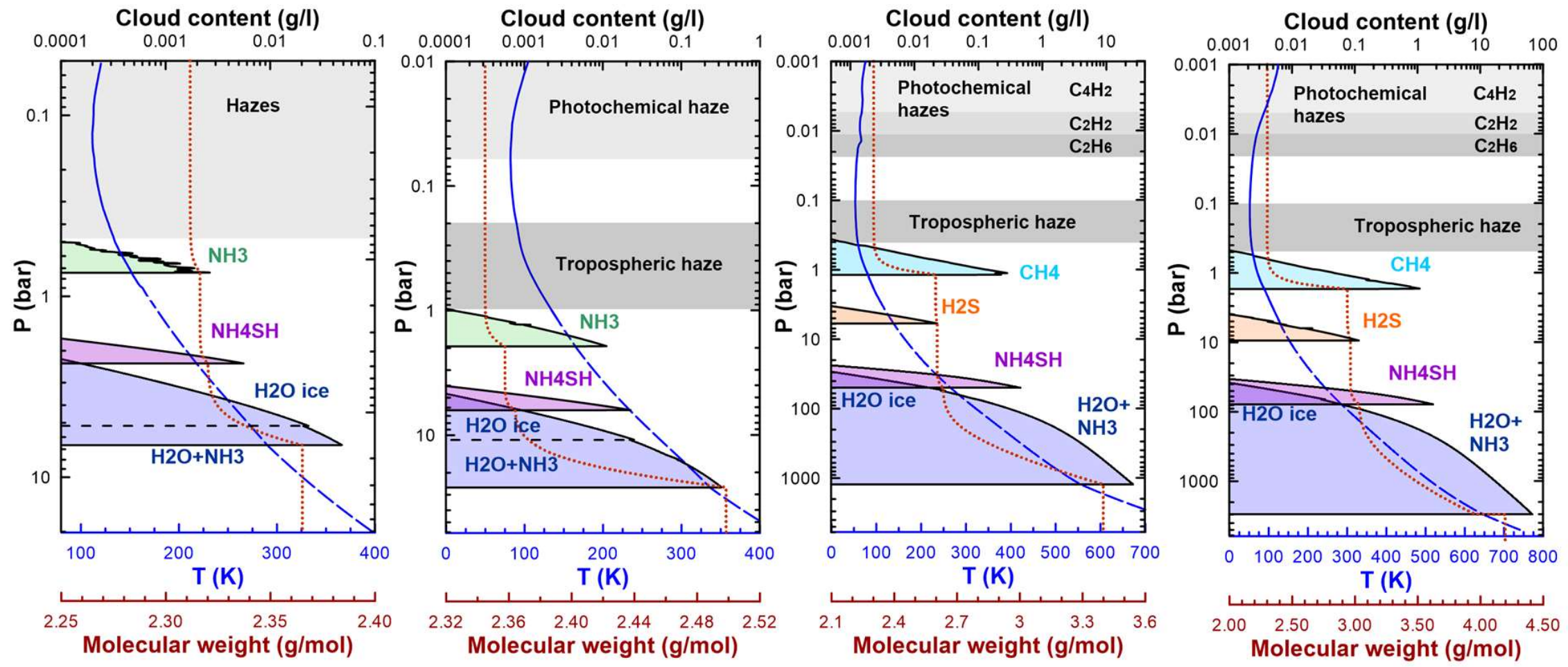

Figure 3.12. Vertical thermal and cloud structure in the Gas and Ice Giants (from left to right, Jupiter, Saturn, Uranus and Neptune), based on moist adiabat extensions (dashed blue lines) of the Voyager thermal profiles (solid blue lines). These profiles assume 2.7 times solar abundance of condensates for Jupiter, compatible with Juno latest measurements (Li et al., 2020), 5.0 times solar abundances for Saturn and 30- and 80-times solar abundances for Uranus and Neptune, respectively, except for ammonia, which is assumed to be 3 and 8 times solar abundance for Uranus and Neptune, respectively, to be consistent with the detection of $H_2S$ in the lower atmosphere and the absence of ammonia clouds. Condensates in the ice giants are very uncertain and ammonia and water could be depleted in a deep

water and ionic/superionic water ocean (Atreya et al., 2020). The upper atmosphere is also home to several photochemical layers. From Mousis et al. (2021).

Integrating the diverse but sparse information we have from multiwavelength observations into a consistent picture describing the longitudinally averaged zonal and meridional circulation at all levels is a challenge in itself. Recent reviews for Jupiter (Ingersoll et al., 2004; West et al., 2004), Saturn (Showman et al., 2018; Fletcher et al., 2018) and the Ice Giants (Hueso & Sánchez-Lavega, 2019; Moses et al., 2020) reveal some of the complexities of studying giant planet atmospheres. Fletcher et al. (2020a, 2020b, 2020c), following Showman & de Pater (2005), presented a recent synthesis of available measurements for the banded structure within the weather layers of giant planets that is most compatible with all available observational constraints. Fig. 3.13 shows their description of Jovian atmospheric circulation in a meridional plane, from the dry convection layer at the bottom of the figure to the tropopause at the top. The three cloud layers, composed from top to bottom of $NH_3$, $NH_4SH$ and $H_2O$, are indicated, as well as their latitude distributions marked by successive clearing zones. Atmospheric circulation is described there as a superposition of "stacked cells", one at shallow pressures that has been studied for decades, and a deeper cell below a "transition altitude" somewhere within the condensate clouds. These cells can be likened to Ferrel-like circulations on Earth, forced by eddy momentum flux convergence, but the sources of the frictional forces 'closing' these circulations near the tropopause and at high pressures deep beneath the weather layer remain a topic of active debate. The representation in Fig. 3.13 is, as explicitly mentioned in Fletcher et al. (2020a), "an attempt to reconcile the observed properties of belts and zones with (i) the meridional overturning inferred from the convergence of eddy angular momentum into the eastward zonal jets at the cloud level on Jupiter and Saturn and the prevalence of moist convective activity in belts; and (ii) the opposing meridional motions inferred from the upper tropospheric temperature structure, which implies decay and dissipation of the zonal jets with altitude above the clouds." This interpretative scheme certainly does not close the issue and cloud microphysics, precipitation and chemical disequilibrium processes could all complicate the picture. Nevertheless, it provides a useful reference for future observers and modellers to validate or disprove some of its aspects and possibly provide a more accurate and consistent description of this complex atmosphere.

While similar descriptions exist for the other gas giant, Saturn, their extrapolation to the Ice Giants atmospheres is even more speculative (see Fletcher et al., 2020a). In particular, Uranus and Neptune both demonstrate fine-scale albedo bands that appear disconnected from the observed winds and temperatures, in addition to significant equator-to-pole gradients in key volatiles (methane and $H_2S$) that are not observed on the gas giants. Furthermore, Uranus and Neptune show stark differences in the strength of vertical mixing and energy balance, possibly related to inhibition of convective processes, which might separate the circulation cells into stacked layers. How and why Ice Giant tropospheric circulation (and associated meteorological features like storms and vortices) differs from the better-studied cases of Jupiter and Saturn remains a topic of active investigation.

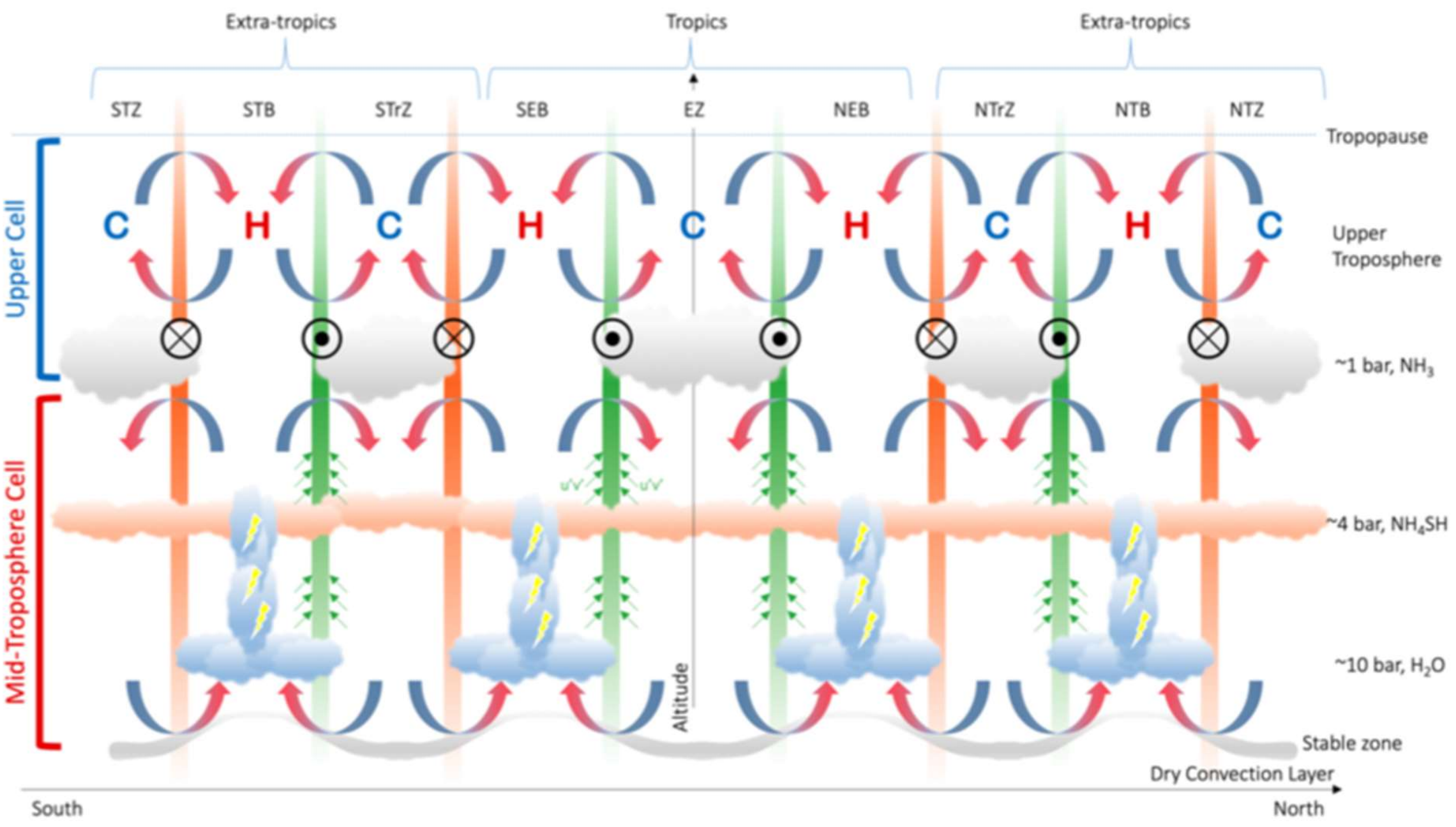

Figure 3.13. A schematic representation of zonally averaged circulation in Jupiter's atmosphere in a meridional plane as a superposition of two "stacked cells", one above and one below a "transition altitude" somewhere within the condensate clouds, based on a synthesis of observational constraints relevant for the different altitude layers. Eastward jets are shown as green bars, with circles with black dots indicating flow 'out of the page.' Westward jets are shown as orange bars, with circles and crosses indicating flow 'into the page.' Small green arrows indicate eddy momentum flux into the eastward jets; plume activity with putative lightning is indicated in the centres of the cyclonic belts; and associated belt-to-zone meridional transport is indicated at the top of this cell. The stable inversion layer beneath the water cloud, separating the moist weather layer from the dry convection lower troposphere, is indicated by the grey shaded line. This schematic expands upon that first shown in Showman & de Pater (2005). From Fletcher et al. (2020a).

Comprehensive knowledge of the common features and diversity of the two kinds of (gas giant and ice giant) planet atmospheres will require combination at each planet of the deep sounding capacities of an IR-submillimetre instrument similar to Juno Microwave Radiometer (MWR) and the high vertical resolution and high-precision determinations of the chemical composition of the gas and condensed phases made possible only by an atmospheric entry probe. In particular, we advocate for:

- Detailed characterisation of atmospheric circulation, meteorology, clouds, chemistry and radiative balance, from the deep atmosphere, through the weather layer, upper troposphere, stratosphere and thermosphere, using multispectral remote sensing;
- In situ measurements via a descent probe to determine bulk atmospheric composition and provide a ground-truth atmospheric profile;
- Gravity science to determine the depth of atmospheric flows and the transitions to deeper interior;
- Comparative planetology between the four giants to understand how different origins, different evolutionary processes and different dynamical/chemical regimes (e.g., size, rotation, enrichment, distance from the Sun) influence the environments that we observe today.

Taken together, these studies will reveal how atmospheres serve to redistribute energy, momentum and material from place to place, as well as connect their ever-changing appearance to (i) the deeper interior described in Section 2.1.1 and the external charged-particle environment of the magnetosphere described in Section 3.5.

## 2.2. Dwarf planets, regular moons, and ocean worlds

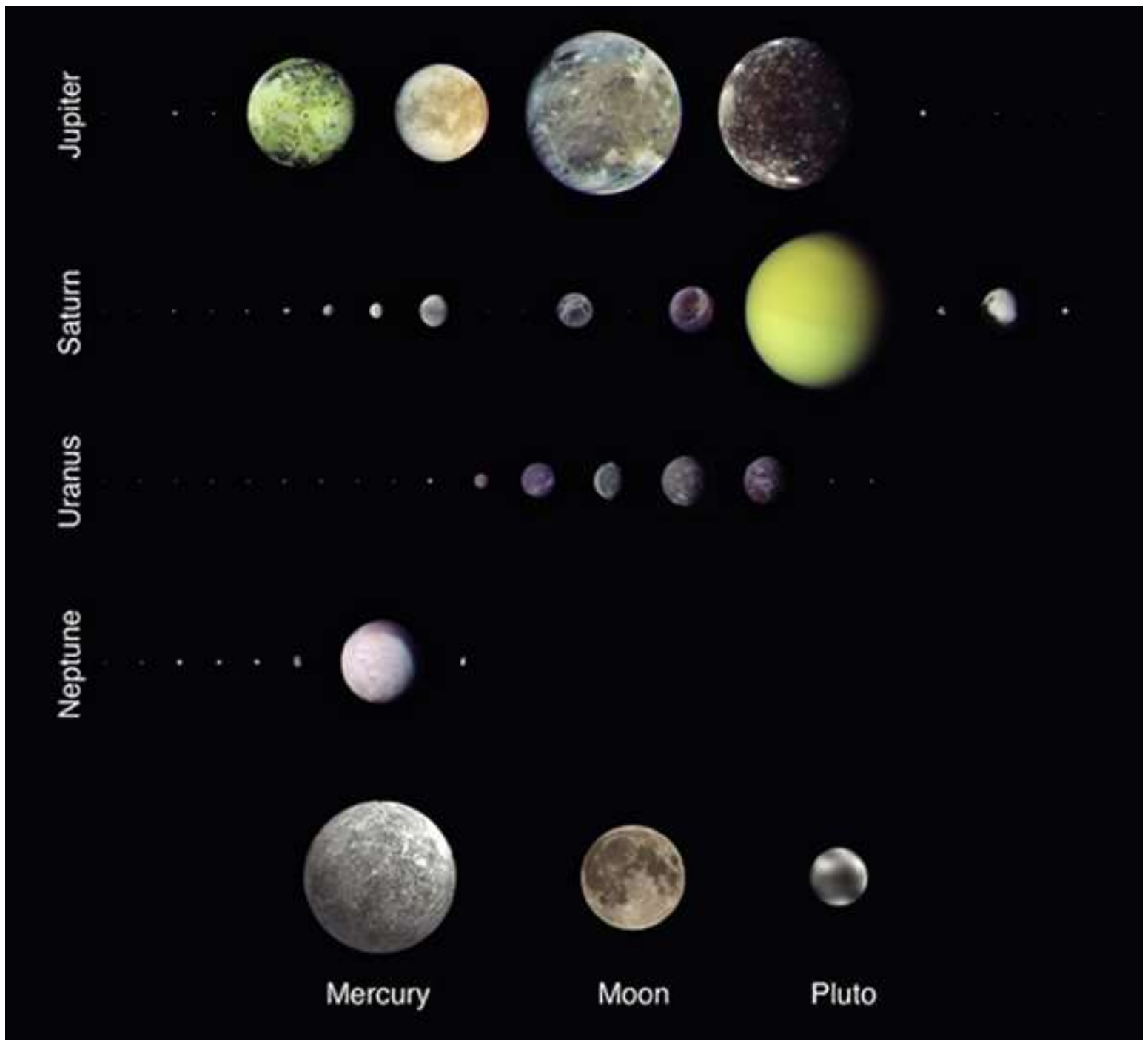

Figure 3.14. An illustration of the diversity of regular moons in the Solar System, compared to a reference dwarf planet, Pluto. The smallest of planets, Mercury, is of a comparable size to the Earth's Moon, to the Galilean moons and to Titan. © NASA.

There is a large diversity of dwarf planets and regular moons (see Fig. 3.14). A number of moons and dwarf planets show evidence for the presence of subsurface oceans. Confirmed ocean worlds include Europa, Ganymede, Callisto, Titan and Enceladus, and bodies such as Pluto, Triton, Dione and Ceres are also considered as candidates (Hendrix et al., 2019). Among them, Enceladus (e.g., Porco et al., 2006) and Europa (e.g., Roth et al., 2014) stand out as objects with direct evidence of interaction between the subsurface, surface and a tenuous, localized and time-variable, atmosphere sourced by plume activity (episodic only at Europa). Io is another example of a (magmatic, in this case) interior ultimately determining the object's atmospheric and surface composition, injecting Sulphur species ($SO_2$, SO, $S_2$) and salts (NaCl, KCl) to form a permanent but time- and space-variable atmosphere that ultimately affects the Io plasma torus and the surface chemistry of both Io and the other Galilean satellites. Cryovolcanism may be powering Triton's plumes, although an internal or external origin remains to be determined.

Given the inherently time-variable character of (sub)surface activity, monitoring observables (atmospheric density and composition, surface composition, surface morphology) on these bodies with long temporal sampling is needed and achievable only from ground-based monitoring.

The atmospheres of Pluto and Triton result more directly from sublimation-condensation exchanges with volatile-rich, spatially heterogeneous surfaces. Such exchanges vary along the orbit with the interplay of seasonal and (in Pluto's case) heliocentric distance effects and are thought to lead to the redistribution of the volatile ices ($N_2$, $CH_4$, CO) with time and to attendant evolution of the atmospheric composition and structure. While the New Horizons encounter has provided an exquisite view of the Pluto system, it was limited to a single point in time in 2015. Temporal monitoring over orbital timescales is needed to address for example the question of the fate of Pluto's atmosphere when it will reach its 49.3 AU perihelion by the year ~ 2110. Stellar occultations (e.g., Sicardy et al., 2016), which do not require large instrumentation, but need multi-telescope campaigns, are ideally suited to monitor the evolving surface pressure. On the other hand, large ground-based facilities that would provide spatially resolved observations such as the Atacama Large Millimetre/submillimetre Array (ALMA), the Very Large Telescope (VLT) array and, in the near future, the James Webb Space Telescope (JWST) and the Extremely Large Telescope (ELT) are needed to follow the detailed aspects of Pluto's complex climatic cycle. The same considerations hold for other KBOs suspected to harbour atmospheres (e.g., Eris, Makemake).

## 2.3. Small Bodies

Small bodies of the Solar System are grouped into a small number of categories of objects orbiting the Sun at a broad variety of distances: by increasing distance to the Sun, the Near-Earth asteroids, the asteroid belt, Jupiter Trojans, a population of objects - the Centaurs - orbiting in the region of the giant planets, Neptune Trojans, Trans-Neptunian Objects (TNOs) and finally the Oort cloud. Comets originate in part from several of these populations.

In complement to the numerous ground-based observations of these populations, space exploration offers the possibility of close-up observations of individual objects and even, for the closest ones, of returning samples from their surface. In fact, detailed properties of their surfaces (e.g., presence of regolith and its properties, boulder distribution, etc.) and interiors can only be accessed through a visit by a spacecraft. In addition, recent sample return missions showed us that inferring the mechanical properties of their surface from images can be misleading and this important knowledge actually requires a direct physical interaction. For instance, while some surface reaction was expected during the OSIRIS-REx spacecraft touchdown on Bennu, the surface behaved like an almost cohesionless one. The same holds true for the impact experiment performed by the Hayabusa-2 spacecraft on Ryugu, whose result can only be understood if cohesion is assumed negligible despite the apparent surface roughness and presence of boulders. Fig. 3.15 shows three of these objects, belonging to three different families of small bodies that have been studied by three different missions: near-Earth asteroid Ryugu visited by JAXA's Hayabusa-2, comet 67/P Churyumov-Gerasimenko visited by ESA's Rosetta mission and 486958 Arrokoth, the target of the second flyby of NASA's New Horizons mission. While only telescope observations can give access to the diversity of objects in each family and across all small bodies, these close-up space observations allow unprecedented investigations of their shapes, multiscale surface properties, masses and densities, in addition to some hints on their chemical composition and pave the way to systematic sample returns from an increasing number of these objects.

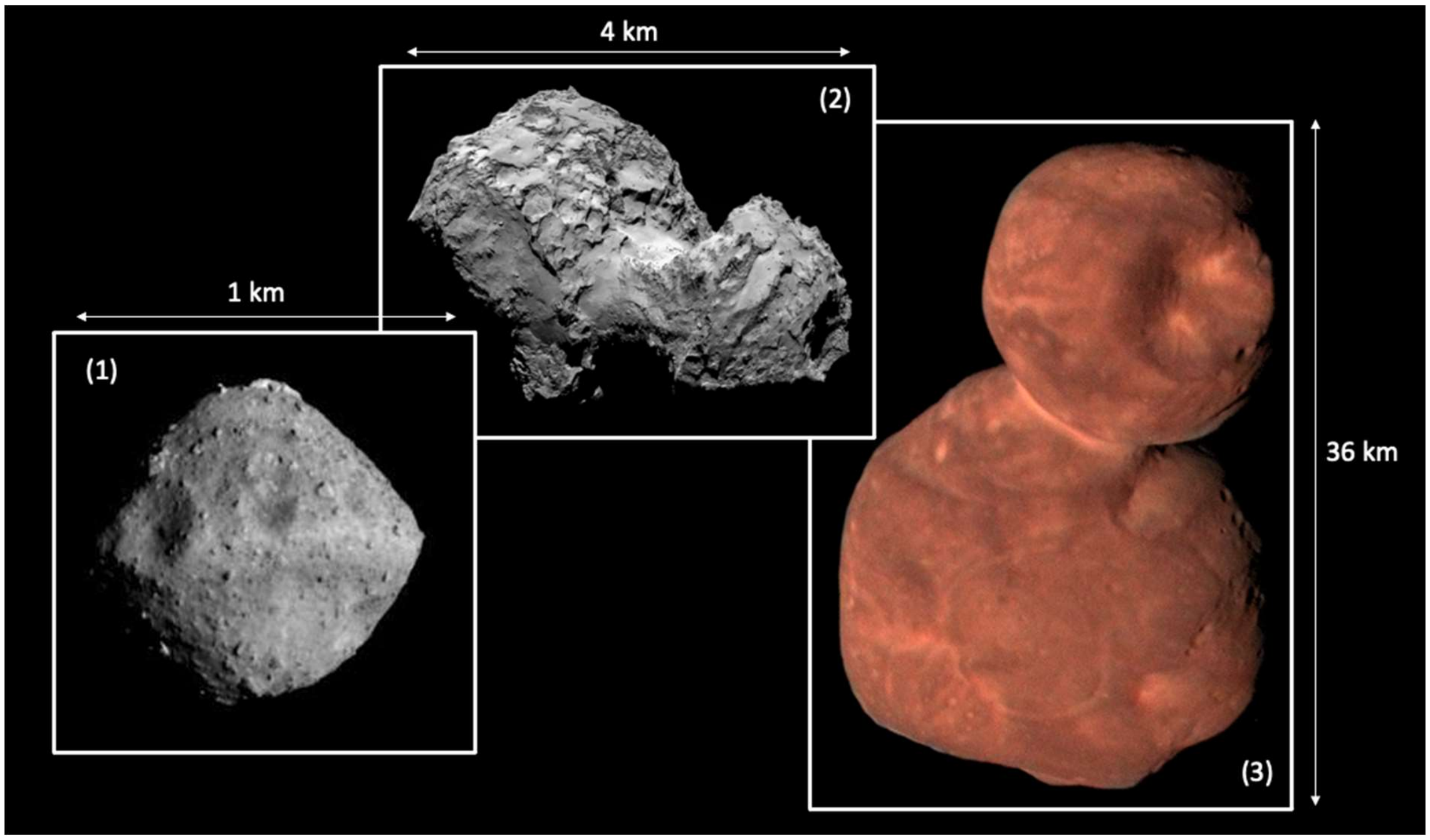

Figure 3.15. Thanks to the complementary Solar System exploration programs of the different agencies, orbital and in-situ investigations or close flybys of objects belonging to the three main categories of Solar System objects have been achieved: (1) a near-Earth asteroid, Ryugu, was explored by JAXA's sample return mission Hayabusa-2; (2) the nucleus of comet 67/P Churyumov-Gerasimenko was orbited by ESA's Rosetta mission; (3) NASA's New Horizons mission, 5 years after its close encounter with the Pluto-Charon system, flew by the Trans-Neptunian Object (TNO) 486958 Arrokoth, nicknamed Ultima Thule during the encounter. Approximate sizes of the different objects are indicated. Credits JAXA, ESA, NASA respectively.

Taken together, observations of the small bodies of our Solar System performed over the last 50 years have allowed us to reach a preliminary understanding of the architecture of our Solar System (see Section 3) and of the distribution of surface compositions across inner Solar System small bodies (≤5A AU; see Vernazza & Beck, 2017, for a review). In particular, they have revealed a number of puzzling features in each dynamical population of small bodies (e.g., the compositional distribution and diversity of the asteroid belt, the inclination distribution of the Jupiter Trojans and the peculiar orbital distribution of TNOs).

It is therefore imperative to collect and interpret a new generation of data for small bodies (in particular with semi major axis ≥5 AU) in order to provide new constraints on Solar System formation models. In particular, whereas preliminary indications exist on the surface composition of main belt asteroids and Jupiter Trojans, little is known regarding their bulk composition, shape, and cratering history. For outer Solar System small bodies (≫5 AU), similar constraints are available for only a handful of bodies. Inevitably, sample return will be the technique to be used for better understanding the small bodies of the Solar System as was done for Hayabusa 1 and 2 missions to Itokawa and Ryugu asteroids by JAXA, and for OSIRIS-REx to Bennu by NASA.

### 2.3.1. Diversity in surface composition

Present day knowledge of the surface compositions of small bodies shows that they are quite diverse, although the true diversity may be even larger as it is biased by the abundant data for the inner Solar System main-belt asteroids (MBAs): the amount of available information regarding the composition of small bodies decreases drastically with heliocentric distance.

Moreover, spectroscopic observations from the Earth provide only limited and sometimes ambiguous information on the actual composition of these objects, even when they are not that far from the Sun (e.g., Near-Earth Asteroids). Many of the irregular satellites of the giant planets were captured from initially heliocentric orbits. Graykowski & Jewitt (2018) have performed an optical colour survey of irregular satellites of the outer planets using three standard colour filters (Blue, Violet and Red) and compared the observed colour ratios B-V and V-R, which are a fair characterisation of more or less blue or red colours of these objects, to other planetary bodies in order to obtain similarities and differences that may reflect upon the origin of the satellites. They found that ultrared matter (with colour index B–R ≥ 1.6) is abundant in the Kuiper Belt and Centaur populations but depleted from the irregular satellites (see Fig. 3.16).

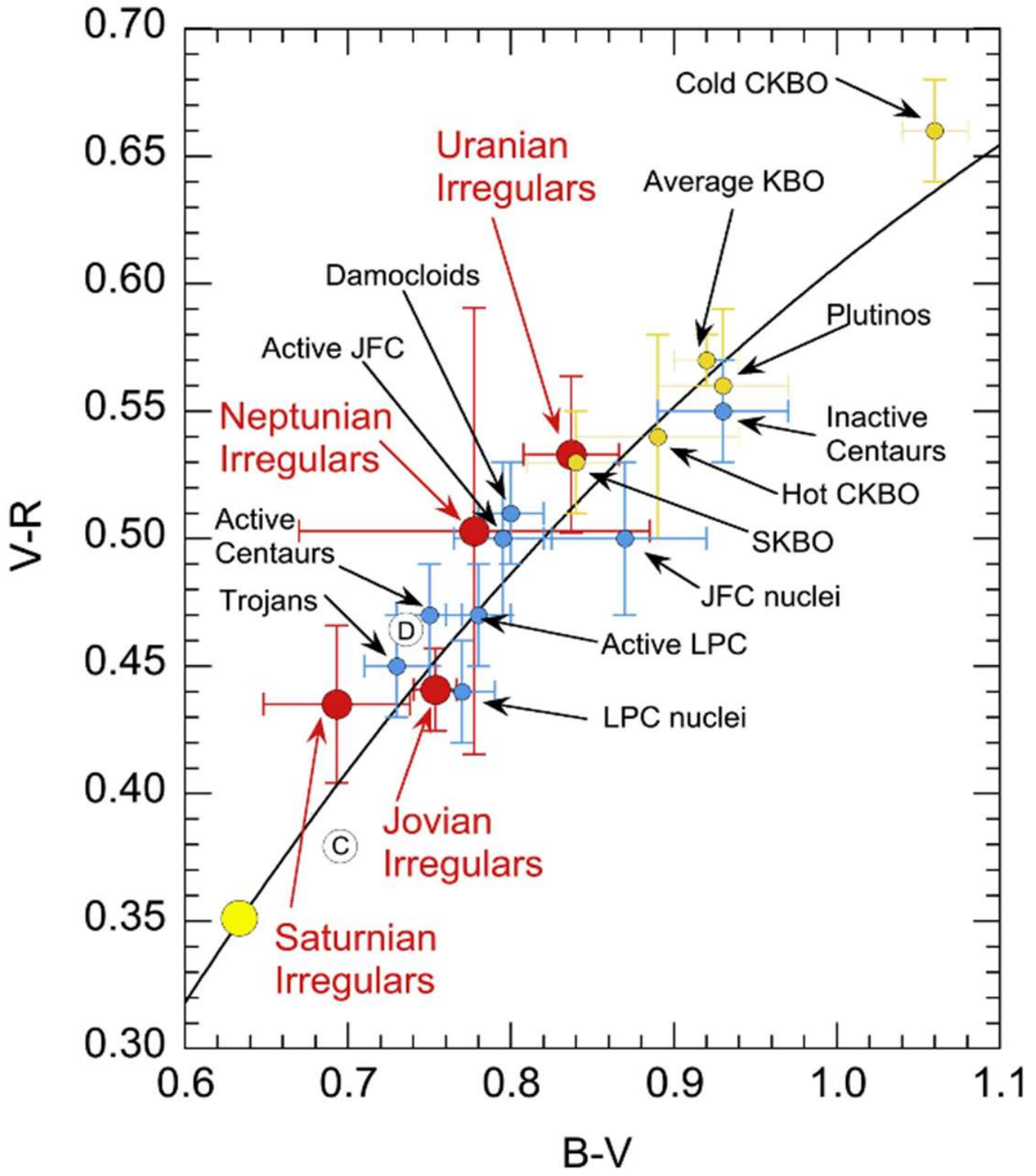

Figure 3.16. Colour vs. colour plot of small-body populations in the Solar System. Yellow data points represent Kuiper Belt objects and blue data points represent comets or comet-like objects. The circles labelled “C” and “D” represent the average colour of the C-class and D-class asteroid populations, respectively. The red data points are the average colours of the irregular satellites for each of the four giant planets. The colour of the Sun is represented by the large yellow circle. From Graykowski & Jewitt (2018).

Available data are limited in wavelength primarily to the near-infrared (λ≤2.5 μm) due to constraints on the brightness of the sources, telescope size, detector performance and transmissivity of Earth’s atmosphere. This wavelength limitation deprives investigators of key spectral features that are highly diagnostic of many ices, minerals and organic chemicals. This explains why basic yet fundamental questions have not been answered so far:

- What is the compositional diversity among outer (>5 AU) Solar System bodies? i.e., what types of volatiles and organics are present among these bodies and what is the composition of their refractory phase (anhydrous versus hydrated; amorphous versus crystalline)?
- What types of volatiles and organics are present on the surfaces of inner Solar System small bodies (main belt asteroids and Jupiter Trojans)?

To make progress in our understanding of the distribution of the surface composition across Solar System small bodies, spectroscopy in spectral regions where the absorption and emission bands diagnostic of ices (useful range: 1-5 μm), silicates (useful ranges: 0.6-3 μm and 6-28 μm) and organic materials (useful range: 3-4 μm and 5-10 μm) can be detected and analysed is required. In addition, constraining the geometric albedo of these bodies would provide useful and complementary information regarding the nature of their surface composition.

Determination of the surface composition of at least one or two representatives of each of the different classes of small bodies is needed before we can have a global view of the chemical diversity of small bodies and of its implications for the building blocks and volatile inventories of the different types of planets.

Measurements of elemental and isotopic compositions of comets will need to be performed in order to reveal whether the elements come from Solar System reservoirs or are interstellar material delivered to the outer Solar System. These analyses will provide constraints on the delivery of cometary volatiles to the terrestrial atmosphere and oceans. Resolving these issues will have important implications for the origin of life on Earth and for the different sources of Solar System material.

By the 2061 horizon, a plan should be implemented for sample returns from representative members of each large class of small bodies, in the asteroid belt, among giant planet Trojans and irregular satellites and ultimately, from TNOs or at least from short-period comets linked to this family.

### 2.3.2. Diversity in bulk composition, shape and cratering history

So far, little is known regarding the density, bulk composition, shape, and cratering history of Solar System small bodies. These physical properties have only been measured for a handful of bodies (mostly via in-situ space missions and via Earth-based observations in the case of the largest main belt asteroids; e.g., Hanus et al., 2019; Fetick et al., 2019; Vernazza et al., 2020; Marsset et al., 2020; Ferrais et al., 2020), which shows as well the diversity in the possibilities. In fact, recent space missions, such as OSIRIS-REx (NASA) and Hayabusa2 (JAXA) showed us that not only is there a wide diversity within the small body population, but also that there is a wide diversity of geological features and rock types on a single body. This geological richness on bodies as small as a few hundred metres in diameter is still not clearly understood and tells us something about their history and the context in which they formed and evolved.

It follows that the following fundamental questions are still unanswered:

- What was the shape of planetesimals at the end of the accretion process? Are the shapes of D>200km small bodies close to equilibrium shapes?
- What are the collisional histories of main belt asteroids, Jupiter Trojans and trans-Neptunian objects and their influence on small bodies' shapes?
- What is the bulk density of large (D>100km) small bodies and is there a relationship with their surface composition? Is there any evidence of differentiation among those bodies?
- Is the density of those bodies that are predicted to be implanted from the outer Solar System (P/D-type asteroids and Jupiter Trojans) compatible with the density of TNOs?

To provide elements of answers to these questions, high angular resolution imaging observations using either giant Earth-based telescopes or close flybys by space probes are needed. Stellar occultations by small bodies are also key sources of information on small body sizes; they should be encouraged, coordinated at the world scale and their data need to be systematically recorded.

## 2.4. Cosmic dust particles

The lowest-mass objects of the Solar System are cosmic dust particles. Cosmic dust can be divided in two categories: interstellar dust (ISD) – 'visible' as the dust blocking the light of the stars of the Milky Way – and interplanetary dust particles (IDPs) – 'visible' as the zodiacal light, which is sunlight scattered by interplanetary dust particles. Interstellar dust particles reside in diffuse or in dense clouds and are the basic building blocks of what eventually become stars, planets and, later on, life. They play a crucial role in astrochemistry for cloud thermodynamics. Characterising ISD is important for astronomical observations as it is the medium we look through to observe the universe and the dust physical properties are needed for interpreting observations of faraway protoplanetary disks, for example. Classical astronomical observations of ISD over long kiloparsec (kpc) scales are used to reveal ISD composition and size distribution using measurements of wavelength-dependent extinction and polarisation of starlight, emission by the dust in the infrared and observations of chemical abundances in the gas (assuming the missing elements, in comparison with the abundances of a reference like the Sun, are locked in the dust). Using this ensemble of observations, models are built for ISD size distribution and composition.

In 1993, a new type of observation became available, providing ground truth information on ISD: for the first time, interstellar dust had been detected in situ in the Solar System with a dust detector onboard the spacecraft Ulysses. This is possible thanks to the relative motion of the Solar System and the Local Interstellar Cloud. Ulysses flew out of the ecliptic plane, and its orbit, being almost perpendicular to the inflow direction of ISD, has facilitated distinguishing interstellar from interplanetary dust. Ulysses has detected between about 500 and 900 particles over 16 years and opened the era of in situ ISD research in the Solar System. More observations followed (Galileo, Helios, Cassini) and in 2016, the Cassini Cosmic Dust Analyser (CDA) measured the composition of 36 ISD particles, whereas the Stardust mission brought back some samples of ISD in its sample return capsule (2006, with analysis in 2014). A comprehensive review on interstellar dust in the Solar System (incl. relevant references) is given in Sterken et al. (2019).

The zodiacal dust has been explored with in situ detectors for more than half a century! In addition to ordinary IDPs, various types of dust 'between' the planets have been examined, such as dust coming from active moons, stream particles, planetary rings, cometary dust and dust clouds around airless bodies.

Enceladus is an example of such an active moon with a subsurface ocean, where water ice particles escape via vents into space. Cassini CDA (a time-of-flight mass spectrometer) measured the composition of the dust particles in Enceladus' plumes and in Saturn's E-ring, illustrating that subsurface ocean compositions can be probed without the need for landing on such a moon. In addition, Io has volcanoes whose tiny particles are accelerated in the Jovian magnetic field and become very fast nanometre-sized 'stream particles'. Their composition was also measured by Cassini CDA. Dust impacts on airless bodies cause ejecta and as such, airless moons are surrounded by an ejecta cloud. Measurements of these ejecta can be used to compositionally map the surfaces of these moons without a landing (Postberg et al., 2011).

Thus, the importance of in situ cosmic dust measurements goes far beyond "just" measuring dust. Furthermore, interplanetary dust, which mostly stems from comet activity and asteroid collisions, provides us with insights into the history of our Solar System; these particles are also a means towards understanding geochemical conditions on subsurfaces of active moons and towards probing the surface composition of airless bodies. In addition, physical processes of the dust and dust as charged probes to investigate the Interplanetary Magnetic Field (IMF) or planetary magnetospheres are subjects of study. A comprehensive review of interplanetary dust is given in Grün et al. (2019) and Koschny et al. (2019).

# 3. Diversity of planetary system architectures within the Solar System (Q2)

## 3.1. Introduction

Observation of extrasolar planetary systems teaches us that the Solar System has a peculiar architecture (Hatzes, 2016; Lammer & Blanc, 2018; see also Chapter 2, Section 3) illustrated by Fig. 3.17 (from Vernazza & Beck, 2017), which shows the radial distances to the Sun and orbital inclinations of the eight planets and of the populations of small bodies (including dwarf planets in this category).

The distribution of planets, as organised by their gravitational interactions, can be described by a few key features (see Winn & Fabrycky, 2015):

- The smaller four ($R_{pl}$ = 0.4 to 1.0 $R_{Earth}$) are all interior to the four larger ones ($R_{pl}$ = 3.9 to 11.2 $R_{Earth}$);
- The small low-mass planets populate the inner part of the system, while giant planets populate the outer Solar System;
- Planetary orbits are nearly circular, with a mean eccentricity of 0.06 and individual eccentricities ranging from 0.0068 to 0.21; they are also nearly-coplanar, with a rms inclination of 1.9° relative to the plane orthogonal to the total angular momentum of the Solar System (the “invariant plane”);
- They carry 99.5% of the angular momentum of the Solar System (compared to 0.5% for the Sun);
- The sizes of neighbouring orbits have ratios in the range of 1.4 to 3.4, but significant gaps exist, first between Mars and Jupiter and farther out beyond Neptune where small bodies dominate.

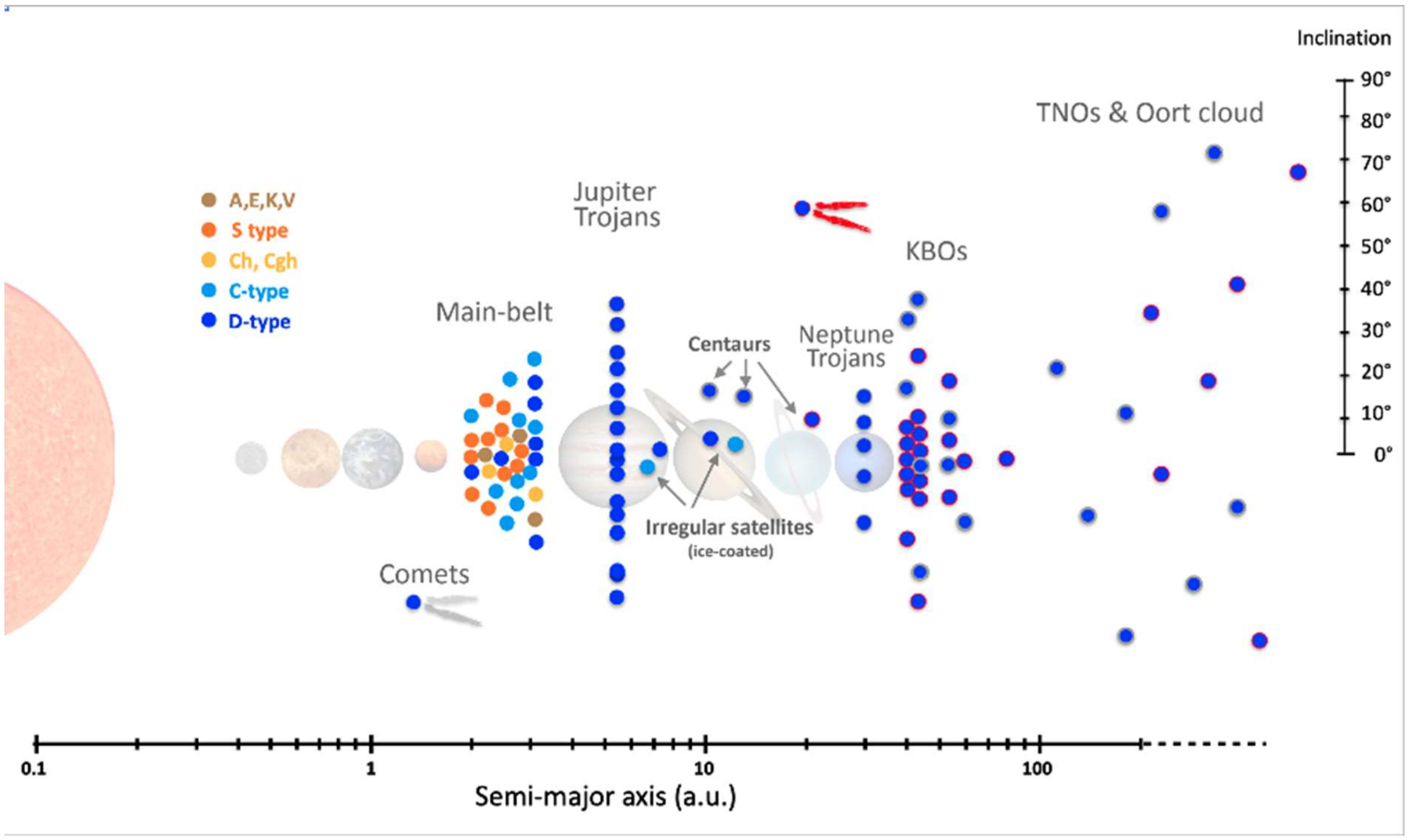

Figure 3.17. A simplified representation of the distribution of Solar System objects in radial distance

(using a logarithmic scale), orbital inclination and object type. The different spectral types of small bodies are indicated in the top left and represented with a colour code. Adapted from Vernazza & Beck (2017).

In contrast to planets, the distribution of these small bodies and of the few embedded dwarf planets can be briefly described as follows:

- They mainly occupy gaps between or beyond planets: the asteroid belt extends mainly between Mars and Jupiter and the TNOs and Oort cloud objects dominate the trans-Neptunian Solar System.
- This population also extends in smaller numbers over the domains occupied by the terrestrial planets (Near-Earth objects) and by giant planets (Jupiter and Neptune Trojans, Centaurs)
- The diversity and radial distribution of the surface composition of these objects, as derived from their spectral colours can be organised in the form of a few spectral (or taxonomic) classes represented by short names and corresponding colour codes in Fig. 3.17. While there is a tendency for “blue” (C-type and D-type) objects, corresponding to the highest levels of ice content, to dominate at and beyond the orbit of Jupiter, and for the other types (likely “more rocky”), to dominate in the inner Solar System, the populations show a high degree of radial mixing, which has to be explained by Solar System formation theories.
- Finally, in contrast to planets that occupy very low inclination orbits, the distribution of small bodies’ inclinations is much broader, from typically 0° to 40° and even isotropic in the case of the Oort cloud.

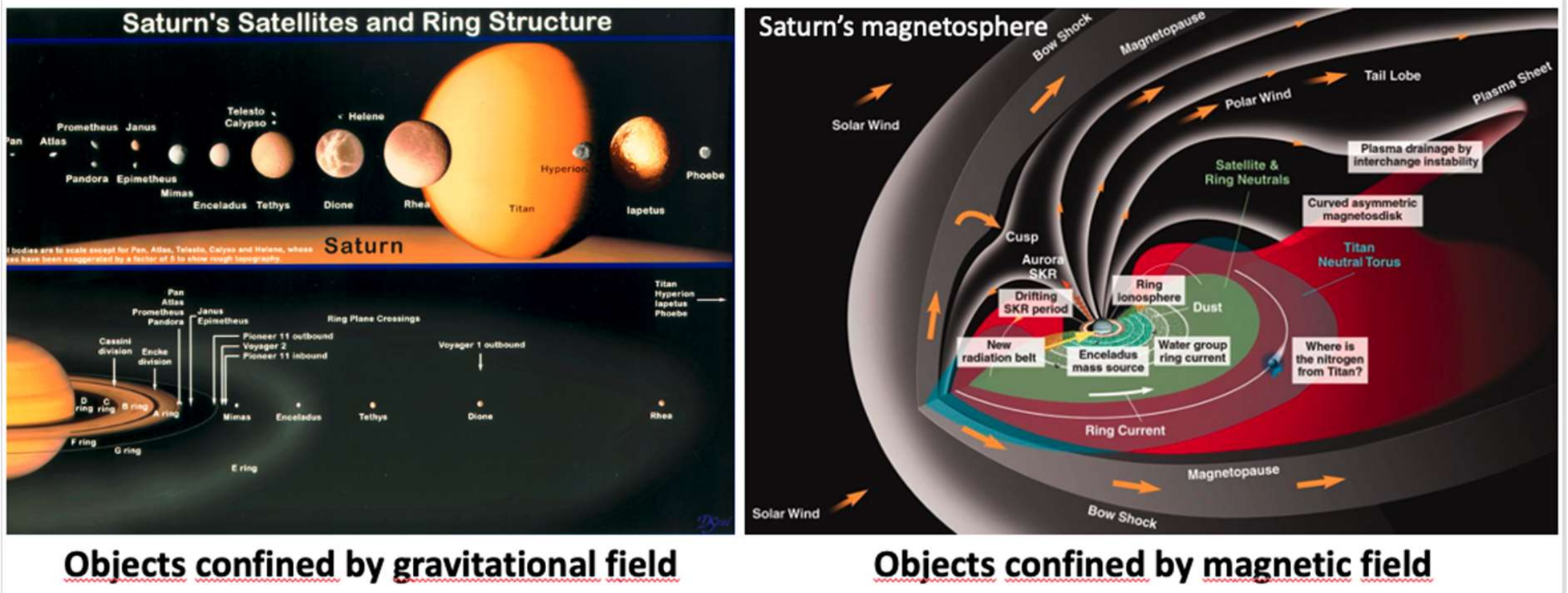

Figure 3.18. (Left-hand panel) Cartoon showing the architecture of the Saturnian system, the most diverse and best explored of secondary systems thanks to the Cassini-Huygens mission: its objects, confined by the gravity field are distributed in radial distance, with the dense rings closest to the planet extending outward up to the Roche limit of ice, then the family of regular moons, including Titan, and irregular moons extending farther outward. The dusty E ring extends along the orbit of Enceladus. (Cartoon credit NASA). (Right-hand panel) The planet’s magnetic field confines plasmas and charged particles inside the Saturnian magnetosphere and induces additional electrodynamic interactions between the planet and most of the objects of the system. Adapted from Blanc et al. (2002, see also Blanc et al., 2021).

The large-scale architecture of the Solar System can also be characterized by its heliosphere, which is the domain of space within which the solar magnetic field controls electrodynamic interactions between its embedded populations of plasmas and charged particles and a large fraction of its objects: planets, small bodies, and dust populations. Its outer boundary, the heliopause, first explored in-situ by the two Voyager probes (Gurnett et al., 2013; Burlaga et al., 2019), has a still poorly known geometry which controls its interactions with the nearby local interstellar medium (as described in Section 5.6). The heliosphere provides a partly opaque barrier to the penetration of galactic cosmic rays into the Solar System and to their interactions with planets and moons, thus playing a role in their habitability. As the heliosphere is the Solar System's astrosphere, its space exploration is of special importance, being the only astrosphere that can be explored in situ.

In addition to this large-scale architecture of the Solar System, each giant planet, as well as Earth and Mars to a lesser extent, is the center of a secondary system of moons, rings, dust, charged particles and plasma populations, illustrated in Fig. 3.18 by the best-studied system, the Saturn system. This section will review the distribution of these different families of objects across the different secondary systems, starting with moon systems, then rings, sometimes forming intricate ring-moon systems in the innermost parts of these systems. It will end with a description of planetary magnetospheres, their interactions with the different objects of the system and their role in the confinement of plasma and charged particle populations (see Section 3.5).

## 3.2. Regular moon systems

A notable characteristic of the giant planets of our Solar System is the presence of systems of regular moons orbiting each of them. These moons are thought to have primarily formed in situ, as will be discussed in Section 4 and as such, their ubiquity has often been considered as indicative of their status as by-products of the formation of their parent planet (e.g., Canup & Ward, 2006). Many of these moons are geologically active today and some of them may harbour subsurface oceans.

The architectures of the regular moon systems of the four giant planets exhibit a large diversity, which likely points to different formation scenarios (see Section 4). The spectacular contrast between the four large Uranian moons that follow near-circular orbits close to the equatorial plane, and the presence of a single large moon at Neptune, Triton, which follows a retrograde orbit with a large inclination to Neptune's equatorial plane, is exceedingly intriguing.

For further complementing the inventory and characterisation of all regular satellites systems, telescope measurements or spacecraft flybys or orbiters will be necessary.

One small object of the Solar System that has regained a lot of interest recently is Triton. Despite being one of the largest and most fascinating natural satellite of our Solar System, Triton is a barely explored body, only visited once by the Voyager 2 mission during a flyby in 1989. Combined with the limited resolution of Earth orbit-based observations due to the large Earth-Neptune distance, there are only few data available today that can be related to the nature and physical properties of this moon (Dale & Cruikshank, 1996; Buratti et al., 2011). Yet, Triton is by many means one of the most puzzling and exciting worlds of our Solar System. Triton has a peculiar orbit, highly inclined and retrograde, which strongly suggests that it was not formed in-situ but captured after the formation of Neptune. Given Triton's composition, it is most likely a Kuiper Belt Object (KBO) that was captured by Neptune. Either, it could have drifted close to Neptune over a billion years ago (Woolfson, 1999; Agnor & Hamilton, 2006), or its capture could have pre-date the capture of Neptune's regular satellites before the dynamical instability of the Nice model (Nogueira et al., 2011). Thus, Triton likely formed orbiting the Sun in a similar region as other icy dwarf planets and primitive bodies in the outer Solar System. This makes Triton unique, in the sense that it would be the only large moon in the Solar System that did not form around its host planet. The physical characteristics of Triton hold the key to understanding the icy dwarf planets of the distant Kuiper Belt, an opportunity that no other planetary system can claim. In fact, it is subject to the tidal, radiolytic and collisional environment like other icy satellites, but with the initial composition of a KBO. Its capture must have left it on an orbit that was much larger and more eccentric than its current one. This early orbit, with a very close pericenter and a very far apocenter, would have raised large tides on Triton causing large tidal heating that may have liquefied its icy internal layers. As a result of this large tidal dissipation, the orbit of Triton would have seen its eccentricity and semi major axis reduced until the orbit reached its current near circular state with a small semi major axis (smaller than the one of the Earth's Moon). Triton's post-capture evolution likely dominated the subsequent evolution of the Neptunian system and subjected the planetary satellite system to extreme processing via catastrophically disruptive collisions, gravitational scattering and tidal heating (Goldreich et al., 1989; McKinnon et al., 1995; Cuk & Gladman, 2005; Nogueira et al., 2011; Crida & Charnoz, 2012).

The present knowledge of Triton is based on very few observations and models and the questions addressed above would need missions to Triton with both orbiting and landing components.

### 3.3. Irregular moon systems.

While regular satellites reside deep inside the potential wells of giant planets, occupying near-circular orbits close to their host planet, a population of small irregular satellites can be found around all four giants at larger distances, up to one half or two thirds of the Hill radius. Fig. 3.19 shows the main characteristics of the distribution of orbital parameters of these populations concerning their orbit inclination and semimajor axis (top part of the figure) as well as their relative mass and semimajor axis (bottom figure). Jewitt and Haghighipour (2007) also mention the prevalence of retrograde orbits, more stable against perturbations by other planets than prograde orbits, the preferred inclinations mainly between 30° and 60° (prograde) and between 130° and 170° (retrograde), and the existence of rather large eccentricities spanning from 0.1 to 0.5. The irregular moons of Neptune stand out with their large inclinations.

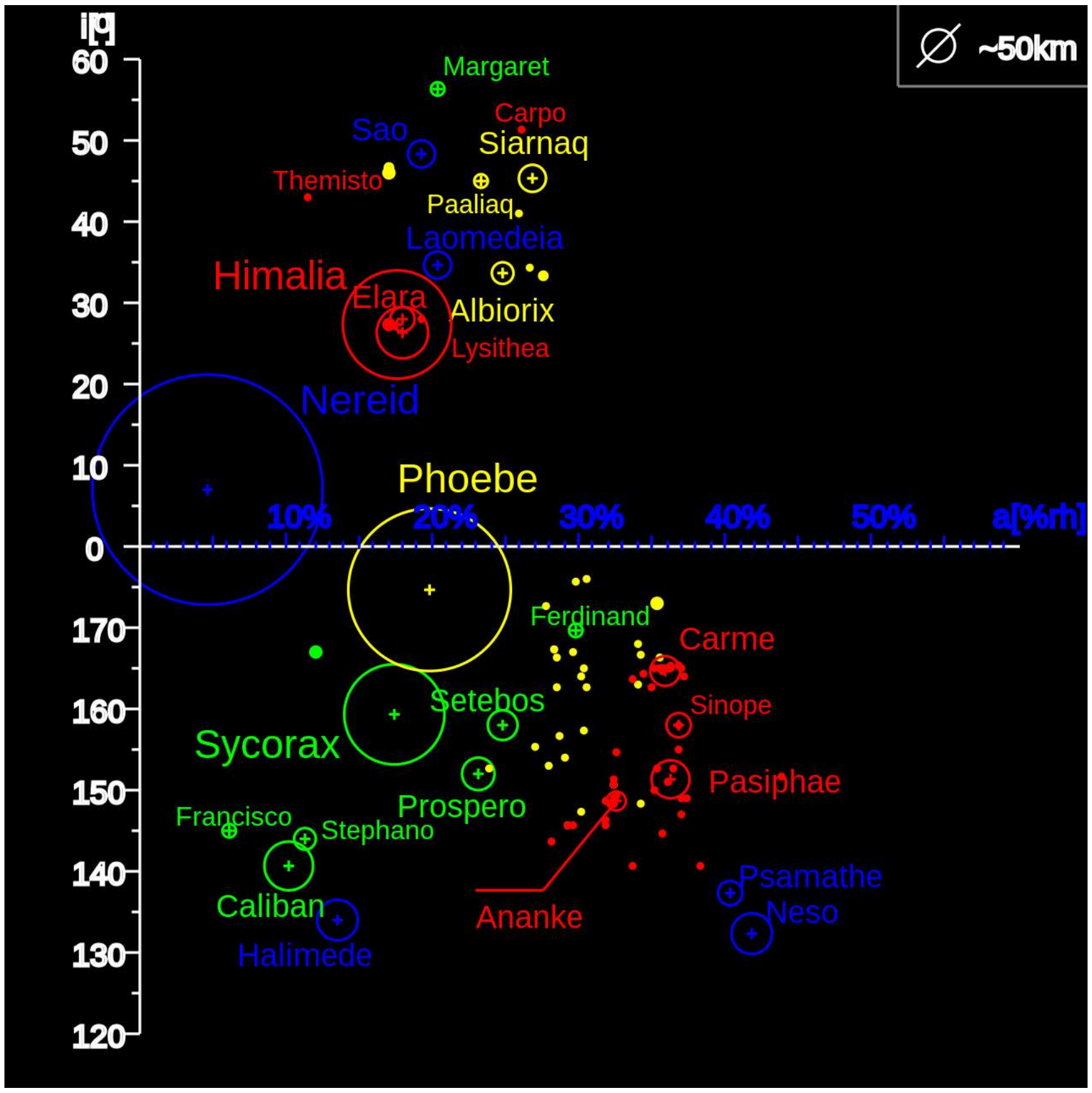

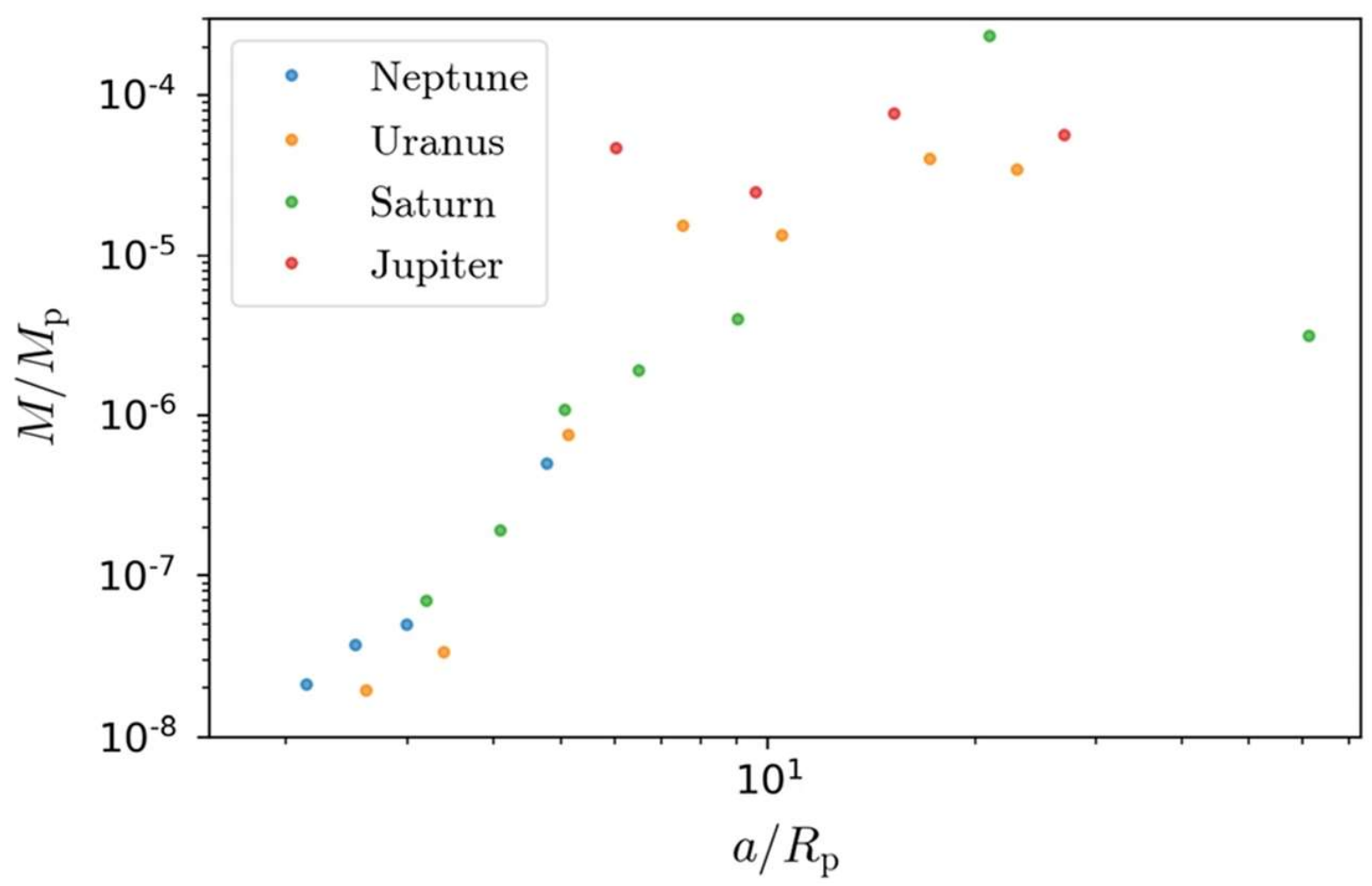

Figure 3.19. Distribution of the main orbital parameters of giant planets satellites: (Top) Irregular satellites of Jupiter (red), Saturn (yellow), Uranus (green) and Neptune (blue) (excluding Triton). The horizontal axis shows their distance from the planet (semimajor axis) expressed as a fraction of the planet's Hill sphere radius. The vertical axis shows their orbital inclination. Points or circles represent their relative sizes. (Bottom) Architecture of the regular moon systems of the four giant planets. Shown are the masses of the moons (normalised to that of their parent planet) against their semimajor axis a (normalised to the radius of their parent planet $R_p$). Only major moons are reported. (Top) From Wikipedia (https://en.wikipedia.org/wiki/Irregular_moon#/media/File:TheIrregulars.svg). (Bottom) From Blanc et al. (2021).

As in the previous paragraph, for further complementing the inventory and characterisation of all irregular moons, telescope measurements or spacecraft flybys or orbiters will be necessary.

## 3.4. Ring-moon systems

All Solar System giant planets possess ring systems (see Esposito, 2006; Tiscareno & Murray, 2018; Showalter, 2020). Despite the fact that the same physical processes are at work that shape all of the planetary ring systems, a surprising variety of ring types and structures exist within each ring system and large differences are observed between the systems (see Fig. 3.20).

Generally, one distinguishes dense rings, formed by macroscopic particles (typically centimetres to metres in size) and dusty rings (with grain sizes in the range from submicron to hundreds of microns), while rings with both dense and dusty components are also known. The best-studied systems to date are the rings of Saturn (that include both dense and dusty components) and Jupiter (that are mostly dusty).

Saturn's main rings (A, B and C) are prototype examples of dense rings (Colwell et al., 2009; Cuzzi et al., 2018), where particle collisions and self-gravity of the ring matter are important physical mechanisms that determine ring evolution and lead to the formation of structures. Collisions of ring particles are dissipative, and the action of cohesion and gravitational forces may lead to the formation of aggregates, which, in turn, can be disrupted by collisions and Keplerian shear (Brilliantov et al., 2015; Salo et al., 2018). Additionally, gravitational interaction with moons that are exterior or embedded in the rings, leads to an exchange of torques between ring matter and these moons, a process that can form resonant structures in the rings (Goldreich & Tremaine, 1978a; 1978b), maintain the sharp edges of the various ring segments (Borderies et al., 1982; Longaretti, 2018; Nicholson et al., 2018b), or confine narrow ringlets (a long-standing hypothesis for Saturn's F ring, e.g., Goldreich & Tremaine, 1979). The ring particles are subject to the external flux of interplanetary projectiles, which erodes them, pollutes their surfaces with exogenic material and leads to a redistribution of angular momentum between neighbouring ring segments (Estrada et al., 2018). Saturn's ring system also comprises dusty components (Horanyi et al., 2009), some of which are embedded in the main rings, while others, like the G and E rings or the Phoebe ring, lie outside. Macroscopic bodies in the system serve as sources and sinks of dust. Important dust generating processes are erosion of surfaces by interplanetary meteoroids, collision, and disruption events (as is the case for Saturn's F ring or for the moonlets of Uranus and Neptune) and volcanic activity of moons (as for Enceladus, forming the E ring). The observed shape of the dust rings is largely determined by the orbital dynamics of individual dust particles, which are perturbed by electromagnetic forces acting on the charged grains, solar radiation forces and the interaction with the magnetospheric plasma (Horanyi, 1996).

Jupiter's rings are dusty in nature, although a population of macroscopic cm to dm size particles has also been inferred (Burns et al., 1999, 2004; Throop et al., 2004; de Pater et al., 2018a). They are directly associated with Jupiter's small, inner moons (Burns et al., 1999; (see Fig. 3.20), which likely serve as sources of the dust, together with a population of smaller, cm sized, unseen bodies. Images of the rings were obtained by the two Voyager flybys, the Galileo orbiter, as well as the Cassini and New Horizons missions. In situ measurements of at least the outer part of the very thin gossamer rings were performed by the Dust Detection Subsystem of the Galileo spacecraft (Krüger et al., 2009), and more recently NASA's Juno star-tracker camera imaged Jupiter's rings from inside their orbits.

Models of the ring systems of Uranus (Nicholson et al., 2018a) and Neptune (de Pater et al., 2018b) are those least constrained by data. The rings of Uranus (Elliot et al., 1977) and the arcs of the Neptune rings (Hubbard, 1986; Hubbard et al., 1986) were discovered in stellar occultations recorded from Earth-based observations. Images of the Uranian and Neptunian rings have been obtained by Voyager 2 (see Fig. 3.20) and Voyager 2 Radio Occultations revealed the structure of the dense Uranian rings (French et al., 1991; Nicholson et al., 2018a). Various observations of the two ring systems were performed with the Hubble Space Telescope and with large Earth-based telescopes with adaptive optics (Nicholson et al., 2018a; de Pater et al., 2018b).

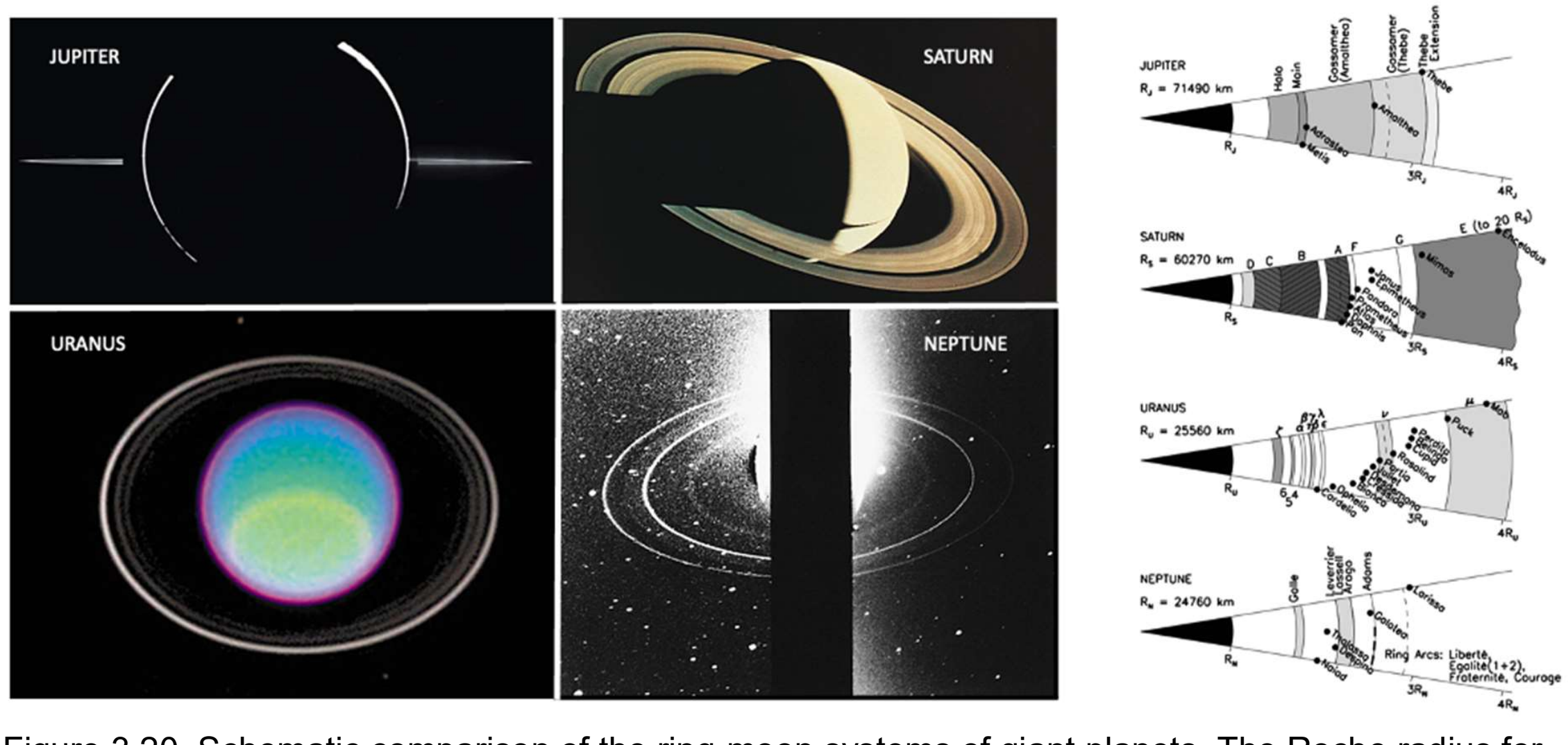

Figure 3.20. Schematic comparison of the ring-moon systems of giant planets. The Roche radius for ice is shown as a dashed line in the right-hand side diagrams. From Blanc et al. (2021).

The dynamics and compositions of rings and their interactions with moons around giant planets need to be further studied, particularly around ice giants where their characterisation is far from complete and where critical information about their variation at timescales of decades can be captured, combining Voyager observations, telescope observations and the future extensive orbiter missions around Uranus and Neptune.

## 3.5. Diversity of planetary magnetospheres

The solar wind propagates from the Sun to the outer boundary of the Heliosphere, encountering all objects within the Solar System. When the solar wind flow encounters planetary magnetic fields, this field digs a long cavity in the solar wind flow, which is filled by the planetary magnetic field and for this reason is called the "magnetosphere". In this case, which applies to Mercury, Earth and all giant planets, one speaks of an "intrinsic magnetosphere". In the case where the solar wind interacts directly with the ionized upper atmosphere of an obstacle (like Venus, Mars and comets), a similar cavity is formed by the effects of the currents induced in the planetary or cometary ionosphere. In the latter case, one speaks of an 'induced magnetosphere', whose size is comparable to the obstacle's size. When the solar wind directly strikes a body that lacks both atmosphere and magnetic field (like some moons and asteroids), just a small void is created in the solar wind on the downstream side (as explicitly mentioned in Baumjohann et al., 2010).

An intrinsic planetary magnetosphere is the planet-dominated magnetic field bubble that surrounds any planet with an intrinsic magnetic field (in the Solar System: Mercury, Earth and the Giant Planets at present day). A planetary magnetosphere interacts with the surrounding supersonic solar wind (at the origin of the bow shock on the dayside external boundary of the magnetosphere) and interplanetary magnetic field (IMF) emanating from the Sun. Every planetary magnetosphere in the Solar System is unique. The Giant Planets have large magnetospheres that can be cross-compared using different criteria. In terms of in-situ exploration with plasma/magnetic field instruments, the most explored magnetosphere is by far that of Earth (see, e.g., Borovsky & Valdivia, 2018, and references therein), then the ones of Saturn (see, e.g., Dougherty et al., 2009, Chapter 9) and Jupiter (see, e.g., Bolton et al., 2015, and references therein) and then, equally poorly visited by only one Voyager II flyby, stand Uranus and Neptune. The intrinsic magnetic field at Mercury was first discovered during Mariner 10's flyby in 1974. The Hermean magnetosphere, the only planetary magnetosphere to be of the same length scale (on dayside) as the planet itself, was explored by the MESSENGER (NASA) mission (see, e.g., Solomon et al., 2018, Chapters 16 and 17) and is now waiting for the two complementary BepiColombo (ESA/JAXA) spacecraft to be better explored with more comprehensive plasma instrumentation and coordinated observations (Milillo et al., 2020). In size and shape, Mercury's magnetosphere is comparable to the one of Ganymede, Jupiter's magnetized moon. On a global scale, the magnetospheres of Giant Planets have different sizes. The typical length scale of a planetary magnetosphere is given by the subsolar distance of the magnetopause, which is the distance at which the magnetic pressure of the planetary magnetic field perfectly balances the dynamic pressure of the solar wind (see Fig. 3.21(a) and Fig. 3.21(b)).

As the four giant planets are much bigger than the Earth, with shorter rotation periods (at first order, a day lasts 10h at Jupiter and Saturn, 18h at Uranus and 16h at Neptune), planetary rotation plays a major part in the global dynamics of those four magnetospheres, much more than solar wind dynamics or reconnection of the planetary magnetic field with the interplanetary magnetic field. The global dynamics of a giant planet magnetosphere also depends on the tilt angle between the magnetic axis and the spin axis (see Fig. 3.21(c)). From this point of view, Jupiter and Saturn belong to the same category, with a small (Jupiter) or a null (Saturn) angle between those two axes. For those two planets, the typical timescales of global magnetospheric physics are determined by the planet rotation period and by the orbital period of satellites that interact with the magnetosphere (Io at Jupiter, Enceladus and Titan at Saturn) through mass loading processes and energetic radiation absorption. These mass loading processes lead to magnetic sub-storms that participate in the evacuation of mass through the magnetotail (the night-side part of the magnetosphere that expands for millions of kilometres in the solar wind direction). The second category in terms of tilt angle between the magnetic and spin axes concerns Uranus and Neptune, whose magnetic fields are tilted from the spin axes by approximately 60 and 47 degrees, respectively. These large tilts produce a dynamic magnetosphere whose entire configuration with respect to the solar wind is changing on a daily basis for both planets, but also on a seasonal scale at Uranus, whose spin axis is almost lying in the ecliptic plane. Those two magnetospheres are still very poorly known.

Internal plasma sources, the role of their moons and rings, and radiation belts are still under debate and cannot be understood with any certainty given the lack of in-situ plasma and magnetic field measurements at Uranus and Neptune.

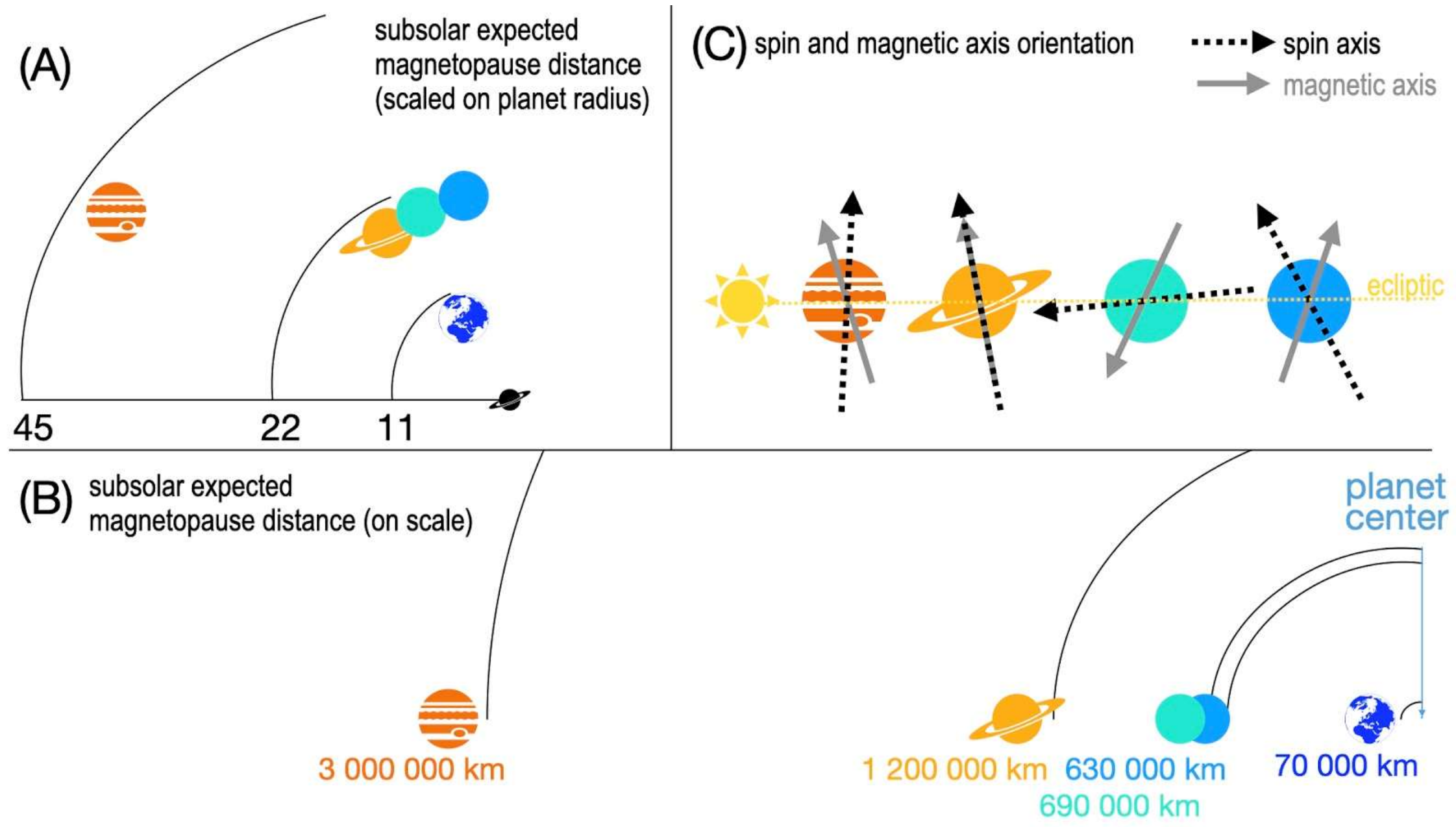

Figure 3.21. Expected subsolar distance of the magnetopause (A) scaled on planetary radius (planet is represented by a black Saturn) and on absolute scale (B) with the planet centre represented by a blue dot, for the Earth and the Giant Planets. Panel (C) shows the respective inclination of spin and magnetic axis with respect to the ecliptic plane (yellow dashed line) for each Giant Planet. Mercury could not be added using the same scaling law since its subsolar magnetopause stands within two Mercury radii, with the radius of Mercury being 2 440 km. Mercury's magnetic and spin axis are perfectly aligned and orthogonal to the ecliptic plane.

# 4. Origin of Planetary Systems

In this section, we review our current understanding of the initial conditions in the ISM (Inter Stellar Medium) before the primordial collapse that led to the formation of the Solar Nebula as well as the initial structure (radial, latitudinal, chemical composition, etc.) of the Solar Nebula. We identify the role of the Sun and possibly nearby stars in Solar Nebula formation and early evolution until dissipation of the gas nebula. We then review our knowledge of the formation of Solar System objects and secondary systems that led to the current architecture of the Solar System (see Fig. 3.1).

## 4.1. Chronology of Solar System formation

The history of the evolution of the Solar System has been described by Coradini et al. (2011) and Turrini et al. (2014) as a succession of three periods, each one dominated by different evolution processes, as illustrated by Fig. 3.22 below. During the first “Solar Nebula” (SN) period, Solar System formation is dominated by gas-solid interactions. Giant planets and at least a first generation of regular satellites form by accretion, respectively inside the SN itself and in the Circum Planetary Disks (CPDs) of the forming giant planets. The second “Primordial Solar System” period opens with dissipation of the gas component of the SN, leaving gravitational dynamics between planets and planetesimals to drive the evolution of the Solar System towards its near-final architecture. Following an episode of large-scale chaotic reconfiguration of the orbits of outer planets and planetesimals, Solar System formation is believed to end with the violent reconfiguration of small bodies that produced the Late Heavy Bombardment (LHB). Note that the LHB, in addition to being the consequence of a prominent spike in the impact rate could also have another origin. It could also correspond to the accretion tail. In this view, the bombardment decayed monotonically since the time of formation of the terrestrial planets (Zellner, 2017; Morbidelli et al., 2016, 2018; Hartmann, 2019).

The end of the LHB opens the period of the “Modern Solar System”, characterised by a slower, less violent secular evolution. It is during this quieter period that most of the cratering record one can read today on planetary and satellite surfaces was written. But this record has been partly erased and overwritten by a variety of processes slowly transforming their surfaces: tectonics, volcanism, weathering, space weathering and even transport of material between different bodies, as discovered for instance by Cassini between Enceladus and the other satellites in the Saturn system, or by New Horizons for the Pluto-Charon system. This late evolution, which partly erased or masked the historical record, introduces an additional difficulty in our search for the origins of solid bodies, moons and rings.

This description encompasses some critical open questions about the different periods and the key mechanisms at play, as reviewed for instance by Morbidelli & Raymond (2016). Nevertheless, altogether, except for the LHB, it offers a sufficiently consensual “story” to serve as a reasonable framework for our analysis of the different periods and events that shaped the Solar System.

In this section, we discuss successively how to determine the initial conditions in the solar nebula, the formation of planetesimals and the assembly of planets and giant planet systems. We then describe how the current distribution and characteristics of small bodies provide key constraints on the migration of planets. At each step, we identify the open questions and the key measurements needed to reconstruct the intricate puzzle of the assembly of the Solar System.

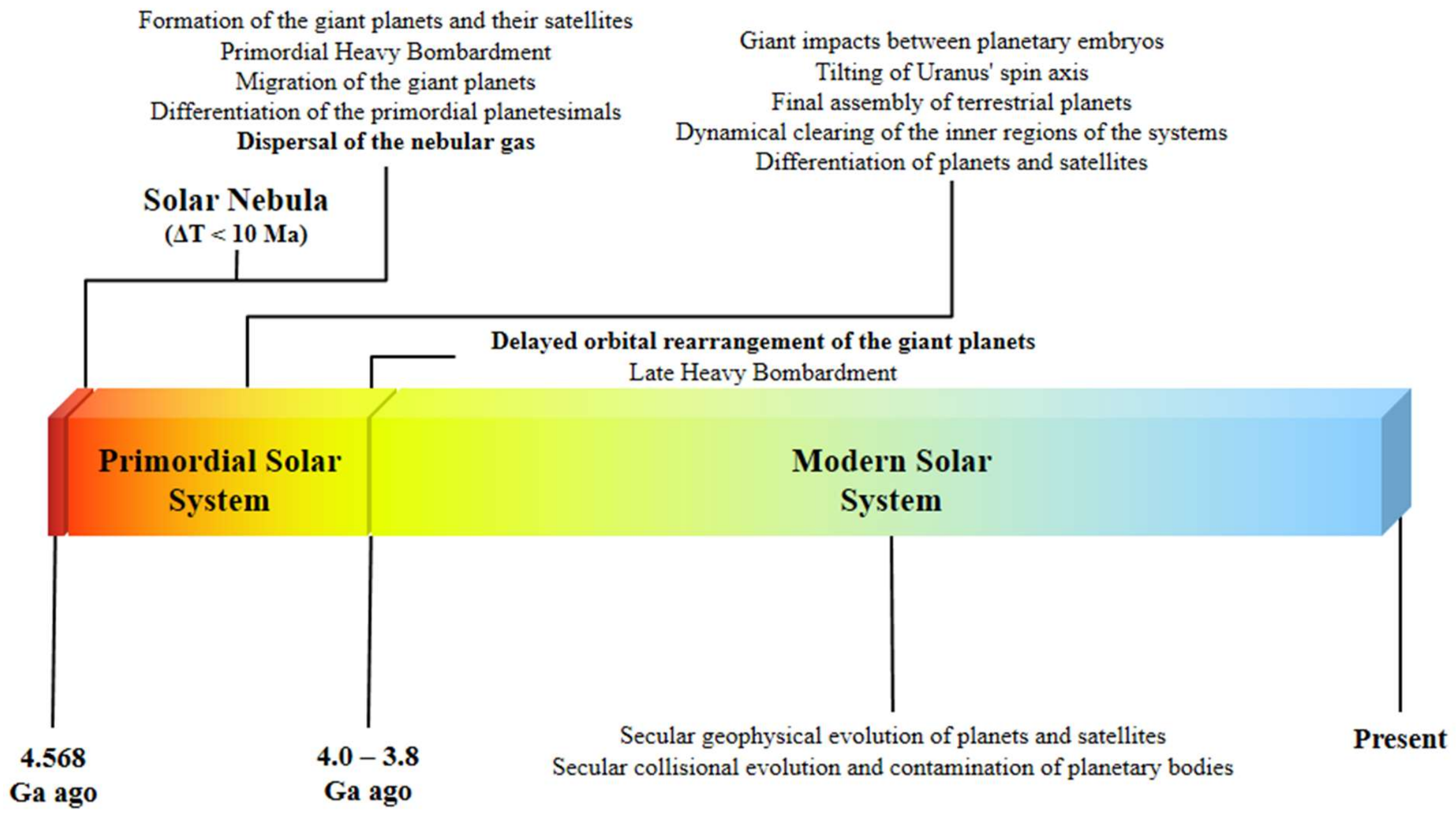

Figure 3.22. Chronology of the formation and evolution of the Solar System, divided into three ages, the “Solar Nebula” age (up to 10 Myr), the Primordial Solar System up to at most 3.8 Gyr ago and the Modern Solar System, according to Coradini et al. (2011). Note that this figure does not account for the reservation about the LHB. The orbital rearrangement of the giant planets might have occurred much earlier than 4 Gy ago. The period 4.0-3.8 Gy ago would mark in this case the gradual end of the late heavy bombardment. Figure adapted from Turrini et al. (2014).

## 4.2. Formation and chemical differentiation of the disk

The solar nebula from which the Solar System formed was a fraction of a molecular cloud made of interstellar $H_2$-rich gas and dust. Dust originated from different stellar sources, whose refractory remnants are found in primitive meteorites, and was also produced as organics from interaction between stellar photons and gaseous species. The solar nebula separated from the main molecular cloud due to gravitational collapse of the cloud or by shockwaves from nearby supernovae. Due to conservation of angular momentum, the solar nebula formed a protoplanetary disk (generally referred to as the disk) rotating around the central protostar. The angular momentum was transported outward, while significant mass was transported inward and accreted onto the protostar or was lost by magnetically driven winds and FUV (Far-Ultraviolet) photoevaporation. Hence, the nascent Solar System was a highly energetic and, at least in some parts, turbulent medium.

### 4.2.1. Reading the messages of primitive meteorites and asteroids.

Our knowledge on the first instants of the Solar System stems mainly from analysis of the so-called primitive meteorites, or chondrites, which originated from asteroids that did not undergo planetary differentiation. As such, they preserve ancient remnants of Solar System material that were present in the protoplanetary disk. Although termed primitive, these objects are by no means a gentle assemblage of dust from the molecular cloud and should be regarded as sedimentary rocks whose components include high-temperature mineral phases (e.g., calcium-aluminium inclusions (CAI), olivine, pyroxene etc. resulting from a condensation sequence of an initially hot gas), as well as low-temperature phases, some of which interacted with water. Hence, the composition of chondrites indicates large scale mixing of matter that evolved in drastically different thermodynamic environments. Further evidence for transport arises from the discovery of high-temperature phases (silicates, metal, CAI) in cometary grains returned by the Stardust mission. Some of the meteoritic groups (e.g., carbonaceous chondrites or CCs) host nucleosynthetic signatures different from others (e.g., noncarbonaceous chondrites or NCs), strongly suggesting that the solar nebula (and by extension the parent molecular cloud) was significantly heterogeneous. CCs are rich in highly volatile elements (O, C, N, noble gases), presumably because they formed in cool environments where water (ice) was present. In contrast, NCs are volatile-poor, suggesting that they formed closer to the nascent Sun where ice and/or organics were not stable. The reason why primordial nucleosynthetic heterogeneities seem to correlate with distance from the central star is presently debated (Nanne et al., 2019).

Overall, the general picture that emerges is a heliocentric zonation of the disk with inner regions being hot and water-poor and outer cool regions having experienced interactions with ice. A zonation in mass and chemistry is now seen in the planets, where the outer planets (Jupiter, Saturn, Uranus, Neptune) are gas- and ice-rich, whereas the inner "rocky" planets (Mercury, Venus, Earth, Mars) are poor in highly-volatile (O, C, N, noble gases) and volatile elements. Yet our understanding of the processes and sources that led to this zonation is incomplete. More specifically, while the zonation due to volatile condensation is driven by the radial temperature gradient, the isotopic zonation is not perfectly understood, e.g., one does not understand yet why enstatite (E-type) chondrites and ordinary chondrites are depleted in refractory elements (e.g., Al) despite the fact that they formed in a warm disk. Of particular importance is the delivery of highly volatile elements to the inner Solar System that permitted the establishment of habitable conditions on Earth and possibly on some of the other inner planets. Furthermore, giant planets that formed before the completion of the inner planets might have also migrated, disturbing drastically the original distribution of disk material including planetesimals. Hence, the present distribution of Solar System objects is poorly representative of the initial structure of the disk.

Ancient material such as chondrites that permits documentation of the first instants of Solar System formation is rare. Available meteorites originated from specific and restricted regions of the disk that are not representative of its overall structure. Furthermore, the composition of primitive meteorites might have been severely altered during their ejection from their parent body, their transport in interplanetary space and their delivery to Earth (e.g., atmospheric entry, terrestrial alteration). To circumvent this problem, several missions have been designed to sample asteroids and return rocks to terrestrial laboratories. So far, JAXA’s Hayabusa mission has sampled an asteroid of the NC S-type (Itokawa), Hayabusa 2 sampled an asteroid of the Cb C-type (Ryugu) and NASA’s OSIRIS-Rex (Origins, Spectral Interpretation, Resource Identification, Security, Regolith Explorer) has sampled a B-type asteroid (Bennu).

Although very valuable, these missions can only address very restricted regions of the disk, probably already sampled by meteorites and it will be necessary to return samples from different classes of asteroids and planetary bodies with increasing heliocentric distances that are unlikely sources of meteorites, such as D-type asteroids, active asteroids (i.e. small Solar System bodies that have asteroid-like orbits but show comet-like visual characteristics), Trojans and comets.

### 4.2.2. Reading the messages of comets

Measurements of elemental and isotopic compositions of comets have revealed large scale heterogeneities of the disk, with strong enrichments in D and $^{15}N$ compared to inner Solar System reservoirs (inner planets, meteorites) suggesting the occurrence of an isotopic gradient with heliocentric distance. These isotopic enrichments could have resulted from photon-gas interactions at the edge of the disk or in the molecular cloud. Alternatively, they could highlight exotic contributions to the outer Solar System. In line with this possibility, the isotopic composition of several key elements analysed on comet 67P/Churyumov-Gerasimenko by the ESA Rosetta spacecraft appears strongly different from that of inner Solar System material. This raises the possibility that cometary ice and organics could partly be interstellar in origin as, on the one hand, sublimation and recondensation of water in the disk may equilibrate the D/H ratio when using the hydrogen in the disk and on the other hand, organic molecules can be derived directly from the interstellar medium or can be synthetized in the disk. These analyses permitted constraints to be set for the first time on the delivery of cometary volatiles to the terrestrial atmosphere and oceans. Resolving these issues will have important implications for the origin of life on Earth and for the different sources of Solar System material.

It will be of utmost importance to document as widely as possible the composition of outer Solar System small bodies and reservoirs. In-situ analysis will allow a first order characterisation and investigation of the diversity of cometary objects. Dedicated sample return missions will give access to invaluable samples and permit the level of analytical precision (on the order of one part per mil to one part per million) necessary to identify stellar sources of Solar System material, to document precisely the outer Solar System environments and to establish a chronology of cometary components.

## 4.3. Formation of planetesimals

Presolar grains and interstellar gas are transformed via thermal processing and, possibly, even sublimation and recondensation. Moreover, submicron and micron sized grains are transformed into ever larger aggregates until they are large enough that their mutual collisional energy overcomes the ability of electrostatic forces to stick grains together. At this stage, mass loss takes place in grain-grain collisions rather than mass gain. The grains of the largest possible size are called “pebbles”

Aerodynamic drag was first considered as a “barrier” in planetary growth. It removes energy from the pebbles so that the pebble spirals into the Sun on very short timescales. Now, it is seen as the potential solution: the back-reaction of the drifting pebbles onto the gas can lead to hydrodynamical instabilities, such as the streaming instability, that form clumps of pebbles. Some of these clumps can be dense enough such that the pebbles are held together by their common gravity (Yang et al., 2017). The contraction of these self-gravitating clumps leads to the formation of planetesimals, with a characteristic size of ~100km (Klahr and Schreiber, 2020). The planetesimals can then grow further by mutual collisions. Once planetesimals reach a mass comparable to a fraction of the lunar mass, they start accreting pebbles efficiently, due to the combination of gravitational attraction and gas-drag (Ormel & Klahr, 2010; Lambrechts & Johansen, 2012). This process, dubbed pebble accretion can lead to the formation of massive planets, such as the cores of the giant planets or the Mars-mass planetary embryos that are supposed to have been the precursors of the terrestrial planets. Both the concepts of streaming instability and pebble accretion still need further testing.

The question of what is the physical nature of the pebbles is still unsolved. Are pebbles chondrules? Are chondrules reprocessed materials? Our meteorite collection is probably not yet representative enough to draw conclusions on that.

The classic view is that some planetesimals formed early and thus they melted and differentiated due to the heat released by radioactive decay of short-lived isotopes. Vesta would be one of these planetesimals. Others would be the parent bodies of iron meteorites and other achondrites (stony meteorite that does not contain chondrules). On the other hand, some planetesimals formed sufficiently late such that the heat released by radioactive decay was not great enough to melt them. These planetesimals would be the parent bodies of chondrites and most of the asteroids and Kuiper belt objects that we observe today.

This classic view is shaken by some evidence from paleomagnetic measurements that even chondritic parent bodies might have undergone internal differentiation. As explained in Elkins-Tanton et al. (2011), “chondritic meteorites are unmelted and variably metamorphosed aggregates of the earliest solids of the Solar System. The variety of metamorphic textures in chondrites motivated the “onion shell” model of the parent body, in which chondrites originated at varying depths within the parent body heated primarily by the short-lived radioisotopes, with the highest metamorphic grade originating nearest the centre. However, a few chondrites possess a unidirectional magnetisation that can be best explained by a core dynamo in their parent body, indicating internal melting and differentiation (a differentiated interior). The parent body could have produced a magnetic field lasting more than 10 Ma. Some chondrites such as the CV chondrites, so named after the Vigarano meteorite, characterized by the presence of lithophile elements and the abundance of presolar isotopes, could originate from the unprocessed crusts of internally differentiated planetesimals. Such planetesimals may exist in the asteroid belt today but are difficult to identify. The asteroid Lutetia, with its chondritic appearance but a high density, may be a candidate” (shortened from Elkins-Tanton et al., 2011).

Thus, one science question to be answered is: Can we find unambiguous evidence of cores in chondritic asteroids?

## 4.4. Formation of planets

As anticipated above, planets are expected to have formed via collisions of planetesimals and pebble accretion. Most of the mass of Jupiter and Saturn is in H and He, captured from the solar nebula around an original 10-20 Earth-mass solid core (Pollack et al., 1996). Neptune and Uranus also contain a few Earth masses of these gases around a ~10 Earth mass core. This unambiguously shows that giant planets formed before the dissipation of gas from the disk. The accretion of giant planets was probably stopped by the dissipation of the gas. Had the nebula lasted longer, giant planets would have been more massive. The satellites of the giant planets are also likely to have formed during the disappearance of gas because this is the only moment when the circum-planetary disk becomes cold enough to allow for the condensation of ice (Lambrechts et al., 2019; Batygin & Morbidelli, 2020). Small moons may not be geologically processed, which makes them very interesting to study as likely witnesses of the early ages of the formation of giant planet satellite systems.

### 4.4.1. Giant planets and their systems

Because they formed in a disk of gas, giant planets should have undergone orbital migration, due to planet-disk interactions. This may have had an effect on the chemistry and composition of the gas accreted. This migration might also have had a significant effect on terrestrial planet formation (see below). However, migration leads to a different orbital configuration than the current one: the migrating planets should have acquired resonant orbits mutually close to each other, which are typically also less eccentric and mutually inclined (Morbidelli et al., 2007). The Nice model (Gomes et al., 2005; Morbidelli et al., 2005; 2007; Tsiganis et al., 2005; Nesvorný & Morbidelli, 2012) explains the current orbit of the giant planets as the result of a phase of dynamical instability that occurred after the disappearance of gas from the protoplanetary disk. This instability does not only reconcile the current orbits of the planets with the original ones achieved via migration, but also explains the formation of the Oort cloud (Brasser & Morbidelli, 2013) and the existence and the orbital structure of populations of small Solar System bodies including the Kuiper belt (Nesvorný et al., 2015; Nesvorný, 2015a,b), the Neptune and Jupiter Trojans (Morbidelli et al., 2005; Nesvorný et al., 2013; Gomes & Nesvorný, 2016), the numerous resonant trans-Neptunian objects dominated by Neptune (Nesvorný & Vokrouhlický, 2016) and the irregular satellites of the giant planets (Nesvorný et al., 2007). See Fig. 3.23. Based on current observational evidence, the Nice scenario best matches many aspects of the observed Solar System (see, e.g., Nesvorný, 2018, for a review).

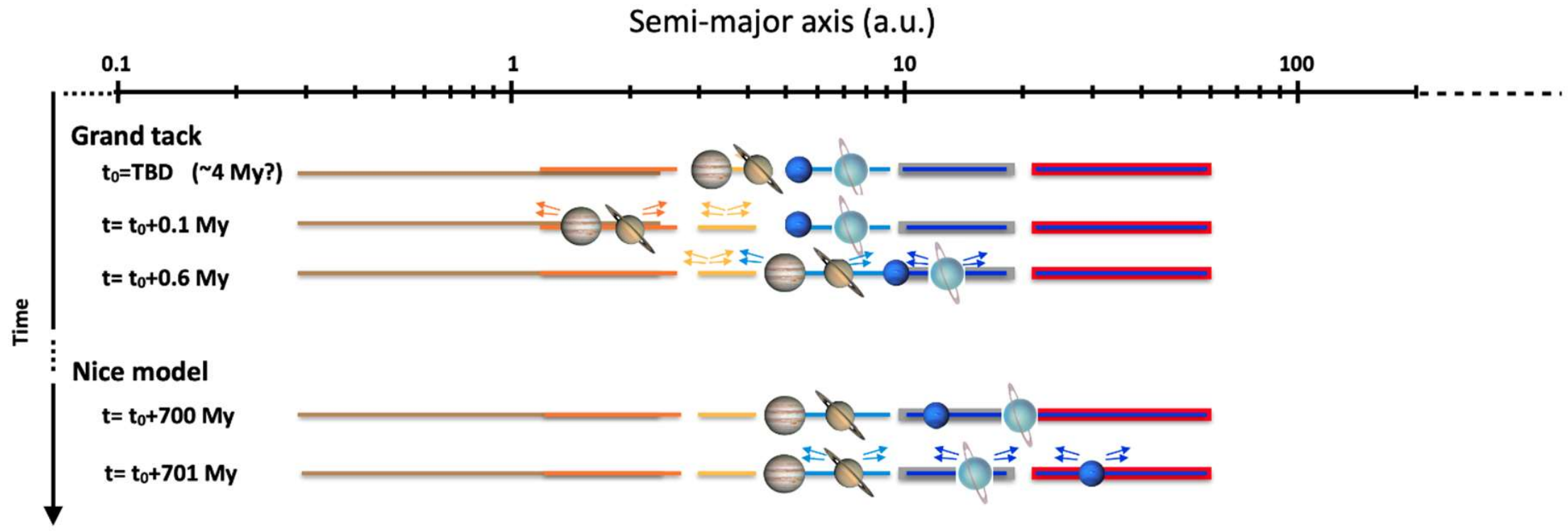

Figure 3.23. A synthetic description of giant planet migrations in the early ages of the Solar System implied by the two main models describing the sculpting of its contemporary architecture: in the Grand Tack model (Walsh et al., 2011) giant planets experience large-amplitude migrations during the solar nebula phase; in the original version of the Nice model (Gomes et al., 2005) a large-scale dynamical instability of the Solar System occurs around 700 My after the dissipation of the solar nebula. The instability is currently believed to have occurred during the first 100My of Solar System history (Nesvorny et al., 2019). From Vernazza & Beck (2017).

### 4.4.2. Terrestrial planets

As explicitly mentioned in Dehant et al. (2019) and written by us based on the references included as well here, “the formation of the terrestrial planets could begin with planetesimals distributed throughout the inner Solar System and the asteroid belt (Rubie et al., 2015). However, in this case it would end with 0.5–1.0 Earth-mass planets near 2 AU, much larger than the actual mass of Mars, or 0.1 Earth-mass (Raymond et al., 2009). Obtaining the order of magnitude mass-contrast between the Earth and Mars requires that the planetesimals were concentrated within 1 AU from the Sun (Hansen, 2009). This planetesimal concentration can be obtained in two ways. The ‘grand tack hypothesis’ (Walsh et al., 2011; Pierens et al., 2014) proposes that, after its formation at ~3.5 AU, Jupiter migrated inward to 1.5 AU, before reversing course due to capturing Saturn in an orbital resonance (Masset & Snellgrove, 2001; Morbidelli & Crida, 2007), eventually halting near its current orbit at 5.2 AU. Due to Jupiter’s migration, the inner planetesimal disk was truncated at 1.0 AU, leaving behind only a mass-depleted and dynamically excited asteroid belt beyond this radius. The second possibility is that the streaming instability was effective only inwards of 1 AU (Drazkowska et al., 2016). Beyond this limit, asteroids could form only later, at the time of the photo-evaporation of the disk (Carrera et al., 2017), producing a low-mass population of objects.

Our planet formed slowly over tens of millions of years, as indicated by radioactive chronometers (Kleine et al., 2009) and explained by models of collisional accretion of a disk of planetesimals and planetary embryos. The Earth formed mostly after the disappearance of the protoplanetary disk of gas, via a sequence of giant impacts. Precursors of the Earth, planetary embryos that formed within the disk lifetime, had a mass presumably smaller than that of Mars. Thus, they did not migrate significantly while in the protoplanetary disk.

It is clear from the above that further examination is needed of different kinds of asteroids, with in-situ missions as well as sample return missions. Additionally, missions to the planets (ice giants, gas giants or terrestrial planets), either in-situ or remote sensing, are necessary to test these evolution models. In situ measurements of elemental and isotopic compositions at Saturn, Uranus, and Neptune would be especially valuable (Mousis et al., 2018; 2021).

## 4.5. Characteristics and distribution of small bodies and captured moons

The observed characteristics and distribution of small bodies throughout the Solar System (described in Section 2) have been exploited to constrain the formation and evolution of the Solar System (e.g., Nice and Grand Tack models; Tsiganis et al., 2005; Gomes et al., 2005; Morbidelli et al., 2005; Levison et al., 2009; Walsh et al., 2011). Based on these models, the idea of a static Solar System history has dramatically shifted to one of dynamic change and mixing (Johnson et al., 2008; DeMeo & Carry, 2013; 2014). Indeed, some predictions from these models have already been confirmed based on other metrics (e.g., the similarity in size distributions between the Jupiter Trojans and TNOs; Fraser et al., 2014).

To date, however, the critical test – that of compositional similarity between now dynamically isolated populations – has remained elusive. While it is clear that important migration episodes did occur during the history of the Solar System that affected the vast majority of the small bodies and the giant planets themselves, a clear understanding of their formation and subsequent evolution, including the timing and nature of the migration episode(s), is currently missing.

# 5. How does the Solar System work?

We here consider the Solar System as a laboratory to study how planetary systems work. First, we shall provide information on exploration of terrestrial planets and the Moon, in relation to phenomena of high relevance for habitability and evolution of these objects, a question that is still of highest importance. Second, we shall explore some of the dynamical processes at work for the other terrestrial objects of the Solar System, which are related to transport processes of mass and energy. We will further look at dynamics involved in the interiors of giant planets and atmospheres across the Solar System. Lastly, we conclude with small body hazards and space awareness, an important part of the dynamics of the Solar System that can jeopardize our own existence.

## 5.1. Exploration of terrestrial planets and the Moon

Terrestrial planets, including Mercury, Venus, the Earth-Moon system, and Mars, constitute the innermost planetary bodies. They developed early in the formation of the Solar System within the ice line, so that they are composed predominantly of silicate minerals, with metallic cores due to differentiation that occurred early in the history of each body. They also cumulatively represent the most studied bodies in the Solar System, having each been visited by several robotic spacecraft and, in the case of the Moon, human explorers. While Earth is the only terrestrial planet that currently has mobile lid tectonics, localized tectonovolcanic activity has occurred on all these bodies (except for the Section 2 Moon) within the past several hundred million years. Despite this, we have only a rudimentary understanding of the dynamical processes in the interior of any of the terrestrial planets, other than Earth.

Better understanding of the interior processes in the terrestrial planets in the Solar System in the coming decades is needed in the following areas:

- the chemical and physical structures of the terrestrial planets;
- geodynamical processes in terrestrial planet interiors, including mechanisms of thermal and chemical transport;
- differentiation processes producing chemically layered planetary bodies;

- establishment and sustainment of magnetic fields and how they are coupled to the solar wind;
- processes by which heat is lost or retained in the interior of planetary bodies;
- interaction and evolution of the myriad of chemical and physical processes in planetary interiors and surfaces, including surface-atmosphere interactions.

In terms of our exploration of the inner planets, we have flown by (all), orbited (all), landed on (Venus, Moon, Mars), roved on (Moon, Mars), returned samples from (Moon) and sent human explorers to (Moon). In the process of these activities, we have built a vast collection of images and data that provide great insight into surface chemistry, geomorphology and processes, as well as some indication of interior structure and processes. Of all the inner Solar System bodies, we know the interior of the Moon the best (other than Earth) thanks to the emplacement of geophysical instrumentation during the Apollo missions and detailed gravitational measurements by GRAIL (Gravity Recovery and Interior Laboratory). The recent arrival of the InSight mission at Mars (Interior Exploration using Seismic Investigations, Geodesy and Heat Transport), bearing a French seismometer, a German thermal probe, and US radioscience antennas to measure the rotation, is greatly expanding our understanding of the interior structure and dynamics of that planet, which is key to understanding the history of volcanism and, hence, habitability of that planet. Mercury has been visited recently by MESSENGER (NASA) and will soon be orbited by BepiColombo (ESA/JAXA). Included in these missions were instruments designed to study Mercury's interior structure and composition, as well as its magnetic field. Shrouded in dense clouds, Earth's twin sister Venus remains enigmatic, largely due to its opaque atmosphere and extreme surface conditions, which, to date, have only allowed landers to survive for a maximum of several hours. While we have some idea about its internal structure and indications that there may be relatively recent resurfacing and volcanism, its tectonic and geodynamic history remains unclear. The recent selection of three missions to Venus by NASA (DAVINCI+ (Deep Atmosphere Venus Investigation of Noble gases, Chemistry, and Imaging) and VERITAS (Venus Emissivity, Radio Science, InSAR, Topography, and Spectroscopy)) and ESA (EnVision) will unveil many of the planet's mysteries.

While much information can be collected remotely using telescopes or from orbit, it is mostly limited to bulk planetary composition and structure. For example, it is possible to measure moments of inertia for planetary bodies to determine interior bulk density distributions, allowing characterization of planetary core size and state. In addition to seismology, other critical instrumentation includes radio-science. Radio-science allows measurement of rotation and orientation of terrestrial bodies, which are both sensitive to deep interior properties, such as the presence or not of a liquid or partially-liquid core.

Nonetheless, detailed understanding of internal thermal and chemical structure and dynamical behaviour really requires surface or shallow emplacement of a variety of geophysical and geothermal tools. Of particular importance, the ability to emplace seismic instrumentation on multiple sites on a planetary surface can provide critical information on planetary structure and geodynamical state. The seismometers needed for planetary characterization have very specific and challenging design criteria. Since Earth is the only terrestrial planetary body with active lid tectonics, seismometers for other planetary bodies must generally be very sensitive and able to withstand and compensate for local conditions. For the Moon, the large thermal day-night variance and the need for surviving the 14-day cold night present a challenge. For Mars, protection from interference due to winds was identified as a potential challenge. For Venus, the high temperatures and pressures at the surface are not suitable for the long-term presence of traditional electronics and communications infrastructure. Mercury also presents challenges for maintaining power in a harsh thermal environment.

Radio frequency probing of planetary interiors, especially in the search of water ice deposits, may enable sustained human presence, and may help us understand the interior structure and chemistry of the terrestrial planets. Such capability has been demonstrated on the Moon and Mars but requires further development if needed to characterize localized ice deposits. As demonstrated by the difficulties in determining thermal fluxes on the Moon using instrumentation emplaced during Apollo and recent challenges for the thermal profiling mole on InSight, measurements of interior heat fluxes require sophisticated instrumentation that is still under development. Additional information can be obtained by electromagnetic sounding of planetary interiors. These and other instrumentation will see significant technical development with planet-specific designs in the coming decades. Techniques such as neutrino tomography are only in their infancy and may offer major new insight in future missions.

In general, major advances in understanding interior structure and processes in terrestrial planets require landed missions and, potentially, surface mobility to obtain 2-D and 3-D information.

## 5.2. Interior Processes in Rocky Planetary Bodies

The evolution of rocky planets, large regular moons and dwarf planets is physically determined by transport processes of mass and energy in their interiors (e.g., Breuer & Moore, 2015). The rates of transport depend crucially on the thermodynamic state variables of pressure and temperature through material transport properties such as viscosity, thermal conductivity and thermal expansion. On a planetary scale, the main interior reservoirs of the bodies of the inner Solar System (Mercury, Venus, Earth, Mars and the Moon) are the silicate crust and mantle and the iron-rich core. In addition, the core and mantle of some of the moons of the outer planets (e.g., Europa, Ganymede and Enceladus) may be surrounded by sizable reservoirs of water. For most planetary bodies, chemical differentiation has led to layering with material density increasing with depth, as proven by measurements of the moment of inertia indicative of mass concentration toward the deep interior (e.g., Sohl & Schubert, 2015). Two notable exceptions are the Jovian satellite Callisto and the Saturnian satellite Titan, which may only be partially differentiated (see Soderlund et al. (2020) for a review).

Whether the planets accreted to form at least partially layered bodies with, for instance, water added late during accretion (e.g., Albarede, 2009), or whether the planets accreted homogeneously (e.g., Halliday, 2013), is a matter of debate (see e.g., the review by Morbidelli et al., 2012). In the latter case, formation of the core required efficient separation of metals from silicates and large-scale melting of the proto-planet, most likely including an early magma ocean. Dissipation of gravitational potential energy into heat and large impacts during the early phase after the formation of the planet likely provided sufficient energy to cause large-scale mantle melting (e.g., Nakajima & Stevenson, 2015) and hence the formation of deep magma oceans. Upon magma crystallisation, certain materials – in particular volatiles, radioactive heat-producing elements and iron oxides – tend to be partitioned preferentially into the remaining liquid phase (e.g., Elkins-Tanton, 2012). As a consequence, the process of magma ocean solidification can lead to the formation of a so-called primary crust, compositionally distinct from the underlying mantle. With the exception of the Moon, whose anorthositic crust has long been recognized as primary, i.e. directly resulting from magma ocean solidification (e.g., Warren, 1985), the crusts of the terrestrial planets are thought to be mostly secondary, i.e. the product of differentiation by partial melting of the mantle (and recycled crust) that may be ongoing for most if not all of the evolution of the planet (Taylor & McLennan, 2009).

Although it is difficult to generalize on the basis of only a handful or so major-sized terrestrial planets in the Solar System and a few tens of silicate-rich moons, following the accretion process and the solidification of putative magma oceans, the long-term evolution of a terrestrial body is largely determined by sub-solidus creep and localized partial melting in its mantle (e.g., Tosi et al., 2014). As long as at least the outer part of the core is liquid, the core can mostly be regarded as mechanically decoupled from the mantle. This is so because of the immense difference in the (effective) viscosity of the solid mantle and liquid core (roughly a difference of 20 orders of magnitude). Although both reservoirs are thermally coupled, the core, because of its mass, plays only a minor role in the global energy balance of the planet. The simple fact that the core is overlain by the mantle causes its rate of cooling to be controlled by that of the mantle. Because a planetary-wide magnetic field is generally held to be generated by dynamo action in the liquid core or potentially a magma ocean at the base of the mantle (Ziegler and Stegman, 2013; Scheinberg et al., 2018), and because a dynamo depends in many ways on the cooling rate of the core, even the generation of the magnetic field is governed by processes in the mantle (e.g., Stevenson et al., 1983).

The importance of the mantle for the evolution of the planet as a whole is underlined when tectonic and volcanic processes are considered. Tectonism and volcanism shape the surface of the planet. Volcanism transports volatiles from the mantle to the atmosphere and oceans (or cryosphere) and hence affects their evolution. On the one hand, in the unique case of Earth's plate tectonics, transport between the interior and atmosphere and oceans occurs in both directions, with subduction transporting volatile-rich crust into the interior and volcanism generating new crust and transporting volatiles to the atmosphere and oceans. What is more, since tectonics and volcanism have a strong influence on the long-term evolution of the planetary climate, which in turn can feed back on interior processes (Lenardic et al., 2016), they may well also affect the biosphere (Tosi et al., 2017; Höning et al., 2019). The way this occurs is complicated but depends largely on exchange processes between the deep interior and the near-surface reservoirs of crust and atmosphere (e.g., Southam et al., 2015). On the other hand, terrestrial bodies other than Earth are characterized by an immobile plate that largely prevents transport from the surface to the interior, leaving volcanism or plume tectonics as the only ways connecting the deep interior with the surface.

While over the past few decades we have been developing some general understanding of the planetary thermo-chemical engine, much fundamental knowledge is still missing. For instance, for most planets, our knowledge is restricted to the gravity and magnetic fields that can be measured by orbiters. In addition, spectroscopy and imaging provide important constraints on the composition and, to some extent, also indirect constraints on the evolution of the planets, e.g., via the inference of the pressure and temperature conditions at which surface lavas of a certain age were generated in the deep mantle (e.g., Filiberto & Dasgupta, 2015; Namur et al., 2016). This is particularly true for Mars due to the wealth of spacecraft missions that visited the planet and for which we even have samples of volcanic rocks in our meteorite collections. In addition, recent missions to Mars have provided the first rovers to investigate rocks in-situ (e.g., Squyres et al., 2003; Grotzinger et al., 2012) and, with the InSight mission, the first geophysical station (Banerdt et al., 2020). Yet, for Venus – our other neighbouring planet – this is much less the case. While we do have radar altimetry and gravity data, the fact that Venus' surface cannot be observed in the visible, together with the prohibitive surface conditions restricting the feasibility of in-situ operations, strongly limits our present knowledge of the surface and interior of the planet. However, with three missions in planning, namely ESA's EnVision (Wideman et al., 2020) and NASA's VERITAS (Smrekar et al., 2019a) and DAVINCI+ (Garvin et al., 2020), significant progress is to be expected over the next two decades. Despite the dramatic advances made by the MESSENGER mission (Solomon et al., 2018) and expected progress from the BepiColombo mission (Benkhoff et al., 2010) on its way to reach Mercury in 2025, the exploration of Mercury is only in its infancy, mostly due to the technical challenges for missions to the planet. In contrast, the Moon has seen multiple human missions. Rock samples returned by the Apollo missions provide insights into the very early evolution of the Solar System. In addition, dating of these samples combined with analysis of the crater density of surface areas allows constraints to be placed on the age of such areas (e.g., Zellner, 2017) – a concept that has been applied to all planetary bodies with cratered surfaces in the Solar System using the lunar record as a gauge. For a recent overview on science of the Moon, see Taylor (2016).

Many satellites of the outer Solar System are largely unexplored. Exceptions are the Jovian satellites, of which at least Europa and Ganymede will be further explored by the upcoming

Europa Clipper and JUpiter ICy moon Explorer (JUICE) missions (Pappalardo et al., 2017; Grasset et al., 2013), the Saturnian satellites Titan and Enceladus, particularly relevant because of their astrobiological potential and the chemical similarity of Titan's atmosphere with that of Earth.

Understanding the processes in the interiors of rocky planets requires a precise knowledge of their structures, as explained in Section 2.1.1. To first order, this involves determination of the mean thickness of the crust and the radii / composition of the outer liquid core and, if present, of the solid inner core. While we have a very accurate picture of the Earth's radial structure and, thanks to seismic tomography, a continuously growing understanding of even the three-dimensional structure of the mantle at multiple spatial scales, we still do not know precisely the basic one-dimensional layering of the interior of the other major terrestrial planets. The size of Mercury's liquid core is relatively well constrained from a variety of geodetic data to be around 2000 km (Margot et al., 2018). The radius of Mars' liquid core is thought to be 1830 ± 40 km (Khan et al., 2018; 2021; Plesa et al., 2018; Smrekar et al., 2019b; Stähler et al., 2021), yet with an uncertainty of at least 100 km. The size of Venus' core was only just determined to be approximately 3500 km with large uncertainties through high-precision measurements of its spin state (Margot et al., 2021). While geodetic data seem to support the existence of a relatively large solid core for Mercury (Genova et al., 2019), we have no data proving or disproving that this is the case for Mars, although this may change thanks to refined rotational data expected from the InSight radio science experiment (Smrekar et al., 2019b). For initial results from the InSight mission on Mars, refer to Johnson et al. (2020) for the crustal and time-varying magnetic fields at the InSight landing site, to Giardini et al. (2020) for the seismicity of Mars, to Lognonné et al. (2020) for the mantle and crustal structure from seismic data and to Kahan et al. (2021) for the radio science data analysis.

The existence of a dynamo-generated magnetic field, at present and/or in the past, provides an indirect, yet powerful constraint for the global evolution of a terrestrial body. Mercury, Earth and the icy moon Ganymede are the only rocky planetary bodies known to possess a dynamo-generated magnetic field today. Laboratory measurements of billion-year-old rock samples suggest that a dynamo has been active for most of Earth's history (e.g., Tarduno et al., 2020). Similarly, analysis of Apollo samples indicates that also the Moon had a magnetic field persisting for a large part of its history (e.g., Tikoo et al., 2017). In the absence of ancient rock samples, information on magnetic fields of the other bodies stems from measurements of crustal magnetisation. The identification of an old region of Mercury's highly magnetised crust during MESSENGER's final low-altitude campaign provided evidence for a magnetic field active ~3.7 Ga (Johnson et al., 2015). Yet, whether or not a magnetic field was active throughout the rest of the planets evolution is unknown. Global low-altitude mapping, which unfortunately is not foreseen by the BepiColombo mission, would be needed to fill this gap. Magnetisation of the Martian crust inferred from orbit suggests that a dynamo was active during the first few hundred million years of the planet's evolution (Acuna et al., 1999). Yet, the latest measurements at the InSight landing site indicate that the local magnetisation is much higher than previously inferred from orbit (Johnson et al., 2020), reinforcing once again the importance of landed missions collecting data in-situ. Venus has no internal magnetic field today and it is unknown whether or not it had one in the past, which prevents us from placing any constraints on the evolution of the interior related to the absence or existence of a dynamo. Although it has long been believed that Venus' crust would be unable to preserve any magnetisation due to its high temperature, recent work suggests instead that this could be actually identified by a low-altitude magnetometer (O'Rourke et al., 2019).

As important as knowledge of the interior structure, but even more difficult to come by, are quantitative assessments of the heat transfer and the rheological properties of the interior. Heat flow measurements on extraterrestrial bodies have been conducted so far only on the Moon by the astronauts of the Apollo 15 and 17 missions (Langseth et al., 1976). The Heat

Flow and Physical Properties Package ($HP^3$) onboard the InSight lander on Mars has not succeeded in measuring the heat flow on Mars. It was equipped with a self-hammering probe – a “mole” – designed to penetrate to a depth of 3 to 5m and measure the thermal conductivity along its way (see Spohn et al., 2018, for details on the experiment). The mole was, however, blocked in the first centimetres.

This calls into question the applicability of the above technique to measure the heat flow, which nevertheless remains a fundamental observational target due to its ability to provide powerful constraints for the thermal evolution of a planet (e.g., Plesa et al., 2015). Alternative techniques may require the drilling of boreholes (Stamenkovic et al., 2019), as was done during the Apollo missions but would have to reach significantly deeper on Mars. The borehole could be used synergistically for other science, for example the search for traces of life, as well.

Direct estimates of the effective viscosity of the Earth’s mantle are traditionally obtained in two ways: by combining static gravity data with models of the planet’s viscous response to internal loading associated with mantle convection, and/or by combining time-dependent gravity data with models of the viscoelastic response of the planet to surface loadings associated with processes occurring over shorter timescales, such as the last deglaciation (e.g., Mitrovica & Forte, 2004). The former relies on assumptions on the Earth’s three-dimensional density distribution driving mantle convection, which can be inferred through seismic tomography. The latter, in addition to accurate measurements of the time-dependence of the Earth’s gravity field, depend on the existence of relatively recent large-scale deformation affecting mass transport in the deep interior. Applying a similar approach to other planets will be difficult. On the one hand, inferring the three-dimensional distribution of internal density anomalies on other bodies would be a major challenge due to the lack of strong seismic sources and widespread distribution of seismometers available on Earth. On the other hand, although measuring time-dependent gravity is possible, the lack of recent events inducing large-scale redistribution of mantle material limits the use of these data to the study of short-term processes, which hardly affect the deep interior. Seasonal changes in the masses of Mars’ polar caps provide a good example in this sense (Smith et al., 2009). Since the assessment of rheology, heat flux and volcanic processes requires measurements of time-dependent processes, future missions to active planetary bodies are essential.

## 5.3. Interior processes in giant planets

The interiors of the four giant planets – Jupiter, Saturn, Uranus, and Neptune – are expected to be fundamentally different from those of the terrestrial planets. First, giant planets contain significant amounts of gas, primarily hydrogen and helium from the protoplanetary disk. The presence of the gas shapes the planetary conditions and processes – it affects the solid accretion process as the planet forms and determines the heat and material transports in the interior later. Second, the pressure in the interiors of the giant planets is much higher, as the planets are much more massive. Under such high pressures, material interaction and properties are poorly known. And third, the giant planets’ formation locations, further out in the protoplanetary disk, indicate volatile-rich composition. As a consequence, the interiors of the giant planets are more uncertain and expected to greatly differ from those of the terrestrial planets. As of today, there are many open questions regarding the chemical composition of each of the planets and its distribution in their interiors (see Section 2.1.1).

The standard model of a giant planet, as described in Section 2.1, is composed of rock-ice core surrounded by a hydrogen-helium dominated envelope. This interior structure is still debated, as several recent studies contradict this simplified core-envelope structure. Studies of planet formation of intermediate mass planets and gas giants show that a substantial amount of the core building blocks (metals) dissipates in the accreting envelope and does not reach the core (Podolak et al., 1988; Hori & Ikoma 2011; Brouwers et al., 2018). Such formation processes lead to interiors with gradual composition distributions, where metal fraction decreases gradually from the deep interior to the gas envelope (Lozovsky et al., 2017; Bodenheimer et al., 2018; Valletta & Helled 2020; Ormel et al., 2021). Miscibility of metals in hydrogen (e.g., Guillot et al., 2004; Wilson & Militzer 2012) is expected to further dissolve the interior metals in the hydrogen-dominated envelope and to inhibit differentiation and the settling of metals to a distinct core. Interestingly, observations in the Solar System agree with this picture of gradual composition distribution in the giant planet interiors. Gradual composition distribution in Jupiter and Saturn is consistent, respectively, with the new Juno mission measurements of gravitational moments (Wahl et al., 2017; Debras & Chabrier 2019) and Cassini's ring seismology observations (Mankovich and Fuller, 2021); a gradual composition distribution in Uranus has also been suggested to explain its low luminosity (Marley et al., 1995; Helled et al., 2011; Vazan & Helled 2020).

The main processes that determine the interior conditions and structure are thermal and material transport. Heat can be transported in the interior by convection, conduction and radiation, depending on how planetary conditions and material properties in the interior vary with time. In the basic model, each layer of a uniform composition is assumed to be adiabatic (i.e., fully convective) and the outermost layer of the planet is taken to be radiative (e.g., Guillot & Gautier 2015; Nettelmann et al., 2012). This simple adiabatic model is good as a first estimate but has difficulties in explaining the observed properties of the giant planets, like, for example, the gradual structure of Jupiter (Wahl et al., 2017; Debras & Chabrier 2019), the high luminosity of Saturn (Stevenson & Salpeter 1977; Fortney & Nettelmann 2010), and the low luminosity of Uranus (Hubbard et al., 1995; Fortney et al., 2011; Nettelmann et al., 2013).

Heat transport in the interiors of the giant planets does not operate only by large-scale (adiabatic) convection. One reason for a nonadiabatic structure is the distribution of metals inside the planet. The existence of a stable compositional gradient can suppress convection (Ledoux 1947) and slow the cooling of the deep interior. This may lead to higher temperatures in the deep interior and super-adiabatic thermal profiles. Not only does this affect the rate at which the planet cools, but the higher internal temperatures also influence the heavy element mass fraction inferred from interior models, as higher temperatures allow for more metals to fit the mass-radius relation at present (e.g., Chabrier & Baraffe 2007; Leconte & Chabrier 2012; Vazan et al., 2015). Such interior structure models can explain the observed properties of the giant planets in our Solar System (Leconte & Chabrier 2013; Vazan et al., 2016; Vazan et al., 2018; Nettelmann et al., 2016; Podolak et al., 2019; Vazan & Helled 2020).

As a giant planet evolves, material transport processes may take place and affect the planet structure and its energy transport. Two main mechanisms are settling (downwards process) and mixing (upwards process). A settling process in the two gas giants is helium rain, whereby under certain pressure-temperature conditions helium separates from hydrogen and settles to a deeper layer (Stevenson & Salpeter 1977). This process leads to the formation of a helium-rich shell above the heavy-element deep interior (Fortney & Hubbard 2003; Hubbard & Militzer 2016). The formation processes of such a helium-rich layer increase the planet luminosity as gravitational energy is being released. The helium-rich layer changes the interior structure and in particular the uniformity of the envelope. This more complex structure can, in turn, change the heat transport from the deeper layers outwards (Vazan et al., 2016; Mankovich et al., 2016). An additional settling process that may be significant in the ice giants is ice condensation. Condensation of volatiles in the planetary envelope has an impact on the energy transport in the outer layers and therefore on its luminosity (e.g., Kurosaki & Ikoma 2017). The second, upwards material transport mechanism, is mixing by convection. The onset of convection in a metal-rich inhomogeneous region may lead to mixing of materials from the deep interior. Efficient convection can act to homogenize the interior and change the distribution of metals as the planet evolves (e.g., Guillot et al., 2004; Stevenson 1982). Moreover, upward mixing enriches the outer envelope with metals, which in turn affects the planet's thermal evolution (Vazan et al., 2015; Muller & Helled 2020) and also drives the generation of their magnetic fields (Soderlund & Stanley, 2020).

In the last decade, the gas giants were explored extensively by dedicated space missions – Cassini to Saturn and Juno to Jupiter. The ice giants, in contrast, have only been visited by single flybys. By their nature and location, they hold keys for understanding Solar System formation. While Jupiter and Saturn went through a massive (runaway) gas accretion phase to become hydrogen-helium dominated planets, Uranus and Neptune did not reach the gas accretion process and therefore remained much smaller. In this context, ice giants can be considered as embryos of potential gas giant planets. As such, they contain clearer imprints of the planet formation processes. Steep composition gradients in their interiors (e.g., Podolak et al., 2019) are more stable against convective mixing and hence keep the envelopes primordial. The complex magnetic fields of the two ice giants (Fig. 3. 6) is additional evidence for their nonuniform interiors.

Space exploration of the ice giants will not only shed light on them, it will also greatly contribute to our understanding of the formation of the Solar System, of planet formation in general and of intermediate mass planetary interiors. A dedicated mission to Uranus and/or Neptune, preferably with atmospheric probe(s), will collect measurements of the planet's gravitational moments, atmospheric abundances and conditions and magnetic field. These complementary data will provide much better constraints on the interior structure of the ice giants in particular, and also allow us to draw conclusions on broader families of giant planets and exoplanets, such as volatile-rich planets, intermediate mass planets and further-out planets.

## 5.4. The Solar System as a fluid dynamics laboratory: studying superrotation in slow and fast rotating planets

### 5.4.1. Atmospheric superrotation

**Introduction**

Superrotation, i.e. the tendency for planetary atmospheres to rotate faster than their parent planets and in the same direction in the equatorial region, is one of the typical dynamical regimes of planetary atmospheres. Observations and supporting theoretical interpretations of superrotation have been reviewed by Imamura et al. (2020), from whom comes most of the elements to follow. Superrotation is found in slow rotators like Venus and Titan as well as in fast rotators like the gas giants (Read & Lebonnois, 2018) (see Fig. 3.24). Whereas at Venus and Titan superrotation has been observed to be global, prevailing to different degrees at all latitudes, it extends from the equator to only about 15° at Jupiter and 30° at Saturn. Signatures of atmospheric superrotation have also been detected at some tidally locked exoplanets orbiting close to their parent star, as an eastward displacement of their IR hot spot on their sunlit side (see Heng & Showman, 2015, for a review). An implication is that superrotation may play a role in the habitability of these planets by allowing some redistribution of heat between their dark and star-lit sides.

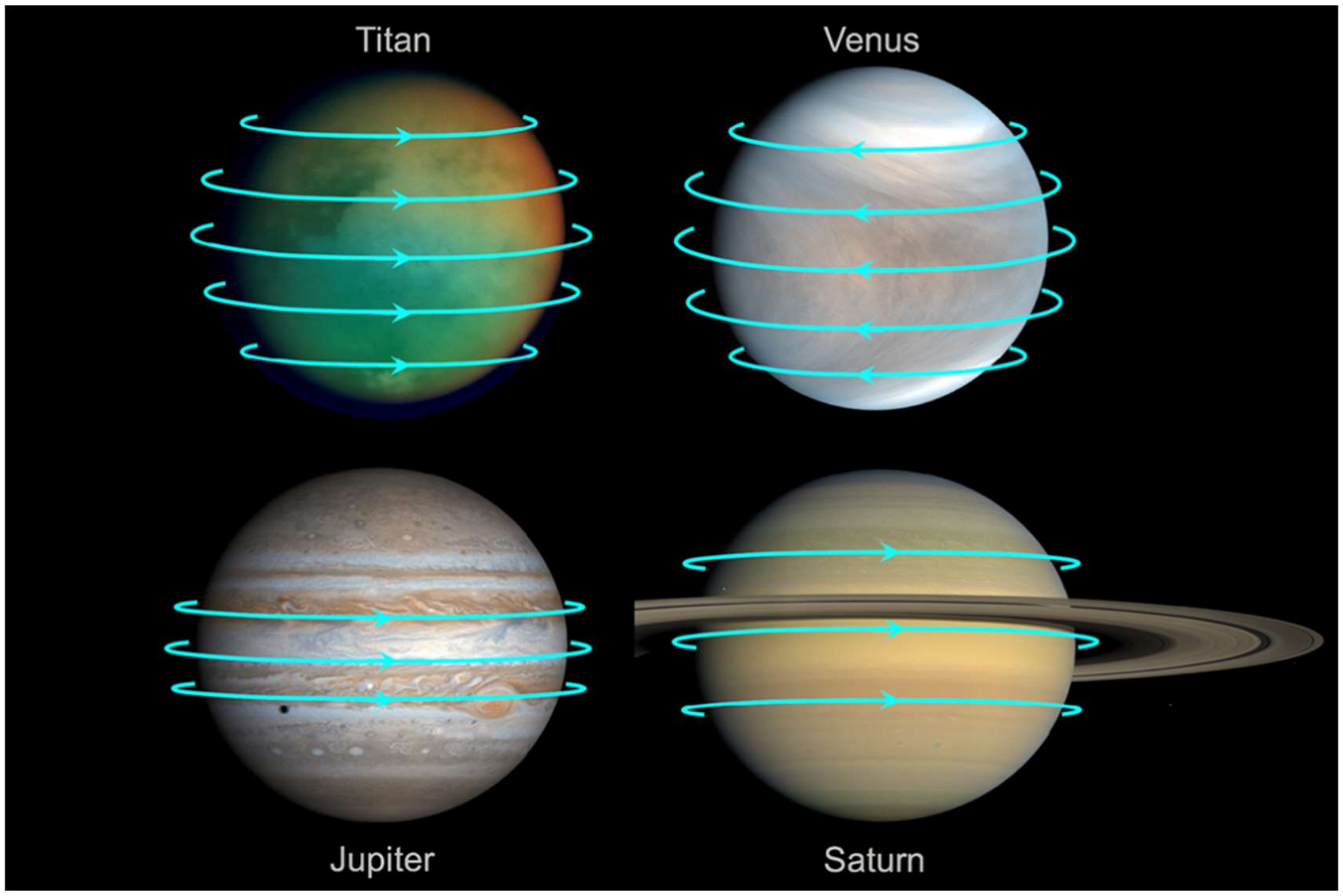

Figure 3.24. Schematic of the direction and the latitudinal extent of superrotation for Titan, Venus, Jupiter and Saturn. Superrotation extends to the high latitudes on Titan and Venus, while it is confined to the low latitudes on Jupiter and Saturn. Superrotation is westward at Venus, eastward at the other planets, in all cases in the same direction as the planetary rotation. Images of Titan, Jupiter and Saturn were provided by NASA. Venus image provided by JAXA. From Imamura et al. (2020).

Table 3.2 displays the key parameters characterising superrotation in these bodies, from Venus (top row) to the exoplanet HD 189733b.

| Planet | Radius (km) | Rotation period (days) | Equatorial rotation speed (m/s) | Equatorial wind speed (m/s) | Superrotation index, *s*, on the equator |
|---|---|---|---|---|---|
| Venus | 6,052 | 243 | 1.81 | 100–120 | 55–66 |
| Titan | 2,576 | 16.0 | 11.7 | 100–180 | 8.5–15 |
| Jupiter | 69,911 | 0.41 | 12,300 | 60–140 | 0.005–0.011 |
| Saturn | 58,232 | 0.44 | 9,540 | 350–430 | 0.037–0.045 |
| HD 189733b | 79,500 | 2.2 | 2600 | 2400 | 0.92 |

Table 3.2. Basic information on the superrotation of the atmospheres (i.e., rotating faster that their parent body) of the Solar System planets and a hot Jupiter (modified from Read & Lebonnois, 2018).

Superrotation appears to be the net result of exchanges of angular momentum between adjacent latitudes and altitudes in a dynamic atmosphere, involving a complex interplay between meridional convection cells, turbulent eddies and a broad spectrum of atmospheric waves, including planetary-scale waves like thermal tides and planetary waves, which redistribute angular momentum globally. The condition for transition from sub-rotation (like Earth and Mars) to superrotation has been studied using idealized general circulation models (GCM) (e.g., Williams, 1988; Dias Pinto & Mitchell, 2014) and deep convection models (e.g., Aurnou et al., 2007; Soderlund et al., 2013; King & Aurnou, 2013). Identifying the key control parameters, such as the Rossby number and thermal inertia of the atmosphere, based on observations and modelling would lead to a general understanding of circulation regimes including superrotation.

Superrotation is found to prevail in planetary atmospheres, not only with very different compositions and condensable gas contents but also with radically different rotation rates and heating sources:

- It prevails on slow rotators like Venus and Titan, where pressure gradients are mainly balanced by inertial forces (large Rossby number) and on fast rotators like Jupiter and Saturn, where it is mainly balanced by the Coriolis force (small Rossby number);
- It prevails in atmospheres where circulation is driven in a shallow layer in the vicinity of the cloud layers by heat input from the Sun, but also in gas giant atmosphere where circulation may be deep, driven by convective transport of heat from the planetary interior (see previous section).

What needs to be understood is why and by which mechanisms superrotation prevails in this diversity of situations, which will be illustrated here by two examples: Venus, a slow rotator with a shallow circulation, and the gas giants Jupiter and Saturn, two fast rotators where, according to recent results, deep circulation may prevail.

### Venus superrotation

Superrotation is by far best documented at Venus, thanks to Earth-based telescope observations combined with several space missions that provided key measurements by orbiters, descent probes and balloons: the Venera and Vega series, Pioneer Venus, Venus Express and most recently JAXA's Akatsuki mission. The data accumulated on atmospheric circulation combine diverse techniques: in situ measurements of winds by atmospheric probes and balloons, temperature maps by IR spectrometers and spectro-imagers, radio occultations and the tracking of cloud motions at various wavelengths (e.g., Sánchez-Lavega et al., 2017; Horinouchi et al., 2018). These diverse data sets have fed the development of general circulation models of increasing complexity, allowing fruitful two-way feedback between data and models. As an illustration, Fig. 3.25 shows the measurements of mean zonal winds performed by Akatsuki's ultraviolet imager (UVI) (top) and the predictions of the Institut Pierre Simon Laplace (IPSL) Venus GCM (bottom).

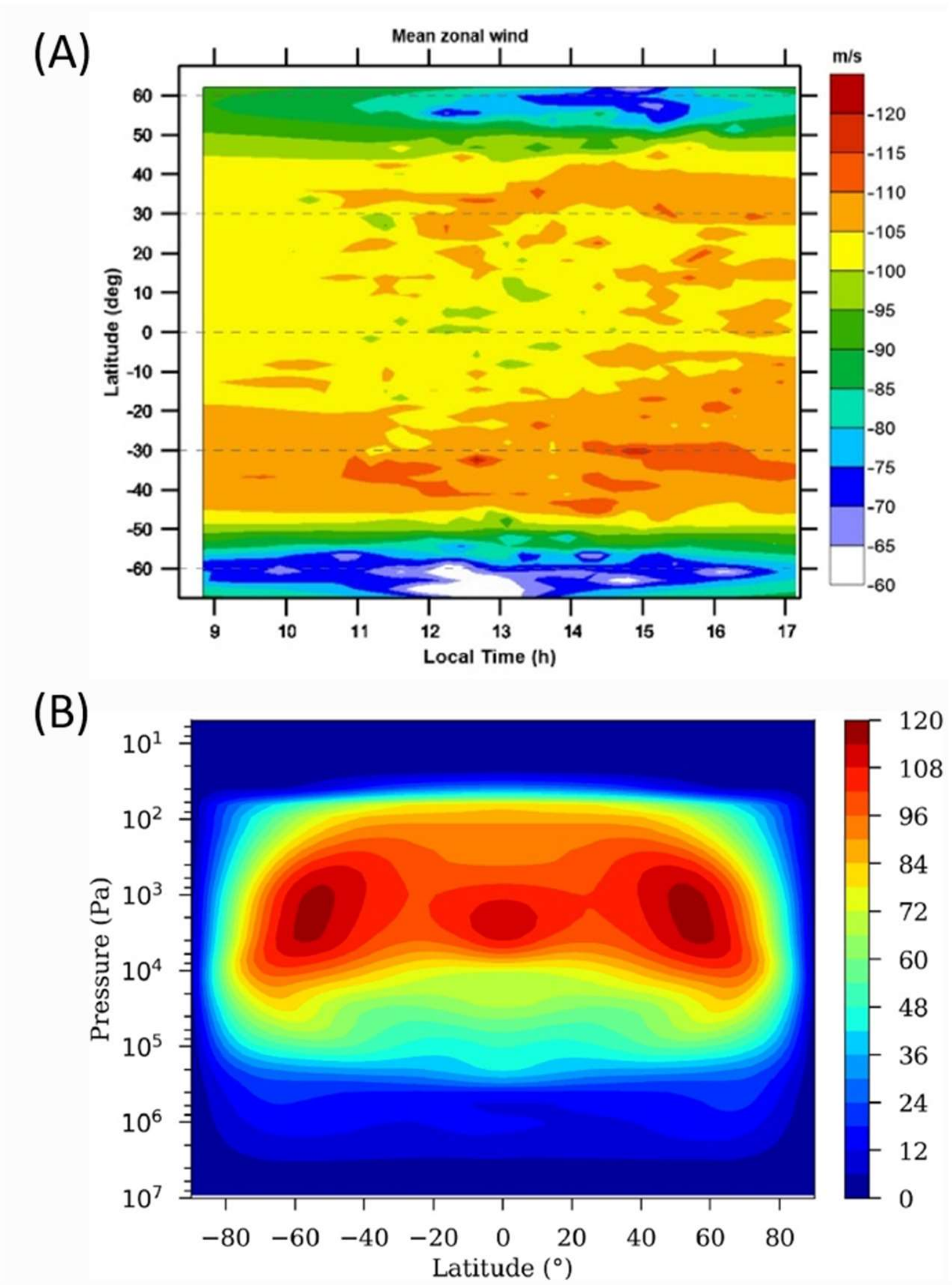

Figure 3.25. Top: Mean winds obtained by cloud tracking using Akatsuki/UVI images during January 2017, as a function of local time and latitude. Bottom: zonal wind distributions in the IPSL Venus GCM, in colour (unit is m/s), with the mean meridional stream function as contours (unit is $10^9$ kg/s). (Top) From Gonçalves et al. (2020). (Bottom) From Garate-Lopez et al. (2018).

Imamura et al. (2020) presented a detailed analysis of the convergences and discrepancies between observations and the most advanced Venus GCMs and concluded that, despite spectacular progress, several features of the observations are not yet fully captured by models. While the superrotation observed at the cloud level has been largely reproduced by recent GCMs, there are still problems in modelling the superrotation in the lower atmosphere. Sugimoto et al. (2019) argued that a very small vertical viscosity is essential for superrotation to occur in the lower atmosphere. While observational evidence for momentum fluxes that sustain the superrotation has been limited, equatorial convergence of eddy angular momentum fluxes associated with thermal tides was discovered at the cloud top by cloud tracking using images taken by Akatsuki UVI (Horinouchi et al., 2020); a GCM by Yamamoto et al. (2019) reproduced such momentum fluxes. At lower levels, equatorward eddy angular momentum fluxes driven by global-scale shear instability might play a role (e.g., Wang & Mitchell, 2014), although observational evidence is lacking.

Future progress in the understanding of the processes involved will require new Venus missions and observations providing a comprehensive spatial and temporal coverage of atmospheric dynamics and temperature fields from the surface to the cloud tops. Observations of near-surface winds, which are extremely weak and not well constrained, are particularly important because the total angular momentum of the atmosphere should be controlled by the exchange of angular momentum between the atmosphere and the solid planet. In addition, comparisons with the Titan case, which is the most similar one but different in key parameters such as the insolation, will be particularly instructive (Read & Lebonnois, 2018).

## Gas giant superrotation

What we know of atmospheric circulation at Jupiter and Saturn has been dominated until recently by the study of circulation at the cloud levels (see Section 2.1.2), leaving open the very nature of the observed circulation, shallow or deep with respect to both penetration depth and the driving energy source (e.g., Sánchez-Lavega and Heimpel, 2018). The only in-situ measurements came from the Galileo probe that descended in 1995 into Jupiter's atmosphere around latitude 6.5◦N. The probe found that the zonal wind velocity increased from 80 m $s^{-1}$ at the cloud level, where the probe entered, to ∼160 m $s^{-1}$ at a depth of 4 bars, below which the zonal velocity remained nearly constant down to 21 bars, 130 km below the clouds, where the signal was lost (Atkinson et al., 1996). The fact that fast zonal winds extend so deep indicates that the zonal flow is not restricted to the thin layer within a few scale heights of the cloud level and opens the question of how deep it extends below the level observed by the Galileo probe.

The key contributions from gravity measurements performed at Jupiter by the Juno radio science experiment and nearly at the same time at Saturn by Cassini during its Grand Finale, have been extensively analysed by Kaspi et al. (2020). Moments of the gravity field were determined with a reasonable accuracy up to order 10 and an inversion technique allowed the determination of the structure of flows deep below the cloud layers of the two planets. The main results of this inversion are displayed in Fig. 3.26. They show for Jupiter (right-hand diagrams) and for Saturn (left-hand diagrams) the vertical profile of these flows (top diagrams) and their distributions in a meridional plane. Two important conclusions were drawn from these measurements: (1) the inferred zonal flows extend more or less exactly for both planets to the $10^5$ bar level, which is known to correspond to the top of the conducting layer, within these planets' interiors, where their magnetic field is believed to be generated by dynamo action; (2) projection of this level from the equator along a cylinder with an axis parallel to the planets rotation crosses the weather layers at latitudes of about 13° for Jupiter and 31° for Saturn, coinciding nearly exactly with the extent of the equator-centered superrotation zone at both planets (see Fig. 3.26).

These conclusions are certainly not final and one must keep in mind the ambiguity inherent to the inversion techniques used. Nevertheless, with the addition of complementary pieces of evidence, Kaspi et al. (2020) provided a strong case in favour of a deep circulation at Jupiter and Saturn extending from the top of the clouds down to the interior conducting layer and organized along cylinders rotating rigidly parallel to the planetary spin axis. Their analysis of the much more limited observations at the ice giants also led to conclude that flows might be much shallower at the ice giants.

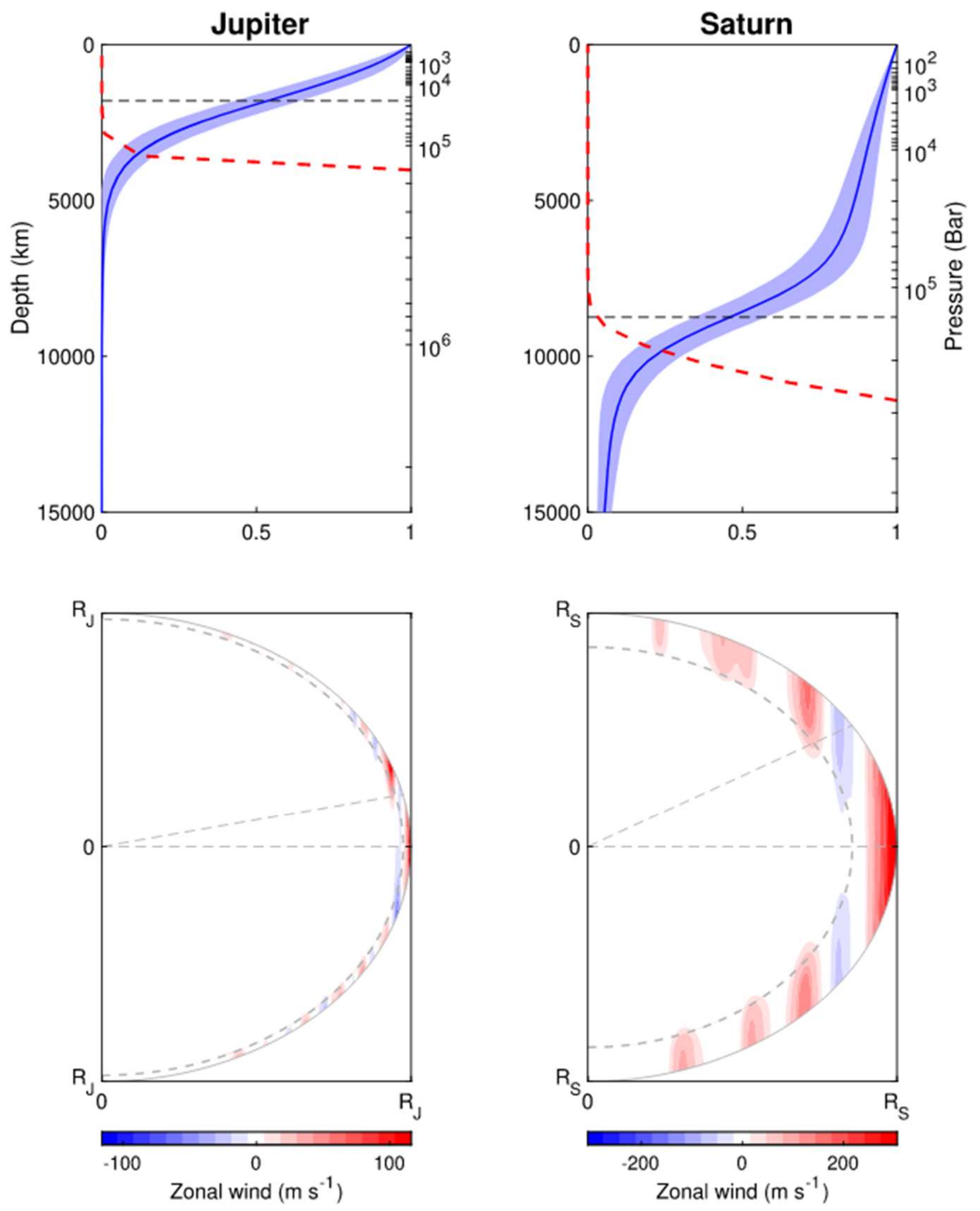

Figure 3.26. (Top panels) Vertical decay profile of the zonal flow on Jupiter (left) and Saturn (right) as a function of depth (blue curve, with uncertainty shown as blue shading), corresponding to the best fits to gravity moment determinations by the radio science experiments onboard Juno and Cassini. The red dashed lines on the same panel show the electrical conductivity profile (red dashed) as given by Liu et al. (2008) for Jupiter and Saturn, in units of S $m^{-1}$, with the scale going linearly from 0 to 100. The middle point in the decay profile, at depths of 1831 km and 8743 km for Jupiter and Saturn, respectively, is marked by the dashed horizontal line. (Bottom panels) Zonal flow profile ($m.s^{-1}$) as a function of latitude and depth in the spherical projection. The middle point shown in the upper panels appears as the curved dashed line. The radial dashed lines show the angle (latitude) derived from extending the depth of the flow along the direction of the spin axis. From Kaspi et al. (2020).

## Key observations for the future

The two cases presented here illustrate the diversity of situations and flow regimes (shallow circulation in a "weather layer", vs. deep circulation driven from below) in which planetary atmospheres display superrotation. Despite spectacular progress in modelling, we are still a long way from understanding both the universal mechanisms driving superrotation and their specific characteristics at each planet. Progress in the coming decades should come from:

(a) a comprehensive description of the circulation regimes at "representative" planets (Venus, one of the gas giants, one of the ice giants, etc.), based on a conjunction of (a) 3-D mapping of the basic fields constraining general circulation (pressure/temperature, velocity, concentration of condensable species) from the stratosphere to the planetary surface or to the top of the internal conducting layers;

(b) a comparison of these cases between planets, with analogue situations in the Solar System (i.e., Venus with Titan, Jupiter with Saturn, Uranus with Neptune) and in the longer term with exoplanets.

For the Solar System, combining the different available techniques at each planet will be mandatory: orbital remote sensing and in situ exploration by atmospheric probes of the shallow atmosphere, deep sounding of the atmosphere and its coupling with the interior via gravity and magnetic field measurements.

### 5.4.2. Terrestrial planet atmospheric and climate evolution

Of the terrestrial planets, Mars and Venus have been the most visited targets of several space missions focused on investigating their atmospheres and the interaction with the solar wind (e.g., for Mars, Jakosky et al., 2015; Jakosky, 2017) and between the different volatiles reservoirs in the (sub)surface or escape to space (see Fig. 3.27 for a comparison of atmospheric temperature profiles of the different planets having an atmosphere). The comparison of surface temperatures without greenhouse effects (-55° for Mars, -43° for Venus, and -17° for the Earth) with the actual surface temperatures reveals how differently this effect works at each planet (e.g., Bennett et al., 2019). Although we have now gathered tremendous collections of data on these bodies and more specifically on their atmospheres, several key questions regarding their evolution and habitability remain unanswered.

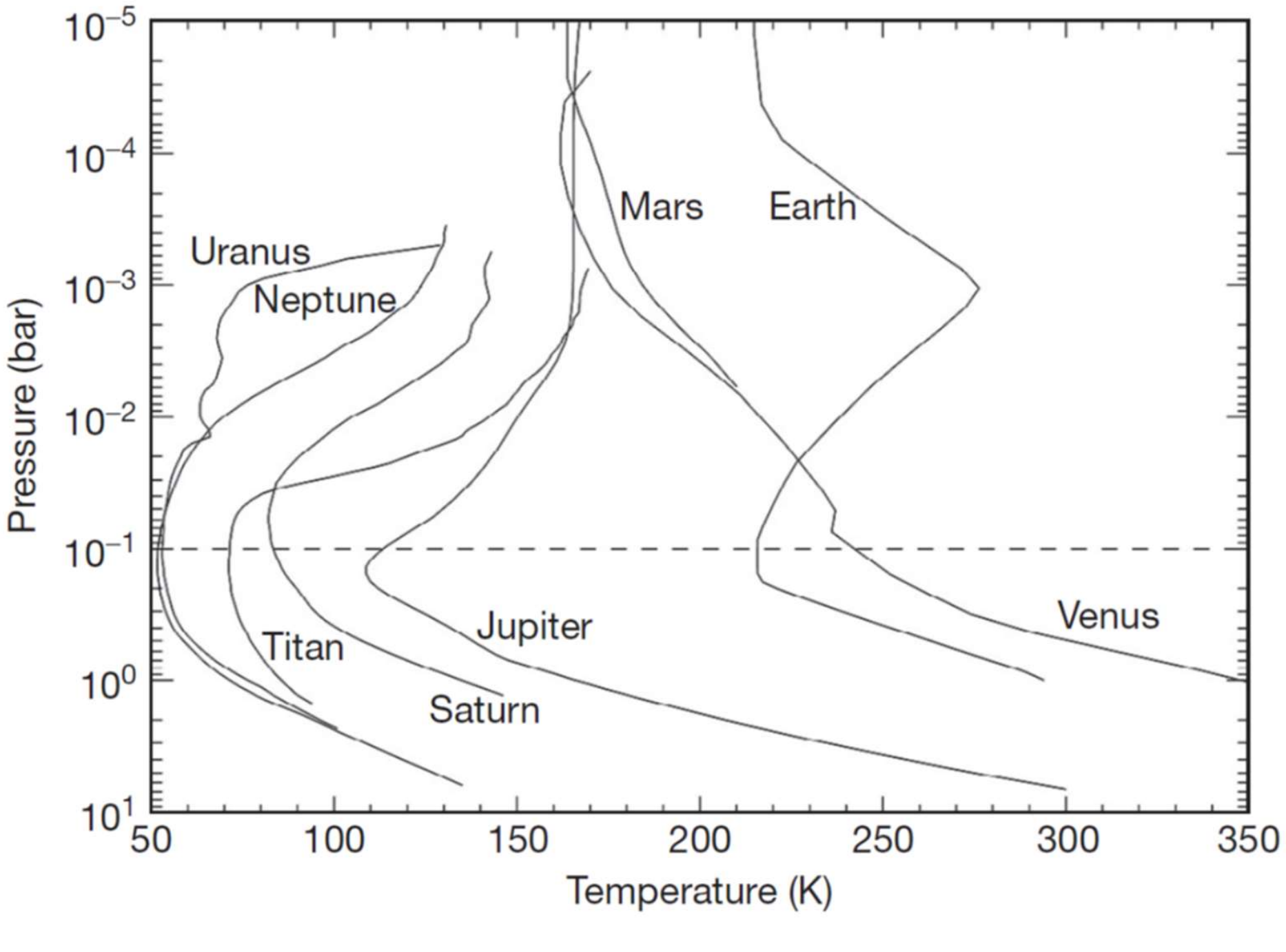

Figure 3.27. This comparison of atmospheric temperature profiles at Uranus, Neptune, Titan, Saturn, Jupiter, Mars, Earth, and Venus reveals striking differences between the planets; in particular for the three terrestrial planets, despite likely similar initial compositions at the time of their formation. (1) At Venus, $CO_2$ accumulation in the atmosphere and the escape of water led to a run-away greenhouse; at Earth, the 30 degrees gained thanks to natural greenhouse gases maintained habitable conditions over most of the planet's life; at Mars, escape of most of the atmosphere leaves the planet without any significant greenhouse effect to warm its surface. (2) The upward temperature gradient in the Earth temperature profile is a signature of the presence of the ozone layer. (3) The dashed line at 0.1 bar indicates that planets with temperature inversions have tropopause minima near this pressure level. From Catling (2015).

Venus, Earth's closest planetary neighbour, is considered as the Earth's twin because both planets share several properties such as mass and size. Moreover, their bulk densities and inventories of carbon and nitrogen are similar. Meteorological and geological phenomena occurring on Mars and Venus are also found on Earth. Our current understanding of planetary formation coupled to the existing observations of their composition, in particular of their isotopic signatures, suggest that all three planets – Mars, Venus and Earth – evolved from comparable geological, surface and atmospheric environments. Yet, despite their close proximity and similar origins, these three planets have evolved into quite different states.

Note that the atmospheres of other bodies of our Solar System, like Triton or Titan, have also been studied through flybys of missions like Voyager and Cassini, or through a lander (Huygens probe) entering, descending in the atmosphere of Titan and landing on the surface. Titan, the second largest moon in our Solar System, is often considered as the largest abiotic organic factory in the Solar System. Titan's atmosphere, like that of the Earth, Mars and Venus, is a place where 'greenhouse' effects warm the surface. It has a nitrogen-based atmosphere like Earth but organics play a much more important role in its composition, for example forming clouds and being precipitated as rain. The extent to which present-day Titan resembles the prebiotic Earth is not clear: for example, the present Titan is more oxygen-poor than is the Earth, but formation of organic haze may nonetheless have taken place on our own planet early in its oxygen-poor history. Thus, while the analogy of Titan to the early Earth is not perfect (e.g., Earth was always warmer than Titan was), it is close enough to advocate for closer scrutiny.

Titan is in a state of rapid loss of volatiles, a situation that occurred also on Mars and Venus in their past. On Venus, a runaway greenhouse effect led to destruction of water via escape of the hydrogen. On Mars, the loss of the protective magnetic field following the shutdown of the planet's internal dynamo has probably accelerated the disappearance of the atmosphere.

Further addressing Titan's atmosphere with missions will help us understand the atmosphere of Titan and the differences and similarities with early Earth.

There is much we can learn about Earth's atmosphere through the investigation of those for other planets. Water is a common building block of these diverse atmospheres: it is or was present, it is a prerequisite for life as we know it to emerge and maintain itself; understanding how much there was throughout the history of the planet, and how and when it disappeared will provide clues to the fundamental question 'How do planetary atmospheres evolve?".

However, planetary atmospheres are a dynamic environment changing both in space and in time. The atmospheres of the different bodies in our Solar System and further away provide us with a wide panel of evolutionary histories. Titan-like planets may be common in the universe: planets around M dwarfs, the most common stellar type, at the distance of the Earth from the Sun will be as cold as Titan. Titan may usefully inform us about their organic chemistry and potential habitability. Numerous Earth or Venus-sized exoplanets have been discovered (Kane et al., 2019). Are any of these habitable? Until now, most of these have been identified as more resembling Venus than Earth. Study of Venus and analogues would improve our understanding of the processes that lead to totally uninhabitable planets.

Improved understanding of the processes occurring on terrestrial planets in the Solar System in the coming decades will help answer the following questions:

- What is the detailed atmospheric composition and chemistry? In particular, what can isotopic ratios tell us about its past, its history or its habitability?
- Where did the water go? What are the processes explaining its current abundances when we know or postulate that water was present in their pasts?
- What is the role of dust and clouds? How is dust exchanged with the surface?
- What is the impact of tectonics, volatile cycling and volcanic resurfacing on the history of an atmosphere? What is the impact of the interior of the planet and exchanges with the atmospheres?

- What are the volatiles fluxes across the surface? What is the impact of surface weathering?
- How is the atmosphere changing and what are the causes of these changes?
- Did any of these planets have a habitable period? Are there regions in the present atmospheres that could support life? How do we detect life, past, extinct or extant?

In order to find some solutions, future missions will need to characterize the atmospheric composition and structure as well as the exchanges with the (sub)surface and space at high enough spatio-temporal resolutions covering global to regional scales. Structure not only involves temperature and total density, but also winds and energy sources (radiative warming/cooling, greenhouse warming, heat fluxes).

Progress in planetary atmosphere modelling in parallel to new space missions, with a systematic use of data assimilation into models, will be critical to efficiently address these challenging questions: comparison between model results and observations help identify gaps in our understanding of the processes at play; using data assimilation, models can directly integrate observations; models can provide forecasts of the state of an atmosphere in support to space and landed missions; they can investigate the response of the system to different changes impacting the atmosphere.

## 5.5. Studying universal processes in planetary magnetospheres

In 1928, Irving Langmuir (1881-1957) introduced the word "plasma" to designate a partially or totally ionised gas (Langmuir, 1928). In 1963, David A. Frank-Kamenezki first referred to plasma as the "fourth state of matter" (Piel, 2010). The solar wind is a plasma, which evolves as it travels away from the Sun, and at the present distance of the Earth is ten billion times less dense than air at sea level on Earth. Its temperature exceeds 100,000 degrees and its average speed is more than one million kilometres per hour (Meyer-Vernet, 2007; Lang, 2011; Russell et al., 2016). Under these extreme conditions, we no longer speak of a gas but of a plasma, because the matter is in a state where the atoms are ionised. A plasma is therefore a collection of charged particles, ions, and electrons that are electrically neutral overall and that nevertheless exhibit collective behaviour (Meyer-Vernet, 2007; Russell et al., 2016; Piel, 2010). The latter means that when a perturbation is applied to the plasma, a large number of particles in the plasma are involved in the macroscopic response to the perturbation. The plasma conducts electricity and interacts with the magnetic field. The magnetic field guides the plasma, but the plasma can, in turn, modify the magnetic field.

Plasma is studied at various scales by different theories. A charged particle subjected to a magnetic field directed in a certain direction performs a circular motion around a straight line aligned with the direction of the magnetic field (see Fig. 3.28), the radius of which depends on the mass of the particle, the local strength of the magnetic field and the speed of the particle. This circular motion is called gyration and the radius of gyration is also known as the Larmor radius. This gyration motion is also associated with a gyration period. In a strong magnetic field that varies slowly in both space and time (relative to gyration), the guiding-center theory is used to describe the motion of a particle relative to the magnetic flux tube. This theory thus allows us to understand how a charged particle evolves with respect to the magnetic field. However, for obvious computational reasons, the equations of motion are impossible to solve for each of the millions or billions of particles involved in the phenomena that interest us at the scale of celestial objects. To understand the behaviour of the particles that make up the plasma, we must therefore rely on statistical calculations and in particular on the probability that there is a particle with a given velocity at a given location. This probability function is called the distribution function.

In the general case of a plasma where collisions between particles are rare (i.e. a "collisionless" plasma), a detailed description of the particle distribution function is necessary, especially when one is interested in small scales, below the particle radius of gyration. This is the domain of plasma kinetic theories, most often based on equations of the type of the Boltzmann equation for gas but with more complex interaction force terms. Their use to simulate a plasma on large scales (a planetary magnetosphere, for example) is unfortunately very difficult because of their heavy demands in terms of numerical computing resources. Finally, when one is primarily interested in large scales, one can neglect (under certain conditions) microscopic phenomena and consider the plasma as a fluid of ions and a fluid of electrons, or even as a single fluid of charged particles. The theory based on the latter assumption is called magnetohydrodynamics (MHD).

A very active domain of current research on planetary magnetospheres looks for the modes of energy storage, transfer, release and dissipation in the collision-less plasmas that populate the largest fraction of magnetospheres. One particular mode of electromagnetic energy transfer, which operates at the interfaces between domains with different magnetic topologies, is magnetic reconnection (Fig. 3.28). This process is believed to be responsible for a significant fraction of the transfer of energy between the solar wind and planetary magnetospheres. Its study in space is or the utmost importance, as it also plays a key role in the dynamics and stability of tokamak fusion plasmas.

Future exploration of the Solar System will require us to push the boundaries of our knowledge of physics, considering the phenomena that have been observed in our most frequently encountered parameter spaces and how we might need to adapt our measurement and analysis techniques for more exotic conditions, spanning a range of heliocentric distances, planetary magnetic field strengths and rotation rates and radiation environments. Measurements have been made in the Earth's magnetosphere for decades and the missions and instrumentation have changed and improved over time, but the basic requirements for in situ measurements remain the same:

- the distribution functions of all constituents of the plasma,
- the DC/AC magnetic and electric field,
- the distribution functions of energetic particles,
- all measured with high spatial, temporal, and directional information.

In situ measurements alone do not give the full picture of how the system works; they need to be combined with ground-based/remote sensing information. Moreover, single spacecraft measurements bring the challenge of separating spatial from temporal effects: are variations seen due to the motion of the spacecraft through a spatial structure or due to changes with time at the location of the spacecraft? The terrestrial system is well served by multispacecraft missions that work in tandem with ground-based monitoring of the space plasma environment (namely the ionosphere and magnetosphere). Multispacecraft missions that have afforded big science breakthroughs at Earth have included: (i) missions to fly in formation to different locations in Earth's magnetosphere (e.g., ISEE1&2, Van Allen Probes, THEMIS), (ii) missions to study small-scale processes (e.g., Cluster, MMS), (iii) and upstream solar wind monitors to provide context for magnetospheric and ionospheric response (e.g., ISEE3, AMPTE). Exploration beyond Earth, such as to the gas giant planets (e.g., Cassini at Saturn, Juno at Jupiter), has faced the challenge of interpreting findings without, for example, an upstream solar wind monitor. Multiple spacecraft missions are thus an inevitable step of future planetary exploration, as the BepiColombo (ESA/JAXA) mission now en route to Mercury.

Another way for improvement is to design instrumentation that simultaneously provides higher performance but with smaller size, lower power and lower cost. A key point for future multiple spacecraft missions to planetary magnetospheres may be miniaturisation of current key instruments while retaining capability. Radiation tolerance is also important. Hopefully, much will be learnt from the Juno mission, for application to future missions to Europa (Europa Clipper and follow-on missions).

Plasmas do not exist in their natural state on Earth but constitute more than 99% of the baryonic matter in the Universe. Plasmas within the Solar System are the only ones accessible for 'in situ' measurements. On Earth, 'artificial' plasmas are being studied in an attempt to produce nuclear energy by fusion as in the Sun (and not by fission as in current nuclear power plants) in order to, among other issues, reduce the abundance of waste currently associated with nuclear energy production. Thus, confronting plasma physics theories with observations in the solar wind and in magnetospheric environments is an excellent way to advance plasma physics itself, with essential future benefits for humanity.

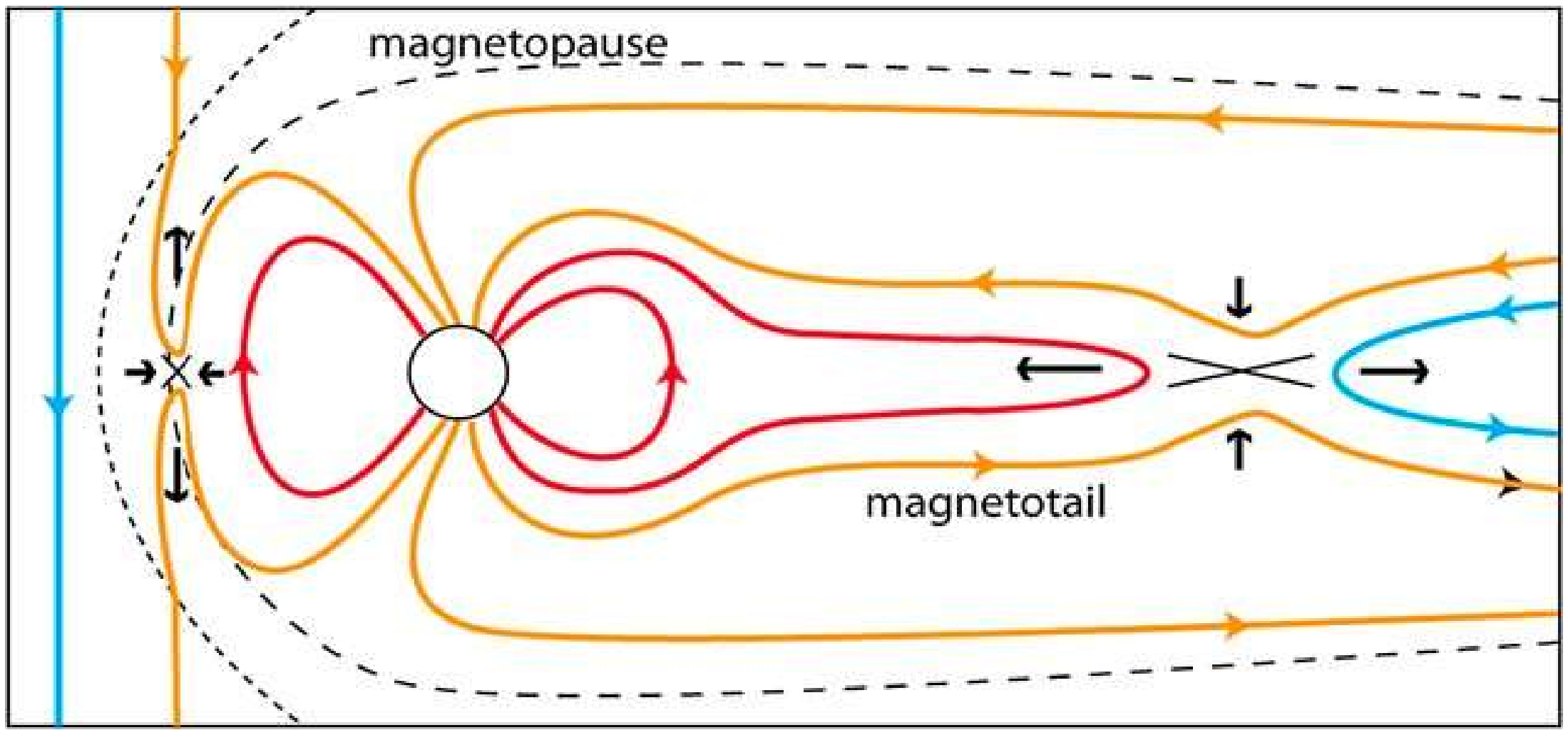

Figure 3.28. An example of energy exchange: magnetic reconnection (typically at location marked by a black cross) between the interplanetary magnetic field (blue field lines) and a planetary magnetic field (red field lines) transfers magnetic energy to kinetic energy and is globally known as an efficient mechanism to allow solar wind plasma to enter a planetary magnetosphere. From Eastwood et al. (2017).

## 5.6. Interfaces and interaction processes between the Heliosphere and the Local Interstellar Medium

The Solar System moves at a relative velocity of 26 km/s (Witte, 2004; McComas et al., 2015) through the local interstellar medium (LISM). Until recently, it was thought that the Solar System is located near the edge of the Local Interstellar Cloud (LIC) – or in a transition region – and moves in the direction of the neighbouring G-cloud (Redfield et al., 2004; Frisch et al., 2011). These two clouds are warm, low-density interstellar clouds of partially ionized gas (H, He) and dust (1% of the total mass). As the Sun expels its solar wind, a bubble called "the heliosphere" is formed around the Solar System that ploughs through the local interstellar environment. This environment is quite complex: most recent research (Linsky et al., 2019) suggests that the heliosphere is actually in contact with four different interstellar clouds: the LIC, the G cloud, the AQL cloud and the Blue cloud (see Fig. 3.29) and that the heliosphere is either already outside of the LIC or is still in a transition zone and will move out of in the next 3000 years. Linsky et al. (2019) also suggest that these adjacent clouds may deliver the $^{60}$Fe that was found in the Antarctic (Koll et al., 2019).

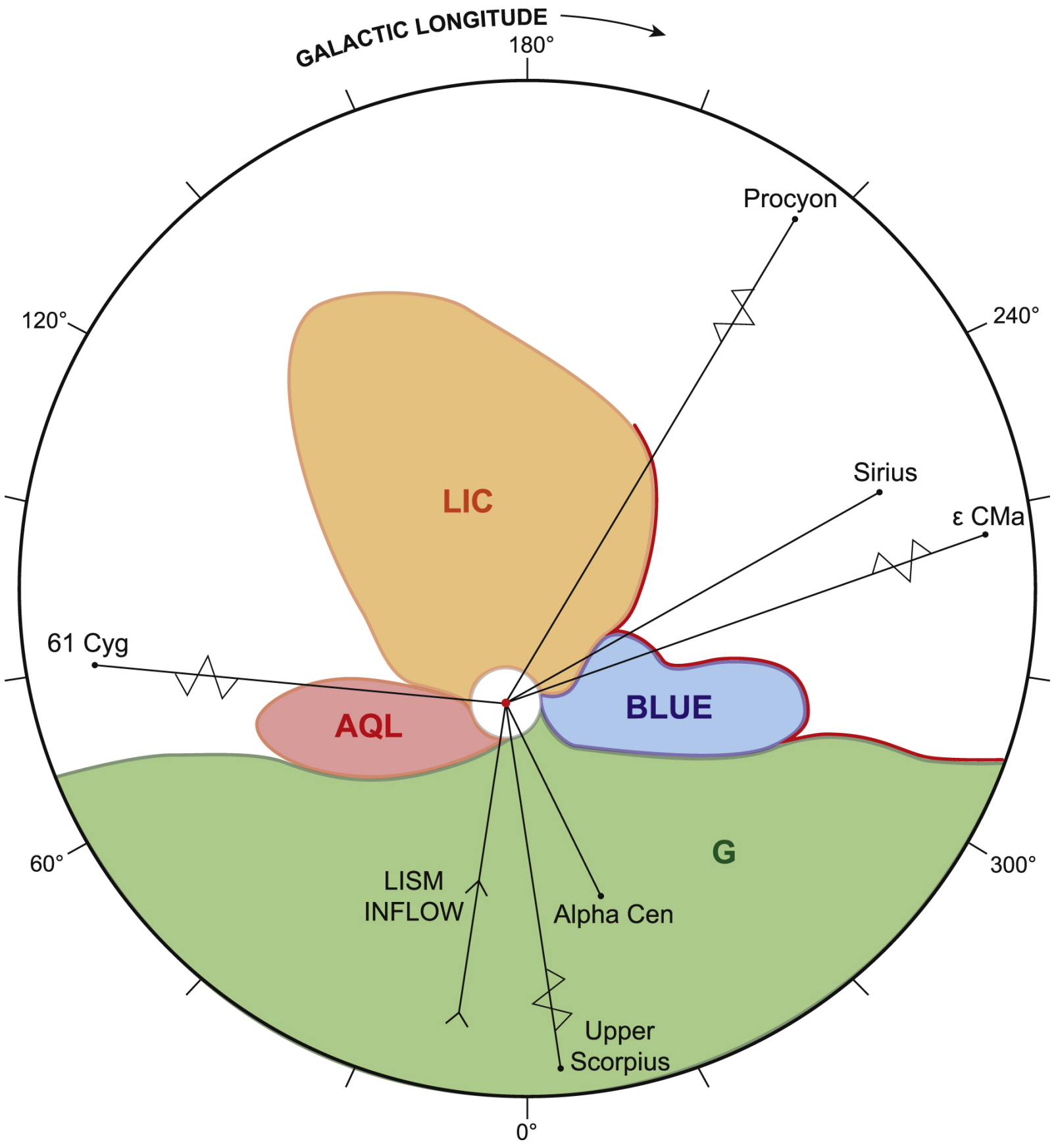

Figure 3.29. The local ISM region within 3 pc of the Sun as viewed from the north Galactic pole, showing the location of the four partially ionised clouds that are in contact with the outer heliosphere. Not shown are other clouds lying outside the four clouds. Shown are the Sun (point), an exaggerated representation of the heliopause (circle around the Sun) and the LIC, G, Aql and Blue clouds. Lines of sight projected onto the Galactic equator are shown for five stars. Red shading shows the Strömgren shells produced by EUV radiation from ε CMa. Also shown are the directions of inflowing interstellar gas as seen from the Sun and the direction to the Upper Scorpius region of the Scorpius–Centaurus Association, where the most recent supernovae likely occurred. From Linsky et al. (2019).

Our immediate local interstellar neighbourhood has so far been largely unexplored: only two spacecraft have crossed the boundary to interstellar space and sent back in-situ data: Voyager 1 and Voyager 2.

The classical view of the heliosphere is a streamlined bubble of ionized gas that interferes with its local surroundings. If the LISM were supersonic, a bow shock would form a few hundred AU upwind of the heliosphere where the partially ionized interstellar gas becomes subsonic. Its existence depends on the LISM parameters and on the velocity of the Solar System through the LISM and is controversial. Scherer & Fichtner (2014) argue for a weak bow shock; Zieger et al. (2013) suggest only a slow shock is present; McComas et al. (2012) think that there is no shock but only a gradual bow wave. The Voyager 1 and 2 lifetimes are far too short to reach the bow shock. The heliopause is the boundary between the region in space dominated by interstellar material and the region dominated by material coming from the Sun. The heliopause is located where the pressures from the ISM and from the solar wind are in balance. It moves inward and outward as the solar wind pressure changes over the solar cycle. Voyager 1 crossed the heliopause in August 2012 at about 122 AU, between solar minimum and solar maximum (Burlaga et al., 2013; Gurnett et al., 2013; Krimigis et al., 2013; Stone et al., 2013), while Voyager 2 crossed it in November 2018 at about 119 AU, between solar maximum to solar minimum (Richardson et al., 2019; Stone et al., 2019; Krimigis et al., 2019;

Burlaga et al., 2019; Gurnett & Kurth, 2019). Before the heliopause is the termination shock, where the solar wind is slowed down to subsonic speeds (see Fig. 3.30). Voyager 1 crossed the Termination Shock at about 94 AU and Voyager 2 at about 84 AU.

Because of the motion of the heliosphere through the interstellar medium, the neutral interstellar H, He and the dust pass through the Solar System. The interstellar neutral gas that flows into the heliosphere forms pick-up ions when it is photo-ionized by UV radiation from the Sun or through charge exchange with solar wind particles. These ionized particles are then accelerated to (on average) the solar wind speed and have thermal energies equal to the solar wind energy. Pick-up ions that cross the termination shock were thought to be accelerated and become Anomalous Cosmic Rays (ACR's), although the Voyagers did not find evidence of this acceleration when they crossed the Termination Shock and the acceleration mechanism remains to be understood (Opher, 2016). When the LISM atoms charge exchange with the solar wind to form pickup ions, the former solar wind ions (now neutral) move out of the Solar System. In a two-step process, they are first ionized by charge exchange with LISM neutrals outside the heliosphere, then charge exchange again with LISM ions to become energetic neutral atoms (ENAs). Some of these ENAs move sunward back into the heliosphere where they could be observed by IBEX (Interstellar Boundary EXplorer) and Cassini and used to probe the plasma conditions (and magnetic fields) beyond the Solar System. It is debated how many ENAs originate from the heliosheath and how many from beyond the heliopause (Opher, 2016). The Earth's magnetic field protects us from the local Galactic Cosmic Rays (GCRs), but the heliosphere magnetic field shields us from a large percentage of these very energetic particles before they reach Earth. Changes in the solar wind with the solar cycle cause variation in both the number of GCRs and interstellar dust particles that pass through the Solar System. Dust particles smaller than about 0.015 micron are filtered out completely from the heliosphere because of Lorentz forces on these dust particles that have high charge-to-mass ratios. Mid-sized (around 0.3 micron) interstellar dust particles have charge-to-mass ratios that are low enough to enter the heliosphere and then be filtered out by Lorentz forces in the solar wind that vary with the solar cycle (see Sterken et al., 2019, for a review of interstellar dust in the Solar System). Large micron-sized particles have low charge-to-mass ratios and are not significantly affected by the magnetic fields in the heliosphere.

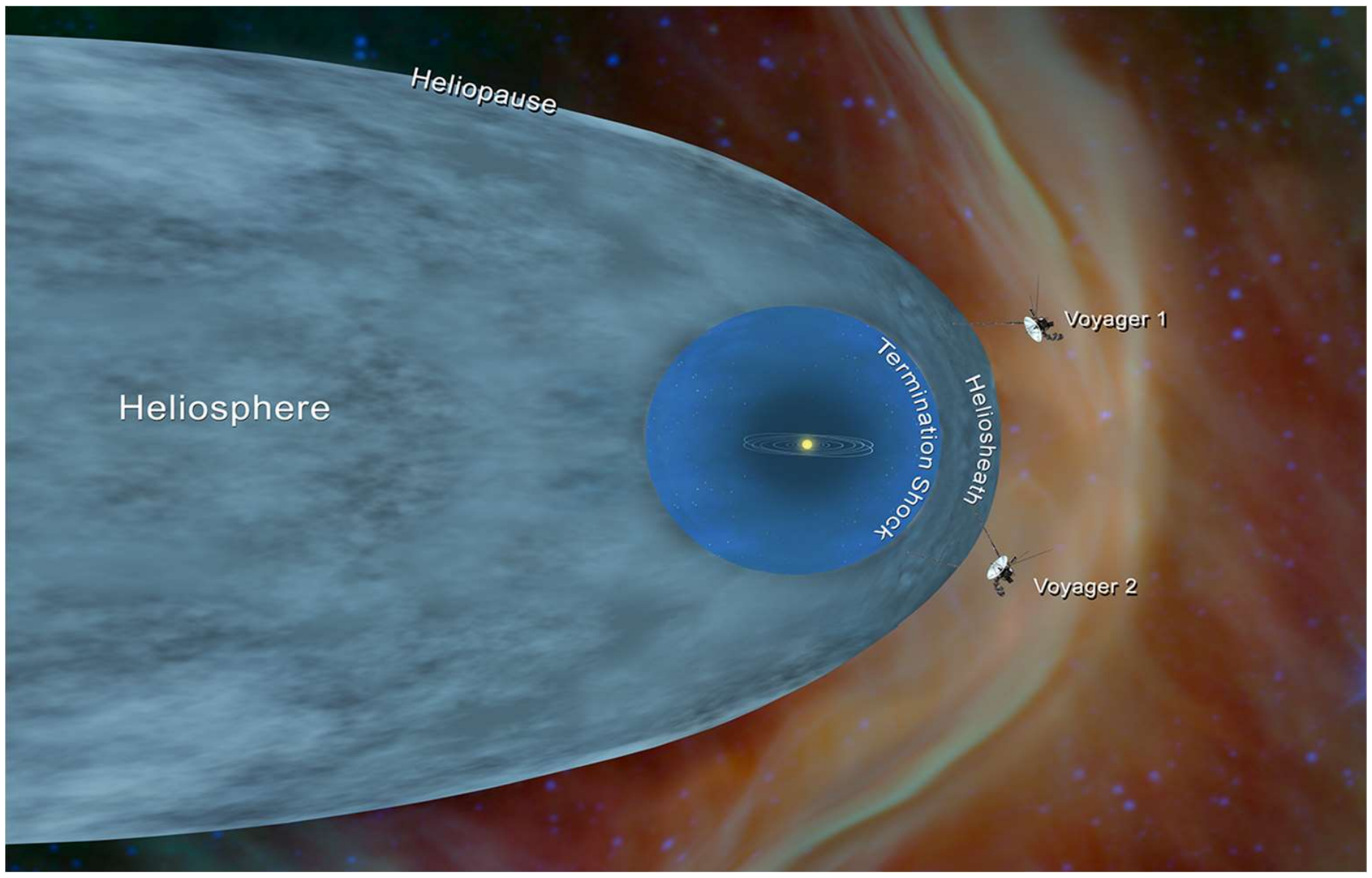

Figure 3.30. Artist impression of the heliosphere, its different regions (Heliosheath, Termination Shock, Heliopause) and the two Voyager missions. The shape of the heliosphere is still under discussion. © NASA/JPL-Caltech.

## Major open questions

Apart from the above-mentioned unresolved questions of the ACR acceleration mechanism, the origins of ENAs and the existence of the bow shock, other major open questions about the heliosphere still exist. The Voyager missions have explored uncharted territory, leading to many new discoveries and to surprising challenges in understanding our immediate interstellar neighbourhood. One such surprise was that (1) the heliosheath turned out to be much thinner than theoretical models would predict, indicating our relatively limited knowledge on the interaction mechanisms between the heliosphere and its surroundings. Also, (2) the magnetic field direction was expected to change dramatically towards the interstellar magnetic field direction as the Voyagers crossed the heliopause, but surprisingly it did not. (3) The interstellar magnetic field strength and the direction of the local interstellar medium are also not yet determined: remote measurements in the Solar System using Energetic Neutral Atom measurements at different energies from the IBEX and Cassini missions, yield a derived field strength and direction that is different from the one derived from models of the Voyager in-situ data. (4) The dust particles from interstellar space have been measured in-situ inside the Solar System using dust detectors on-board of the Ulysses, Galileo and Cassini missions. The dust mainly consists of silicates. The presence of carbon in the local interstellar dust is subject of debate. The dust dynamics have not yet been completely understood: the heliosphere magnetic fields block the smallest particles from entering the heliosphere and filter the particles from a few tens of nanometres to half a micron intermittently in a solar-cycle dependent way. However, the complete dynamics are not yet understood as computer simulations of interstellar dust moving through the heliosphere cannot yet fully explain the in-situ observations of the dust (Sterken et al., 2019). Probably the most intriguing unresolved research question is (5) the shape of the heliosphere in which we live. The current benchmark model is a comet-shaped streamlined heliosphere (e.g., Parker, 1961; Baranov & Malama, 1993). However, ENA data from Cassini's Imaging Neutral Camera (INCA) suggest a spherical shape for the heliosphere (Dialynas et al., 2017), while modelling studies, taking into account the solar wind magnetic field, lead to the prediction of a croissant-shaped heliosphere (Opher & Drake, 2013; Opher et al., 2015; Opher et al., 2020) and finally MHD/neutrals models (Pogorelov et al., 2015) produce a more elongated, jellyfish-shaped heliosphere. Other open questions are why the plasma flows are so different at the two Voyager spacecraft, what is the nature (e.g., porosity) of the heliopause, why and to what extent the disturbances in the plasma from the Sun are still present in the outer heliosphere and LISM and what the role of the heliosphere and the VLISM is in the evolution of stellar and planetary systems?

These research questions ultimately lead to fundamental considerations in space plasma physics and are ultimately important for studies of astrospheres as well. Opher (2016) provides a more in-depth review of these challenges (except for the dust, see Sterken et al., 2019).

## Future missions

The Voyager missions and their instruments were tailored primarily for the (inter)planetary research phase of the mission, they represent 'only' two measurement 'lines' in space at two specific time periods in the solar cycle and their energy provision will be depleted around 2028, so they will not provide observations of the "undisturbed" interstellar space beyond the bow shock/wave. The Voyager 1 plasma instrument failed at 10 AU (but some plasma densities in the LISM were derived using the plasma wave instrument) and the two Voyagers do not have an instrument suite onboard to measure the important particle energy range for particles from 6 keV to 30-40 keV, which constrains the pick-up ions. Also, the ambient spacecraft field makes it difficult for the magnetometer to measure the weak fields in the outer heliosphere and interstellar space. Finally, yet importantly, although some impacts of dust on the spacecraft body have been registered by the plasma wave instrument (Gurnett et al., 1997), the Voyagers do not carry a dedicated dust instrument. There are no in-situ measurements of dust impacts near or beyond the Termination Shock to provide reliable estimates for the dust mass and impact velocity, or for obtaining the chemical composition of interstellar and outer Solar System dust with time-of-flight mass spectrometry.

New in-situ explorations of the immediate interstellar neighbourhood inside and outside of the heliosphere are necessary to solve these open questions. To do so, a synergy of measurements beyond the heliopause as well as 'remote' observations from spacecraft within the Solar System of material coming from interstellar space should be obtained and exploited. Remote measurements of our own heliosphere by a spacecraft that has reached far distances may also open up surprising new views of our own 'home' in interstellar space. Such missions should ideally have instruments onboard that are tailored for the in-situ exploration of the heliosphere, the interaction regions between the heliosphere and the interstellar medium and interstellar space itself. Key measurements on such a mission are plasma properties in all energy ranges (including 1 keV–40 keV), magnetic field measurements with sufficiently high sensitivity and precision for the weak fields in the outer heliosphere and interstellar medium, measurements of galactic cosmic rays and especially dust particle number, mass, velocity, and composition. Small (a few tens of nanometres) interstellar dust particles cannot enter the heliosphere but are likely most abundant in number density outside of the heliosphere. Carrying a dust detector out of our home "bubble" would thus provide new compelling astrophysical information about our local interstellar neighbourhood. Measuring dust in interstellar space beyond the heliopause would thus be a tremendous step forward to study the interstellar dust in a diffuse cloud thanks to unique in-situ, or 'ground truth' information. Also, measuring the time variation of the dust flux throughout the mission's lifetime would provide understanding on the modulation of the dust flow by the heliospheric magnetic fields. Such a mission should aim to at least reach undisturbed interstellar space (~ 500 AU), implying strict requirements for telemetry and power provision.

## 5.7. Small body hazards and space awareness

Asteroids, when their orbits become close to the Earth's orbit, present an important threat to humankind that systematic observations can help to predict and possibly to mitigate: this is the field of Space Awareness (Planetary Defense). When the 21 fragments of the Comet Shoemaker Levy were observed to impact Jupiter in July 1994, this was the first time that human eyes could directly observe such an event and have a direct evidence that impacts keep occurring in the Solar System, with potentially major consequences.

Impactors of terrestrial planets are small bodies belonging to the Near-Earth Object (NEO) population. Most of them come from the asteroid belt, between Mars and Jupiter, where dynamical resonances can increase their orbital eccentricity so that they are transported from a circular orbit in this region to an orbit crossing that of terrestrial planets, including Earth. The majority go directly into the Sun as their eccentricities keep increasing on a Myr timescale. A small fraction of these bodies is captured on their way from the resonance by a planetary close approach and can evolve deeply into the Near-Earth space where they are mostly perturbed by planetary encounters and resonances with the terrestrial planets. Some of them ultimately collide with a planet, while others are eventually re-injected in a main resonance that drives them into the Sun or beyond Jupiter if they encounter the giant planet at their aphelion. The median lifetime for NEOs is about 10 Myr.

Studies of lunar craters indicate that the impact flux in the inner Solar System has been constant on average over the last 3 billion years, although it was higher prior to that. The impact rate on Earth can be calibrated from that on the Moon (e.g., Zellner et al., 2017) since that body retains the best record of the impact history in the inner Solar System as, contrary to the Earth, the Moon does not have oceans, or plate tectonics, winds and other active processes that erase craters over time. This constant impact flux means that despite the short lifetime of individual NEOs, the whole population is maintained in a steady state so that the impact flux does not decrease over time. In fact, in the asteroid belt, asteroid collisions continuously occur, which generates new bodies, and dynamical mechanisms such as the Yarkovsky effect can make these fragments slowly diffuse into one of the resonances that can transport them to the near-Earth space.

Analysis of the impact flux and numerical modelling of the NEO population has allowed estimation of the average impact frequencies of those bodies on the terrestrial planets. It is thus estimated that an object 10 km in diameter (threshold for extinction of species) impacts our planet every few 100 Myr, while an object of 1 km in diameter (global damage threshold) impacts Earth every Myr and an object of about 140 metres in diameter (regional damage threshold) collides with Earth every few 10,000 years (Granvik et al., 2018). On June 30$^{th}$, 1908, an object exploded over the Tunguska forest in Siberia, which flattened 2000 square kilometres of forest. Given the estimated size of the object (about 50 metres), the frequency of this kind of event is estimated about 1000 years. More recently, on February 15, 2013, a 17 m-diameter object exploded over the city of Chelyabinsk with an energy equivalent to 500 kilotons of TNT (about 30 Hiroshima bombs), causing injuries to 1000 people. Such an event occurs on average every century and yet, this is the first time in recorded history that it has occurred over an area where humans were present. This is because most of the Earth is covered by water and desert and, therefore, the likelihood that such a local event occurring over a populated area remains extremely small.

Thus, the risk of asteroid impacts is one of the least likely natural disasters. However, contrary to other more likely disasters, such as earthquakes, tsunamis, and volcanoes, we can predict them and prevent them with means that are feasible and reasonable. Moreover, given the impact frequencies, we know that on the more or less long term, we will face the reality of this threat. In addition, to date, we know only ~30% of the population of NEOs larger than 140 m and we cannot guarantee that there is no object coming our way, even if the probability remains very small. As for all risks with low probability but huge consequences, it is wise to be prepared before we need it, especially since we have the capability to do so.

In complement to the scientific study of asteroids and other potential Earth-grazing small bodies, deployment, maintenance and operation of a combined Earth-based and space-based observation system capable of identifying long in advance collisions with these near-Earth bodies is mandatory. Even more, studies of potential space-based techniques dedicated to the mitigation of small-body threats need to be pursued. By 2061, one should expect the full deployment of a space-based monitoring and mitigation system. The current perspectives of this space system will be described in Chapter 4.

In order to be prepared for a real mitigation attempt, significant further studies are required. In parallel with ground-based detection programs, space-based observatories (optical and/or infrared) are required to detect potential impactors, particularly those in orbits largely interior to the Earth. Follow-up observations are required to provide basic characteristics (size, shape, spin, spectra diagnostic of composition) but properties relevant to impact mitigation such as variation of physical and thermal properties and surface structure, require in-situ spacecraft, as that knowledge is not measurable from Earth. For instance, a fleet of low-cost spacecraft could provide multiple flybys to characterize the full range of taxonomic and physical types, in different size regimes, inferred from observations. A dedicated tomography and physical properties mission, including deployable lander(s), would provide ground-truth for subsurface structure and physical properties. Both kinds of missions were already described in the ESA NEOMAP (Near-Earth Object Mission Advisory Panel) report (Harris et al., 2004). A second kinetic impactor mitigation test mission would allow a direct test of predictive capability (i.e. the ability to deflect an object, with characteristics defined by remote observations, by a specified amount, based on models validated with NASA's Double Asteroid Redirect Test (DART), ESA's Hera mission and laboratory studies). We additionally note that missions whose requirements are driven by planetary defense objectives also have a high science return, as they contribute greatly to the scientific characterisation of small asteroids and to the scientific understanding of their response to external actions. In particular, collisions play a major role in all phases of the Solar System history and need an improved understanding at the actual scale of those objects, which is by far larger than the scale of objects used in terrestrial laboratories. Deflection tests using the kinetic impactor technique offer fully documented impact experiments at asteroid scale that can feed with more reliable parameters for collisional evolution models of small body populations.

# 6. Potential habitats in the Solar System

A simple concept of habitability considers three major requirements: an energy source, the availability of nutrients and the presence of a solvent (e.g., Southam et al., 2015). The last requirement is often reduced to the necessity of liquid water since water is the essential solvent sustaining life as-we-know-it on Earth (e.g., Cockell et al., 2016). The three requirements may be met on the surface of a planet and/or at shallow depths beneath the surface. They may also be met in an ocean beneath an ice layer, which has led to speculation about life in Enceladus and Europa (see e.g., Lammer et al., 2009; Coustenis & Encrenaz, 2013, for reviews). A combination of tidal heating and radioactive decay may provide sufficient energy to keep both satellite's subsurface oceans liquid (e.g., Spohn & Schubert, 2002; Moore & Hussmann, 2009, and references therein; Iess et al., 2014). Furthermore, the internal heat flow may result in hydrothermal vents, which could provide an energy source for potential ocean life (Barge & White, 2017).

Although the processes leading to the emergence of life are still debated, models suggest that thermodynamic disequilibrium, as may be found at or near active volcanoes, is essential (e.g., Westall et al., 2018), making the subsurface oceans of icy moons particularly interesting candidates for potential life in the Solar System.

Enceladus is the moon of Saturn on which water plumes have been observed with the Cassini mission (Porco et al., 2006). Evaluation of Enceladus' habitability involves understanding nearly all other aspects of Enceladus related to the potential presence of liquid water, either in a subsurface ocean or in the plume vent regions. A dedicated mission to Enceladus would enable a better understanding its habitability and dynamics as well as search for evidence of life.

Missions to Europa and Enceladus that probe the subsurface have been considered (e.g., Weiss et al., 2011; Konstantinidis et al., 2015) but would require landers or penetrators, which are typically associated with high-cost missions. However, chemical analysis of particles that have been ejected by cryovolcanism may also allow inferences about biochemical processes within these oceans (McKay et al., 2008; Postberg et al., 2018). The search for life at icy ocean worlds is more challenging than at planets in the habitable zone, if only for the large distances to these bodies and the difficulties of identifying biomarkers in ocean-derived materials, yet offset by Europa and Enceladus being the best known targets for extant life in our Solar System.

Liquid water on a planetary surface requires suitable temperature and pressure conditions. These are mainly set by the stellar luminosity, the orbital distance to the star and by the atmosphere and its greenhouse gases (Kasting et al., 1993). The evolution of the atmosphere is closely linked to the planetary interior structure and dynamics, with atmospheric carbon dioxide that has been outgassed by volcanoes making up the largest fraction of molecules in the atmospheres of Mars and Venus (de Pater & Lissauer, 2015). On Earth, the long-term carbonate-silicate cycle includes the transfer of carbon from the Earth's mantle to the surface reservoirs of the oceans and the atmosphere and the recycling back into the mantle in subduction zones (e.g., Kasting & Catling, 2003). For efficient recycling, a low temperature gradient geotherm along the subducting plate is key (Kerrick & Connolly, 2001) to keep the carbonates from decomposing before they become mixed into the mantle. Subduction zones on early Earth were probably hot and recycling was therefore rare, so that $CO_2$ accumulated in the atmosphere (Dasgupta & Hirschmann, 2010). The existence of a plate-tectonics planet (Earth) in the Solar System suggests that there should be similar planets in other planetary systems but the likelihood of their occurrence is not at all clear. Although the rheology of rock and its dependence on temperature and volatile presence and concentration seems to be key, the exact conditions for plate tectonics to occur are debated (e.g., O'Neill et al., 2007; Noack & Breuer, 2014), even on Earth, and even the effect of planetary size has been discussed controversially (Valencia et al., 2007; Kite et al., 2009).

In addition to ensuring climate stability via the long-term carbonate silicate cycle, there are other reasons why plate tectonics should be beneficial for the evolution of life. For example, this tectonic mode implies a large mantle heat flux when compared with stagnant lid planets that allows for a more effective cooling of the core and the maintenance of the magnetic field (e.g., Stevenson et al., 1983). The presence of a magnetic field is generally held to be important for the protection of life against harmful radiation and to protect the atmosphere from erosion by the solar wind (Driscoll, 2018). Another potential consequence of plate tectonics is Earth's bimodal hypsometry with a substantial reservoir of water in ocean basins and continents that provide life with easy access to solar energy (e.g., Höning & Spohn, 2016). Moreover, plate tectonics could recycle crust and thereby continuously rejuvenate the most fundamental element – rock – of the crust.

Active surface recycling into the mantle by plate tectonics potentially implies that the surface biosphere plays a role in subduction zone processes and the evolution of the solid Earth (Sleep et al., 2012). It has even been speculated (e.g., Rosing et al., 2006) that the mere presence of continents may be due to bioactivity and, hence, be a biosignature, while Höning & Spohn (2016) maintain that bioactivity may have prevented the Earth from becoming either a desert planet largely covered by continental rock or an ocean planet largely covered by water, even in the presence of active plate tectonics. Clearly, the surface biosphere plays a significant role in the energy budget of the Earth's system by capturing solar energy through photosynthesis (Kleidon, 2016). But it takes plate tectonics to then cycle the captured energy within the bulk of the planet.

A major effect of the biosphere on planetary habitability is biological enhancement of silicate weathering, as a key component of the carbonate-silicate cycle that helps to regulate the atmospheric CO2 concentration (Schwartzman & Volk, 1989, Berner, 1992). Since biological primary productivity increases with temperature, a negative feedback to increasing incident insolation is established, which results in a prolongation of the habitable period of inhabited planets (Lenton & von Bloh, 2001, von Bloh et al., 2007). Chopras & Lineweaver (2016) argued that a particular early ability of Earth's biosphere to shape its environment is necessary in order to prevent a planet from becoming uninhabitable, Höning (2020) showed that an active biosphere on land and in oceans does not only stabilize the climate against increasing incident insolation but furthermore weakens climate oscillations over a large range of timescales (see Fig. 3.31).

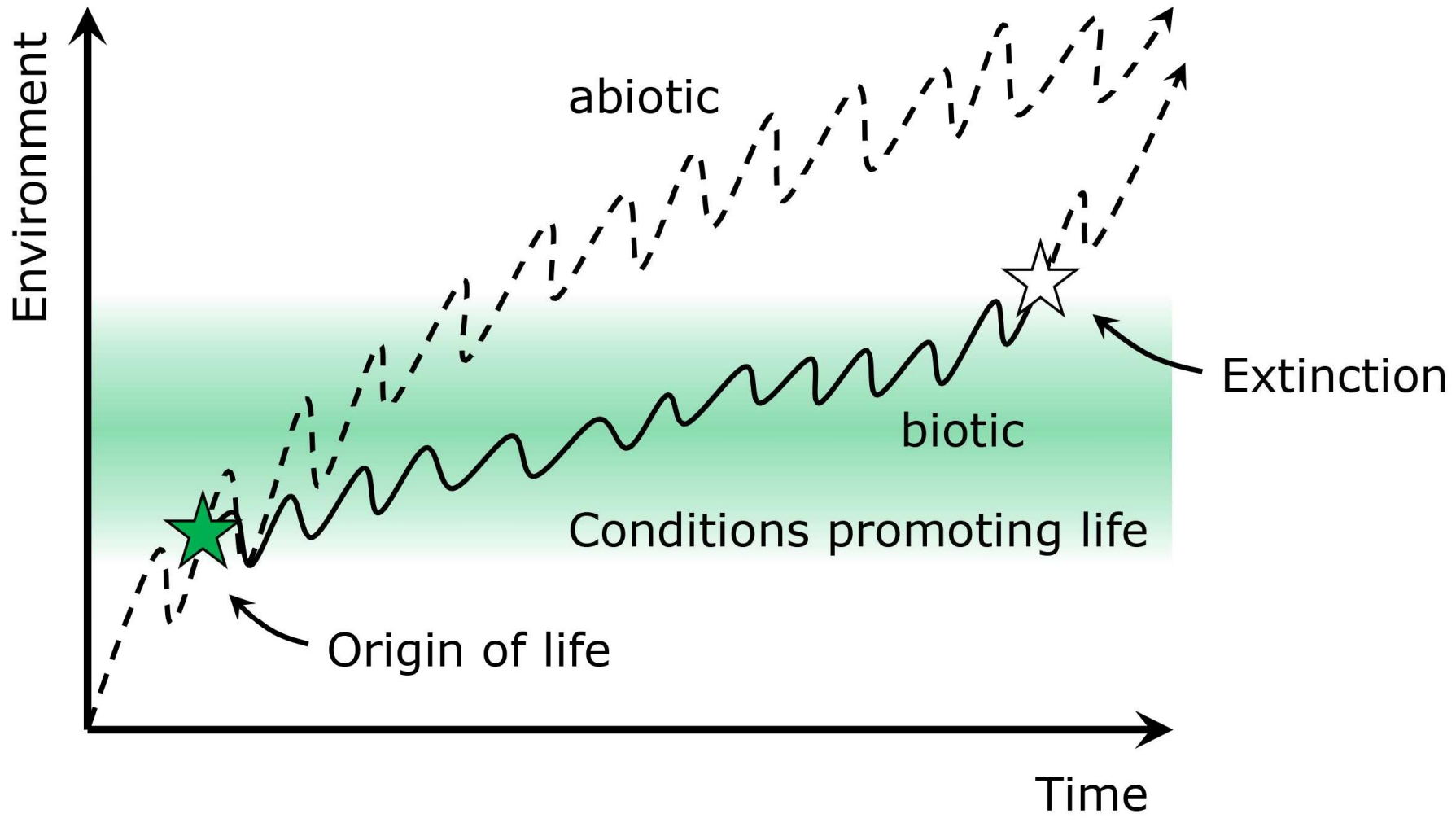

Figure 3.31. Gaian regulation of the habitability of a planet. While the conditions for habitability (green area) stay constant over time, environmental conditions will change and may evolve away from habitability (dashed curve, Venus, Mars). This will happen if the active biomass does not become widespread and massive enough to affect the planetary environment and stabilize habitability. On the contrary, an inhabited planet will remain habitable for a longer period of time (solid curve). From Chopras & Lineweaver (2016).

The conjecture of plate tectonics being essential for life has been challenged. Primitive life may find local (small-scale) energy and chemical cycles that could be capable of sustaining it, at least for primitive life forms (e.g., Westall et al., 2013, 2015) in local habitats. After all, models such as those of Tosi et al. (2017) and Foley (2019) have suggested that volcanic outgassing of carbon dioxide could result in a temperate climate for Earth-sized stagnant-lid planets. Temperature-dependent and $CO_2$-dependent weathering may regulate the climates of these planets to some extent (Höning et al., 2019), although recycling of carbon back into the mantle would be strongly limited.

Two questions pose themselves in that context:
1. Could life flourish or even exist for extended periods of time on a planet without plate tectonics? If so, what are the necessary tectonic components or processes that would still be required?
2. What causes a planet to have plate tectonics?

Solving the question for life on Mars would help to answer the first question. It can be hoped that establishing whether or not there is extinct or extant life on Mars will be possible through the planned ExoMars in situ rover mission (Vago et al., 2017), as well as ESA/NASA's sample return mission scheduled for returning samples in ~2032 (e.g., Mattingly and May, 2011) and/or through missions by other space faring nations. There is no other planet in the Solar System that would be as well suited to answer this question as Mars, where the surface environmental conditions are the closest of all Solar System bodies to the conditions on the Earth. Mars will help us to understand whether or not plate tectonics is required to maintain habitable conditions in the long term. The latter problem would require a determination of the duration for which life has been active on Mars as well as its evolutionary stage. Even if the planned missions fail to find traces of life at the surface of the planet (or down to a depth of 2m, the length of the sampling drill on the ExoMars rover), it will be difficult to prove the absence of life definitively, especially since we know that there is a huge microbial biomass existing within the crust of the Earth (Magnabosco et al., 2017) and, potentially, within Mars' crust (Michalski et al., 2013).

Should life – extinct or extant – be found on Mars, one would, of course, be interested in its detailed interaction with the planet, requiring fieldwork as is done on Earth. Projects such as the German project "Earth Shape" (Oeser et al., 2018) exploring the interaction of microbial life with rock would be important. Since on Mars the habitat – if any – might be found underground, the technical hurdles would be even higher. The fieldwork could be done using rovers or crawlers that would need to be capable of entering difficult terrain such as steep slopes – where, for instance, liquid water can be maintained for some time in "gullies" despite its thermodynamic instability – and caves.

A better understanding of Venus' climate evolution would also provide insights into the necessity of plate tectonics for climate regulation. Even though Venus' surface is too hot today, it may have been habitable earlier in its history (e.g., Way et al., 2016; Way and Del Genio, 2020). A knowledge of rock alterations of Venus' crust could be helpful in order to spot potential rock-fluid interactions (e.g., Zolotov, 2019) and therefore to better constrain its climate evolution.

Determining what causes the emergence of plate tectonics and its maintenance over geological time is an issue of fundamental importance not only to understand the way our planet operates, but also to assess whether or not this process, with all its crucial consequences, can occur on other rocky bodies. From a mechanical standpoint, Earth's surface is characterized by wide, thin and largely rigid plates, separated by narrow and weak boundaries where most of the deformation is concentrated. These ultimately allow subduction and surface recycling to occur. The strong temperature dependence of the viscosity of rocks naturally leads to the formation of a stagnant lid, the continuous rigid plate that currently occupies the upper layers of all known terrestrial bodies other than Earth, with the possible exception of Venus. The latter can also have a stagnant lid as Earth or other planets when considering the sluggish lid hypothesis (Foley, 2018). In order to generate tectonic plates, one or more mechanisms inducing weakness in the otherwise rigid surface and lithosphere are necessary. The physics underlying such weakening mechanisms is still not fully understood. Nonlinear pseudo-plastic rheologies are often considered for their simplicity and ability to generate plate-tectonics-like behaviour (e.g., Trompert & Hansen, 1998; Tackley, 2000). These assume the lithosphere to have a finite strength – the so-called yield stress – and to fail when such a critical stress is exceeded, for example, because of the arrival of a plume at the base of the lithosphere (Gerya et al., 2015) or because of sub-lithospheric stresses generated by mantle convection (e.g., Crameri & Tackley, 2016). Yet, the critical yield stress that allows plate-like behaviour is typically much lower than expected from laboratory experiments, suggesting that further weakening is necessary. The presence of water has been considered as a potentially important factor to reduce lithospheric strength and favour the generation of plates (Regenauer-Lieb et al., 2001; Korenaga, 2007). A more sophisticated theory based on the interplay between multiple solid phases and on the weakening effects induced by grain size reduction in shear zones has been developed during the past two decades through a vast amount of work by Bercovici and Ricard (see e.g., Bercovici & Ricard, 2014, for a synthesis; Foley, 2018). This theory successfully explains not only important first-order features of Earth's plate tectonics, but also why Venus' high surface temperature tends to cause rapid healing of damaged lithosphere, preventing development of Earth-like plate tectonics. Nevertheless, the theory relies on a number of parameters that are difficult to constrain in the laboratory, suggesting that more experimental and theoretical work is needed to make progress in the search for the ultimate causes of the origin of plate tectonics.

Earth's geochemical record also provides insight into suitable conditions for plate tectonics. Note, however, that much of the interpretations are based on analyses of ancient zircons, the majority of which are not as old as previously thought, i.e. inferences regarding Hadean crustal composition and formation of protocontinental crust need to be taken with caution (Whitehouse et al., 2017). In particular, felsic continental crust is thought to be a product of subduction (e.g., Tang et al., 2016) and reconstructing Earth's history of continental growth can place constraints on the onset of plate tectonics. In this respect, Rosas & Korenaga (2018) argue for an early rapid production of continental crust, which would imply that surface recycling was already present in the Hadean. Similarly, Arndt & Nisbet (2012) claim that early plate tectonics had already produced a significant fraction of today's continental crust by the Mid-Archean. These two studies, however, do not take into account the reservations noted above (Whitehouse et al., 2017). On the other hand, Cawood et al. (2018) among others find geochemical evidence for a change in the tectonic style between 3.2 and 2.5 Ga, which would imply that a cooler mantle might be required for plate tectonics to initiate.

Certainly, a comparison between Venus and the Earth will provide additional insights into the causes of plate tectonics. Subduction has been proposed for Venus (Ghail, 2015; Davaille et al., 2017). However, obtaining details on the subduction style, such as the steadiness of potential subducting plates, would require seismic measurements. Geophysical exploration of Venus as well as continued study of the Earth would help to improve our understanding of the requirements for plate tectonics. On the longer term, it may be possible to identify signatures of plate tectonics or large-scale recycling processes on exoplanets. For example, the atmospheric spectra of an exoplanet have been argued to depend on the presence of geochemical cycles (Kaltenegger & Sasselov, 2009). Furthermore, it is possible that an active magnetic field would be promoted by plate tectonics (e.g., Nimmo, 2002). If future telescopes will allow for a characterisation of surface albedo features on exoplanets (Stam, 2008; Fan et al., 2019; Kawahara, 2020), the presence of continents – if detectable – would be a strong indication of plate tectonics.

In summary, if plate tectonics is important for the long-term presence and evolution of life on a terrestrial planet and, as long as plate tectonics is known only for Earth, much research needs to focus on our home planet to study planetary interactions with life and the causes for this tectonic mode. This must include the interplay between the biosphere and the planetary interior, biogeochemical studies of biologically produced and altered sediments, as well as petrologic and numerical studies of their role in subduction zones. Nevertheless, it will be important to investigate other planets and moons that appear to show either lateral tectonic movement, such as Europa and Venus, or that are highly energetic, such as Io, to determine how they differ from Earth or are similar in one or other respect. What is more, as long as the question of extraterrestrial life remains unanswered, the search must continue on all bodies that are considered. Understanding of their tectonics is also necessary with respect to long-term habitability.

As explained in the paragraph concerning interaction processes and evolution of terrestrial objects of the Solar System, habitability must be considered in a general sense. The role of interior evolution and its interaction with potential atmospheres is important for bodies having an atmosphere with the possibility of surface liquid water. Water can also be found in the subsurface. This is the case for Earth and might also be the case for colder bodies like Mars. Radar subsurface sounding has potentially observed liquid reservoirs below the surface of Mars (see Fig. 3.32) and will be a key technique used to characterize the ice shells of Europa and Ganymede, including the search for shallow liquid water, by the Europa Clipper and JUICE missions.

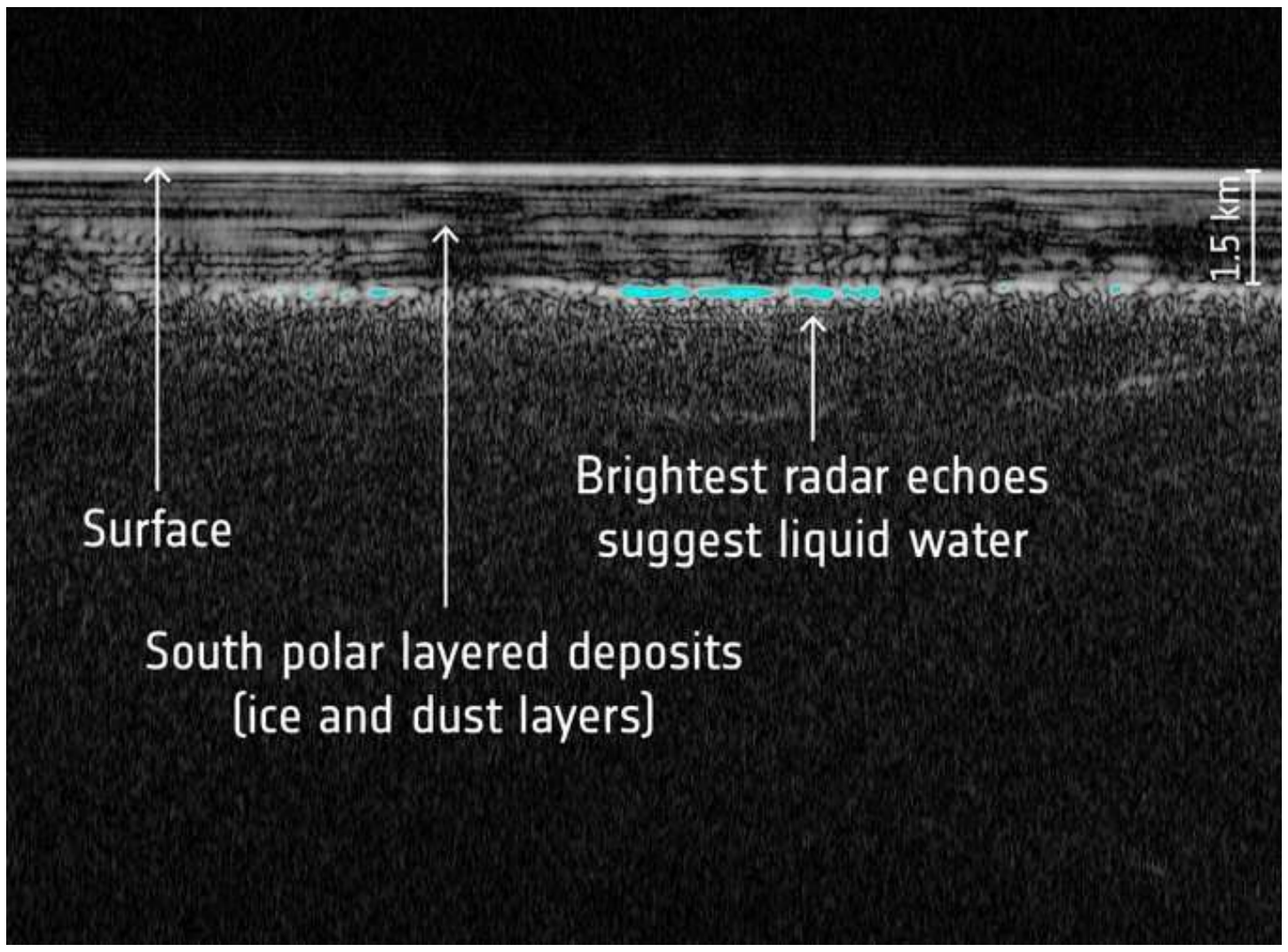

Figure 3.32. Radar sounding of planetary interiors can detect subsurface structure and deposits. Ground-penetrating radar indicates a layer of liquid water below ice and sediment near the south pole of Mars. Credit: ESA/NASA/JPL/ASI/Univ. Rome

# 7. Detection of life – Strategies for the detection of biosignatures in the Solar System

A series of missions from ESA and NASA are planned that will open a new era of astrobiology exploration of the Solar System. They include missions such as Mars 2020-Perseverance, ExoMars-Rosalyn Franklin, Dragonfly and the still under discussion NASA Europa lander and Enceladus Orbiter. While the Mars 2020 mission is tasked with taking sample that might contain traces of life, the other missions have explicit objectives to search for life considering the principal characteristics of the potential habitable environment of each body. Half a century ago, the ambiguous results from the Mars Viking landers demonstrated the difficulty in designing experiments and interpreting findings related to the search for life. The great importance of understanding the environment was recognized and, as a consequence, a reorientation to a more cautious period of exploration was initiated, in which characterising the context was given priority. Life signatures based on carbon chemistry in water-based media defined the parameters that could be detected by the available technology for astrobiology missions to Mars and icy moons (see previous sections). Carbon is ubiquitous in the Universe and has unique capabilities to establish stable covalent bonds with other abundant elements (particularly HONPS), leading to the synthesis of millions of organic molecules. Water is a perfect solvent and also very abundant. It plays an active role in a number of chemical reactions, offers protection against radiation and provides a favourable environment for carbon-based chemistry to develop (Westall and Brack, 2018).

The current search for life is availed by the evolving knowledge regarding the limits of life on Earth, the outcomes of science related to the biosignatures, and the technological developments of analytical instruments qualified for space use. In 1994, the NASA Exobiology Discipline group adopted a working definition of life that provide some perspective for space exploration: "Life is a self-sustaining chemical system capable of Darwinian evolution". Having in mind that the main property of this particularly complex chemical system is the combination of compartmentalisation, self-replication and metabolism, many astrobiologists suggest basing the search for life on these functions that characterize biochemical processes (see Fig. 3.33). To find traces of life in past environments can be very challenging as there exist morphological or chemical abiotic processes that mimic or alter them (see Javaux, 2019), but clay minerals provide the highest fidelity of preservation (see, e.g., Wacey et al., 2014).

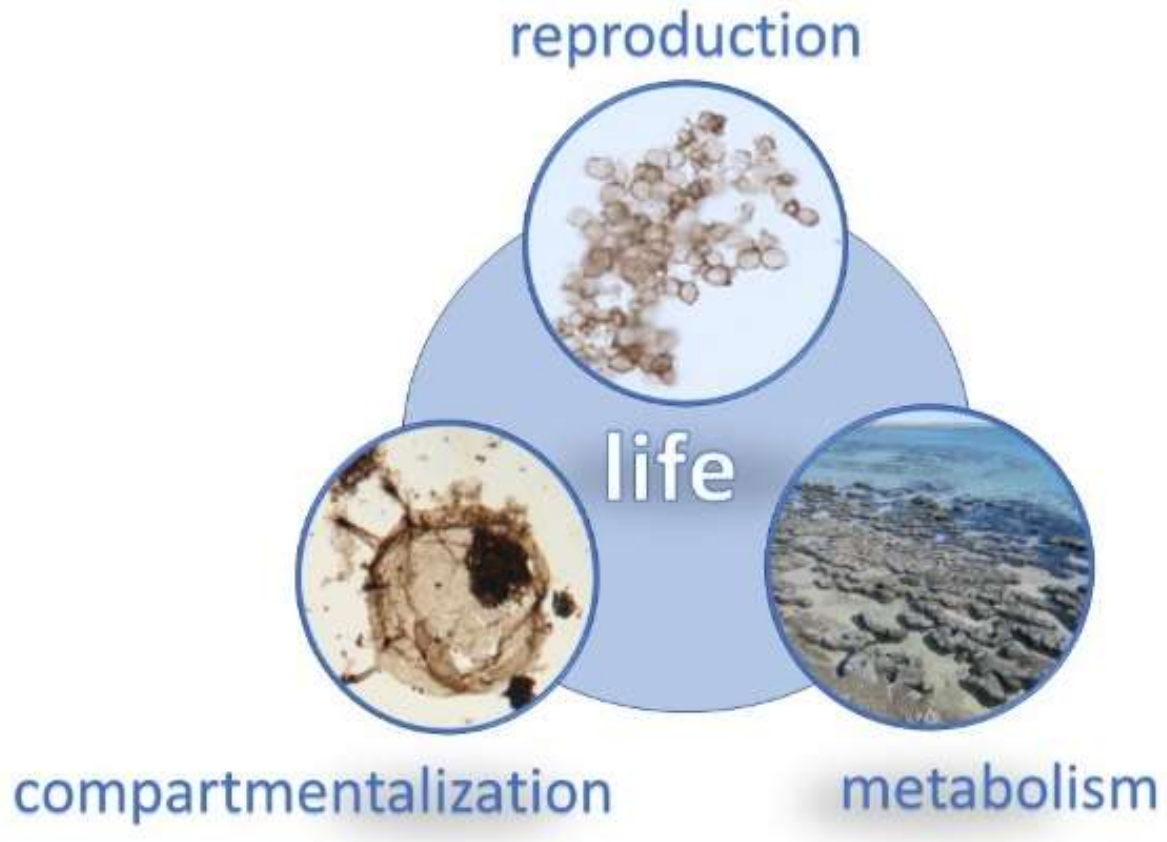

Figure 3.33. The basic elements of life: compartmentalisation, metabolism and reproduction help to define signs of life when it operates in the environment. Compartmentalisation accounts for chemical organisation and system individualisation, metabolism allows gathering material resources and energy from the environment and transforming them to self-maintenance of the whole system and reproduction transmits the blueprint of the system to the progeny including mutations, which leads to diversity and evolution. The figure has been provided to us by Emmanuelle Javaux.

In 2003, Des Marais et al. (2003a; see also Des Marais et al., 2003b; Des Marais, 2003; Cady et al., 2003; Des Marais et al., 2008) established that a biosignature is any object, substance, and/or pattern whose origin specifically requires a biological agent, such as chemical disequilibria, distribution of patterns of structurally related compounds, isotopic signatures of the dominance of catalysis in biochemistry, or the concentration of chemical monomers that are dictated by adaptability and utility. Biosignatures are closely connected to the environment, so the control of the prevailing ambient parameters is key to identify false positives or negatives. More recently, Neveu et al. (2018) discussed the main rungs to advance in the search for life from suspicious biological materials to strong evidence of life as we know it. The criteria include parameters related to the instrumentation and the context such as detectability, specificity and ambiguity. Interestingly, the group of molecules and structures conferring function was assigned the lowest ambiguity in the ladder of life organigram (e.g., polymers, structural preferences in organic molecules, enantiomeric excess, pigments). Although not all can be considered generic and, in some cases, their detection requires further technological developing for space uses, this is a promising group of biosignatures for future planetary exploration. The detection of higher levels of chemical complexity and structural organisation will allow deduction of the presence of highly advanced prebiotic chemistry or active biochemistry manifesting on other planets (Fairén et al., 2020).

The complex chemistry based on polymers can be characterized, for instance, by the combination of several techniques dedicated to measure functional properties. Fairén et al. (2020) propose a micro-fluidic based instrument with three detectors to search for life: the microscopic unit, the Raman spectrometer unit, and the biomarker detector unit. The microscope identifies ultra-structures and cell-like morphologies. The Raman spectrometer detects universal intramolecular complexity and resolves 3D secondary and tertiary polymeric structures. And the biomarker sensor contains multiple bio-affinity probes (antibodies and aptamers) for up to 200 life-related and nonlife-related chemical compounds, able to identify the nature and structure of the molecules detected, at least the part of the molecule that has been captured by the receptor molecule (antibody or antigen), following the lock-and-key principle.

As previously mentioned, ocean worlds are major targets for future astrobiological exploration, particularly Europa, Enceladus and Titan. In addition to search for signs of extant/extinct life, they enable the quest for a potential second genesis in the Solar System and alternative biochemistries. Europa shows evidence of a sub-surface ocean in contact with a rocky seafloor, which could be geothermally active. Indeed, the moon seems to be active now since water plumes emerging from the surface have been detected by the Hubble Telescope and the Hawaii observatory. In the case of the Saturnian satellite Enceladus, the Cassini spacecraft detected and analysed the plume materials directly and found that they contain organic molecules, water, salt and silica particles that could be the product of the alteration of a rocky layer. Titan, on the other hand, possesses a rich collection of organic prebiotic molecules in the atmosphere and in the solid and liquid surfaces.

Future missions to these moons should explore their habitable environments. This could be achieved by a series of missions with increasing risk and technological challenges to eventually reach and study the oceans in-situ (Sherwood et al., 2018; Prieto-Ballesteros et al., 2019) (see Fig. 3.34).

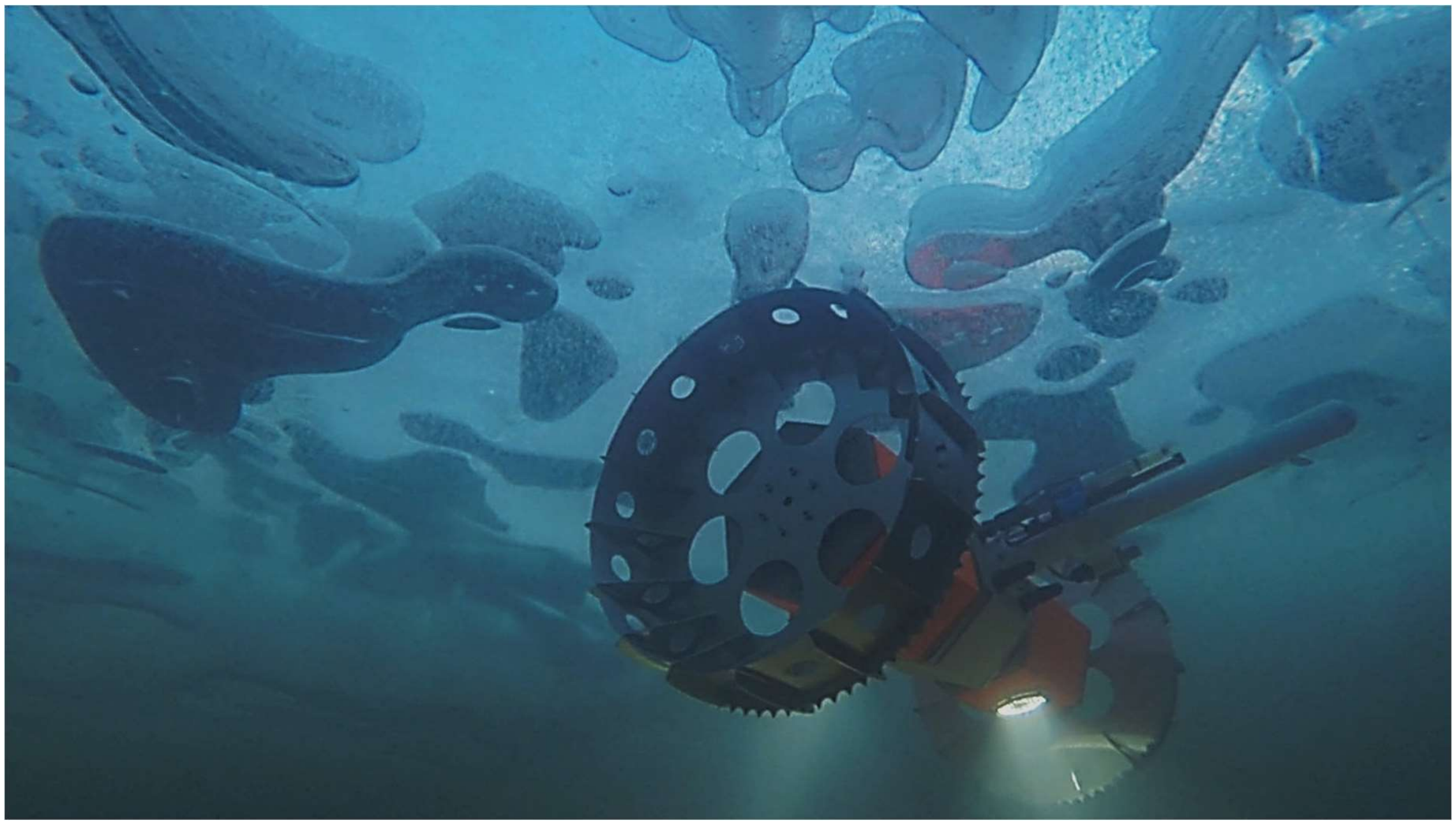

Figure 3.34. Picture of the robotic submersible BRUIE (Buoyant Rover for Under-Ice Exploration) developed at NASA-JPL to explore ice-covered oceans on moons like Europa and Enceladus. The picture was taken in an Arctic lake near Barrow during the 2015 campaign. Credit: NASA.

The existence of life on other planetary bodies and satellites is conditioned by our understanding of the conditions for the emergence of life on Earth. This is a clearly a terrestrial bias but it is not our remit here to delve into hypotheses about other forms of life. Already, from the outset, there are many different theories concerning the modalities for the emergence of life during Hadean era (4.5-4.0 Ga) on Earth. Those invoking some form of hydrothermal activity, either subaerial or subaqueous, are the most favoured since these environments combine prebiotic organic molecules, water, reactive mineral surfaces, and gradients of different kinds (pH, temperature, ionic composition etc.) (Baross & Hoffman, 1985; Russell et al., 1988; 1997; Martin et al., 2008; Westall & Brack, 2018; Westall et al., 2018; Barge et al., 2019; Damer & Deamer, 2020).

Looking more closely at the subaerial versus subaqueous scenarios has important implications for the possibility of the emergence of life on other bodies in the Solar System. The common denominator of hypotheses concerning the emergence of life on land is the phenomenon of wet-dry cycling, a useful method of concentration and physical juxtaposition of prebiotic molecules in a volcanic environment with reactive minerals (Damer & Deamer, 2020; Pearce et al., 2017; Xu et al., 2020). While Xu et al. (2020) require UV radiation to fuel photochemical reactions leading to the formation and selection of nucleosides RNA pyrimidine and DNA purine nucleosides, Pearce et al. (2017) underline the negative effects of UV radiation during dry cycles on nucleobases, concluding that they must have emerged extremely rapidly in order to avoid being almost immediately broken down. Damer & Deamer (2020), however, hypothesize the emergence of protocells – primitive cells before the evolution of organisms capable of photosynthesis in the subaerial environment – and the adaptation to and colonisation of the oceans from the freshwater realm (see Fig. 3.35; see also Li & Kusky, 2007).

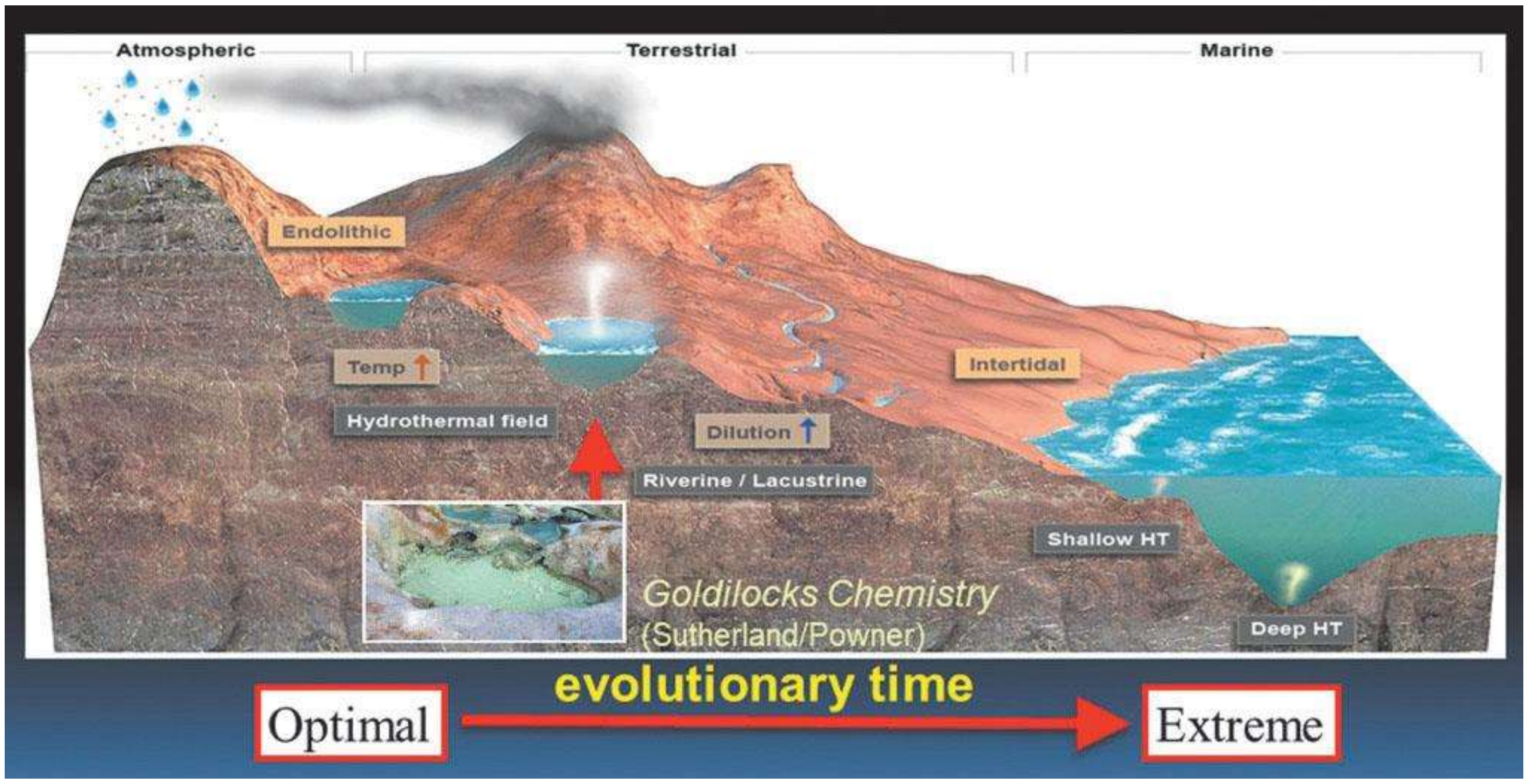

Figure 3.35. The emergence of life in a volcanic land mass interacting with freshwater and saltwater conditions environment. From Damer & Deamer (2020).

Submarine hydrothermal environments include not only the vents and vent edifices but also the surrounding sediments, as demonstrated by field and geochemical studies of the oldest well-preserved volcanic environments (Westall et al., 2018; see Fig. 3.36). Importantly, these environments concentrate molecular components in favourable microenvironments characterized by significant gradients fuelling the prebiotic reactions. An important component of these submarine environments is the ubiquity of silica gel in the early seawater, functioning as a porous molecular sieve and as support for the reactions (Dass et al., 2018). While hydrothermal environments and activity were abundant on the early Earth in all the preserved environments, it is necessary to keep in mind that only certain portions of the early Earth's crust have been preserved, namely relatively shallow water environments (going from below the wave base, i.e. 10s-100s m, to the littoral environment) representing lava and sediment-filled basinal structures (Nijman et al., 2014). Deep-water early terrestrial environments have not been preserved. These environments were largely protected from UV radiation, especially when intermixed with a protective mineral matrix, except in the littoral environment where radiation could penetrate up to several metres (Fleischmann, 1989). In the scenario of life emerging in the submarine environment, the hypothesis is that it could then colonize the exposed terrestrial surfaces. In fact, there is good evidence for the exposure of phototrophic microbial mats to the atmosphere in the littoral environment by 3.5-3.33 Ga (Westall et al., 2006), as well as subaerial springs (Djokic et al., 2017).

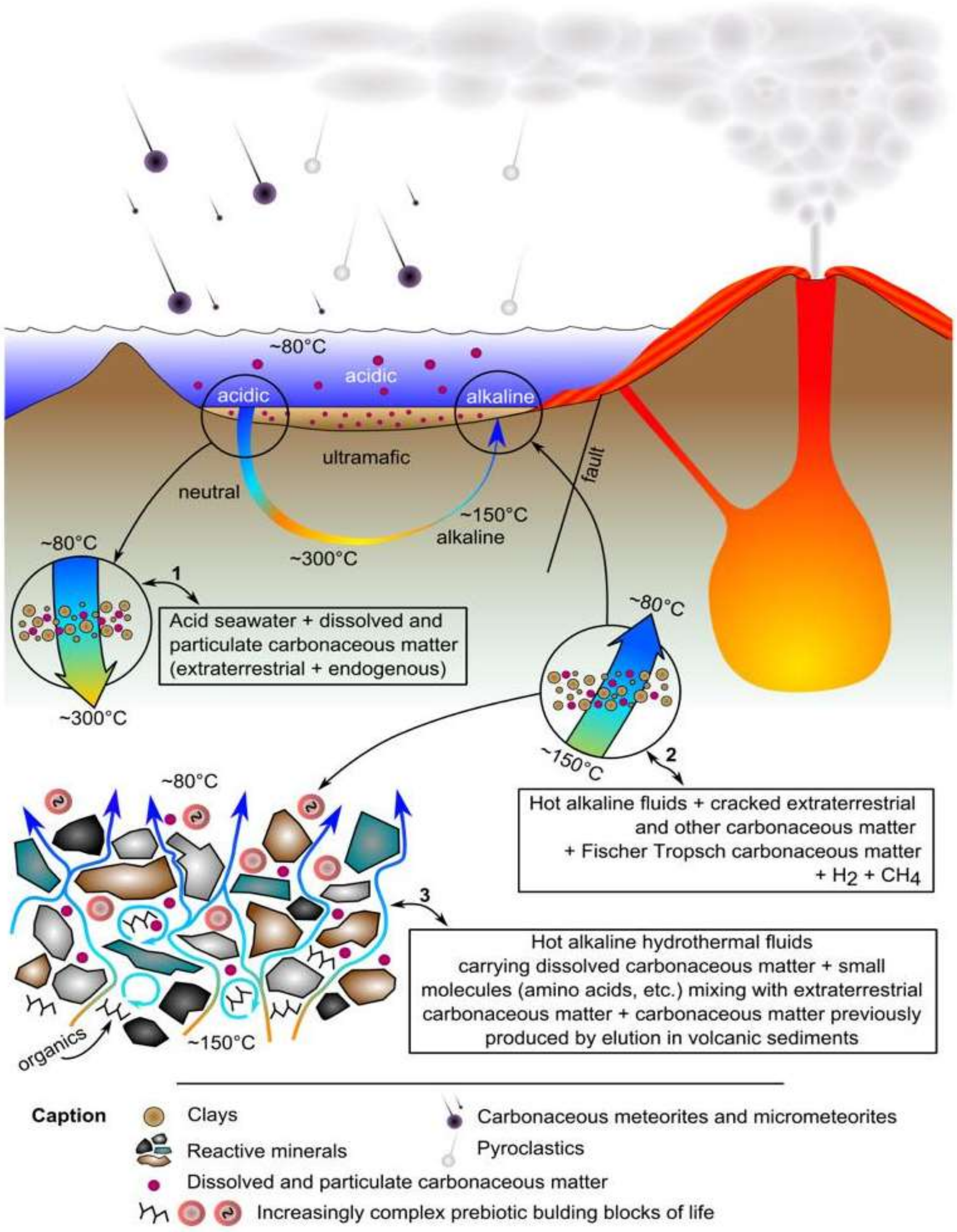

Figure 3.36. Subaqueous hydrothermal environments and the emergence of life. From Westall et al. (2018).

We may never know exactly how life emerged on the Earth; perhaps parts of both scenarios are correct; perhaps useful complex molecules formed both in freshwater and in salty seawater and were mixed in the littoral zone (but in some UV-protected environment). However, if one or the other scenario is correct, this has significant consequences for the emergence of life elsewhere in the Solar System and beyond. If life emerged only in freshwater on land, then it is clear that it could not have emerged on any of the icy satellites where no exposed landmasses were formed. If, on the other hand, it emerged in the subaqueous environment, life could have appeared on other terrestrial planets, such as Venus and Mars, as well as the water/ice covered satellites of Jupiter and Saturn. We will take the latter scenario further to address the existence of extraterrestrial life in the Solar System in general and how to search for it.

If we accept that life could have emerged elsewhere in the Solar System, we now need to address what kind of life we might be looking for. Earth has had the advantage of being continuously habitable with variable distribution of landmasses and associated shallow water environments throughout much of its history. The situation for Venus, apart from the fact that the planet is presently not habitable at the surface or sub-surface, is less clear. Was it ever habitable in the sense of having significant bodies of water? That depends on its geological history and development, factors that are difficult to penetrate although various models support its habitability for varying lengths of geological time, depending upon the model (Way & Del Genio, 2020).

On Mars, the situation is certainly clearer: we understand that there was a major climate change likely related to loss of its atmosphere and decline of its dynamo and the protective magnetic field that it produced, which could have resulted in stripping of much of its volatile inventory, including water, early in Mars' history. Nevertheless, the presence of water might not have been continuous in the planet's early history. There is abundant evidence for large floods that carved the landscape until about 3.7 Ga (see Section 2.1.2). Together with the ubiquitous rain of prebiotically-significant organic molecules in carbonaceous meteorites and micrometeorites from space (as demonstrated by the SAM instrument on MSL's rover Curiosity, Freissinet et al., 2015; Eigenbrod et al., 2018) and having similar volcanic components, hydrothermal activity and micro-environments as on the early Earth (Westall et al., 2015), the hypothesis that life could have appeared on the red planet is entirely plausible – plausible enough that quite a few missions are already on the planet or heading there to search for traces of (past) life. The MSL mission to Gale Crater has demonstrated the potential for this formerly lake-filled crater to have been able to host life (Grotzinger et al., 2014), although life would not likely have originated there since the crater does not fulfil the requirement of the long-term co-existence of hydrothermal activity and the elemental ingredients of life that would lead to life emerging ab initio (several tens to hundreds of thousands of years at minimum) that is presently understood to be necessary (Westall et al., 2015, 2018). However, given its great age (~3.6 Ga), existing life elsewhere could have been transported into the crater, perhaps through groundwater and/or aquifer circulation or even through impact transport. If this were not the case, Gale Crater would be an example of an uninhabited habitat (Cockell et al., 2012). Such inhabited or uninhabited habitats were probably common on early Mars. Indeed, life could have appeared in different locations at different times, since the potential habitats would have been isolated from each other (Mars was never an ocean-covered planet as has been the Earth since the Hadean era). Whatever habitat on Mars that might have permitted life to emerge or to colonize, the planet eventually became uninhabitable.

One of the important consequences of the lack of connectivity between habitable locations and their eventual disappearance is that there could be no continuous evolution over long geological periods, as there was on Earth (Westall et al., 2015). By 3.5 Ga on Earth, the planet was teeming with anaerobic organisms that had colonized all subaqueous environments and some subaerial areas. Life had diversified from the initial primitive chemotrophs into phototrophs whose readily identifiable remains are abundant in shallow water environments.

Without being able to state categorically that phototrophs could not have evolved on Mars, the probability of their appearance is low owing to lack of opportunity. The significance of this situation is demonstrated in the ancient terrestrial sediments, analogues of early Mars. The chemotrophic colonies are only well-developed in the vicinity of hydrothermal activity although they occur also in oligotrophic environments, but only as poorly developed colonies on the surfaces of detrital volcanic grains. While the remains of phototrophic mats are relatively readily identifiable in sedimentary rocks and at the same time may also be macroscopically visible, chemotrophic colonies, even if well-developed, have cryptic expressions and their identification is far more controversial (Hickman-Lewis et al., 2020). In-situ identification of such organisms will likely have to rely on analysis of their carbonaceous remains to find molecular compositions and patterns indicative of life (Vago et al., 2017). A sample return mission would be a necessity in order to verify any trace of life.

Further out in the Solar System, the very nature of the icy satellites precludes the development of phototrophic types of microorganisms. This means that, there also, life forms would be chemotrophic, feeding off carbon and inorganic nutrients present under the icy crusts.

Considering life forms based on carbon molecules and water, we can conclude from the above that rocky planets with water and hydrothermal activity are most probably a prerequisite. Other stellar systems may have any number of variants on the rocky planet theme, but so long as the essential volatiles are present, i.e. liquid water and organic molecules, and the conditions for the emergence of life are sufficiently long lasting as to permit cellular life to form, life will be there, either as primitive chemotrophic colonies, possibly as photo-synthesizers and probably rarely as more evolved forms, oxygenic photo-synthesizers and complex forms, such as eucaryotes.

Missions to discover life elsewhere in the Solar System should utilize multiple detectors to search for life, such as a microscopic unit, a Raman spectrometer unit, and a biomarker detector unit, which must all be developed further to survive in harsh environments. Of course, a sample return mission is the best option to discover life or traces of past life in samples.

# 8. Summary

This Chapter explored the Solar System, except the Sun itself, its objects and its secondary systems in the light of six key science questions concerning the diversity, origins, workings and habitability of planetary systems. This “scientific tour” of the Solar System showed that these questions can be addressed in complementary ways by observing many different objects with a diversity of measurement techniques.

## 8.1 Detailed scientific objectives of the exploration of Solar System objects.

Table 3.3 summarizes the detailed scientific objectives to be given at each of the main Solar System destinations to address these questions. They are summarized below the table for each science question.

Table 3.3. This matrix displays for each of the six key science questions and for each of the main destinations in the Solar System the scientific objectives to be assigned to space and Earth-based observations.

**Question 1- How well do we understand the diversity of planetary systems objects?**

We are still far from having established a comprehensive characterisation of all classes of Solar System objects because of their extreme diversity and different distances from Earth. A global strategy to capture both the complexity and diversity of Solar System objects should take into account the differences in accessibility of the different objects and the number of objects in each class.

For planets, it is essential to know, with reasonably comparable accuracies, their key internal, surface, atmosphere, and plasma envelope properties. Accurate characterisation of each class of planet represented in the Solar System (terrestrial planets, gas giants and ice giants) is essential not only to understand their diversity, but also to be able to use them as “nearby” templates of the different types of exoplanets found around other stars. Given current knowledge, the most urgent task is to fill the “knowledge gap” we have concerning ice giants.

Characterising dwarf planets and regular moons around giant planets in the same terms as planets is another urgent task, particularly for those that are likely ocean worlds.

The different classes of small bodies, down to the size of cosmic dust, need to be better characterized, first by visiting or observing many more of them and then by characterising in sufficient detail a representative sample of each class of these objects and by better establishing their connection to meteorites and interplanetary dust particles collected in the Earth environment.

Finally, exploring the outer “frontiers” of the Solar System, i.e. the huge population of objects orbiting at its outskirts, heliosphere boundaries and, beyond, the nearby interstellar space in different directions, should be one of the priorities of the coming decades.

### Question 2- How well do we understand the diversity of planetary system architectures?

Chapter 2 reviewed the diversity of extra-solar planetary systems. Section 3 of this present Chapter illustrated the many contributions Solar System studies can make to a deeper understanding of the diversity of planetary system architectures.

First, taken as a whole together with its different secondary systems, the Solar System offers the ability to examine closely a large diversity of architectures: the system as a whole and its peculiar architecture that needs to be understood better for a useful comparison to other multiple-planet systems and to the debris disks one can observe around other stars; and its diverse secondary systems, from the Earth-Moon system, a particularly important example of a tidally-locked binary system, through the four giant planet systems, to the uniquely interesting secondary systems found among the TNOs. Exploring again with a comprehensive orbiter mission, first the Uranian and Neptunian systems and later, the Pluto-Charon system that were visited only once by mission flybys, are “must-dos” prior to the 2061 horizon.

Second, the Solar System will remain for quite some time the only system where the dynamical interplay between the components of secondary systems (regular and irregular moons, rings, energetic particles and plasma populations) can be observed in situ. The four giant planets must be explored with similar degrees of accuracy, which means in practice raising our knowledge of ice giant systems to the same level as what the Galileo, Cassini and Juno missions taught us about gas giant systems.

Finally, the diversity of planetary magnetospheres, their geometries and extension in space, their interactions with their central planet, with the objects orbiting around it, and with the solar wind, call for future missions carrying comprehensive but compact particles-and-fields instrumentation and a multipoint diagnostic capability to study them. Exploring this diversity is also the best possible preparation for future observations of exo-magnetospheres when they will become available.

## Question 3- What are the origins and formation scenarios for planetary systems?

Understanding how an object as complex as the Solar System was formed, with its multi-scale and multi-object structure, and its embedded secondary systems, is a huge scientific challenge. Meeting this challenge requires a multi-target, multi-instrument, and multi-mission approach and enough time to accomplish it. Looking ahead to the 2061 horizon, a global plan to address this question can be designed and implemented: something like a "mission to the origins of the Solar System". Adequate measurements need to be made at all objects that have recorded a memory, even tiny, of the initial conditions of the solar nebula and of the succession of events that sculpted the current architecture of the Solar System during its early ages: small bodies (which can be prioritized with respect to the other questions addressed here or in Chapter 4 (Lasue et al., 2022)), particularly those related to the external reservoirs like comets and TNOs, giant planet atmospheres and the fraction of dwarf planets and giant planet moons where these records have not been erased by chemical differentiation and surface/sub-surface activity. Missions to all these bodies should include well-designed combinations of instruments on orbiters and in situ probes capable of reading these records and, wherever possible, return samples to Earth for deeper analysis.

## Question 4- How do planetary systems work?

The Solar System is a multiscale system accessible to in situ observation of all its objects at all these different scales: it is a unique laboratory for learning and understanding how planets and planetary systems work. This Chapter touched only partly on this subject, by selecting a few examples. The rich diversity of studies undertaken covers the diversity of scales. From the smallest to the largest one:

- Developing an integrative understanding of how planets, dwarf planets and their moons (i.e., differentiated objects) work is a very important ongoing task. This requires better understanding of the dynamical and energy transfer processes within each layer (core, envelope or mantle, surface when it exists, atmosphere when it exists, plasma envelope and interface with the space environment). Understanding the role played by their working and coupling processes in the generation and maintenance of habitability conditions and how the climates of terrestrial planets work and evolve, is of special importance.
- In the same spirit, building with time an integrative understanding of the Earth-Moon and giant planets systems is a very exciting objective for the planetary exploration programme. This requires planned study of the diversity of coupling processes between all the objects of each system. Future missions to these systems should be designed with the necessary combinations of orbital and in situ observations of these objects specifically designed to allow the study of their couplings.
- Finally, time is ripe to try and to understand how the Solar System interacts with its galactic environment and what consequences these interactions have on its past, present and future habitability. This exciting goal points to a much deeper exploration of the boundaries of the heliosphere and of their 3-D shape, of the distant poorly known objects orbiting there and of the local interstellar medium extending beyond the heliopause, on our way to nearby stars.

## Question 5- Do planetary systems host potential habitats?

The search for habitable worlds should reach a mature stage in the coming decades, with at least three directions of research:

- Trace the "flow of water" from the solar nebula and the early Solar System to the different "candidate" habitable world: to identify the different water reservoirs, and understand the contributions of each class to the delivery of water to planets and moons.

- At each place where there is sufficient water to fulfil the first condition of habitability, to check whether the other habitability conditions are met. Knowing that the Mars exploration program and Cassini-Huygens led to a consensus on the habitability of Mars and Enceladus, the focus now should be on characterising the habitability of other ocean worlds: first Europa, then Ganymede, Titan, Triton and Ceres.
- Finally, to understand the role played by the environments of candidate habitats and the effects of the dynamics of their different layers, such as climate evolution and plate tectonics, in their habitability. Concerning plate tectonics, obtaining details on the subduction style, such as the steadiness of potential subducting plates, would require seismic measurements. Furthermore, an active magnetic field (that could also be promoted by plate tectonics) may play an important role in the exchanges between interior, subsurface, surface, atmosphere and space.

The future missions to be flown to characterise the habitability of planet or moon destinations should combine descent imaging, in-situ analyses and sample return. They should also characterise the atmospheric composition and structure as well as the exchanges with the (sub)surface and space at high spatio-temporal resolution.

### Question 6- Where and how to search for life?

Once a planet or ocean moon's habitability has been assessed, the next logical step is to send new missions to the most promising habitable worlds to search for life, starting with Mars and Enceladus whose habitability has already been assessed and then extending this search of signatures of life to newly characterised habitats.

In order to detect life – extinct or extant – on Mars, one would need to go to places where the habitability conditions are met, which includes underground as well as the oldest terranes in the Southern Highlands, involving significant technical hurdles, notably including planetary protection requirements on forward contamination of the Mars environment and resultant confusion should living organisms be discovered. The fieldwork could be done using rovers or crawlers that have been treated to minimize the possibility of contamination by Earth microbes and that would need to be capable of entering difficult terrain such as steep slopes – where, for instance, liquid water can be maintained for some time in "gullies" despite its thermodynamic instability – and caves. Ocean worlds of the outer Solar System are very challenging destinations that will require the landing of probes bringing a dedicated astrobiology instrumentation package to their surface and sub-surface (with a drill system or a penetrator) after a phase of characterisation of their habitability from their orbit. Further measurements need to be done in situ or in plumes or via sample return wherever possible. Preservation of biosignatures is the key question for the detection of biosignatures in the Solar System.

## 8.2 A diversity of measurement techniques and types of missions to address the key science questions.

The different science questions and more detailed scientific objectives just reviewed are listed in the left-hand side of Fig. 3.37. They can be addressed by the broad diversity of measurement techniques shown on the right-hand side of this figure.

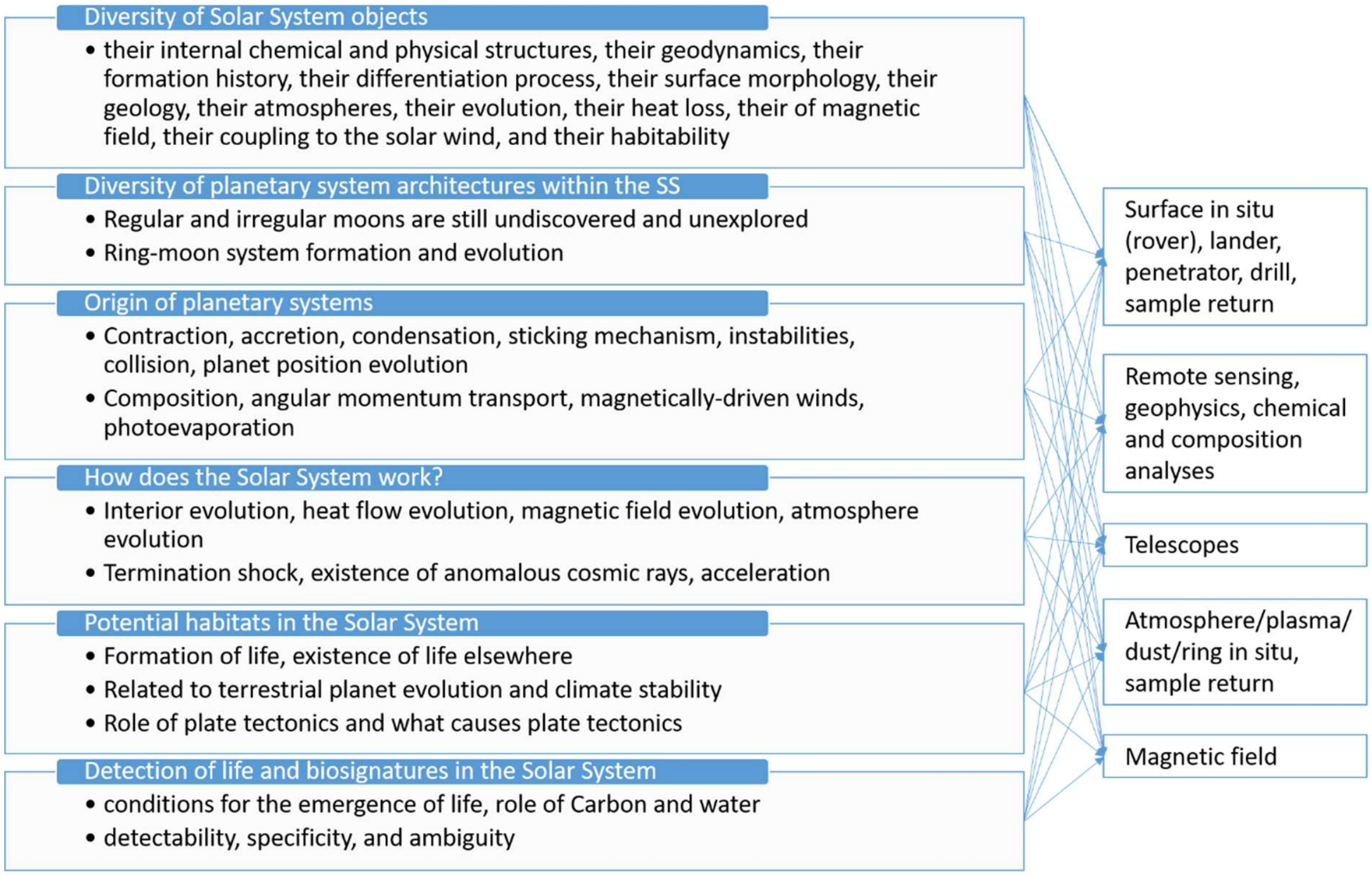

Figure 3.37. Sketch of the measurement needs for the different questions.

To operate these instruments and to provide answers to science questions, many different types of Earth-based observatories and Solar System exploration missions are needed. They can be ordered along a “scale of complexity” of planetary missions. From bottom to top of this ranking:

(1) Observations from Earth’s surface;
(2) Flybys;
(3) Orbital surveys of a small body, of a planet and/or of its system;
(4) Orbital surveys of a moon;
(5) In situ probes/landers/rovers to atmospheres and surfaces;
(6) Sample return from these objects to Earth;
(7) Human missions.

Even given the projected progress in technology development, the last two types of missions are likely to remain limited to the closest destinations even by the 2061 horizon: up to the asteroid belt for Human exploration, up to Jupiter’s Trojans and perhaps some moons of Jupiter and Saturn for sample return missions. The next Chapter explores the diversity of missions to the different destinations of the Solar System that will have to be flown to address the six key science questions of the Horizon 2061 foresight exercise.

## Acknowledgement

**VJS** received funding from the European Union’s Horizon 2020 research and innovation programme under grant agreement N°851544. VJS thanks Dr. J. Richardson for his helpful comments on the manuscript. This project has received funding from the European Research Council (ERC) under the European Union’s Horizon 2020 research and innovation programme (grant agreement n° 679030 /WHIPLASH – **JL** and grant agreement n° 855677 / GRACEFUL – **VD**). **MB** thanks CNES for its support to the preparation, operation and scientific exploitation of space missions to many destinations in the Solar System. He expresses his gratitude to CNES and CNRS for all their support to planetary research at the Institut de Recherche en Astrophysique et Planétologie (IRAP) and his warm thanks to the directors and staff of IRAP and Observatoire Midi-Pyrénées for hosting in September 2019 in Toulouse the “Horizon 2061 synthesis workshop” thanks to which the material presented in this Chapter and in the book were assembled and debated with the community.

## References

Acuna, M.H., Connerney J.E.P., Ness N.F., Lin R.P., Mitchell D., Carlson C.W., McFadden J., Anderson K.A., Reme H., Mazelle C., Vignes D., Wasilewski P., Cloutier P., 1999, Global distribution of crustal magnetization discovered by the Mars Global Surveyor MAG/ER experiment. Science, 284(5415), 790-793, DOI: 10.1126/science.284.5415.790.

Agnor, C.B., Hamilton D.P., 2006, Neptune’s capture of its moon Triton in a binary–planet gravitational encounter, Nature, 441, 192-194, DOI: 10.1038/nature04792.

Albarede, F., 2009, Volatile accretion history of the terrestrial planets and dynamic implications. Nature, 461(7268), 1227-1233, DOI: 10.1038/nature08477.

Arndt, N.T., Nisbet E.G., 2012, Processes on the young Earth and the habitats of early life. Annual Review of Earth and Planetary Sciences, 40, 521-549, DOI: 10.1146/annurev-earth-042711-105316.

Atkinson, D.H., Pollack J.B., Seiff A., 1996, Galileo Doppler Measurements of the Deep Zonal Winds at Jupiter. Science, 272, 5263, 842-843, DOI: 10.1126/science.272.5263.842.

Atreya, S.K., Hofstadter M.H., In J.H., Mousis O., Reh K., Wong M.H., 2020, Deep atmosphere composition, structure, origin, and exploration, with particular focus on critical in situ science at the Icy Giants. Space Science Reviews, 216, 1, Id.18, DOI: 10.1007/s11214-020-0640-8.

Aurnou, J., Heimpel M., and Wicht J., 2007, The effects of vigorous mixing in a convective model of zonal flow on the ice giants. Icarus, 190(1), 110-126.

Banerdt, W.B., Smrekar S.E., Banfield D., Giardini D., Golombek M., Johnson C., Lognonné P., Spiga A., Spohn T., Perrin C., Stähler S., Antonangeli D., Asmar S., Beghein C., Bowles N., Bozdag E., Chi P., Christensen U., Clinton J., Collins G., Daubar I., Dehant V., Drilleau M., Fillingim M., Folkner W., Garcia R., Garvin J., Grant J., Grott M., Grygorczuk J., Hudson T., Irving J., Kargl G., Kawamura T., Kedar S., King S., Knapmeyer-Endrun B., Knapmeyer M., Lemmon M., Lorenz R., Maki J., Margerin L., McLennan S., Michaut C., Mimoun D., Mittelholz A., Mocquet A., Morgan P., Mueller N., Murdoch N., Nagihara S., Newman C., Nimmo F., Panning M., Pike W., Plesa A.C., Rodriguez S., Rodriguez-Manfredi J., Russell C., Schmerr N., Siegler M., Stanley S., Stutzmann E., Teanby N., Tromp J., van Driel M., Warner N., Weber R., Wieczorek M., 2020, Initial results from the InSight mission on Mars. Nature Geoscience, 13, 183-189, DOI: 10.1038/s41561-020-0544-y.

Baranov, V.B., Malama Y.G., 1993, Model of the solar wind interaction with the local interstellar medium numerical solution of self-consistent problem. Journal of Geophysical Research, 98(A9), 15157-15164, DOI: 10.1029/93JA01171.

Barge, L.M., White L.M., 2017, Experimentally testing hydrothermal vent origin of life on Enceladus and other icy/ocean worlds. Astrobiology, 17(9), 820-833, DOI: 10.1089/ast.2016.1633.

Barge, L.M., Flores E., Baum M.M., VanderVelde D.G., Russell M.J., 2019, Redox and pH gradients drive amino acid synthesis in iron oxyhydroxide mineral systems. Proc. Natl Acad. Sci. USA 116, 4828-4833, DOI: 10.1073/pnas.1812098116.

Baross, J.A., Hoffman S.E., 1985, Submarine hydrothermal vents and associated gradient environments as sites for the origin and evolution of life. Origins Life Evol Biosphere, 15, 327-345, DOI: 10.1007/BF01808177.

Basilevsky, A.T., Nikolaeva O.V., Weitz C.M., 1992, Geology of the Venera 8 landing site region from Magellan data: Morphological and geochemical considerations. J. Geophys. Res., 97(E10), 16315-16335, DOI: 10.1029/92JE01557.

Batygin, K., Morbidelli A., 2020, Formation of giant planet satellites. The Astrophysical Journal, 894(2), Id.143, 23 pages, DOI: 10.3847/1538-4357/ab8937.

Baumjohann, W., Blanc, M., Fedorov, A., Glassmeier, K.-H., 2010, Current Systems in Planetary Magnetospheres and Ionospheres. Space Science Reviews 152, 99-134.

Benkhoff, J., Van Casteren J., Hayakawa H., Fujimoto M., Laakso H., Novara M., Ferri P., Middleton H.R., Ziethe R., 2010, BepiColombo—Comprehensive exploration of Mercury: Mission overview and science goals. Planetary and Space Science, 58(1-2), 2-20, DOI: 10.1016/j.pss.2009.09.020.

Bennett, J.O., Donahue, M.O., Schneider, N., Voit, M., 2019. The Cosmic Perspective: The Solar System. Pearson, pp. 1e832. ISBN-10: 0134874366, ISBN-13: 978-0134874364.

Bercovici, D., Ricard Y., 2014, Plate tectonics, damage, and inheritance. Nature, 508, 513-516, DOI: 10.1038/nature13072.

Berner, R.A., 1992, Weathering, plants, and the long-term carbon cycle. Geochimica et Cosmochimica Acta, 56 (8), 3225-3231, DOI: 10.1016/0016-7037(92)90300-8.

Blanc, M., Bolton, S., Bradley, J. et al., 2002, Magnetospheric and Plasma Science with Cassini-Huygens.Space Science Reviews, 104, 253-346, DOI: 10.1023/A:102360511071

Blanc, M., Mandt K., Mousis O., André N., Bouquet A., Charnoz S., Craft K.L., Deleuil M., Griton L., Helled R., Hueso R., Lamy L., Louis C., Lunine J., Ronnet T., Schmidt J., Soderlund K., Turrini D., Turtle E., Vernazza P., Witasse O., 2021, Science Goals and Mission Objectives for the Future Exploration of ice Giant Systems: a Horizon 2061 perspective. Space Sci. Rev., 217(3), DOI: 10.1007/s11214-020-00769-5.

Blanc, M., Ammannito E., Bousquet P., Capria M.-T., Dehant V., Foing B., Grande M., Guo L., Hutzler, A., Lasue J., Lewis J., Perino M.A., Rauer H., 2022, Planetary Exploration, Horizon 2061 Report - Chapter 1: Introduction to the “Planetary Exploration, Horizon 2061” Foresight Exercise. Science Direct, Elsevier.

Bodenheimer P., Stevenson D.J., Lissauer J.J., and D'Angelo G., 2018, New Formation Models for the Kepler-36 System. The Astrophysical Journal, 868(2), Id. 138, 17 pages, DOI: 10.3847/1538-4357/aae928.

Bolton, S.J., Bagenal F., Blanc M., Cassidy T., Chané E., Jackman C., Jia X., Kotova A., Krupp N., Milillo A., Plainaki C., Smith H.T., Waite H., 2015, Jupiter’s Magnetosphere: Plasma Sources and Transport. Space Sci Rev 192, 209-236, DOI: 10.1007/s11214-015-0184-5.

Borderies, N., Goldreich P., Tremaine S., 1982, Sharp edges of planetary rings. Nature, 299, 209-211, DOI: 10.1038/299209a0.

Borovsky, J.E., and Valdivia J.A., 2018, The Earth's magnetosphere: a systems science overview and assessment. Surveys in geophysics, 39(5), 817-859, DOI: 10.1007/s10712-018-9487-x.

Brasser, R., Morbidelli A., 2013, Oort cloud and scattered disc formation during a late dynamical instability in the Solar System. Icarus, 225(1), 40-49, DOI: 10.1016/j.icarus.2013.03.012.

Breuer, D., Moore W.B., 2015, Dynamics and thermal history of the terrestrial planets, the Moon and Io. In Treatise in Geophysics, 2nd ed., Vol 10, eds. T. Spohn and G. Schubert, Elsevier, Amsterdam, 255-305.

Brilliantov, N.V., Pimenova A.V., Goldobin D.S., 2015, A dissipative force between colliding viscoelastic bodies: Rigorous approach. Europhysics Letters, 109, 1, Id. 14005, DOI: 10.1209/0295-5075/109/14005.

Brouwers, M.G., Vazan A., and C.W. Ormel, 2018, How cores grow by pebble accretion. I. Direct core growth. Astronomy and Astrophysics, 611, Id. A65, 12 pages, DOI: 10.1051/0004-6361/201731824.

Buratti, B.J., Bauer J.M., Hicks M.D., Hillier J.K., Verbiscer A., Hammel H., Schmidt B., Cobb B., Herbert B., Garsky M., Ward J., 2011, Photometry of Triton 1992–2004: Surface volatile transport and discovery of a remarkable opposition surge. Icarus, 212(2), 835-846.

Burlaga, L.F., Ness N.F., Stone E.C., 2013, Magnetic Field Observations as Voyager 1 Entered the Heliosheath Depletion Region. Science, 341(6142), 147-150, DOI: 10.1126/science.1235451.

Burlaga, L.F., Ness N.F., Berdichevsky D.B., Park J., Jian L.K., Szabo A., Stone E.C., Richardson J.D., 2019, Magnetic field and particle measurements made by Voyager 2 at and near the heliopause. Nature Astronomy, 3, 1007-1012, DOI: 10.1038/s41550-019-0920-y.

Burns, J.A., Showalter M.R., Hamilton Do.P., Nicholson P.D., de Pater I., Ockert-Bell M.E., Thomas P.C., 1999, The Formation of Jupiter's Faint Rings. Science, 284, 5417, 1146, DOI: 10.1126/science.284.5417.1146.

Burns, J.A., Simonelli D.P., Showalter M.R., Hamilton D.P., Porco C.D., Throop H., Esposito L.W., 2004, Jupiter's ring-moon system. In: Jupiter. The planet, satellites and magnetosphere. Eds. F. Bagenal, T.E. Dowling, and W.B. McKinnon, Cambridge planetary science, Vol. 1, Cambridge, UK: Cambridge University Press, ISBN 0-521-81808-7, 241-262.

Cady, S.L., Farmer J.D., Grotzinger J.P., Schopf J.W., Steele A., 2003, Morphological Biosignatures and the Search for Life on Mars. Astrobiology, 3(2), 351-368, DOI: 10.1089/153110703769016442.

Canup, R.M., W.R. Ward, 2006, A common mass scaling for satellite systems of gaseous planets. Nature, 441, 7095, 834-839, DOI: 10.1038/nature04860.

Carrera, D., Gorti U., Johansen A., Davies M.B., 2017, Planetesimal Formation by the Streaming Instability in a Photoevaporating Disk. The Astrophysical Journal, 839(1), 16, DOI: 10.3847/1538-4357/aa6932.

Catling, D.C., 2015. Planetary atmospheres, in G. Schubert (ed.) Treatise on Geophysics, 2nd Ed., vol. 10, Oxford, Elsevier, 429-472.

Cawood, P.A., Hawkesworth C.J., Pisarevsky S.A., Dhuime B., Capitanio F.A., Nebel O., 2018, Geological archive of the onset of plate tectonics. Philosophical Transactions of the Royal Society A: Mathematical, Physical and Engineering Sciences, 376(2132), 20170405, DOI: 10.1098/rsta.2017.0405.

Chabrier, G., and Baraffe I., 2007, Heat Transport in Giant (Exo)planets: A New Perspective. The Astrophysical Journal, 661(1), L81-L84, DOI: 10.1086/518473.

Chopras, A., Lineweaver C., 2016, The Case for a Gaian Bottleneck: The Biology of Habitability. Astrobiology, 16(1), DOI: 10.1089/ast.2015.1387.

Cockell, C.S., Balme M., Bridges J.C., Davilad A., Schwenzer S.P., 2012, Uninhabited habitats on Mars. Icarus, 217, 184-193, DOI: 10.1016/j.icarus.2011.10.025.

Cockell, C.S., Bush T., Bryce C., Direito S., Fox-Powell M., Harrison J.P., Lammer H., Landenmark H., Martin-Torres J., Nicholson N., Noack L., O'Malley-James J., Payler S.J., Rushby A., Samuels T., Schwendner P., Wadsworth J., Zorzano M.P., 2016, Habitability: a review. Astrobiology, 16(1), 89-117, DOI: 10.1089/ast.2015.1295.

Colwell J.E., Nicholson P.D., Tiscareno M.S., Murray C.D., French R.G., Marouf E.A., 2009, The Structure of Saturn's Rings. In: Saturn from Cassini-Huygens, Eds. M.K. Dougherty, L.W. Esposito, and S.M. Krimigis, ISBN 978-1-4020-9216-9, 375, DOI: 10.1007/978-1-4020-9217-6_13.

Coradini, A., Turrini D., Federico C., Magni G., 2011, Vesta and Ceres: crossing the history of the Solar System. Space Science Reviews, 163, 1-4, 25-40, DOI: 10.1007/s11214-011-9792-x.

Coustenis, A., Encrenaz Th., 2013, Life Beyond Earth, Cambridge University Press, DOI: 10.1017/CBO9781139206921.

Crameri, F., Tackley P.J., 2016, Subduction initiation from a stagnant lid and global overturn: new insights from numerical models with a free surface, Progress in Earth and Planetary Science, 3(1), 1-19, DOI: 10.1186/s40645-016-0103-8.

Crida, A., Charnoz S., 2012, Formation of regular satellites from ancient massive rings in the Solar System, Science, 338(6111), 1196-1199, DOI: 10.1126/science.1226477.

Cuk, M., Gladman B.J., 2005, Constraints on the orbital evolution of Triton, Astrophys. J., 626, L113-L116, DOI: 10.1086/431743.

Cuzzi, J.N., Filacchione G., Marouf E.A., 2018, The Rings of Saturn. In: Planetary Ring Systems – Properties, Structure, and Evolution, Eds. M.S. Tiscareno and C.D. Murray, Cambridge University Press, 51-92, DOI: 10.1017/9781316286791.003.

Dale, A.M.S., Cruikshank P., 1996, Mildred Shapley Matthews, Neptune and Triton, edited by D. P. Cruikshank (University of Arizona Press).
Damer, B., Deamer D., 2020, The Hot Spring Hypothesis for an Origin of Life. Astrobiology, 20(4), 429-452, DOI: 10.1089/ast.2019.2045.
Dasgupta, R., Hirschmann M.M., 2010, The deep carbon cycle and melting in Earth's interior. Earth and Planetary Science Letters, 298(1-2), 1-13, DOI: 10.1016/j.epsl.2010.06.039.
Dass, A.V., Jaber M., Brack A., Foucher F., Kee T.P., Georgelin T., Westall F., 2018, Potential Role of Inorganic Confined Environments in Prebiotic Phosphorylation. Life, 8(1), 7, DOI: 10.3390/life8010007.
Davaille, A., Smrekar S.E., Tomlinson S., 2017, Experimental and observational evidence for plume-induced subduction on Venus. Nature Geoscience, 10(5), 349-355, DOI: 10.1038/ngeo2928.
De Pater, I., Lissauer J., 2015, Planetary Sciences. Cambridge: Cambridge University Press. DOI: 10.1017/CBO9781316165270.
De Pater, I., Hamilton D.P., Showalter M.R., Throop H.B., Burns J.A., 2018a, The Rings of Jupiter. n: Planetary Ring Systems. Properties, Structure, and Evolution, Eds. M.S. Tiscareno and C.D. Murray, ISBN: 9781316286791, Cambridge University Press, 112-124, DOI: 10.1017/9781316286791.006.
De Pater, I., Renner S., Showalter M.R., Sicardy B., 2018b, The Rings of Neptune. A review of the jovian ring system. In: Planetary Ring Systems. Properties, Structure, and Evolution, Eds. M.S. Tiscareno and C.D. Murray, ISBN: 9781316286791, Cambridge University Press, 125-134, DOI: 10.1017/9781316286791.005.
Debras, F., and Chabrier G., 2019, New Models of Jupiter in the Context of Juno and Galileo. The Astrophysical Journal, 872(1), Id. 100, 22 pages, DOI: 10.3847/1538-4357/aaff65.
DeMeo, F.E., Carry B., 2013, The taxonomic distribution of asteroids from multi-filter all-sky photometric surveys. Icarus, 226, 1, 723-741, DOI: 10.1016/j.icarus.2013.06.027.
DeMeo, F.E., Carry B., 2014, Solar System evolution from compositional mapping of the asteroid belt. Nature, 505, 7485, 629-634, DOI: 10.1038/nature12908.
Des Marais, D.J., 2003, Biogeochemistry of hypersaline microbial mats illustrates the dynamics of modern microbial ecosystems and the early evolution of the biosphere. Biol. Bull. 204, 160-167, DOI: 10.2307/1543552.
Des Marais, D.J., Allamandola L.J., Benner S.A., Boss A.P., Deamer D., Falkowski P.G., Farmer J.D., Hedges S.B., Jakosky B.M., Knoll A.H., Liskowsky D.R., Meadows V.S., Meyer M.A., Pilcher C.B., Nealson K.H., Spormann A.M., Trent J.D., Turner W.W., Woolf N.J., Yorke H.W., 2003a, The NASA Astrobiology Roadmap. Astrobiology, 3(2), 219-235, DOI: 10.1089/153110703769016299.
Des Marais, D.J., Beard B., Canfield D., 2003b, Stable isotopes. In Biosignatures for Mars Exploration, edited by J.F. Kerridge, NASA publication SP-XXX, U.S. Government Printing Office, Washington DC.
Des Marais, D.J., Jakosky B.M., Hynek B.M., 2008, Astrobiological implications of Mars surface composition and properties. In The Martian Surface: Composition, Mineralogy and Physical Properties, ed. J.A. Bell, Cambridge planetary science series 9, Cambridge University Press, New York, 599-623, DOI: 10.1017/CBO9780511536076.027.
Dialynas, K., Krimigis S.M., Mitchell D.G., Decker R.B., Roelof E.C., 2017, The bubble-like shape of the heliosphere observed by Voyager and Cassini. Nature Astronomy, Volume 1, Id. 0115, DOI: 10.1038/s41550-017-0115.
Dias Pinto, J.R., Mitchell J.L., 2014. Atmospheric superrotation in an idealized GCM: Parameter dependence of the eddy response, Icarus, 238, 93-109, DOI: 10.1016/j.icarus.2014.04.036.
Djokic, T., Van Kranendonk M.J., Campbell K.A., Walter M.R., Ward C.R., 2017, Earliest signs of life on land preserved in ca. 3.5 Ga hot spring deposits. Nature Communications, 8, Id. 15263.
Dougherty, M.K., Esposito L., Krimigis S. (Eds.), 2009, Saturn from Cassini-Huygens, DOI 10.1007/978-1-4020-9217-6_9, Springer Science+Business Media B.V. (Chapter 9).

Drazkowska, J., Alibert Y., Moore B., 2016, Close-in planetesimal formation by pile-up of drifting pebbles. Astronomy and Astrophysics, 594(A105), 1-12, DOI: 10.1051/0004-6361/201628983.

Driscoll, P.E., 2018, Planetary interiors, magnetic fields, habitability. Handbook of Exoplanets, part of Springer Nature, Id.76, DOI: 10.1007/978-3-319-55333-7_76.

Eastwood, J.P., Nakamura R., Turc L., Hesse M., 2017, The Scientific Foundations of Forecasting Magnetospheric Space Weather. Space Sci. Rev., 212, 1221-1252, DOI: 10.1007/s11214-017-0399-8.

Eigenbrod, J.L., Summons R.E., Steele A., Freissinet C., Millan M., Navarro-González R., Sutter B., McAdam A.C., Franz H.B., Glavin D.P., Archer Jr P.D., Mahaffy P.R., Conrad P.G., Hurowitz J.A., Grotzinger J.P., Gupta S., Ming D.W., Sumner D.Y., Szopa C., Malespin C., Buch A., Coll P., 2018, Organic matter preserved in 3-billion-year-old mudstones at Gale crater, Mars. Science, 360, 1096-1101, DOI: 10.1126/science.aas9185.

Elkins-Tanton L.T., Weiss B.P., Zuber M.T., 2011, Chondrites as samples of differentiated planetesimals. Earth Planet. Sci. Lett., 305(1-2), 1-10, DOI: 10.1016/j.epsl.2011.03.010.

Elkins-Tanton, L.T., 2012, Magma oceans in the inner Solar System. Annual Review of Earth and Planetary Sciences, 40, 113-139, DOI: 10.1146/annurev-earth-042711-105503.

Elliot, J L., Dunham E., Mink D., 1977, The rings of Uranus. Nature, 267, 5609, 328-330, DOI: 10.1038/267328a0.

Esposito, L.W., 2006, Planetary Rings, ISBN 0521362229, Cambridge University Press.

Estrada, P.R., Durisen R.H., Latter H.N., 2018, Meteoroid Bombardment and Ballistic Transport in Planetary Rings. In: Planetary Ring Systems. Properties, Structure, and Evolution, Eds. M.S. Tiscareno and C.D. Murray, ISBN: 9781316286791, Cambridge University Press, 198-224, DOI: 10.1017/9781316286791.009.

Fairén, A.G., Gómez-Elvira J., Briones C., Prieto-Ballesteros O., Rodríguez-Manfredi J.A., López Heredero R., Belenguer T., Moral A.G., Moreno-Paz M., Parro V., 2020, The Complex Molecules Detector (CMOLD): A fluidic-based instrument suite to search for (bio)chemical complexity on Mars and icy moons. Astrobiology, 20(9), 1076-1096, DOI: 10.1089/ast.2019.2167.

Fan, S., Li C., Li J.Z., Bartlett S., Jiang J.H., Natraj V., Crisp D.,Yung Y.L., 2019, Earth as an exoplanet: A two-dimensional alien map. The Astrophysical Journal Letters, 882(1), L1, DOI: 10.3847/2041-8213/ab3a49.

Ferrais, M., Vernazza P., Jorda L., et al. 2020, Asteroid (16) psyche's primordial shape: a possible Jacobi ellipsoid. Astron. Astrophys. 638, 1-9, L15, DOI: 10.1051/0004-6361/202038100.

Fetick, R.J.L., Jorda, L., Sevecek, P., et al., 2020, A basin-free spherical shape as an outcome of a giant impact on asteroid Hygiea. Nature Astronomy, 4, 136-141, DOI: 10.1038/s41550-019-0915-8.

Filiberto, J., Dasgupta R., 2015, Constraints on the depth and thermal vigor of melting in the Martian mantle. Journal of Geophysical Research: Planets, 120(1), 109-122, DOI: 10.1002/2014JE004745.

Fleischmann, E.M., 1989, The measurement and penetration of ultraviolet radiation into tropical marine water. Limnol. Oceanogr., 34(8), 1623-1629, DOI: 10.4319/lo.1989.34.8.1623.

Fletcher, L.N., Greathouse T.K., Guerlet S., Moses J.I., and West R.A., 2018, Saturn's seasonally changing atmosphere. Thermal structure composition and aerosols. In: Saturn in the 21st Century, Eds. K.H. Baines, F.M. Flasar, N. Krupp, and T. Stallard, Cambridge University Press, Cambridge Planetary Science, ISBN: 9781316227220, DOI: 10.1017/9781316227220, Chapter 10, 251-294.

Fletcher, L.N., Kaspi Y., Guillot T., Showman A.P., 2020a, How well do we understand the belt/zone circulation of Giant Planet atmospheres? Space Sci. Rev. 216, 2, Id. 30, DOI: 10.1007/s11214-019-0631-9.

Fletcher, L.N., Simon A.A., Hofstadter M.D., Arridge C.S., Cohen I.J., Masters A., Mandt K., and Coustenis A., 2020b, Ice giant system exploration in the 2020s: an introduction.

Philosophical Transactions of the Royal Society of London Series A, 378(2187), DOI: 10.1098/rsta.2019.0473.

Fletcher, L.N., Helled R., Roussos E., Jones G., Charnoz S., André N., Andrews D., Bannister M., Bunce E., Cavalié T., Ferri F., Fortney J., Grassi D., Griton L., Hartogh P., Hueso R., Kaspi Y., Lamy L., Masters A., Melin H., Moses J., Mousis O., Nettleman N., Plainaki C., Schmidt J., Simon A., Tobie G., Tortora P., Tosi F., and Turrini D., 2020c, Ice Giant Systems: The scientific potential of orbital missions to Uranus and Neptune. Planetary and Space Science, 191, Id. 105030, DOI: 10.1016/j.pss.2020.105030.

Foley, B.J., 2018, The dependence of planetary tectonics on mantle thermal state: applications to early Earth evolution. Phil. Trans. R. Soc. A, 376(2132), DOI: 10.1098/rsta.2017.0409.

Foley, B.J., 2019, Habitability of Earth-like Stagnant Lid Planets: Climate Evolution and Recovery from Snowball States. The Astrophysical Journal, 875, 72, 1-20, DOI: 10.3847/1538-4357/ab0f31.

Forget, F., Leconte J., 2014, Possible climates on terrestrial exoplanets. Philos. Trans. R. Soc. A, 372, 20130084, DOI: 10.1098/rsta.2013.0084.

Forget, F., Korablev O., Venturini J., Imamura T., H Lammer., Blanc M., 2021, Understanding the Diversity of Planetary Atmospheres, Space Science Reviews Topical collection – Space Science Series of ISSI #81, ISSN: 0038-6308; editorial: 217, 51, 1572-9672, DOI: 10.1007/s11214-021-00820-z.

Fortney, J.J., and Hubbard W.B., 2003, Phase separation in giant planets: inhomogeneous evolution of Saturn. Icarus, 164(1), 228-243, DOI: 10.1016/S0019-1035(03)00130-1.

Fortney, J.J., and Nettelmann N., 2010, The interior structure, composition, and evolution of Giant Planets. Space Science Reviews, 152(1-4), 423-447, DOI: 10.1007/s11214-009-9582-x.

Fortney, J.J., Ikoma M., Nettelmann N., Guillot T., and Marley, M.S., 2011, Self-consistent Model Atmospheres and the Cooling of the Solar System's Giant Planets. The Astrophysical Journal, 729(1), Id. 32, 14 pages, DOI: 10.1088/0004-637X/729/1/32.

Fraser, W.C., Brown M.E., Morbidelli A., Parker A., Batygin K., 2014, The Absolute Magnitude Distribution of Kuiper Belt Objects. The Astrophysical Journal, 782, 2, Id. 100, 14 pages, DOI: 10.1088/0004-637X/782/2/100.

Freissinet, C., Glavin D.P., Mahaffy P.R., Miller K.E., Eigenbrode J.L., Summons R.E., Brunner A.E., Buch A., Szopa C., Archer P.D., Franz Jr. H. B., Atreya S.K., Brinckerhoff W. B., Cabane M., Coll P., Conrad P.G., Des Marais D.J., Dworkin J.P., Fairén A.G., François P., Grotzinger J.P., Kashyap S., ten Kate I.L., Leshin L.A., Malespin C.A., Martin M.G., Martin-Torres F.J., McAdam A.C., Ming D.W., Navarro-González R., Pavlov A.A., Prats B.D., Squyres S.W., Steele A., Stern J.C., Sumner D.Y., Sutter B., Zorzano M.-P., the MSL Science Team, 2015, Organic molecules in the Sheepbed Mudstone, Gale Crater, Mars, J. Geophys. Res. Planets, 120, 495-514, DOI: 10.1002/2014JE004737.

French, R.G., Nicholson P.D., Porco C.C., Marouf E.A., 1991, Dynamics and structure of the Uranian rings. In: Uranus (A92-18701 05-91). Tucson, AZ, University of Arizona Press, 1991, 327-409.

Frisch, P.C., Redfield S., Slavin J.D., 2011, The Interstellar Medium Surrounding the Sun. Annual Review of Astronomy and Astrophysics, 49(1), 237-279, DOI: 10.1146/annurev-astro-081710-102613.

Garate-Lopez, I., Lebonnois S., 2018, Latitudinal variation of clouds' structure responsible for Venus' cold collar. Icarus, 314, 1-11, DOI: 10.1016/j.icarus.2018.05.011.

Garvin, J.B., Arney G., Getty S., Johnson N., Kiefer W., Lorenz R., Ravine M., Malespin C., Webster C., Campbell B., Izenberg N., 2020, March. DAVINCI+: Deep Atmosphere of Venus Investigation of Noble Gases, Chemistry, and Imaging Plus. In Lunar and Planetary Science Conference (No. 2326, p. 2599).

Genova, A., Goossens S., Mazarico E., Lemoine F.G., Neumann G.A., Kuang W., Sabaka T.J., Hauck S.A. II, Smith D.E., Solomon S.C., Zuber M.T., 2019, Geodetic evidence that Mercury has a solid inner core. Geophysical Research Letters, 46(7), 3625-3633, DOI: 10.1029/2018GL081135.

Gerya, T, Stern R, Baes M, Sobolev S, Whattam S., 2015, Plate tectonics on the Earth triggered by plume-induced subduction initiation. Nature 527, 221-225, DOI: 10.1038/nature15752.

Ghail, R., 2015, Rheological and petrological implications for a stagnant lid regime on Venus. Planetary and Space Science, 113, 2-9, DOI: 10.1016/j.pss.2015.02.005.

Giardini, D., Lognonné P., Banerdt W.B., Pike W.T., Christensen U., Ceylan S., Clinton J.F., van Driel M., Stähler S.C., Böse M., Garcia R.F., Khan A., Panning M., Perrin C., Banfield D., Beucler E., Charalambous C., Euchner F., Horleston A., Jacob A., Kawamura T., Kedar S., Mainsant G., Scholz J.-R., Smrekar S.E., Spiga A., Agard C., Antonangeli D., Barkaoui S., Barrett E., Combes P., Conejero V., Daubar I., Drilleau M., Ferrier C., Gabsi T., Gudkova T., Hurst K., Karakostas F., King S., Knapmeyer M., Knapmeyer-Endrun B., Llorca-Cejudo R., Lucas A., Luno L., Margerin L., McClean J.B., Mimoun D., Murdoch N., Nimmo F., Nonon M., Pardo C., Rivoldini A., Rodriguez Manfredi J.A., Samuel H., Schimmel M., Stott A.E., Stutzmann E., Teanby N., Warren T., Weber R.C., Wieczorek M., Yana C., 2020, The seismicity of Mars. Nature Geoscience, 13(3), 205-212, DOI: 10.1038/s41561-020-0539-8.

Gillmann, C., Tackley P., 2014, Atmosphere/mantle coupling and feedbacks on Venus. Journal of Geophysical Research: Planets, 119(6), 1189-1217, DOI: 10.1002/2013JE004505.

Guillot, T., and Gautier D., 2015, Giant Planets., In: Treatise on Geophysics, 2d edition, Eds. T. Spohn, G. Schubert, Volume 10 – Planets and Moons, 42 pages, DOI: 10.1016/B978-0-444-53802-4.00176-7.

Guillot, T., Stevenson, D.J., Hubbard, W.B., and Saumon D., 2004, The interior of Jupiter. In: Jupiter. The planet, satellites and magnetosphere. Eds. F. Bagenal, T.E. Dowling, and W.B. McKinnon, Cambridge University Press, Cambridge planetary science, ISBN 0-521-81808-7, Vol. 1, 35-57.

Goldreich, P., Tremaine S.D., 1978a, The velocity dispersion in Saturn's rings. Icarus, 34, 2, 227-239, DOI: 10.1016/0019-1035(78)90164-1.

Goldreich, P., Tremaine S.D., 1978b, The formation of the Cassini division in Saturn's rings. Icarus, 34, 2, 240-253, DOI: 10.1016/0019-1035(78)90165-3.

Goldreich, P., Tremaine S., 1979, The rings of Saturn and Uranus. In: Instabilities in dynamical systems: Applications to celestial mechanics; Proceedings of the Advanced Study Institute, held July 30-August 12, 1978 in Cortina d'Ampezzo, Italy. NATO ASI Series C, Vol. 47. Dordrect: D. Reidel Publishing Co., 129-133.

Goldreich, P., Murray N., Longaretti P.Y., Banfield D., 1989, Neptune's story, Science, 245(4917), 500 500-504, DOI: 10.1126/science.245.4917.500.

Gomes, R., Levison H.F., Tsiganis K., Morbidelli A., 2005, Origin of the cataclysmic Late Heavy Bombardment period of the terrestrial planets. Nature, 435, 466-469, DOI: 10.1038/nature03676.

Gomes, R., Nesvorný D., 2016, Neptune Trojan formation during planetary instability and migration. Astronomy & Astrophysics, 592, Id. A146, 8 pages, DOI: 10.1051/0004-6361/201527757.

Gonçalves, R., Machado P., Widemann T., Peralta J., Watanabe S., Yamazaki A., Satoh T., Takagi M., Ogohara K., Lee Y.-J., Harutyunyan A., Silva J., 2020. Venus' cloud top wind study: Coordinated Akatsuki/UVI with cloud tracking and TNG/HARPS-N with Doppler velocimetry observations, Icarus 335, 113418, DOI: 10.1016/j.icarus.2019.113418.

Grasset, O., Dougherty M.K., Coustenis A., Coustenis A., Bunce E., Erd C., Titov D., Blanc M., Coates A., Drossart P., Fletcher L., Hussmann H., Jaumann R., Krupp N., Lebreton J.P., Prieto-Ballesteros O., Tortora P., Tosi F., Van Hoolst T., 2013, JUpiter ICy moons Explorer (JUICE): An ESA mission to orbit Ganymede and to characterise the Jupiter system. Planetary and Space Science, 78, 1-21, DOI: 10.1016/j.pss.2012.12.002.

Granvik, M., Morbidelli A., Jedicke R., Bolin B., Bottke W.F., Beshore E., Vokrouhlický D., Nesvorný D., Michel P., 2018, Debiased orbit and absolute-magnitude distributions for near-Earth objects. Icarus, 312, 181-207, DOI: 10.1016/j.icarus.2018.04.018.

Graykowski, A., Jewitt D., 2018, Colors and Shapes of the Irregular Planetary Satellites.The Astronomical Journal, 155, 4, 184, DOI: 10.3847/1538-3881/aab49b.

Grenfell, J. L., Leconte J., Forget F., Godolt M., Carrión-González Ó., Noack L., Tian F., Rauer H., Gaillard F., Bolmont A., Charnay B., Turbet M., 2020, Possible Atmospheric Diversity of Low Mass Exoplanets – Some Central Aspects. Space Sci. Rev., 216(5), 98, DOI: 10.1007/s11214-020-00716-4.

Grotzinger, J.P., Crisp J., Vasavada A.R., Anderson R.C., Baker C.J., Barry R., Blake D.F., Conrad P., Edgett K.S., Ferdowski B., Gellert R., Gilbert J.B., Golombek M., Gómez-Elvira J., Hassler D.M., Jandura L., Litvak M., Mahaffy P., Maki J., Meyer M., Malin M.C., Mitrofanov I., Simmonds J.J., Vaniman D., Welch R.V., Wiens R.C., 2012, Mars Science Laboratory mission and science investigation. Space Sci. Rev., 170, 5-56, DOI: 10.1007/s11214-012-9892-2.

Grotzinger, J.P., Sumner D.Y., Kah L.C., Stack K., Gupta S., Edgar L., Rubin D., Lewis K., Schieber J., Mangold N., Milliken R., Conrad P.G., DesMarais D., Farmer J., Siebach K., Calef F. III, Hurowitz J., McLennan S.M., Ming D., Vaniman D., Crisp J., Vasavada A., Edgett K.S., Malin M., Blake D., Gellert R., Mahaffy P., Wiens R.C., Maurice S., Grant J.A., Wilson S., Anderson R.C., Beegle L., Arvidson R., Hallet B., Sletten R.S., Rice M., Bell J. III, Griffes J., Ehlmann B., Anderson R.B., Bristow T.F., Dietrich W.E., Dromart G., Eigenbrode J., Fraeman A., Hardgrove C., Herkenhoff K., Jandura L., Kocurek G., Lee S., Leshin L.A., Leveille R., Limonadi D., Maki J., McCloskey S., Meyer M., Minitti M., Newsom H., Oehler D., Okon A., Palucis M., Parker T., Rowland S., Schmidt M., Squyres S., Steele A., Stolper E., Summons R., Treiman A., Williams R., Yingst A., MSL Science Team, 2014, A Habitable Fluvio-Lacustrine Environment at Yellowknife Bay, Gale Crater, Mars. Science 343(6169), 1242777, DOI: 10.1126/science.1242777.

Grün, E., Krüger H., Srama R., 2019, The Dawn of Dust Astronomy. Space Science Reviews, 215, 46, 97 pages, DOI: 10.1007/s11214-019-0610-1.

Gurnett, D.A., Ansher J.A., Kurth W.S., Granroth L.J., 1997, Micron-sized dust particles detected in the outer Solar System by the Voyager 1 and 2 plasma wave instruments. Geophysical Research Letters, 24(24), 3125-3128, DOI: 10.1029/97GL03228.

Gurnett, D.A., Kurth W.S., Burlaga L.F., Ness N.F., 2013, In-situ Observations of Interstellar Plasma with Voyager 1, Science, 341(6153), 1489-1492, DOI: 10.1126/science.1241681.

Gurnett, D.A., Kurth W.S., 2019, Plasma densities near and beyond the heliopause from the Voyager 1 and 2 plasma wave instruments. Nature Astronomy, 3, 1024-1028, DOI: 10.1038/s41550-019-0918-5.

Halliday, A.N., 2013, The origins of volatiles in the terrestrial planets. Geochimica et Cosmochimica Acta, 105, 146-171, DOI: 10.1016/j.gca.2012.11.015.

Hansen, B.M.S., 2009, Formation of the Terrestrial Planets from a Narrow Annulus. The Astrophysical Journal, 703(1), 1131-1140, DOI: 10.1088/0004-637X/703/1/1131.

Hanus, J., Marsset M., Vernazza P., et al., 2019, The shape of (7) Iris as evidence of an ancient large impact. Astronomy & Astrophysics, 624, 1-17, A121, DOI: 10.1051/0004-6361/201834541.

Harris, A. W., Benz W., Fitzsimmons A., Galvez A., Green S.F., Michel P., and Valsecchi G.B., 2004, The near-earth object impact hazard: space mission priorities for risk assessment and reduction. ESA NEOMAP (Near-Earth Object Mission Advisory Panel) report. In: Proc. International Seminar on Nuclear War and Planetary Emergencies — 32nd Session, 185-185, DOI: 10.1142/9789812701787_0021.

Hartmann, W.K., 2019, History of the Terminal Cataclysm Paradigm: Epistemology of a Planetary Bombardment That Never (?) Happened. Geosciences, 9(7), 285, DOI: 10.3390/geosciences9070285.

Hatzes, A.P., 2016, The Architecture of Exoplanets. Space Science Reviews, 205(1-4), DOI: 10.1007/s11214-016-0246-3.

Helled, R., Anderson J.D., Podolak M., and Schubert G., 2011, Interior Models of Uranus and Neptune. The Astrophysical Journal, 726(1), Id. 15, 7 pages, DOI: 10.1088/0004-637X/726/1/15.

Hendrix A.R., Hurford T.A., Barge L.M., Bland M.T., Bowman J.S., Brinckerhoff W., Buratti B.J., Cable M.L., Castillo-Rogez J., Collins G.C., Diniega S., German C.R., Hayes A.G., Hoehler T., Hosseini S., Howett C.J.A., McEwen A.S., Neish C.D., Neveu M., Nordheim

T.A., Patterson G.W., Patthoff D.A., Phillips C., Rhoden A., Schmidt B.E., Singer K.N., Soderblom J.M., Vance S.D., 2019, Astrobiology, DOI: 10.1089/ast.2018.1955.
Heng, K., Showman A.P., 2015, Atmospheric Dynamics of Hot Exoplanets. Annual Review of Earth and Planetary Sciences, 43, 509-540, DOI: 10.1146/annurev-earth-060614-105146.
Hewins, R.H., Zanda B., Humayun M., Nemchin A., Lorand J.-P., Pont S., Deldicque D., Bellucci J.J., Beck P., Leroux H., Marinova M., Remusat L., Göpel C., Lewin E., Grange M., Kennedy A., Whitehouse M.J., 2017, Regolith Breccia Northwest Africa 7533: Mineralogy and Petrology With Implications For Early Mars, Meteoritics Planetary Science, 52, 89-124, DOI: 10.1111/maps.12740.
Hickman-Lewis, K., Cavalazzi B., Sorieul S., Gautret P., Foucher F., Whitehouse M.J., Jeon H., Georgelin T., Cockell C.S., Westall F., 2020, Metallomics in deep time and the influence of ocean chemistry on the metabolic landscapes of Earth's earliest ecosystems. Sci. Rep., 10, 4965, DOI: 10.1038/s41598-020-61774-w.
Höning, D., Spohn T., 2016, Continental growth and mantle hydration as intertwined feedback cycles in the thermal evolution of Earth. Physics of the Earth and Planetary Interiors, 255, 27-49, DOI: 10.1016/j.pepi.2016.03.010.
Höning, D., Tosi N., Spohn T., 2019, Carbon cycling and interior evolution of water-covered plate tectonics and stagnant-lid planets. Astronomy Astrophysics, 627, A48, 1-15, DOI: 10.1051/0004-6361/201935091.
Höning, D., 2020, The impact of life on climate stabilization over different timescales. Geochemistry, Geophysics, Geosystems, 21 (9), e2020GC009105, DOI: 10.1029/2020GC009105.
Horanyi, M., 1996, Charged Dust Dynamics in the Solar System. Annual Review of Astronomy and Astrophysics, 34, 383-418, DOI: 10.1146/annurev.astro.34.1.383.
Horanyi, M., Burns J.A., Hedman M.M., Jones G.H., Kempf S., 2009, Diffuse Rings. In: Saturn from Cassini-Huygens, Eds. M.K. Dougherty, L.W. Esposito, and S.M. Krimigis, ISBN 978-1-4020-9216-9, Springer Science+Business Media B.V., 511, DOI: 10.1007/978-1-4020-9217-6_16.
Hori, Y. and Ikoma M., 2011, Gas giant formation with small cores triggered by envelope pollution by icy planetesimals. Monthly Notices of the Royal Astronomical Society, 416, 1419-1429, DOI: 10.1111/j.1365-2966.2011.19140.x.
Horinouchi, T., Kouyama T., Lee Y.J., Murakami S.-Y, Ogohara K., Takagi M., Imamura T., Nakajima K., Peralta J., Yamazaki A., Yamada M., Watanabe S., 2018, Earth, Planets and Space, 70, 1, Id. 10, 19 pages, DOI: 10.1186/s40623-017-0775-3.
Horinouchi, T., Y.-Y. Hayashi, Watanabe S., Yamada M., Yamazaki A., Kouyama T., Taguchi M., Fukuhara T., Takagi M., Ogohara K., Murakami S.-Y., Peralta J., Limaye S.S., Imamura T., Nakamura M., Sato T.M., Satoh T., 2020, How waves and turbulence maintain the super-rotation of Venus' atmosphere. Science, 368, 405-409, DOI: 10.1126/science.aaz4439.
Hubbard, W.B., and Militzer B., 2016, A Preliminary Jupiter Model. The Astrophysical Journal, 820(1), Id. 80, 13 pages, DOI: 10.3847/0004-637X/820/1/80.
Hubbard, W.B., Pearl J.C., Podolak M., and Stevenson D.J., 1995. The Interior of Neptune, In: Neptune and Triton, University of Arizona Press, Ed. D.P. Cruishank, 109-138.
Hueso, R., and Sánchez-Lavega A., 2019, Atmospheric Dynamics and Vertical Structure of Uranus and Neptune's Weather Layers. Space Science Reviews, 215, Id. 52, DOI: 10.1007/s11214-019-0618-6.
Hubbard, W.B., Brahic A., Sicardy B, Elicer L.-R., Roques F., Vilas F., 1986, Occultation detection of a neptunian ring-like arc. Nature, 319, 6055, 636-640, DOI: 10.1038/319636a0.
Iess, L., Stevenson D.J., Parisi M., Hemingway D., Jacobson R.A., Lunine J.I., Nimmo F., Armstrong J.W., Asmar S.W., Ducci M., Tortora P., 2014, The gravity field and interior structure of Enceladus, Science, 344(6179), 78-80, DOI: 10.1126/science.1250551.
Imamura, T., Mitchell J., Lebonnois S., Kaspi Y., Showman A.P., Korablev O., 2020, Superroration in Planetary Atmospheres. Space Sci. Rev., 216, 87, DOI: 10.1007/s11214-020-00703-9.

Ingersoll, A.P., Dowling T.E., Gierasch P.J., Orton G.S., Read P.L., Sánchez-Lavega A., Showman A.P., Simon-Miller A.A., and Vasavada A.R., 2004, Dynamics of Jupiter's atmosphere. In: Jupiter. The Planet, Satellites and Magnetosphere, Eds. F. Bagenal, T.E. Dowling, W.B. McKinnon, ISBN: 9780521035453, Cambridge University Press, Cambridge Planetary Science, Chapter 6, 105-128.
Jakosky, B.M., Grebowsky J.M., Luhmann J.G., Connerney J., Eparvier F, Ergun R., Halekas J., Larson D., Mahaffy P., McFadden J., Mitchell D.L., Schneider N., Zurek R., Bougher S., Brain D., Ma Y.J., Mazelle C., Andersson L., Andrews D., Baird D., Baker D., Bell J.M., Benna M., Chaffin M., Chamberlin P., Chaufray Y.-Y., Clarke J., Collinson G., Combi M., Crary F., Cravens T., Crismani M., Curry S, Curtis D, Deighan J, Delory G, Dewey R, DiBraccio G, Dong C, Dong Y, Dunn P, Elrod M, England S, Eriksson A., Espley J, Evans S, Fang X, Fillingim M, Fortier K., Fowler C.M, Fox J, Gröller H, Guzewich S, Hara T, Harada Y, Holsclaw G, Jain S.K, Jolitz R, Leblanc F, Lee C.O, Lee Y., Lefevre F, Lillis R, Livi R., Lo D, Mayyasi M, McClintock W, McEnulty T, Modolo R, Montmessin F, Morooka M, Nagy A, Olsen K, Peterson W, Rahmati A, Ruhunusiri S, Russell C.T, Sakai S., Sauvaud J.-A, Seki K, Steckiewicz M, Stevens M, Stewart A.I.F, Stiepen A, Stone S., Tenishev V, Thiemann E, Tolson R, Toublanc D, Vogt M, Weber T, Withers P, Woods T, Yelle R., 2015, MAVEN observations of the response of Mars to an interplanetary coronal mass ejection. Science 350(6261), DOI: 10.1126/science.aad0210.
Jakosky, B.M., 2017, MAVEN observations of the Mars upper atmosphere, ionosphere, and solar wind interactions, Journal of Geophysical Research Space Physics, 122(9), 9552-9553, DOI: 10.1002/2017ja024324.
Janssen, M.A., Ingersoll, A.P., Allison, M.D., Gulkis, S., Laraia, A.L., Baines, K.H., Edgington, S.G., Anderson, Y.Z., Kelieher, K., Oyafuso, F.A., 2013, Saturn’s thermal emission at 2.2-cm wavelength as imaged by the Cassini RADAR radiometer. Icarus 226, 522-535, DOI: 10.1016/j.icarus.2013.06.008.
Janssen, M. A.; Oswald, J. E.; Brown, S. T.; Gulkis, S.; Levin, S. M.; Bolton, S. J.; Allison, M. D.; Atreya, S. K.; Gautier, D.; Ingersoll, A. P.; Lunine, J. I.; Orton, G. S.; Owen, T. C.; Steffes, P. G.; Adumitroaie, V.; Bellotti, A.; Jewell, L. A.; Li, C.; Li, L.; Misra, S. ; ..., 2017 MWR: Microwave Radiometer for the Juno Mission to Jupiter. Space Science Reviews, 213, 1-4, 139-185, DOI: 10.1007/s11214-017-0349-5.
Javaux, E.J., 2019. Challenges in evidencing the earliest traces of life. Nature 572, 451-460, DOI: 10.1038/s41586-019-1436-4.
Jewitt, D., Haghighipour, N., 2007, Irregular satellites of the planets: products of capture in the early Solar System, Ann. Rev. Astron. Astrophys. 45, 261-295, DOI: 10.1146/annurev.astro.44.051905.092459.
Johnson, T.V., Castillo-Rogez J.C., Matson D.L., Morbidelli A., Lunine J.I., 2008, Constraints on outer Solar System early chronology. Early Solar System Impact Bombardment conference, Lunar and Planetary Institute Science Conference Abstracts, March 1, 2008, 39, 2314.
Johnson, C.L., Phillips R.J., Purucker M.E., Anderson B.J., Byrne P.K., Denevi B.W., Feinberg J.M., Hauck S.A. II, Head J.W. III, Korth H., James P.B., Mazarico E., Neumann G.A., Philpott L.C., Siegler M.A., Tsyganenko N.A., Solomon S.C., 2015, Low-altitude magnetic field measurements by MESSENGER reveal Mercury’s ancient crustal field. Science, 348(6237), 892-895, DOI: 10.1126/science.aaa8720.
Johnson, C.L., Mittelholz A., Langlais B., Russell C.T., Ansan V., Banfield D., Chi P.J., Fillingim M.O., Forget F., Fuqua Haviland H., Golombek M., Joy S., Lognonné P., Liu X., Michaut C., Pan L., Quantin-Nataf C., Spiga A., Stanley S., Thorne S.N., Wieczorek M.A., Yu Y., Smrekar S.E., Banerdt W.B., 2020, Crustal and time-varying magnetic fields at the InSight landing site on Mars. Nature Geoscience, 13(3), 199-204, DOI: 10.1038/s41561-020-0537-x.
Kahan, D.S., Folkner W.M., Buccino D.R., Dehant V., Le Maistre S., Rivoldini A., Van Hoolst T., Yseboodt M., Marty J.C., 2021, Mars Precession Rate Determined from Radiometric Tracking of the InSight Lander. Planet. Space Sci., 199, 105208, DOI: 10.1016/j.pss.2021.105208.

Kaltenegger, L., Sasselov D., 2009, Detecting planetary geochemical cycles on exoplanets: atmospheric signatures and the case of SO2. The Astrophysical Journal, 708(2), 1162-1167, DOI: 10.1088/0004-637X/708/2/1162.

Kane, S.R., Arney G., Crisp D., Domagal-Goldman S., Glaze L.S., Goldblatt C., Grinspoon D., Head J.W., Lenardic A., Unterborn C., Way M.J., Zahnle K.J., 2019, Venus as a Laboratory for Exoplanetary Science. J. Geophys. Res. Planets, 124(8), 2015-2028, DOI: 10.1029/2019JE005939.

Kaspi, Y., Galanti E., Showman A.P., Stevenson D.J., Guillot T., Iess L., Bolton S.J., 2020, Comparison of the deep atmospheric dynamics of Jupiter and Saturn in light of the Juno and Cassini gravity measurements. Space Sci. Rev., 216, 84, DOI: 10.1007/s11214-020-00705-7.

Kasting, J.F., Catling D., 2003, Evolution of a habitable planet. Annual Review of Astronomy and Astrophysics, 41(1), 429-463, DOI: 10.1146/annurev.astro.41.071601.170049.

Kasting, J.F., Whitmire D.P., Reynolds R.T., 1993, Habitable zones around main sequence stars. Icarus, 101(1), 108-128, DOI: 10.1006/icar.1993.1010.

Kawahara, h., 2020, global mapping of the surface composition on an exo-Earth using color variability. The Astrophysical Journal, 894(1), Id. 58, DOI: 10.3847/1538-4357/ab87a1.

Kerrick, D.M., Connolly J.A.D., 2001, Metamorphic devolatilization of subducted oceanic metabasalts: implications for seismicity, arc magmatism and volatile recycling. Earth and Planetary Science Letters, 189(1-2), 19-29, DOI: 10.1016/S0012-821X(01)00347-8.

Khan, A., Liebske C., Rozel A., Rivoldini A., Nimmo F., Connolly J.A.D., Plesa A.-C., Giardini D., Lognonné P., Samuel H., Schmerr N.C., Stähler S.C., Duran A.C., Huang Q., et al., 2018, A Geophysical Perspective on the Bulk Composition of Mars. Journal of Geophysical Research Planets, 123(2), 575-611, DOI: 10.1002/2017JE005371.

Khan, A., Ceylan S., van Driel M., Giardini D., Lognonné P., Samuel H., Schmerr N.C., Stähler S.C., Duran A.C., Huang Q., et al., 2021, Upper mantle structure of Mars from InSight seismic data.. Science, 373(6553), 434-438, DOI: 10.1126/science.abf2966.

King, E.M. and Aurnou J.M., 2013, Turbulent convection in liquid metal with and without rotation. Proceedings of the National Academy of Sciences, 110(17), 6688-6693.

Kite, E.S., Manga M., Gaidos E., 2009, Geodynamics and rate of volcanism on massive Earth-like planets. The Astrophysical Journal, 700(2), 1732-1749, DOI: 10.1088/0004-637X/700/2/1732.

Klahr, H., and Schreiber A., 2020, Turbulence Sets the Length Scale for Planetesimal Formation: Local 2D Simulations of Streaming Instability and Planetesimal Formation. The Astrophysical Journal, 901, DOI: 10.3847/1538-4357/abac58.

Kleidon, A., 2016, Thermodynamic foundations of the Earth system. Cambridge University Press.

Kleine, T., Touboul M., Bourdon B., Nimmo F., Mezger K., Palme H., Jacobsen S.B., Yin Q.-Z., Halliday A.N., 2009, Hf-W chronology of the accretion and early evolution of asteroids and terrestrial planets. Geochimica et Cosmochimica Acta, 73, 5150-5188, DOI: 10.1016/j.gca.2008.11.047.

Koll, D., Korschinek G., Faestermann T., Gómez-Guzmán J.M., Kipfstuhl S., Merchel S., Welch J.M., 2019, Interstellar 60Fe in Antarctica. Physical Review Letters, 123(7), Id. 072701, DOI: 10.1103/PhysRevLett.123.072701.

Konstantinidis, K., Martinez C.L.F., Dachwald B., Ohndorf A., Dykta P., Bowitz P., Rudolph M., Digel I., Kowalski J., Voigt K., Förstner R., 2015, A lander mission to probe subglacial water on Saturn′ s moon Enceladus for life. Acta astronautica, 106, 63-89, DOI: 10.1016/j.actaastro.2014.09.012.

Korenaga, J., 2007, Thermal cracking and the deep hydration of oceanic lithosphere: a key to the generation of plate tectonics? Journal of Geophysical Research: Solid Earth, 112(B5), Id. B05408, DOI: 10.1029/2006JB004502.

Koschny, D., Soja R.H., Engrand C., Flynn G.J., Lasue J., Levasseur-Regourd A.C., Malaspina D., Nakamura T., Poppe A.R., Sterken V.J., Trigo-Rodríguez J.M., 2019, Interplanetary dust, meteoroids, meteors and meteorites. Space Science Reviews, 215, 4, Id. 34, 62 pages, DOI: 10.1007/s11214-019-0597-7.

Krimigis, S.M., Decker R.B., Roelof E.C., Hill M.E., Armstrong T.P., Gloeckler G., Hamilton D.C., Lanzerotti L.J., 2013, Search for the Exit: Voyager 1 at Heliosphere's Border with the Galaxy. Science, 341(6142), 144-147, DOI: 10.1126/science.1235721.

Krimigis, S.M., Decker R.B., Roelof E.C., Hill M.E., Bostrom C.O., Dialynas K., Gloeckler G., Hamilton D.C., Keath E.P., Lanzerotti L.J., 2019, Energetic charged particle measurements from Voyager 2 at the heliopause and beyond. Nature Astron., 3, 997-1006, DOI: 10.1038/s41550-019-0927-4.

Krüger, H., Hamilton D.P., Moissl R., Grün E., 2009, Galileo in-situ dust measurements in Jupiter's gossamer rings. Icarus, 203, 1, 198-213, DOI: 10.1016/j.icarus.2009.03.040.

Kurosaki, K., and Ikoma M., 2017, Acceleration of cooling of ice giants by condensation in early atmospheres. The Astronomical Journal, 153(6), Id. 260, 9 pages, DOI: 10.3847/1538-3881/aa6faf.

Lambrechts, M., Johansen A., 2012, Rapid growth of gas-giant cores by pebble accretion. Astronomy & Astrophysics, 544, Id. A32, 13 pages, DOI: 10.1051/0004-6361/201219127.

Lambrechts, M., Lega E., Nelson R.P., Crida A., Morbidelli A., 2019, Quasi-static contraction during runaway gas accretion onto giant planets. Astronomy & Astrophysics, 630, Id. A82, 10 pages, DOI: 10.1051/0004-6361/201834413.

Lammer, H., Bredehöft, J.H., Coustenis, A., Khodachenko M.L., Kaltenegger L., Grasset O., Prieur D., Raulin F., Ehrenfreund P., Yamauchi M., Wahlund J.-E., Grießmeier J.-M., Stangl G., Cockell C.S., Kulikov Y.N., Grenfell J.L., Rauer H., 2009, What makes a planet habitable? The Astronomy and Astrophysics Review, 17(2), 181-249, DOI: 10.1007/s00159-009-0019-z.

Lammer, H., Blanc M., 2018, From Disks to Planets: The making of planets and their early atmospheres. An Introduction. Space Science Reviews, 214, 2, Id. 60, 35 pages, DOI: 10.1007/s11214-017-0433-x.

Lang, K. R., 2011, The Cambridge Guide to the Solar System. Cambridge University Press, 2nd edition, March 2011.

Langmuir, I., 1928, Oscillations in Ionized Gases. Proceedings of the National Academy of Science, 14, 627–637, Aug. DOI: 10.1073/pnas.14.8.627.

Langseth, M.G., Keihm S.J., Peters K., 1976, Revised lunar heat-flow values. In Lunar and Planetary Science Conference Proceedings, 7, 3143-3171.

Lasue, J., Bousquet P., Blanc M., André N., Beck P., Berger G., Bolton S., Bunce E., Chide B., Foing B., Hammel H., Lellouch E., Griton L., Mcnutt R., Maurice S., Mousis O., Opher M., Sotin C., Senske D., Spilker L., Vernazza P., Zong Q., 2022. Planetary Exploration, Horizon 2061", Report - Chapter 4: From Planetary Exploration Goals to Technology Requirements. Science Direct, Elsevier.

Lawrence, D.J., Feldman W.C., Goldsten J.O., Maurice S., Peplowski P.N., Anderson B.J., Rodgers D.J., 2013, Evidence for water ice near Mercury's north pole from MESSENGER Neutron Spectrometer measurements. Science, 339(6117), 292-296, DOI: 10.1126/science.1229953.

Leconte, J., and Chabrier G., 2012, A new vision of giant planet interiors: Impact of double diffusive convection. Astronomy Astrophysics, 540, Id. A20, 13 pages, DOI: 10.1051/0004-6361/201117595.

Leconte, J., and Chabrier G., 2013, Layered convection as the origin of Saturn's luminosity anomaly. Nature Geoscience, 6(5), 347-350, DOI: 10.1038/ngeo1791.

Ledoux, P., 1947, On stellar models with convection and discontinuity of the mean molecular weight. Astronomical Journal, 52, 155, DOI: 10.1086/105977.

Lenardic, A., Jellinek A.M., Foley B., O'Neill C., Moore W.B., 2016, Climate-tectonic coupling: Variations in the mean, variations about the mean, and variations in mode. Journal of Geophysical Research: Planets, 121(10), 1831-1864, DOI: 10.1002/2016JE005089.

Lenton, T.M., von Bloh W., 2001, Biotic feedback extends the life span of the biosphere. Geophysical research letters, 28 (9), 1715-1718, DOI: 10.1029/2000GL012198.

Levison, H.F., Bottke W.F., Gournelle M., Morbidelli A., Nesvorný D., Tsiganis K., 2009, Contamination of the asteroid belt by primordial trans-Neptunian objects. Nature, 460, 7253, 364-366, DOI: 10.1038/nature08094.

Levison, H.F., Morbidelli, A., Van Laerhoven, C., Gomes, R.S., Tsiganis, K., 2007, Origin of the structure of the Kuiper belt during a dynamical instability in the orbits of Uranus and Neptune. Icarus 196 (1), 258, DOI: 10.1016/j.icarus.2007.11.035.

Li, J., Kusky T.M., 2007, World's largest known Precambrian fossil black smoker chimneys and associated microbial vent communities, North China: Implications for early life, Gondwana Research, 12(1-2), 84-100, DOI: 10.1016/j.gr.2006.10.024

Li, C., Ingersoll A., Bolton S., Levin S., Janssen M., Atreya S., Lunine J., Steffes P., Brown S., Guillot T., Allison M., Arballo J., Bellotti A., Adumitroaie V., Gulkis S., Hodges A., Li L., Misra S., Orton G., Oyafuso F., Santos-Costa D., Waite H., and Zhang Z., 2020, The water abundance in Jupiter's equatorial zone. Nature Astron., 4(6), 609-616, DOI: 10.1038/s41550-020-1009-3.

Linsky, J.L., Redfield S., Tilipman D., 2019, The Interface between the Outer Heliosphere and the Inner Local ISM: Morphology of the Local Interstellar Cloud, Its Hydrogen Hole, Strömgren Shells, and 60Fe Accretion. The Astrophysical Journal, 886(1), Id. 41, 1-19, DOI: 10.3847/1538-4357/ab498a.

Liu, J., Goldreich P.M., Stevenson D.J., 2008, Constraints on deep-seated zonal winds inside Jupiter and Saturn. Icarus, 196 (2), 653-664, DOI: 10.1016/j.icarus.2007.11.036.

Lognonné, P., Banerdt W.B., Pike W.T., et al., 2020, Constraints on the shallow elastic and anelastic structure of Mars from InSight seismic data. Nat. Geosci., 13, 213-220, DOI: 10.1038/s41561-020-0536-y.

Longaretti, P. -Y., 2018, Theory of Narrow Rings and Sharp Edges. In: Planetary Ring Systems. Properties, Structure, and Evolution, Eds. M.S. Tiscareno and C.D. Murray, ISBN: 9781316286791, Cambridge University Press, 225-275, DOI: 10.1017/9781316286791.010.

Lozovsky, M., Helled R., Rosenberg E.D., and Bodenheimer P., 2017, Jupiter's Formation and Its Primordial Internal Structure. The Astrophysical Journal, 836(2), Id. 227, 16 pages, DOI: 10.3847/1538-4357/836/2/227.

Magnabosco, C., Lin LH., Dong H. et al., 2018, The biomass and biodiversity of the continental subsurface. Nature Geosci., 11, 707-717, DOI: 10.1038/s41561-018-0221-6.

Mangold, N., Dromart G., Ansan V., Massé M., Salese F., Kleinhans M., 2020, Fluvial Regimes, Age and Duration of Jezero Crater Paleolake and its Significance for the 2020 Rover Mission Landing Site, Astrobiology, 20(8), 994-1013, DOI: 10.1089/ast.2019.2132.

Mankovich, C., Fortney J.J., and Moore K.L., 2016, Bayesian Evolution Models for Jupiter with Helium Rain and Double-diffusive Convection. The Astrophysical Journal, 832(2), Id. 113, 13 pages, DOI: 10.3847/0004-637X/832/2/113.

Mankovich, C., and Fuller J., 2021, A diffuse core in Saturn revealed by ring seismology, arXiv, arXiv:2104.13385.x.

Margot, J.-L., Hauck S.A., Mazarico E., Padovan, Peale S.J., 2018, Mercury's internal structure. In Solomon, S.C., Anderson, B.J., Nittler, L.R., editors, Mercury, the view after MESSENGER. Cambridge University Press, Cambridge, UK.

Margot, J.L., Campbell D.B., Giorgini J.D., Jao J.S., Snedeker L.G., Ghigo F.D., Bonsall A., 2021, Spin state and moment of inertia of Venus. Nature Astronomy, 1-8.

Marley, M.S., Gómez P., and Podolak M., 1995, Monte Carlo interior models for Uranus and Neptune. Journal of Geophysical Research, 100(E11), 23349-23354, DOI: 10.1029/95JE02362.

Marsset, M., Broz M., Vernazza P., et al., 2020, The violent collisional history of aqueously evolved (2) Pallas. Nature Astronomy, 4, 569-576, DOI: 10.1038/s41550-019-1007-5.

Martin, W., Baross J., Kelley D., Russell M.J., 2008, Hydrothermal vents and the origin of life. Nat. Rev. Microbiol. 6, 805-814, DOI: 10.1038/nrmicro1991.

Masset, F., Snellgrove M., 2001, Reversing type II migration: resonance trapping of a lighter giant protoplanet. Monthly Notices of the Royal Astronomical Society, 320, L55-L59, DOI: 10.1046/j.1365-8711.2001.04159.x.

Mattingly, R., and May L., 2011, March. Mars sample return as a campaign. In 2011 IEEE Aerospace Conference, 1-13.

McComas, D.J., Alexashov D., Bzowski M., Fahr H., Heerikhuisen J., Izmodenov V., Lee M.A., Möbius E., Pogorelov N., Schwadron N.A., Zank G.P., 2012, The Heliosphere's Interstellar Interaction: No Bow Shock. Science, 336(6086), 1291-1293, DOI: 10.1126/science.1221054.
McComas, D.J., Bzowski M., Frisch P., Fuselier S.A., Kubiak M.A., Kucharek H., Leonard T., Möbius E., Schwadron N.A., Sokół J.M., Swaczyna P., Witte M., 2015, Warmer Local Interstellar Medium: A Possible Resolution of the Ulysses-IBEX Enigma. The Astrophysical Journal, 801(1), Id. 28, 1-7, DOI: 10.1088/0004-637X/801/1/28.
McKay, C.P., Porco C.C., Altheide T., Davis W.L., Kral T.A., 2008, The possible origin and persistence of life on Enceladus and detection of biomarkers in the plume. Astrobiology, 8(5), 909-919, DOI: 10.1089/ast.2008.0265.
McKinnon, W. B., Lunine J., Banfield D., 1995, Origin and evolution of Triton. In: Neptune and Triton, 807-877.
Meyer-Vernet, N., 2007, Basics of the Solar Wind. Cambridge University Press, Jan 2007.
Michalski, J., Cuadros J., Niles P., et al., 2013, Groundwater activity on Mars and implications for a deep biosphere. Nature Geosci., 6, 133-138, DOI: 10.1038/ngeo1706.
Milillo, A., Fujimoto M., Murakami G., et al., 2020, Investigating Mercury's Environment with the Two-Spacecraft BepiColombo Mission. Space Sci Rev 216, 93, DOI: 10.1007/s11214-020-00712-8.
Mitrovica, J.X., Forte A.M., 2004, A new inference of mantle viscosity based upon joint inversion of convection and glacial isostatic adjustment data. Earth and Planetary Science Letters, 225(1-2), 177-189, DOI: 10.1016/j.epsl.2004.06.005.
Molter, E.M., de Pater I., Luszcz-Cook S., Tollefson J., Sault R.J., Butler B., de Boer D., 2021, Tropospheric composition and circulation of Uranus with ALMA and the VLA. Planet. Sci. J., 2(1), DOI: 10.3847/psj/abc48a.
Moore, W.B., Hussmann H., 2009, Thermal evolution of Europa's silicate interior. In: Europa, eds. R.T. Pappalardo, W.B. McKinnon, K. Khurana, University of Arizona Press, 369-380.
Morbidelli, A., Levison H.F., Tsiganis K., Gomes R., 2005, The chaotic capture of Jovian Trojan asteroids during the early dynamical evolution of the Solar System., Nature, 435, 462-465, DOI: 10.1038/nature03540.
Morbidelli, A., Crida, A. 2007, The dynamics of Jupiter and Saturn in the gaseous protoplanetary disk. Icarus, 191, 158-171, DOI: 10.1016/j.icarus.2007.04.001.
Morbidelli, A., Tsiganis, K., Crida, A., Levison, H.F., Gomes, R. 2007, Dynamics of the giant planets of the Solar System in the gaseous protoplanetary disk and their relationship to the current orbital architecture, Astronomical Journal, 134, 1790-1798, DOI: 10.1086/521705.
Morbidelli, A., Lunine J.I., O'Brien D.P., Raymond S.N., Walsh K.J., 2012, Building terrestrial planets. Annual Review of Earth and Planetary Sciences, 40, 251-275, DOI: 10.1146/annurev-earth-042711-105319.
Morbidelli, A., Raymond S.N., 2016, Challenges in planet formation. Journal of Geophysical Research: Planets, 121, 10, 1962-1980, DOI: 10.1002/2016JE005088.
Morbidelli, A., Nesvorny D., Laurenz V., Marchi S., Rubie D.C., Elkins-Tanton L., Wieczorek M., Jacobson S., 2018, The timeline of the lunar bombardment: Revisited. Icarus, 305, 262-276, DOI: 10.1016/j.icarus.2017.12.046.
Morbidelli, A., Libourel G., Palme H., Jacobson S.A., Rubie D.C., 2020, Subsolar Al/Si and Mg/Si ratios of non-carbonaceous chondrites reveal planetesimal formation during early condensation in the protoplanetary disk. Earth and Planetary Science Letters, 538, DOI: 10.1016/j.epsl.2020.116220.
Moses, J.I., Cavalié T., Fletcher L.N., and Roman M.T., 2020, Atmospheric chemistry on Uranus and Neptune. Philosophical Transactions of the Royal Society of London Series A, 378(2187), DOI: 10.1098/rsta.2019.0477.
Mousis, O., Atkinson D.H., Cavalié T., Fletcher L.N., Amato M.J., Aslam S., Ferri F., Renard J.B., Spilker T., Venkatapathy E., Wurz P., 2018, Scientific rationale for Uranus and Neptune in situ explorations. Planetary and Space Science, 155, 12-40.
Mousis, O., Atkinson D.H., et al., 2021, In Situ Exploration of the Giant Planets. White Paper submitted in response to ESA's Call for Voyage 2050 Science Themes.

https://www.cosmos.esa.int/documents/1866264/3219248/MousisO_WP_final.pdf/f60f9c82-aa40-8c67-788a-81212dbb5291?t=1565184649007.
Müller, S., Ben-Yami M., and Helled R., 2020, Theoretical versus Observational Uncertainties: Composition of Giant Exoplanets. The Astrophysical Journal, 903(2), Id. 147, 13 pages, DOI: 10.3847/1538-4357/abba19.
Nakajima, M., Stevenson D.J., 2015, Melting and mixing states of the Earth's mantle after the Moon-forming impact. Earth and Planetary Science Letters, 427, 286-295, DOI: 10.1016/j.epsl.2015.06.023.
Namur, O., Collinet M., Charlier B., Grove T.L., Holtz F., McCammon C., 2016, Melting processes and mantle sources of lavas on Mercury. Earth and Planetary Science Letters, 439, 117-128, DOI: 10.1016/j.epsl.2016.01.030.
Nanne, J.A.M.; Nimmo F., Cuzzi J.N., Kleine T., 2019, Origin of the non-carbonaceous-carbonaceous meteorite dichotomy. Earth and Planetary Science Letters, 511, 44-54, DOI: 10.1016/j.epsl.2019.01.027.
Nesvorný, D., Vokrouhlický D., Morbidelli A., 2007, Capture of Irregular Satellites during Planetary Encounters. The Astronomical Journal, 133(5), 1962-1976, DOI: 10.1086/512850.
Nesvorný, D., Morbidelli A., 2012, Statistical Study of the Early Solar System's Instability with Four, Five, and Six Giant Planets. The Astronomical Journal, 144(4), Id. 117, 20 pages, DOI: 10.1088/0004-6256/144/4/117.
Nesvorný, D., Vokrouhlický D., Morbidelli A., 2013, Capture of Trojans by Jumping Jupiter. The Astronomical Journal, 768(1), Id. 45, 8 pages, DOI: 10.1088/0004-637X/768/1/45.
Nesvorný, D., 2015a, Evidence for Slow Migration of Neptune from the Inclination Distribution of Kuiper Belt Objects. The Astronomical Journal, 150(3), Id. 73, 18 pages, DOI: 10.1088/0004-6256/150/3/73.
Nesvorný, D., 2015b, Jumping Neptune Can Explain the Kuiper Belt Kernel. The Astronomical Journal, 150(3), Id. 68, 14 pages, DOI: 10.1088/0004-6256/150/3/68.
Nesvorný, D., Brož M., Carruba V., 2015, Identification and Dynamical Properties of Asteroid Families. In: Asteroids IV, Eds. P. Michel, F.E. DeMeo, and W.F. Bottke, University of Arizona Press, Tucson, 895 pages. ISBN: 978-0-816-53213-1, 297-321, DOI: 10.2458/azu_uapress_9780816532131-ch016.
Nesvorný, D., Vokrouhlický D., 2016, Neptune's orbital migration was grainy, not smooth. The Astrophysical Journal, 825(2), Id. 94, 18 pages, DOI: 10.3847/0004-637X/825/2/94.
Nesvorný, D., 2018, Dynamical Evolution of the Early Solar System. Annual Review of Astronomy and Astrophysics, 56, 137-174, DOI: 10.1146/annurev-astro-081817-052028.
Nettelmann, N., Becker A., Holst B., and Redmer R., 2012, Jupiter Models with Improved Ab Initio Hydrogen Equation of State (H-REOS.2). The Astrophysical Journal, 750(1), Id. 52, 10 pages, DOI: 10.1088/0004-637X/750/1/52.
Nettelmann, N., Helled R., Fortney J.J., and Redmer R., 2013, New indication for a dichotomy in the interior structure of Uranus and Neptune from the application of modified shape and rotation data. Planetary and Space Science, 77, 143-151, DOI: 10.1016/j.pss.2012.06.019.
Nettelmann, N., Wang K., Fortney J.J., Hamel S., Yellamilli S., Bethkenhagen M., and Redmer R., 2016, Uranus evolution models with simple thermal boundary layers. Icarus, 275, 107-116, DOI: 10.1016/j.icarus.2016.04.008.
Neveu, M., Hays L.E., Voytek M.A., New M.H., Schulte M.D., 2018, The Ladder of Life Detection. Astrobiology, 18(11), 1375-1402, DOI: 10.1089/ast.2017.1773.
Nicholson, P. D.; De Pater, I.; French, R. G.; Showalter, M. R., 2018a, The rings of Uranus. In: Planetary Ring Systems. Properties, Structure, and Evolution, Eds. M.S. Tiscareno and C.D. Murray, ISBN: 9781316286791, Cambridge University Press, 2018, 93-111, DOI: 10.1017/9781316286791.004.
Nicholson, P.D., French R.G., Spitale J.N., 2018b, Narrow rings, gaps, and sharp edges. In: Planetary Ring Systems. Properties, Structure, and Evolution, Eds. M.S. Tiscareno and C.D. Murray, ISBN: 9781316286791, Cambridge University Press, 276-307, DOI: 10.1017/9781316286791.011.

Nijman, N., Kloppenburg A., de Vries S.T., 2014, Archaean basin margin geology and crustal evolution: an East Pilbara traverse. Journal of the Geological Society, 174, 1090-1112, 29 June 2017, DOI: 10.1144/jgs2016-127.

Nimmo, F., 2002, Why does Venus lack a magnetic field? Geology, 30(11), 987-990, DOI: 10.1130/0091-7613(2002)030<0987:WDVLAM>2.0.CO;2.

Nittler, L.R., Starr R.D., Weider S.Z., McCoy T.J., Boynton W.V., Ebel D.S., Ernst C.M., Evans L.G., Goldsten J.O., Hamara D.K., Lawrence D.J., McNutt Jr R.L., Schlemm 2nd C.E., Solomon S.C., Sprague A.L., 2011, The major-element composition of Mercury's surface from MESSENGER X-ray spectrometry. Science, 333(6051), 1847-1850, DOI: 10.1126/science.1211567.

Noack, L., Breuer D., 2014, Plate tectonics on rocky exoplanets: influence of initial conditions and mantle rheology. Planetary and Space Science, 98, 41-49, DOI: 10.1016/j.pss.2013.06.020.

Nogueira, E., Brasser R., Gomes R., 2011, Reassessing the origin of Triton, Icarus, 214(1), 113-130, DOI: 10.1016/j.icarus.2011.05.003.

O'Neill, C., Jellinek A.M., Lenardic A., 2007, Conditions for the onset of plate tectonics on terrestrial planets and moons. Earth and Planetary Science Letters, 261(1-2), 20-32, DOI: 10.1016/j.epsl.2007.05.038.

O'Rourke, J.G., Buz J., Fu R.R., Lillis R.J., 2019, Detectability of remnent magnetism in the crust of Venus. Geophysical Research Letters, 46(11), 5768-5777, DOI: 10.1029/2019GL082725.

Oeser, R.A., Stroncik N., Moskwa L.M., Bernhard N., Schaller M., Canessa R., van den Brink L., Köster M., Brucker E., Stock S., Fuentes J.P., Godoy R., Matus F.J., Pedraza R.O., McIntyre P.O., Paulino L., Seguel O., Bader M.Y., Boy J., Dippold M.A., Ehlers T.A., Kühn P., Kuzyakov Y., Leinweber P., Scholten T., Spielvogel S., Spohn M., Übernickel K., Tielbörger K., Wagner D., von Blanckenburg F., 2018, Chemistry and microbiology of the Critical Zone along a steep climate and vegetation gradient in the Chilean Coastal Cordillera, CATENA, 170, 183-203, DOI: 10.1016/j.catena.2018.06.002.

Ormel, C.W., Klahr H.H., 2010, The effect of gas drag on the growth of protoplanets. Analytical expressions for the accretion of small bodies in laminar disks. Astronomy and Astrophysics, 520, Id. A43, 15 pages, DOI: 10.1051/0004-6361/201014903.

Ormel, C.W., Vazan A., and Brouwers M.G., 2021, How planets grow by pebble accretion. III. Emergence of an interior composition gradient. Astronomy Astrophysics, 647, Id. A175, 19 pages, DOI: 10.1051/0004-6361/202039706.

Opher, M., 2016, The heliosphere: what did we learn in recent years and the current challenges. Space Science Reviews, 200(1-4), 475-494, DOI: 10.1007/s11214-015-0186-3.

Opher, M., Drake J.F., 2013, On the rotation of the magnetic field across the heliopause. The Astrophysical Journal Letters, 778(2), Id. L26, 1-6, DOI: 10.1088/2041-8205/778/2/L26.

Opher, M., Drake J.F., Zieger B., Gombosi T.I., 2015, Magnetized jets driven by the Sun: the structure of the heliosphere revisited. The Astrophysical Journal Letters, 800(2), Id. L28, 1-7, DOI: 10.1088/2041-8205/800/2/L28.

Opher, M., Loeb A., Drake J., Toth G., 2020, A small and round heliosphere suggested by magnetohydrodynamic modelling of pick-up ions. Nature Astronomy, 4, 675-683, DOI: 10.1038/s41550-020-1036-0.

Pappalardo, R.T., Senske D.A., Korth H., Klima R., Vance S.D., Craft K., Phillips C.B., Europa Science Team, 2017, The planned Europa Clipper Mission and its role in investigating ice shell exchange processes. In: Europa Deep Dive 1: Ice-Shell Exchange Processes, Proceedings of the conference held 1-2 November, 2017 in Houston, Texas, LPI Contribution No. 2048, Id.7003.

Parker, E.N., 1961, The Stellar-Wind Regions. Astrophysical Journal, 134, 20, DOI: 10.1086/147124.

Pearce, B.K.D., Pudritz R.E., Semenov D.A., Henning T.K., 2017, Origin of the RNA World: the fate of nucleobases in warm little ponds. Proc. Natl. Acad. Sci. USA, 114, 11327-11332, DOI: 10.1073/pnas.1710339114.

Peplowski, P.N., Klima R.L., Lawrence D.J., Ernst C.M., Denevi B.W., Frank E.A., Goldsten J.O., Murchie S.L., Nittler L.R., Solomon S.C., 2016, Remote sensing evidence for an ancient carbon-bearing crust on Mercury. Nature Geoscience, 9(4), 273-276, DOI: 10.1038/ngeo2669.

Piel, A., 2010, Definition of the Plasma State. In: Plasma Physics. Springer, Berlin, Heidelberg, DOI : 10.1007/978-3-642-10491-6_2.

Pierens, A., Raymond S.N., Nesvorný D., Morbidelli A., 2014, Outward Migration of Jupiter and Saturn in 3:2 or 2:1 Resonance in Radiative Disks: Implications for the Grand Tack and Nice models. The Astrophysical Journal Letters, 795(1), Id. L11, 1-6, DOI: 10.1088/2041-205/795/1/L11.

Pitjeva, E.V., Pitjev N.P., 2016, Masses of asteroids and total mass of the main asteroid belt. Proceedings IAU Symposium No. 318, 2015, International Astronomical Union, Eds. S.R. Chesley, A. Morbidelli, R. Jedicke, and D. Farnocchia, DOI: 10.1017/S1743921315008388.

Pitjeva, E.V., Pitjev N.P., 2018, Mass of the Kuiper belt. Celest. Mech. Dyn. Astr., 130, 57, DOI: 10.1007/s10569-018-9853-5.

Plesa, A.C., Tosi N., Grott M., Breuer D., 2015, Thermal evolution and Urey ratio of Mars. Journal of Geophysical Research: Planets, 120(5), 995-1010, DOI: 10.1002/2014JE004748.

Plesa, A.C., Padovan S., Tosi N., Breuer D., Grott M., Wieczorek M.A., Spohn T., Smrekar S.E., Banerdt W.B., 2018, The thermal state and interior structure of Mars. Geophysical Research Letters, 45(22), 12-198, DOI: 10.1029/2018GL080728.

Podolak, M., Pollack J.B., and Reynolds R.T., 1988, Interactions of planetesimals with protoplanetary atmospheres. Icarus, 73(1), 163-179, DOI: 10.1016/0019-1035(88)90090-5.

Podolak, M., Helled R., and Schubert G., 2019, Effect of non-adiabatic thermal profiles on the inferred compositions of Uranus and Neptune. Monthly Notices of the Royal Astronomical Society, 487(2), 2653-2664, DOI: 10.1093/mnras/stz1467.

Pogorelov, N.V., Borovikov S.N., Heerikhuisen J., Zhang M., 2015, The Heliotail. The Astrophysical Journal Letters, 812(1), Id. L6, 1-7, DOI: 10.1088/2041-8205/812/1/L6.

Pollack, J.B., Hubickyj O., Bodenheimer P., Lissauer J.J., Podolak M., Greenzweig Y., 1996, Formation of the Giant Planets by Concurrent Accretion of Solids and Gas. Icarus, 124(1), 62-85, DOI: 10.1006/icar.1996.019.

Porco, C.C., Helfenstein P., Thomas P.C., Ingersoll A.P., Wisdom J., West R., Neukum G., Denk T., Wagner R., Roatsch T., Kieffer S., Turtle E., McEwen A., Johnson T.V., Rathbun J., Veverka J., Wilson D., Perry J., Spitale J., Brahic A., Burns J.A., DelGenio A.D., Dones L., Murray C.D., Squyres S., 2006, Cassini Observes the Active South Pole of Enceladus. Science, 311(5766), 1393-1401, DOI: 10.1126/science.1123013.

Postberg, F., Grün E., Horanyi M., Kempf S., Krüger H., Schmidt J., Spahn F., Srama R., Sternovsky Z., Trieloff M., 2011, Compositional mapping of planetary moons by mass spectrometry of dust ejecta. Planetary and Space Science, 59(14), 1815-1825, DOI: 10.1016/j.pss.2011.05.001.

Postberg, F., Khawaja N., Abel B., Choblet G., Glein C.R., Gudipati M.S., Henderson B.L., Hsu H.-W., Kempf S., Klenner F., Moragas-Klostermeyer G., Magee B., Nölle L., Perry M., Reviol R., Schmidt J., Srama R., Stolz F., Tobie G., Trieloff M., Waite J.H., 2018, Macromolecular organic compounds from the depths of Enceladus. Nature, 558(7711), 564-568, DOI: 10.1038/s41586-018-0246-4.

Prieto-Ballesteros, O., et al., 2019, Searching for (bio)chemical complexity in icy satellites, with a focus on Europa. White paper ESA Voyage 2050, https://www.cosmos.esa.int/documents/1866264/3219248/Prieto-BallesterosO_WP-complex.pdf/9be87764-b230-5497-d98f-77693b0b9add?t=1565184652275.

Quantin-Nataf C., Carter J., Mandon L., Thollot P., Balme M., Volat M., Pan L., Loizeau D., Millot C., Breton S., Dehouck E., Fawdon P., Gupta S., Davis J., Grindrod P.M., Pacifici A., Bultel B., Allemand P., Ody A., Lozach L., Broyer J., 2021, Oxia Planum: The Landing Site for the ExoMars "Rosalind Franklin" Rover Mission: Geological Context and Prelanding Interpretation. Astrobiology, 21(3), 345-366, DOI: 10.1089/ast.2019.2191.

Rauer, H., Blanc M., Venturini J., Dehant V., Demory B., Dorn C., Domagal-Goldman S., Foing B., Gaudi S., Helled R., Heng K., Kitzman D., Kokubo E., Le Sergeant d'Hendecourt L., Mordasini C., Nesvorny D., Noack L., Opher M., Owen J., Paranicas C., Qin L., Snellen I., Testi L., Udry S., Wambganss J., Westall F., Zarka P., Zong Q., 2022, Planetary Exploration, Horizon 2061 Report - Chapter 2: Solar System/Exoplanet Science Synergies in a Multi-Decadal Perspective. Science Direct, Elsevier.

Raymond, S.N., O'Brien D.P., Morbidelli A., Kaib N.A., 2009, Building the terrestrial planets: Constrained accretion in the inner Solar System. Icarus, 203, 644-662, DOI: 10.1016/j.icarus.2009.05.016.

Read, P.L., Lebonnois S., 2018, Superrotation on Venus, on Titan, and Elsewhere. Annual Review of Earth and Planetary Sciences, 46, 175-202, DOI: 10.1146/annurev-earth-082517-010137.

Redfield, S., Wood B.E., Linsky J.L., 2004, Physical structure of the local interstellar medium. Advances in Space Research, 34(1), 41-45, DOI: 10.1016/j.asr.2003.02.053.

Regenauer-Lieb, K., Yuen D.A., Branlund J., 2001, The initiation of subduction: criticality by addition of water? Science, 294(5542), 578-580, DOI: 10.1126/science.1063891.

Richardson, J.D., Belcher J.W., Garcia-Galindo P., Burlaga L.F., 2019, Voyager 2 plasma observations of the heliopause and interstellar medium. Nature Astronomy, 3, 1019-1023, DOI: 10.1038/s41550-019-0929-2.

Rosas, J.C., Korenaga J., 2018, Rapid crustal growth and efficient crustal recycling in the early Earth: Implications for Hadean and Archean geodynamics. Earth and Planetary Science Letters, 494, 42-49, DOI: 10.1016/j.epsl.2018.04.051.

Rosing, M.T., Bird D.K., Sleep N.H., Glassley W.E., Albarède F., 2006, The rise of continents—an essay on the geologic consequences of photosynthesis. Palaeogeography Palaeoclimatology Palaeoecology, 232(2-4), 99-113, DOI: 10.1016/j.palaeo.2006.01.007.

Roth, L., Saur J., Retherford K.D., Strobel D.F., Feldman P.D., McGrath M.A., Nimmo F., 2014, Transient Water Vapor at Europa's South Pole. Science, 343, 6167, 171-174, DOI: 10.1126/science.1247051.

Rubie, D.C., Jacobson S.A., Morbidelli A., O'Brien D.P., Young E.D., de Vries J., Nimmo F., Palme H., Frost D.J., 2015, Accretion and differentiation of the terrestrial planets with implications for the compositions of early-formed Solar System bodies and accretion of water. Icarus, 248, 89-108, DOI: 10.1016/j.icarus.2014.10.015.

Russell, M.J., Hall A.J., Cairns-Smith A.G., Braterman P.S., 1988, Submarine hot springs and the origin of life. Nature, 336, 117, DOI: 10.1038/336117a0.

Russell, MJ, Hall A.J., 1997, The emergence of life from iron monosulphide bubbles at a submarine hydrothermal redox and pH front. J. Geol. Soc., 154, 377-402, DOI: 10.1144/gsjgs.154.3.0377.

Russell, C.T., Luhmann J. G., Strangeway R. J., 2016, Space Physics: an Introduction. Cambridge University Press, Jul 2016.

Salo, H., Ohtsuki K., Lewis M.C., 2018, Computer Simulations of Planetary Rings. In: Planetary Ring Systems – Properties, Structure, and Evolution, Eds. M.S. Tiscareno and C.D. Murray, ISBN: 9781316286791, Cambridge University Press, 434-493, DOI: 10.1017/9781316286791.016.

Sánchez-Lavega, A., Lebonnois S., Imamura T., Read P., Luz D., 2017, The atmospheric dynamics of Venus. Space Sci. Rev., 212, 1541-1616, DOI 10.1007/s11214-017-0389-x.

Sánchez-Lavega, A., and Heimpel M., 2018, Atmospheric dynamics of giants and icy planets. Handbook of exoplanets, p. 51.

Saunders, R.S., Pettengill G.H., 1991, Magellan: Mission summary, Science, 252(5003), 247-249, DOI: 10.1126/science.252.5003.247.

Sautter, V., Toplis M., Wiens R., Cousin A., Fabre C., Gasnault O., Maurice S., Forni O., Lasue J., Ollila A., Bridges J., Mangold N., Le Mouélic S., Fisk M., Meslin P.-Y., Beck P., Pinet P., Le Deit L., Rapin W., Stolper E., Newsom H., Dyar D., Lanza N., Vaniman D., Clegg S., Wray J., 2015, In-situ evidence for continental crust on early Mars, Nature Geosciences, 8, 605-609, DOI: 10.1038/ngeo2474.

Scheinberg, A.L., Soderlund K.M., Elkins-Tanton L.T., 2018, A basal magma ocean dynamo to explain the early lunar magnetic field. Earth and Planetary Science Letters, 492, 144-151.
Scherer, K., Fichtner H., 2014, The return of the bow shock. The Astrophysical Journal, 782(1), Id. 25, 1-5, DOI: 10.1088/0004-637X/782/1/25.
Schwartzman, D.W., Volk T., 1989, Biotic enhancement of weathering and the habitability of Earth. Nature 340 (6233), 457-460, DOI: 10.1038/340457a0.
Sherwood B., Lunine J., Sotin C., Cwik T., Naderi F., 2018, Program options to explore ocean worlds. Acta Astronautica, 143, 285-296, DOI: 10.1016/j.actaastro.2017.11.047.
Showalter, M.R., 2020, The rings and small moons of Uranus and Neptune. Philosophical Transactions of the Royal Society A, 378, 2187, Id.20190482, DOI: 10.1098/rsta.2019.0482.
Showman, A.P., and de Pater I., 2005, Dynamical implications of Jupiter's tropospheric ammonia abundance. Icarus, 174(1), 192-204, DOI: 10.1016/j.icarus.2004.10.004.
Showman, A.P., Ingersoll A.P., Achterberg R., and Kaspi Y., 2018, The global atmospheric circulation of Saturn. In: Saturn in the 21st Century, Eds. K.H. Baines, F.M. Flasar, N. Krupp, and T. Stallard, Cambridge University Press, Cambridge Planetary Science, ISBN: 9781316227220, DOI: 10.1017/9781316227220, Chapter 11, 295-336.
Sicardy, B., Talbot J., Meza E., Camargo J. I. B., Desmars J., Gault D., Herald D., Kerr S., Pavlov H., Braga-Ribas F., Assafin M., Benedetti-Rossi G., Dias-Oliveira A., Gomes-Júnior A. R., Vieira-Martins R., Bérard D., Kervella P., Lecacheux J., Lellouch E., Beisker W., et al., 2016, Pluto's Atmosphere from the 2015 June 29 Ground-based Stellar Occultation at the Time of the New Horizons Flyby. The Astrophysical Journal Letters, 819, 2, Id. L38, 8 pages, DOI: 10.3847/2041-8205/819/2/L38.
Sleep, N.H., Bird D.K., Pope E., 2012, Paleontology of Earth’s mantle. Annual Review of Earth and Planetary Sciences, 40, 277-300, DOI: 10.1146/annurev-earth-092611-090602.
Smith, D.E., Zuber M.T., Torrence M.H., Dunn P.J., Neumann G.A., Lemoine F.G., Fricke S.K., 2009, Time variations of Mars’ gravitational field and seasonal changes in the masses of the polar ice caps. Journal of Geophysical Research Planets, 114(E5), DOI: 10.1029/2008JE003267.
Smrekar, S.E., Stofan E.R., Mueller N., Treiman A., Elkins-Tanton L., Helbert J., Piccioni G., Drossart P., 2010, Recent hotspot volcanism on Venus from VIRTIS emissivity data. Science, 328(5978), 605-608, DOI: 10.1126/science.1186785.
Smrekar, S.E., Hensley S., Dyar D., Helbert J., 2019a, VERITAS (Venus Emissivity, Radio Science, InSAR, Topography And Spectroscopy): A Proposed Discovery Mission. EPSC-DPS Joint Meeting 2019, Id. EPSC-DPS2019-1124.
Smrekar, S.E., Lognonné P., Spohn T., Banerdt W.B., Breuer D., Christensen U., Dehant V., Drilleau M., Folkner W., Fuji N., Garcia R.F., Giardini D., Golombek M., Grott M., Gudkova T., Johnson C., Khan A., Langlais B., Mittelholz A., Mocquet A., Myhill R., Panning M., Perrin C., Pike T., Plesa A.C., Rivoldini A., Samuel H., Stähler S.C., van Driel M., Van Hoolst T., Verhoeven O., Weber R., Wieczorek M., 2019b, Pre-mission InSights on the Interior of Mars. Space Science Reviews, 215(3), 1-72, DOI: 10.1007/s11214-018-0563-9.
Soderlund, K.M., Heimpel M.H., King E.M., Aurnou J.M., 2013, Turbulent models of ice giant internal dynamics: Dynamos, heat transfer, and zonal flows. Icarus, 224(1), 97-113.
Soderlund, K.M., Kalousová K., Buffo J.J., Glein C.R., Goodman J.C., Mitri G., Patterson G.W., Postberg F., Rovira-Navarro M., Rückriemen T., Saur J., 2020, Ice-Ocean Exchange Processes in the Jovian and Saturnian Satellites. Space Science Reviews, 216(5), 1-57.
Soderlund, K.M., Stanley S., 2020, The underexplored frontier of ice giant dynamos. Philosophical Transactions of the Royal Society A: Mathematical, Physical and Engineering Sciences, 378(2187), DOI: 10.1098/rsta.2019.0479.
Sohl, F., Schubert G., 2015, Interior Structure, Composition, and Mineralogy of the Terrestrial Planets. In: Treatise in Geophysics, 2nd ed, Vol. 10, eds. T. Spohn and G. Schubert, 23-64, Elsevier, Amsterdam.

Solomon, S.C., Nittler L.R., Anderson B.J. (Eds.), 2018, Mercury: The view after MESSENGER. Cambridge University Press, DOI: 10.1017/9781316650684 (Chapters 16 and 17).

Southam, G., Westall F., Spohn T., 2015, Geology, life, and habitability. In Treatise in Geophysics, 2nd ed., Vol 10, eds. T. Spohn and G. Schubert eds., 473-486, Elsevier, Amsterdam.

Spohn, T., Schubert G., 2002, Oceans in the icy Galilean satellites of Jupiter? Icarus, 161, 456-467, DOI: 10.1016/S0019-1035(02)00048-9.

Spohn, T., Grott M., Smrekar S.E., Knollenberg J., Hudson T.L., Krause C., Müller N., Jänchen J., Börner A., Wippermann T., Krömer O., Lichtenheldt R., Wisniewski L., Grygorczuk J., Fittock M., Rheershemius S., Spröwitz T., Kopp E., Walter I., Plesa A.C., Breuer D., Morgan P., Banerdt W.B., 2018, The heat flow and physical properties package (HP3) for the InSight mission. Space Science Reviews, 214(5), Id. 96, DOI: 10.1007/s11214-018-0531-4.

Squyres, S.W., 1989, Urey prize lecture: Water on Mars, Icarus, 79, 229-288, 1989, DOI: 10.1016/0019-1035(89)90078-X.

Squyres, S.W., Arvidson R.E., Baumgartner E.T., Bell J.F. III, Christensen P.R., Gorevan S., Herkenhoff K.E., Klingelhöfer G., Madsen M.B., Morris R.V., Rieder R., Romero R.A., 2003, Athena Mars Rover science investigation. J. Geophys. Res., 108(E12), DOI: 10.1029/2003JE002121.

Stähler, S.C., Khan A., Banerdt W.B., Lognonné P., Giardini D., Ceylan S., Drilleau M., Duran A.C., Garcia R.F., Huang Q., Kim D., Lekic V., Samuel H., Schimmel M., Schmerr N., Sollberger D., Stutzmann E., Xu Z., et al., 2021, Seismic detection of the martian core. Science 373, 443-448, DOI: 10.1126/science.abi7730.

Stam, D.M., 2008, Spectropolarimetric signatures of Earth-like extrasolar planets. Astronomy Astrophysics, 482(3), 989-1007, DOI: 10.1051/0004-6361:20078358.

Stamenković, V., Beegle L. W., Zacny K., Arumugam D. D., Baglioni P., Barba N., Baross J., Bell M. S., Bhartia R., Blank J. G., Boston P. J., Breuer D., Brinckerhoff W., Burgin M. S., Cooper I., Cormarkovic V., Davila A., Davis R. M., Edwards C., Etiope G. Fischer W. W., Glavin D. P., Grimm R. E., Inagaki F., Kirschvink J. L., Kobayashi A., Komarek T., Malaska M., Michalski J., Ménez B., Mischna M., Moser D., Mustard J., Onstott T. C., Orphan V. J., Osburn M. R., Plaut J., Plesa A. -C., Putzig N., Rogers K. L., Rothschild L., Russell M., Sapers H., Lollar B. Sherwood, Spohn T., Tarnas J. D., Tuite M., Viola D., Ward L. M., Wilcox B., Woolley R., 2019, The next frontier for üplanetary and human exploration. Nature Astronomy, 3, 116-120, DOI: 10.1038/s41550-018-0676-9.

Sterken, V.J., Westphal A.J., Altobelli N., Malaspina D., Postberg F., 2019, Interstellar dust in the Solar System. Space Science Reviews, 215(7), Id. 43, 1-32, DOI: 10.1007/s11214-019-0607-9.

Stevenson, D.J., and Salpeter E.E., 1977, The dynamics and helium distribution in hydrogen-helium fluid planets. Astrophysical Journal Supplement Series, 35, 239-261, DOI: 10.1086/190479.

Stevenson, D.J., 1982, Interiors of the Giant Planets. Annual Review of Earth and Planetary Sciences, 10, 257, DOI: 10.1146/annurev.ea.10.050182.001353.

Stevenson, D.J., Spohn T., Schubert G., 1983, Magnetism and thermal evolution of the terrestrial planets. Icarus, 54(3), 466-489, DOI: 10.1016/0019-1035(83)90241-5.

Stone, E.C., Cummings A.C., McDonald F.B., Heikkila B.C., Lal N., Webber W.R., 2013, Voyager 1 observes low-energy galactic cosmic rays in a region depleted of heliospheric ions. Science, 341(6142), 150-153, DOI: 10.1126/science.1236408.

Stone, E.C., Cummings A.C., Heikkila B.C., Lal N., 2019, Cosmic ray measurements from Voyager 2 as it crossed into interstellar space. Nature Astronomy, 3, 1013-1018, DOI: 10.1038/s41550-019-0928-3.

Sugimoto, N., Takagi M., Matsuda Y., 2019, Fully developed superrotation driven by the mean meridional circulation in a Venus GCM. Geophys. Res. Lett., 46, 1776-1784, DOI: 10.1029/2018GL080917.

Surkov, Y.A., Moskaleva L.P., Kharyukova V.P., Dudin A.D., Smimov G.G., Zaitseva S.E., 1986, Venus rock composition at Vega 2 landing site. J. Geophys. Res., 91(B13), E215-E218, DOI: 10.1029/JB091iB13p0E215.

Tackley, P., 2000, Self-consistent generation of tectonic plates in time-dependent, three-dimensional mantle convection simulations, 1. Pseudoplastic yielding. Geochemistry Geophysics Geosystems, 1(8), DOI: 10.1029/2000GC000036.

Tang, M., Chen K., Rudnick R.L., 2016, Archean upper crust transition from mafic to felsic marks the onset of plate tectonics. Science, 351(6271), 372-375, DOI: 10.1126/science.aad5513.

Tarduno, J.A., Cottrell R.D., Bono R.K., Oda H., Davis W.J., Fayek M., van ’t Erve O., Nimmo F., Huang W., Thern E.R., Fearn S., Mitra G., Smirnov A.V., Blackman E.G., 2020, Paleomagnetism indicates that primary magnetite in zircon records a strong Hadean geodynamo. Proceedings of the National Academy of Sciences, 117(5), 2309-2318, DOI: 10.1073/pnas.1916553117.

Taylor, S.R., 2016, Lunar Science: A post-Apollo view. Pergamon Press, 392 pages.

Taylor, S.R., McLennan S., 2009, Planetary crusts: their composition, origin and evolution, Vol. 10, Cambridge University Press.

Teanby, N.A., Irwin P.G.J., Moses J.I., Helled R., 2020, Neptune and Uranus: ice or rock giants?. Philosophical Transactions of the Royal Society A, 378(2187), 20190489.

Throop, H.B., Porco C.C., West R.A., Burns J.A., Showalter M.R., Nicholson P.D., 2004, The jovian rings: new results derived from Cassini, Galileo, Voyager, and Earth-based observations. Icarus, 172, 1, 59-77, DOI: 10.1016/j.icarus.2003.12.020.

Tikoo, S.M., Weiss B.P., Shuster D.L., Suavet C., Wang H., Grove T.L., 2017, A two-billion-year history for the lunar dynamo. Science Advances, 3(8), Id. e1700207, DOI: 10.1126/sciadv.1700207.

Tiscareno, M.S., Murray C.D., 2018, The Future of Planetary Rings Studies. In: Planetary Ring Systems – Properties, Structure, and Evolution, Eds. M.S. Tiscareno and C.D. Murray, ISBN: 9781316286791, Cambridge University Press, 577-579, DOI: 10.1017/9781316286791.021.

Tosi, N., Breuer D., Spohn T., 2014, Evolution of planetary interiors. In: Encyclopedia of the Solar System (3rd Edition), Eds. T. Spohn, D. Breuer, and T. Johnson, Elsevier, Ch. 9, 185-208.

Tosi, N., Godolt M., Stracke B., Grenfell J. L., Höning D., Nikolaou A., Plesa A.-C., Breuer D. Spohn T., 2017, The habitability of a stagnant-lid Earth. Astronomy Astrophysics, 605, A71, 1-21, DOI: 10.1051/0004-6361/201730728.

Trompert, R., Hansen U., 1998, Mantle convection simulations with rheologies that generate plate-like behaviour. Nature 395, 686-689, DOI: 10.1038/27185.

Tsiganis, K., Gomes R., Morbidelli A., Levison H.F., 2005, Origin of the orbital architecture of the giant planets of the Solar System. Nature, 435(7041), 459-461, DOI: 10.1038/nature03539.

Turrini, D., Politi R., Peron R., Grassi D., Plainaki C., Barbieri M., Lucchesi D.M., Magni G., Altieri F., Cottini V., Gorius N., Gaulme P., Schmider F.-X., Adriani A., Piccioni G., 2014, The comparative exploration of the ice giant planets with twin spacecraft: Unveiling the history of our Solar System. Planetary and Space Science, 104, 93-107, DOI: 10.1016/j.pss.2014.09.005.

Vago, J.L., Westall F., Pasteur Teams, Landing site selection Working Group, and other contributors, 2017, Habitability on early Mars and the search for biosignatures with the ExoMars rover. Astrobiology 17, 471-510, DOI: 10.1089/ast.2016.1533.

Valencia, D., O’connell R.J., Sasselov D.D., 2007, Inevitability of plate tectonics on super-Earths. The Astrophysical Journal Letters, 670(1), L45, DOI: 10.1086/524012.

Valletta, C., and Helled R., 2020, Giant Planet Formation Models with a Self-consistent Treatment of the Heavy Elements. The Astrophysical Journal, 900(2), Id. 133, 17 pages, DOI: 10.3847/1538-4357/aba904.

Vazan, A., and Helled R., 2020, Explaining the low luminosity of Uranus: a self-consistent thermal and structural evolution. Astronomy Astrophysics, 633, Id. A50, 10 pages, DOI: 10.1051/0004-6361/201936588.

Vazan, A., Helled R., Kovetz A., and Podolak M., 2015, Convection and Mixing in Giant Planet Evolution. The Astrophysical Journal, 803(1), Id. 32, 11 p ages, DOI: 10.1088/0004-637X/803/1/32.

Vazan, A., Helled R., Podolak M., and Kovetz A., 2016, The Evolution and Internal Structure of Jupiter and Saturn with Compositional Gradients. The Astrophysical Journal, 829(2), Id. 118, 11 pages, DOI: 10.3847/0004-637X/829/2/118.

Vazan, A., Helled R., and Guillot T., 2018, Jupiter's evolution with primordial composition gradients. Astronomy Astrophysics, 610, Id. L14, 5 pages, DOI: 10.1051/0004-6361/201732522.

Vernazza, P., Beck P., 2017, Composition of Solar System Small Bodies. Chapter 13 of Planetesimals – Early differentiation and consequences for planets, Eds. L.T. Elkins-Tanton and B.P. Weiss, Cambridge University Press, DOI: 10.1017/9781316339794.013, 269-297.

Viswanathan, V., Rambaux N., Fienga A., Laskar J., Gastineau M., 2019, Observational constraint on the radius and oblateness of the lunar core-mantle boundary. Geophysical Research Letters, 46, 7295-7303, DOI: 10.1029/2019GL082677.

Von Bloh, W., Bounama C., Cuntz M., Franck S., 2007, The habitability of super-Earths in Gliese 581. Astronomy & Astrophysics, 476 (3), 1365-1371, DOI: 10.1051/0004-6361:20077939.

Wacey, D., Saunders M., Roberts M., et al., 2014, Enhanced cellular preservation by clay minerals in 1 billion-year old lakes. Sci. Rep. 4, 5841, DOI: 10.1038/srep05841.

Wahl, S.M., Hubbard W.B., Militzer B., Guillot T., Miguel Y., Movshovitz N., Kaspi Y., Helled R., Reese D., Galanti E., Levin S., Connerney J.E., and Bolton S.J., 2017, Comparing Jupiter interior structure models to Juno gravity measurements and the role of a dilute core. Geophys. Res. Let., 44, 4649-4659. DOI: 10.1002/2017GL073160.

Walsh, K.J., Morbidelli A., Raymond N., O'Brien D.P., Mandell A.M., 2011, A low mass for Mars from Jupiter's early gas-driven migration. Nature, 475 (7355), 206-209, DOI: 10.1038/nature10201.

Wang, P., Mitchell J.L., 2014, Planetary ageostrophic instability leads to superrotation. Geophys. Res. Lett., 41, 4118-4126.

Warren, P.H., 1985, The magma ocean concept and lunar evolution. Annual Review of Earth and Planetary Sciences, 13(1), 201-240, DOI: 10.1146/annurev.ea.13.050185.001221.

Way, M.J., Del Genio A.D., Kiang, Sohl L.E., Grinspoon D.H., Aleinov I., Kelley M., Clune T., 2016, Was Venus the first habitable world of our Solar System?. Geophysical Research Letters, 43(16), 8376-8383, DOI: 10.1002/2016GL069790.

Way, M.J., Del Genio A.D., 2020, Venusian habitable climate scenarios: modeling Venus through time and applications to slowly rotating Venus-Like exoplanets. JGR Planets 25(5), May 2020, e2019JE006276, DOI: 10.1029/2019JE006276.

Weber, R.C., Lin P.Y., Garnero E.J., Williams Q., Lognonné P., 2011, Seismic Detection of the Lunar Core. Science, DOI: 10.1126/science.1199375.

Weiss, P., Yung K.L., Kömle N., Ko S.M., Kaufmann E., Kargl G., 2011, Thermal drill sampling system on board high-velocity impactors for exploring the subsurface of Europa. Advances in Space Research, 48(4), 743-754, DOI: 10.1016/j.asr.2010.01.015.

West, R.A., Baines K.H., Friedson A.J., Banfield D., Agent B.R., and Taylor F.W., 2004, Jovian Clouds and Haze. In: Jupiter. The Planet, Satellites and Magnetosphere, Eds. F. Bagenal, T.E. Dowling, W.B. McKinnon, ISBN: 9780521035453, Cambridge University Press, Cambridge Planetary Science, Chapter 5, 79-104.

Westall, F, de Ronde C.E.J., Southam G., Grassineau N., Colas M., Cockell C., Lammer H., 2006, Implications of a 3.472-3.333 Ga-old subaerial microbial mat from the Barberton greenstone belt, South Africa for the UV environmental conditions on the early Earth. Phil. Trans. Roy. Soc. Lond. Series B., 361, 1857-1875, DOI: 10.1098/rstb.2006.1896.

Westall, F., Loizeau D., Foucher F., Bost N., Betrand M., Vago J., Kminek G., 2013, Habitability on Mars from a microbial point of view. Astrobiology, 13(9), 887-897, DOI: 10.1089/ast.2013.1000.

Westall, F., Foucher F., Bost N., Bertrand M., Loizeau D., Vago J.L., Kminek G., Gaboyer F., Campbell K.A., Bréhéret J-B., Gautret P., Cockell C.S., 2015, Biosignatures on Mars: What, Where, and How? Implications for the Search for Martian Life. Astrobiology, 15(11), 998-1029, DOI: 10.1089/ast.2015.1374.

Westall, F., Brack A., 2018, The Importance of Water for Life. Space Science Reviews, 214, 50, DOI: 10.1007/s11214-018-0476-7.

Westall, F., Hickman-Lewis K., Hinman N., Gautret P., Campbell K.A., Bréhéret J.-G., Foucher F., Hubert A., Sorieul S., Kee T.P., Dass A.V., Georgelin T., Brack A., 2018, A hydrothermal-sedimentary context for the origin for life. Astrobiology 18, 259-293, DOI: 10.1089/ast.2017.1680.

Whitehouse, M.J., Nemchinb A.A., Pidgeonb R.T., 2017, What can Hadean detrital zircon really tell us? A critical evaluation of their geochronology with implications for the interpretation of oxygen and hafnium isotopes. Gondwana research, 51, 78-91.

Widemann, T., Ghail R.C., Wilson C.F., Titov D.V., 2020, EnVision: Europe's Proposed Mission to Venus. Exoplanets in Our Backyard: Solar System and Exoplanet Synergies on Planetary Formation, Evolution, and Habitability, held 5-7 February 2020 in Houston, TX. LPI Contribution No. 2195, Id.3024.

Wieczorek, M.A., Neumann G.A., Nimmo F., Kiefer W.S., Taylor G.J., Melosh H.J., Phillips R.J., Solomon S.C., Andrews-Hanna J.C., Asmar S.W., Konopliv A.S., Lemoine F.G., Smith D.E., Watkins M.M., Williams J.G., Zuber M.T., 2013, The crust of the Moon as seen by GRAIL. Science, 339, 6120, 671-675, DOI: 10.1126/science.1231530.

Williams, G.P., 1988, The dynamical range of global circulations–I. Clim. Dyn., 2, 205-260, DOI: 10.1007/BF01371320.

Williams, J.G., Konopliv A.S., Boggs D.H., Park R.S., Yuan D.-N., Lemoine F.G., Goossens S., Mazarico E., Nimmo F., Weber R.C., Asmar S.W., Melosh H.J., Neumann G.A., Phillips R.J., Smith D.E., Solomon S.C., Watkins M.M., Wieczorek M.A., Andrews-Hanna J.C., Head J.W., Kiefer W.S., Matsuyama I., McGovern P.J., Taylor G.J., Zuber M.T., 2014, Lunar interior properties from the GRAIL mission. J. Geophys. Res. Planets, 119, 1546–1578, DOI: 10.1002/2013JE004559.

Wilson, H.F., and Militzer B., 2012, Rocky Core Solubility in Jupiter and Giant Exoplanets. Physical Review Letters, 108, 11, Id. 111101, DOI: 10.1103/PhysRevLett.108.111101.

Winn, J.N.. Fabrycky D.C., 2015, The Occurrence and Architecture of Exoplanetary Systems. Annual Review of Astronomy and Astrophysics, 53, 409-447, DOI: 10.1146/annurev-astro-082214-122246.

Witte, M., 2004, Kinetic parameters of interstellar neutral helium. Review of results obtained during one solar cycle with the Ulysses/GAS-instrument. Astronomy and Astrophysics, 426(3), 835-844, DOI: 10.1051/0004-6361:20035956.

Woolfson, M.M., 1999, Neptune-Triton-Pluto system, Monthly Notices of the Royal Astronomical Society, 304(1), 195-198, DOI: 10.1046/j.1365-8711.1999.02304.x.

Wu, W., Li, C., Zuo, W., 2019, Lunar farside to be explored by Chang'e-4. Nat. Geosci., 12, 222–223, DOI: 10.1038/s41561-019-0341-7.

Xu, J., Guo J., Xu M., Chen X., 2020, Enhancement of microbial redox cycling of iron in zero-valent iron oxidation coupling with deca-brominated diphenyl ether removal. Science of the Total Environment, 748, Id. 141328, DOI: 10.1016/j.scitotenv.2020.141328.

Yamamoto, M., Ikeda K., Takahashi M., Horinouchi T., 2019, Solar-locked and geographical atmospheric structures inferred from a Venus general circulation model with radiative transfer. Icarus 321, 232-250, DOI: 10.1016/j.icarus.2018.11.015.

Yang, C.-C., Johansen A., Carrera D., 2017, Concentrating small particles in protoplanetary disks through the streaming instability. Astronomy and Astrophysics, 606, A80, 1-16, DOI: 10.1051/0004-6361/201630106.

Yewdall, N.A., Mason A.F., van Hest J.C.M., 2018, The hallmarks of living systems: towards creating artificial cells. Interface Focus, Review article of the Royal Soc., 8(5), DOI: 10.1098/rsfs.2018.0023.

Zellner, N.E.B., 2017, Cataclysm No More: New Views on the Timing and Delivery of Lunar Impactors. Origins of Life and Evolution of Biospheres, 47(3), 261-280, DOI: 10.1007/s11084-017-9536-3.

Zieger, B., Opher M., Schwadron N.A., McComas D.J., Tóth G., 2013, A slow bow shock ahead of the heliosphere. Geophysical Research Letters, 40(12), 2923-2928, DOI: 10.1002/grl.50576.

Ziegler, L.B., and Stegman D.R., 2013, Implications of a long-lived basal magma ocean in generating Earth's ancient magnetic field. Geochemistry, Geophysics, Geosystems, 14(11), 4735-4742.

Zolotov, M., 2019, Chemical weathering on Venus. In: Oxford Research Encyclopedia of Planetary Science, DOI: 10.1093/acrefore/9780190647926.013.146.

Zuber, M.T., Smith D.E., Watkins M.M., Asmar S.W., Konopliv A.S., Lemoine F.G., Melosh H.J., Neumann G.A., Phillips R.J., Solomon S.C., Wieczorek M.A., Williams J.G., Goossens S.J., Kruizinga G., Mazarico E., Park R.S., Yuan D.-N., 2013, Gravity Field of the Moon from the Gravity Recovery and Interior Laboratory (GRAIL) Mission. Science, 339(6120), 668-671, DOI: 10.1126/science.1231507.